%% file: beyond2pt_apj.tex
\pdfoutput=1
\documentclass{aastex631}
\usepackage{amsmath}
\usepackage{fontawesome5}
\usepackage[T1]{fontenc}
\usepackage{orcidlink}
\usepackage{rotating}

\shorttitle{Beyond-2pt mock challenge}
\shortauthors{Beyond-2pt Collaboration}

\makeatletter
\newcommand{\github}[1]{%
   \href{#1}{\faGithub}%
}
\makeatother

\definecolor{RoyalBlue}{rgb}{0.25,.41,.88}
\definecolor{DeepPurple}{rgb}{.45,.1,.75}
\definecolor{WildStrawberry}{HTML}{EE2967}
\definecolor{RedWine}{rgb}{0.743,0,0}
\definecolor{BitterSweet}{rgb}{1.0, 0.44, 0.37}
\definecolor{BurntOrange}{rgb}{0.8, 0.33, 0.0}
\definecolor{MidnightGreen}{rgb}{0.0, 0.29, 0.33}

\definecolor{cobalt}{rgb}{0.0, 0.28, 0.67}

\newcommand{\Challenge}{\textit{Challenge}}
\newcommand{\Analysis}{\textit{analysis}}

\let\vec\mathbf

\newcommand{\data}{\widehat{\vec d}}
\newcommand{\model}{\vec m}
\newcommand{\lik}{\mathcal{L}}
\newcommand{\post}{\mathcal{P}}
\newcommand{\Par}{\vec\Omega}
\newcommand{\cosPar}{\vec\Theta}
\newcommand{\nuisPar}{\vec\Phi}
\newcommand{\gauss}{\mathcal{N}}    
\newcommand{\unif}{\mathcal{U}}
\newcommand{\cov}{\mathsf{C}}
\newcommand{\identity}{\mathsf{I}}

\newcommand{\code}[1]{\texttt{#1}}

\newcommand{\beq}{\begin{eqnarray}}
\newcommand{\eeq}{\end{eqnarray}}
\newcommand{\hMpc}{h\,\text{Mpc}^{-1}}
\newcommand{\Mpch}{h^{-1}\text{Mpc}}
\usepackage[normalem]{ulem}
\usepackage{enumitem}

\newcommand{\Plin}{P_{\mathrm{L}}}

\newcommand{\<}{\langle}
\renewcommand{\>}{\rangle}

\newcommand{\vk}{\bm{k}}

\newcommand{\shat}{\hat{s}}

\newcommand{\kmax}{k_{\mathrm{max}}}
\newcommand{\eps}{\epsilon}
\newcommand{\op}{O}

\newcommand{\refeq}[1]{Eq.~(\ref{eq:#1})}          
\newcommand{\refeqs}[2]{Eqs.~(\ref{eq:#1})--(\ref{eq:#2})}          
\newcommand{\reffig}[1]{Fig.~\ref{fig:#1}}  
\newcommand{\reffigs}[2]{Fig.~\ref{fig:#1}--\ref{fig:#2}}  
          
\newcommand{\reftab}[1]{Tab.~\ref{tab:#1}}          
\newcommand{\refsec}[1]{Sec.~\ref{sec:#1}}

\newcommand{\be}{\begin{equation}}
\newcommand{\ee}{\end{equation}}

\newcommand\numberthis{\addtocounter{equation}{1}\tag{\theequation}}

\newcommand{\lcdm}{$\Lambda$CDM}

\newcommand{\Om}{\Omega_\mathrm{m}}

\newcommand{\void}{\mathrm{v}}

\newcommand{\gal}{\mathrm{g}}

\newcommand*{\diff}{\ensuremath{{\rm d}}}
\newcommand{\xibar}{\overline{\xi}}

\usepackage{bm}

\graphicspath{{./}{figures/}}

\begin{document}

\reportnum{MIT-CTP/5708; LCTP-24-06}

\title{A Parameter-Masked\footnote{We use \emph{to mask} (and derived forms) in place of the formerly common \emph{to blind} throughout this manuscript.} Mock Data Challenge for Beyond-Two-Point Galaxy Clustering Statistics}

\collaboration{30}{The Beyond-2pt Collaboration}
\email{krausee@arizona.edu, yosukekobayashi@arizona.edu, ansalcedo@arizona.edu}
\author[0000-0001-8356-2014]{Elisabeth Krause}
\affiliation{Department of Astronomy/Steward Observatory, The University of Arizona, 933 North Cherry Avenue, Tucson, AZ 85721, USA}

\author[0000-0002-6633-5036]{Yosuke Kobayashi}
\affiliation{Department of Astronomy/Steward Observatory, The University of Arizona, 933 North Cherry Avenue, Tucson, AZ 85721, USA}
\affiliation{Department of Astrophysics and Atmospheric Sciences, Faculty of Science, Kyoto Sangyo University, Motoyama, Kamigamo, Kita-ku, Kyoto 603-8555, Japan}

\author[0000-0003-1420-527X]{Andr\'{e}s N. Salcedo}
\affiliation{Department of Astronomy/Steward Observatory, The University of Arizona, 933 North Cherry Avenue, Tucson, AZ 85721, USA}

\author[0000-0002-6745-984X]{Mikhail M. Ivanov}
\affiliation{Center for Theoretical Physics, Massachusetts Institute of Technology, 
Cambridge, MA 02139, USA}

\author[0000-0002-5969-1251]{Tom Abel}
\affiliation{Kavli Institute for Particle Astrophysics and Cosmology, Stanford University, 452 Lomita Mall, Stanford, CA 94305, USA}
\affiliation{Department of Physics, Stanford University, 382 Via Pueblo Mall, Stanford, CA 94305, USA}
\affiliation{SLAC National Accelerator Laboratory, 2575 Sand Hill Road, Menlo Park, CA 94025, USA}

\author[0000-0001-6473-3420]{Kazuyuki Akitsu}
\affiliation{Theory Center, Institute of Particle and Nuclear Studies, High Energy Accelerator Research Organization(KEK), Tsukuba, Ibaraki 305-0801, Japan}

\author[0000-0003-2953-3970]{Raul E. Angulo}
\affiliation{Donostia International Physics Center (DIPC), Paseo Manuel de Lardizabal 4, 20018 Donostia-San Sebastian, Spain}
\affiliation{IKERBASQUE, Basque Foundation for Science, E-48013, Bilbao, Spain}

\author[0000-0001-9487-702X]{Giovanni Cabass}
\affiliation{Division of Theoretical Physics,  Ru\dj er Bo\v{s}kovi\'{c} Institute, Zagreb HR-10000, Croatia}

\author[0000-0002-9843-723X]{Sofia Contarini}
\affiliation{Dipartimento di Fisica e Astronomia "Augusto Righi" - Alma Mater Studiorum Universit\`{a} di Bologna, via Piero Gobetti 93/2, I-40129 Bologna, Italy}
\affiliation{INFN-Sezione di Bologna, Viale Berti Pichat 6/2, I-40127 Bologna, Italy}
\affiliation{INAF-Osservatorio di Astrofisica e Scienza dello Spazio di Bologna, Via Piero Gobetti 93/3, I-40129 Bologna, Italy}

\author[0000-0002-6069-2999]{Carolina Cuesta-Lazaro}
\affiliation{The NSF AI Institute for Artificial Intelligence and Fundamental Interactions Massachusetts Institute of Technology, Cambridge, MA 02139, USA}
\affiliation{Department of Physics, Massachusetts Institute of Technology, Cambridge, MA 02139, USA}
\affiliation{Harvard-Smithsonian Center for Astrophysics, 60 Garden Street, Cambridge, MA 02138, USA}

\author[0000-0003-1197-0902]{ChangHoon Hahn}
\affiliation{Department of Astrophysical Sciences, Princeton University, Princeton, NJ 08544, USA}

\author[0000-0002-0876-2101]{Nico Hamaus}
\affiliation{Universit\"ats-Sternwarte M\"unchen, Fakult\"at f\"ur Physik, Ludwig-Maximilians-Universit\"at, Scheinerstr. 1, 81679 M\"unchen, Germany}
\affiliation{Excellence Cluster ORIGINS, Bolzmannstr. 2, 85748 Garching, Germany}

\author[0000-0002-8434-979X]{Donghui Jeong}
\affiliation{Department of Astronomy and Astrophysics and Institute for Gravitation and the Cosmos, 
The Pennsylvania State University, University Park, PA 16802, USA}
\affiliation{School of Physics, Korea Institute for Advanced Study, Seoul, South Korea}

\author[0000-0002-1670-2248]{Chirag Modi}
\affiliation{Center for Computational Astrophysics, Flatiron Institute, 162 5th Avenue, 10010, New York, NY, USA}
\affiliation{Center for Computational Mathematics, Flatiron Institute, 162 5th Avenue, 10010, New York, NY, USA}

\author[0000-0002-2542-7233]{Nhat-Minh Nguyen}
\affiliation{Leinweber Center for Theoretical Physics, University of Michigan, 450 Church St, Ann Arbor, MI 48109-1040}
\affiliation{Department of Physics, College of Literature, Science and the Arts, University of Michigan, 450 Church St, Ann Arbor, MI 48109-1040}

\author[0000-0002-9664-0760]{Takahiro Nishimichi}
\affiliation{Department of Astrophysics and Atmospheric Sciences, Faculty of Science, Kyoto Sangyo University, Motoyama, Kamigamo, Kita-ku, Kyoto 603-8555, Japan}
\affiliation{Center for Gravitational Physics and Quantum Information, Yukawa Institute for Theoretical Physics, Kyoto University, Kyoto 606-8502, Japan}
\affiliation{Kavli Institute for the Physics and Mathematics of the Universe (WPI), The University of Tokyo Institutes for Advanced Study (UTIAS), The University of Tokyo, Kashiwa, Chiba 277-8583, Japan}

\author[0000-0002-4637-2868]{Enrique Paillas}
\affiliation{Waterloo Centre for Astrophysics, University of Waterloo, Waterloo, ON N2L 3G1, Canada}
\affiliation{Department of Physics and Astronomy, University of Waterloo, ON N2L 3G1, Canada}

\author[0000-0003-4680-7275]{Marcos Pellejero Iba\~nez}
\affiliation{Institute for Astronomy, University of Edinburgh, Royal Observatory, Blackford Hill, Edinburgh, EH9 3HJ , UK}

\author[0000-0002-3033-9932]{Oliver H.\,E. Philcox}
\affiliation{Department of Physics, Columbia University, New York, NY 10027, USA}
\affiliation{Simons Society of Fellows, Simons Foundation, New York, NY 10010, USA}

\author[0000-0002-6146-4437]{Alice Pisani}
\affiliation{Aix-Marseille University, CNRS/IN2P3, CPPM, 163 Av. de Luminy, 13009, Marseille, France}
\affiliation{Center for Computational Astrophysics, Flatiron Institute, 162 5th Avenue, 10010, New York, NY, USA}
\affiliation{The Cooper Union for the Advancement of Science and Art, 41 Cooper Square, New York, NY 10003, USA}
\affiliation{Department of Astrophysical Sciences, Princeton University, Princeton, NJ 08544, USA}

\author[0000-0002-6807-7464]{Fabian Schmidt}
\affiliation{Max–Planck–Institut f\"ur Astrophysik, Karl–Schwarzschild–Straße 1, 85748 Garching, Germany}

\author[0000-0003-2442-8784]{Satoshi Tanaka}
\affiliation{Center for Gravitational Physics and Quantum Information, Yukawa Institute for Theoretical Physics, Kyoto University, Kyoto 606-8502, Japan}

\author[0000-0002-1886-8348]{Giovanni Verza}
\affiliation{Center for Cosmology and Particle Physics, Department of Physics, New York University, 726 Broadway, New York, NY 10003, USA}
\affiliation{Center for Computational Astrophysics, Flatiron Institute, 162 5th Avenue, 10010, New York, NY, USA}

\author[0000-0002-5992-7586]{Sihan Yuan}
\affiliation{Kavli Institute for Particle Astrophysics and Cosmology, Stanford University, 452 Lomita Mall, Stanford, CA 94305, USA}
\affiliation{SLAC National Accelerator Laboratory, 2575 Sand Hill Road, Menlo Park, CA  94025, USA}

\author[0000-0002-4458-1754]{Matteo Zennaro}
\affiliation{University of Oxford, Astrophysics, Denys Wilkinson Building, Keble Rd, Oxford OX1 3RH, Regno Unito}

\begin{abstract}
The last few years have seen the emergence of a wide array of novel techniques for analyzing high-precision data from upcoming galaxy surveys, which aim to extend the statistical analysis of galaxy clustering data beyond the linear regime and the canonical two-point (2pt) statistics.
We test and benchmark some of these new techniques in a community data challenge ``Beyond-2pt'', initiated during the Aspen 2022 Summer Program ``Large-Scale Structure Cosmology beyond 2-Point Statistics,'' whose first round of results we present here. The challenge dataset consists of high-precision mock galaxy catalogs for clustering in real space, redshift space, and on a light cone. Participants in the challenge have developed end-to-end pipelines to analyze mock catalogs and extract unknown (``masked'') cosmological parameters of the underlying $\Lambda$CDM models with their methods. The methods represented are density-split clustering, nearest neighbor statistics, \code{BACCO} power spectrum emulator, void statistics, \code{LEFTfield} field-level inference using effective field theory (EFT), and joint power spectrum and bispectrum analyses using both EFT and simulation-based inference.
In this work, we review the results of the challenge, focusing on problems solved, lessons learned, and future research needed to perfect the emerging beyond-2pt approaches. The unbiased parameter recovery demonstrated in this challenge by multiple statistics and the associated modeling and inference frameworks supports the credibility of cosmology constraints from these methods. The challenge data set is publicly available, and we welcome future submissions from methods that are not yet represented. 
\end{abstract}

\keywords{Cosmological parameters from large-scale structure (340) --- Large-scale structure of the universe(902) --- Astrostatistics techniques(1886)}

\section{Introduction} \label{sec:intro}
Cosmic large-scale structure---the large-scale distribution of galaxies in the late universe---is shaped by matter density fluctuations in the early universe, the growth rate of the fluctuations, and the expansion rate of the smooth cosmic background.
Ongoing and upcoming large-scale structure surveys will collectively map the distribution of galaxies over an extensive fraction of the sky at unprecedented depth \citep{DESI:2016fyo,2011arXiv1110.3193L,PFS:whitepaper2014}.
These observations will enable precision tests of fundamental physics, placing tight constraints on inflation, neutrino masses, and novel properties of dark matter, gravity, and dark energy.

On large scales, matter and galaxy distributions preserve the almost perfectly Gaussian statistics from the initial conditions according to the standard inflationary models.
The two-point (2pt) correlation function or power spectrum captures all information in a Gaussian field.
They therefore have been the linchpin of cosmological inference and interpretation.

The situation, however, is beginning to change. Larger and higher-resolution simulations \citep[e.g.][]{Nishimichi:2018etk,Heitmann:2019ytn,Uuchu,Maksimova:2021ynf,Schaye:2023jqv,aemulusnu} and more realistic galaxy mocks \citep[e.g.][]{Behroozi_et_al_2019,Buzzard,Balaguera-Antolinez:2022xko,Yuan:2022ibz,Stevens:2023ozg,Cardinal} have been developed to extend the inferences to smaller scales \citep[e.g.][]{Reid_et_al_2014,Reddick2014,Lange_et_al_2022,2022bYuan,Lange:2023khv,2024ApJ...961..208S}. Furthermore, the gain from small-scale information will become less limited by signal-to-noise ratios in galaxy samples from upcoming galaxy surveys.

On smaller scales, 2pt statistics cannot capture all the information available in the data: even with Gaussian initial perturbations, non-Gaussian features naturally emerge in the galaxy distribution as a result of nonlinear gravitational evolution and galaxy formation. Beyond-2pt statistics aim to extract information from non-Gaussian features arising from nonlinear clustering. Examples of the first cosmological constraints from galaxy clustering using beyond-2pt statistics include bispectrum analyses of PSCz \citep{Feldman:2000vk}, 2dF \citep{Verde:2001sf}, VVDS \citep{Marinoni:2008wx}, SDSS DR6-DR7 \citep{Gaztanaga:2008sq}, and SDSS-III BOSS galaxies \citep{Gil-Marin:2016wya}. 
More recent examples include analyses of the galaxy bispectrum \citep{DAmico:2019fhj,Philcox:2021kcw,DAmico:2022osl,Ivanov:2023qzb,Hahn:2023kky}, three-point function \citep{2023MNRAS.523.3133S}, skew spectrum \citep{2024arXiv240115074H}, density-split clustering \citep{Paillas:2023cpk}, cosmic voids \citep{Hamaus2020,Contarini2023BOSS}, and wavelet scattering transform of the galaxy density field \citep{Valogiannis:2023mxf,Blancard:2023iab}.
While the discussion here focuses on galaxy clustering, beyond-2pt statistics are similarly gaining traction in weak lensing analyses \citep[e.g.,][]{2015PhRvD..91j3511P,2021MNRAS.506.1623H,2022PhRvD.106h3509G,2022A&A...667A.125H,2024arXiv240416085C}.

In most cases, these analyses provided competitive or complementary constraints to those derived from standard 2-pt analyses applied to the same data. Beyond-2pt statistics are poised to mature into a prominent role in cosmological inference from forthcoming galaxy clustering data as nonlinear clustering becomes more accessible and statistical power increases, owing to improvements in the inference models and the clustering signal-to-noise, respectively.

 With great statistical power comes great systematic responsibility. How (in)sensitive are the beyond-2pt statistics to their modeling choices? Do they respond in known ways to observational systematics? Modeling beyond-2pt statistics and thus physics on the associated nonlinear scales adds considerable complexities in both models of matter density perturbation and matter--galaxy connection, \textit{i.e.}\ galaxy bias. Current beyond-2pt statistics adopt a wide array of approaches for each modeling step: the matter field is modeled with a variety of perturbative approaches or N-body simulations, and the matter--galaxy connection is usually encoded via bias expansions, Halo Occupation Distribution (HOD) models, or Sub-Halo Abundance Matching (SHAM) schemes.
 
In addition to modeling systematics, how well do we understand the likelihoods and covariances of these novel statistics? Can we evaluate and estimate them at the precision required by current and future surveys?
 Specifically, intractable non-Gaussian likelihoods\footnote{See, e.g., \citet{Hahn:2018zja} or \citet{Nguyen:2020hxe} for recent examples of non-Gaussian likelihoods and their impacts on cosmological inference from galaxy clustering.} must be approximated with simulations. Furthermore, analytic covariances are not always available, hence simulations and mocks are again required to estimate the covariances.
 The difficulties inherent in the generation of numerous high-resolution simulations and mock data add another layer of complexity to such analyses, which could potentially compromise their robustness. Consequently, the first aim of this data challenge is to validate a variety of beyond-2pt statistics and their modeling approaches.

Our second aim is to address the following question: How much information can we (robustly) extract from galaxy clustering? Although individual comparisons of information content between standard and beyond-2pt statistics have been made elsewhere, no direct comparison between the beyond-2pt methods presented here was attempted with the same survey volume and with the same galaxy density.
This challenge therefore aims to facilitate a community exercise to study the information content in beyond-2pt statistics by providing a standard set of simulated mock data. A critical quantity that controls the amount of beyond-linear information is the maximum cutoff scale in each analysis, $\kmax$. As each beyond-2pt analysis might have a different sensitivity to $\kmax$, we let each group determine their own $\kmax$ up to which they can trust their model against model misspecifications. However, to avoid fine-tuning while simultaneously encouraging better uncertainty quantification and systematic control in each analysis, we mask the parameters of the mock catalogs, which include both the cosmological parameters and the galaxy HOD parameterization(s) plus their parameter values.

For galaxy clustering, only a single public, parameter-masked mock challenge exists prior to this work---the \code{PTchallenge}, introduced in \citet{Nishimichi:2020tvu}---that specifically targets 2pt statistics.
\code{PTchallenge} was cornerstone 
for the establishment of EFT as a 
mature tool to analyze the full shape of the galaxy power spectrum.
In particular, it demonstrates the ability of one-loop EFT theories to recover 
\emph{masked} cosmological 
parameters from both amplitude and shape of the (nonlinear) power spectrum at a 
sub-percent precision.

Important technical lessons from 
\code{PTchallenge} are:
(1) learning parameter degeneracies and optimizing analysis choices with respect to them,
(2) developing a methodology to select scale cuts, 
(3) understanding the role of EFT parameters and their priors.
Since the original publication, the true cosmology of the \code{PTchallenge} has been kept masked and the \code{PTchallenge} continues to serve as a testing ground for new galaxy power spectrum models, e.g., Lagrangian EFT \citep{Chen:2020fxs,Chen:2020zjt}, a simulation- plus HOD-based emulator \citep{Kobayashi:2020zsw,Kobayashi:2021oud}, and a perturbation-based shape-fitting approach \citep{Brieden:2021edu,Brieden:2022ieb}.
However, the \code{PTchallenge} only made available data in the form of redshift-space power spectrum measurements. 

Our ``Beyond-2pt'' mock challenge extends the \code{PTchallenge} \citep{Nishimichi:2020tvu} in several ways.
First, we present the data directly at the level of mock galaxy catalogs, rather than summary statistics thereof. Estimator validation is therefore part of the challenge for each clustering statistics.
Second, we present mock catalogs at three different complexity levels: real-space or redshift-space snapshots and a light cone. These mocks cover a redshift range around $z\sim1$ with different masked flat $\Lambda$CDM cosmologies and different mock galaxy populations with masked HOD parameterization(s). The range of complexity levels enables participation by statistics and methods at different maturity levels, while also facilitating robustness tests of different modeling prescriptions. Third, prior to this, many prescriptions were only tested on mock galaxies resembling those found in the BOSS CMASS sample within a Planck-like cosmology. This challenge---extending both galaxy and cosmology models---therefore marks a step forward in this regard as well.

Finally, we extend the challenge to beyond-2pt statistics---showing their constraints on cosmological parameters, $[\Omega_m,\sigma_8]$, side-by-side with the corresponding constraints from \code{BACCO} P, a representative of 2pt-statistics, for all setups---and dissecting the information content they capture.
The present challenge presents a first of such benchmark and sets the stage for more detailed comparisons in future.
With eight independent analyses submitted from seven international teams, this challenge represents the first community effort toward developing optimal strategies to extract cosmological information from upcoming galaxy surveys.

The mocks and analyses presented in this paper assume a flat $\Lambda\rm{CDM}$ cosmology, which can be described by five parameters: (1) the dimensionless Hubble parameter $h$, (2) the matter fluctuation or primordial power spectrum amplitude, parameterized by either $\sigma_8$ or $A_{\mathrm{s}}$, (3) the power spectrum spectral index $n_{\mathrm{s}}$, plus (4-5) the density parameter of cold dark matter $\Omega_{\mathrm{cdm}}$ and baryons $\Omega_{\mathrm{b}}$. Different analyses adopt priors on either the density parameters $\Omega_{\mathrm{x}}$ or the physical density parameters $\omega_{\mathrm{x}}=\Omega_{\mathrm{x}}\,h^2$. Though no analysis varies neutrino mass, some assume massless ($\omega_\nu = 0$) while others adopt minimal-mass neutrinos ($\omega_\nu=0.0006442$).
The main results of this challenge are summarized and presented in marginalized constraints on $\sigma_8$ and the total matter density, $\Omega_{\rm{m}} = \Omega_{\rm{cdm}}+\Omega_{\rm{b}}+\Omega_\nu$.

To set the stage for the \emph{Summary of Results} in \refsec{results}, we briefly review the information content in galaxy clustering, in the context of the $\Lambda\rm{CDM}$ model (with Gaussian initial conditions), as considered throughout this paper.
In linear theory, all cosmological information encoded in galaxy clustering is captured by the linear power spectrum, particularly its scale dependence and amplitude. The scale dependence is sensitive to both the initial density perturbations, captured by the spectral index $n_{\mathrm{s}}$, and the stress-energy components that determine the background evolution and growth of perturbations. The physical density parameters $\omega_{\mathrm{m,b}}$ determine two characteristic scales: the Hubble horizon at matter-radiation equality and the sound horizon at photon-baryon decoupling, which are imprinted on the broadband shape of the matter power spectrum and baryonic acoustic oscillation (BAO) wiggles, respectively. All the challenge mocks include this information.

In real space, i.e., without accounting for observational effects of galaxy peculiar motion, the amplitude of the (linear) galaxy power spectrum at redshift $z$ is proportional to the product of the amplitude of primordial fluctuations, the linear growth factor $D(z)$ and linear galaxy bias $b_1$. That is, the cosmology parameter $\sigma_8$ is completely degenerate with the unknown linear galaxy bias parameter. 
Observational effects encode additional information: the redshift space distortion (RSD) effect contains information on the growth rate $f$, which breaks the degeneracy between $\sigma_8$ and $b_1$, and the Alcock-Paczynski (AP) effect contains geometry information, which constrains $\Omega_{\mathrm{m}}$ (within $\Lambda$CDM). Growth information from RSD is included in redshift-space and light-cone mocks, while geometry information from AP is only available on the light cone.

Beyond the linear regime, quasi- and nonlinear clustering and collapsed structures encode additional information with different parameter degeneracies. The results from this challenge underscore the robust extraction of such nonlinear information either through the nonlinear matter power spectrum (\code{BACCO} P) or through statistics that have access to higher-order $n$-point functions (all other teams). At the power spectrum level, nonlinear evolution introduces additional smearing of the BAO feature and an enhancement of small-scale power \citep{Rimes:2005xs,Crocce:2007dt,Angulo_2021}. The distinct dependencies of beyond-2pt statistics on cosmology and astrophysics break parameter degeneracies and improve constraints on cosmological parameters. We refer to individual analysis sections, and references therein, for detailed discussions on the sources of nonlinear information.

\begin{figure}
\centering \includegraphics[width=\textwidth]{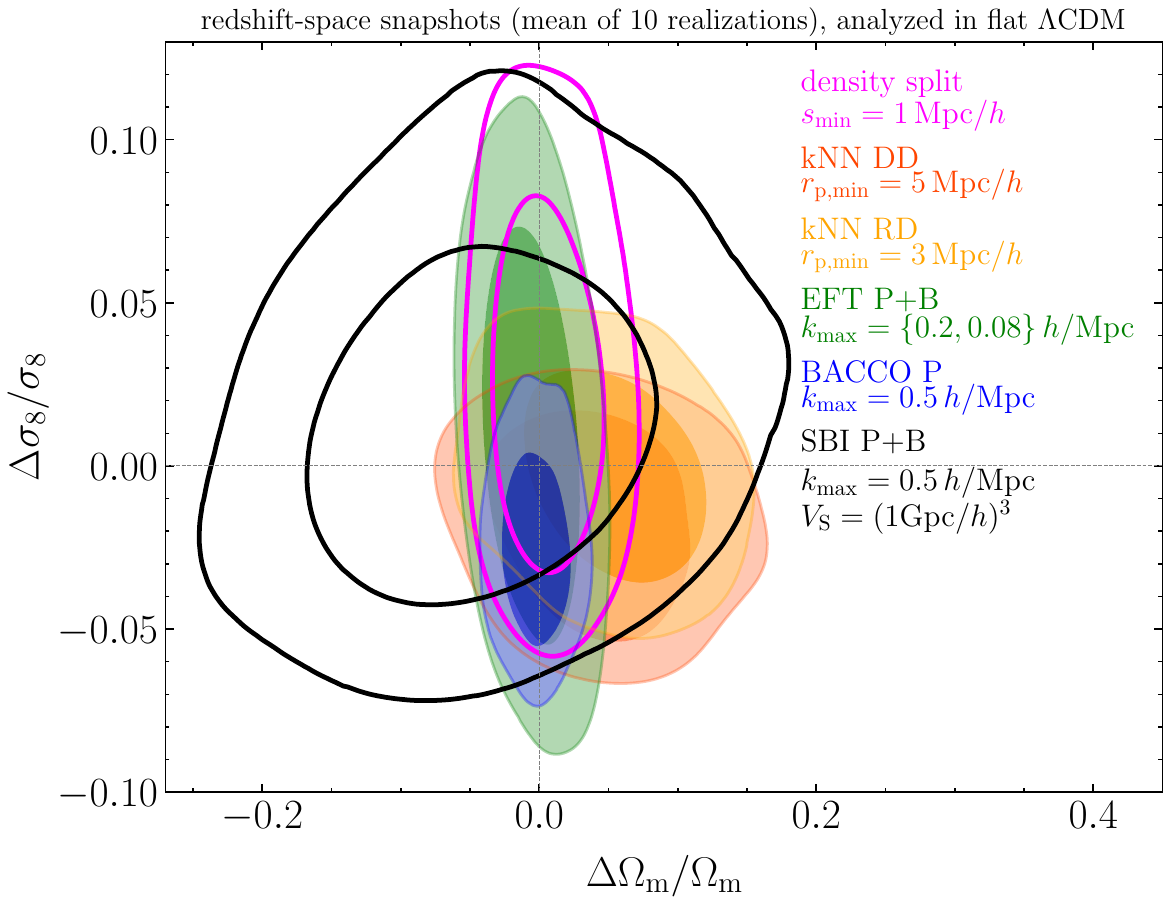}
    \caption{
    2D marginalized constraints on $\Omega_{\rm{m}}$ and $\sigma_8$ for parameter-masked analyses of redshift-space mocks (mean of 10 realizations, errors of 1 box), marginalized over the remaining cosmological parameters of flat $\Lambda\rm{CDM}$ and nuisance parameters specific to each method. We list the scale cuts due to nonlinear modeling for different analyses, which is specified in terms of redshift-space separation ($s$), projected radius ($r_\mathrm{p}$) or Fourier mode ($k$), depending on the analysis method.} 
\label{fig:2D_summary}
\end{figure}

The remainder of the paper is organized as follows. To help readers navigate the following sections, we provide tables summarizing the different analyses (\reftab{methods}) and the common notation (\reftab{variables}).
In \refsec{results}, we summarize and discuss the results of eight parameter-masked analyses.
\reffigs{2D_summary}{1D_summary_real} present the key results, as constraints on the parameter combination $[\Om,\sigma_8]$ from the redshift-space mocks in \reffigs{2D_summary}{1D_summary_rsd} or from the light-cone mock in \reffig{1D_summary_lc}, and as constraints on the parameter $\sigma_8$ from the real-space mocks in \reffig{1D_summary_real}. We introduce the suite of mock catalogs---the main data product of the ``Beyond-2pt'' challenge---in \refsec{mocks}, and the parameter (un)masking procedure in \refsec{mask}.
In \refsec{methods}, each team describes their analysis method. We describe post-unmasking reanalyses and resulting lessons for method refinement and optimization of constraining power in \refsec{discussion}. We summarize the results of this challenge and discuss implications for future mock challenges and beyond-2pt analyses in \refsec{conclusion}.

\section{Summary of Results}
\label{sec:results}

In this section, we summarize the main results of the paper, shown in \reffigs{2D_summary}{1D_summary_real}, and provide an overview of the eight participating analyses in \reftab{methods}.

This challenge is based on a series of mock galaxy catalogs created from N-body simulations with a flat $\Lambda$CDM cosmology and HOD galaxy--halo connection models. It contains three levels of increasing realism: 
\begin{enumerate}
    \item \textbf{real-space} galaxy distribution from a simulation snapshot with periodic boundary conditions (10 realizations with the same cosmology + HOD parameters at $z=1$),
    \item \textbf{redshift-space} galaxy distribution from a simulation snapshot with periodic boundary conditions (10 realizations with the same cosmology + HOD parameters at $z=1$),
    \item \textbf{light-cone} galaxy mock emulating the observational data most closely (one realization, uniform coverage across $0.8<z<1.3$ and with a simple footprint bounded by lines of constant right ascension and declination).
\end{enumerate}

Initially, the organizers communicated to the analysis teams only that the challenge catalogs are HOD-based galaxy mocks in flat $\Lambda$CDM cosmologies without observational systematics, hiding the true cosmological parameters and the HOD model used to generate the simulations from the analysis teams.
Each analysis team then analyzed these parameter-masked mocks while documenting their model, inference, and analysis choices including scale cuts and parameter priors, before submitting their parameter posteriors to the organizers, who subsequently unmasked the fractional distances between posterior means and the ground truth.

The challenge results presented here indicate whether a given analysis returns parameters consistent with the ground truth given their reported error bars.
For real-space and redshift-space analyses, analysis teams report results corresponding to the mean of 10 realizations analyzed with the covariance of a single realization volume to minimize parameter biases due to cosmic variance. Assuming a Gaussian data likelihood, in this setup the probability of a $\ge 1 \sigma$ fluctuation due to cosmic variance is 0.16\%. Hence, biases in the inferred parameters likely indicate model misspecification or incomplete uncertainty modeling. The challenge results do not directly establish whether the reported parameter uncertainties fully reflect the true uncertainties (measurement and model). This is particularly relevant for small-scale analyses and simulation-based methods, based on empirical galaxy--halo connection parameterizations and priors. Hence, the results should not be regarded as a direct quantitative comparison between different analysis methods or summary statistics.

We report results ordered by number of participating analyses for the different mocks. All results on the redshift-space and light-cone mock catalogs are from analyses in flat $\Lambda$CDM with priors that are much broader than current-generation observational constraints \citep{Planck2018,2021PhRvD.103h3533A,2024arXiv240403002D}. However, the cosmology priors differ between all analyses, as illustrated in \reffig{prior}. We note that the void size function (VSF) analysis of the light-cone mock adopts informative priors on three parameters that are weakly constrained by the VSF, with width of three times the current Planck2018~\citep{Planck2018} uncertainties. The analyses 
on the real-space mock
reported here fix all cosmological parameters except $\sigma_8$ to their true values.

\subsection{Results on redshift-space mocks} \label{sec:resultsredshift}

The redshift-space snapshots were analyzed by most teams, comprising of the following analyses: density-split clustering (Section~\ref{subsec:DSC}), two different flavors of nearest neighbor statistics (kNN; Section~\ref{subsec:Abacus_kNN}), a joint power spectrum plus bispectrum analysis using effective field theory (EFT P+B; Section~\ref{subsec:PTPB}), a hybrid-EFT power spectrum analysis with the \code{BACCO} emulator (\code{BACCO} P; Section~\ref{subsec:bacco}), and a simulation-based-inference analysis of the power spectrum plus bispectrum (SBI P+B; Section~\ref{subsec:SBI}). 
The SBI P+B analysis was trained on simulations with volume $(1\,h^{-1}{\rm Gpc})^3$, while all other analyses assume a covariance matrix corresponding to the sample variance of a $(2\,h^{-1}{\rm Gpc})^3$ box, resulting in degradation in constraining power. While it would be convenient to assume a simple survey volume rescaling to compare SBI P+B results with other analyses, we refrain from making such a rescaling as constraining power is a function of both survey volume and scale cuts ($k_{\rm{max}}$), which need to be calibrated anew when changing the survey volume. 

Figure~\ref{fig:2D_summary} shows the marginalized constraints in the $\Om$--$\sigma_8$ plane inferred from the different summary statistics. Remarkably, all analysis teams successfully recover the input cosmology within their $1{-}\sigma$ confidence region. This result from a parameter-masked challenge further demonstrates the maturity of these
reportedly ``novel'' statistics and their potential for analyses of near-term data. 

\begin{figure}
\centering \includegraphics[width=.9\textwidth]{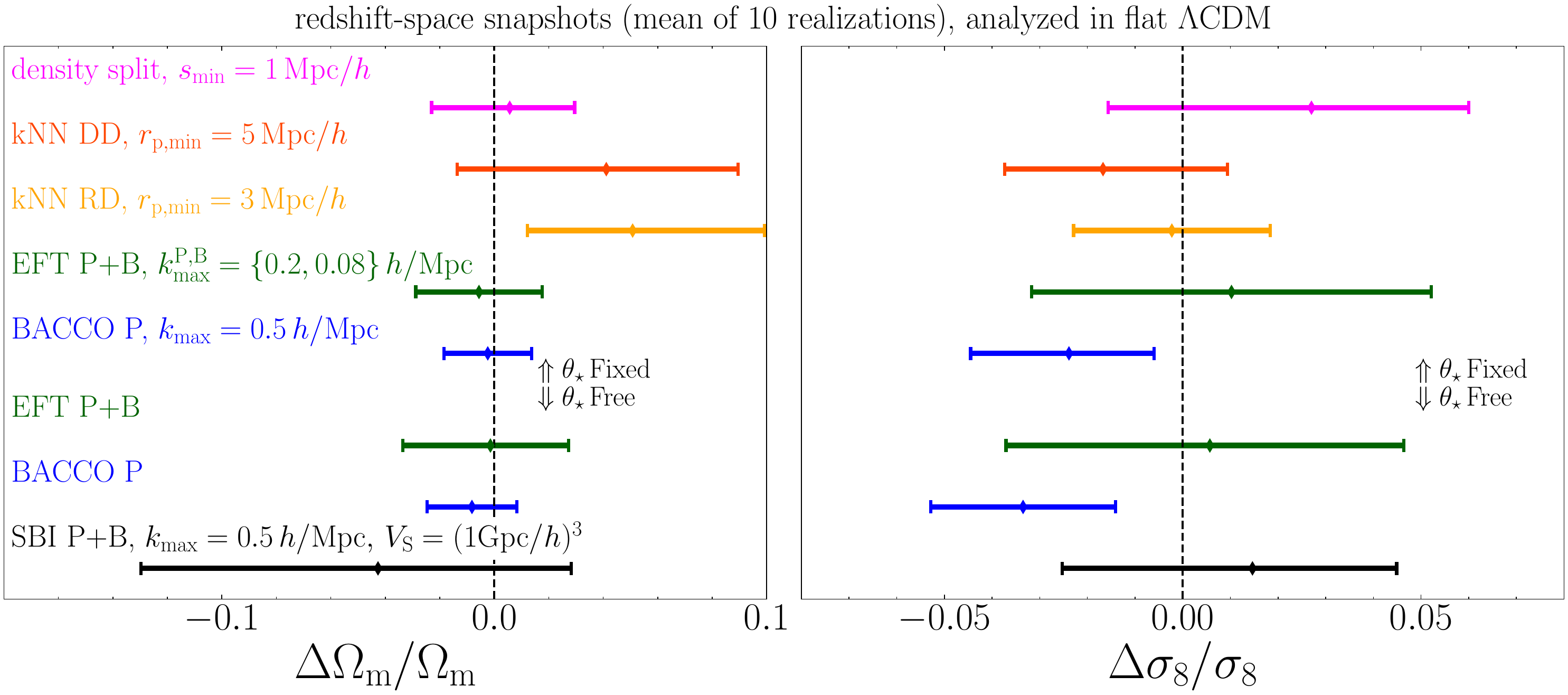}
    \caption{1D marginalized constraints on $\Omega_{\rm{m}}$ and $\sigma_8$ for parameter-masked analyses of redshift-space mocks (mean of 10 realizations, errors of 1 box), marginalized over the remaining cosmological parameters of flat $\Lambda\rm{CDM}$ and nuisance parameters specific to each method. 
    } 
\label{fig:1D_summary_rsd}
\centering \includegraphics[width=.9
\textwidth]{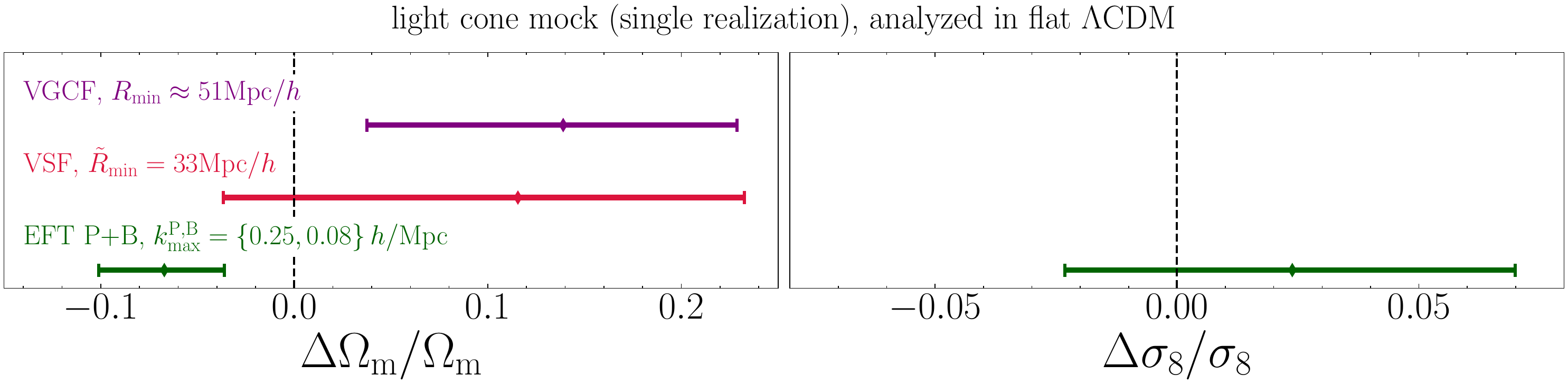}
    \caption{
    1D marginalized constraints on $\Omega_{\rm{m}}$ and $\sigma_8$ for parameter-masked analyses of light-cone mock (single realization), marginalized over the remaining cosmological parameters of flat $\Lambda\rm{CDM}$ and nuisance parameters specific to each method. Void-size cuts are specified in terms of the effective radius $R$ and the cleaned (re-scaled) radius $\tilde{R}$.\\
    } 
\label{fig:1D_summary_lc}
\centering \includegraphics[width=0.45\textwidth]{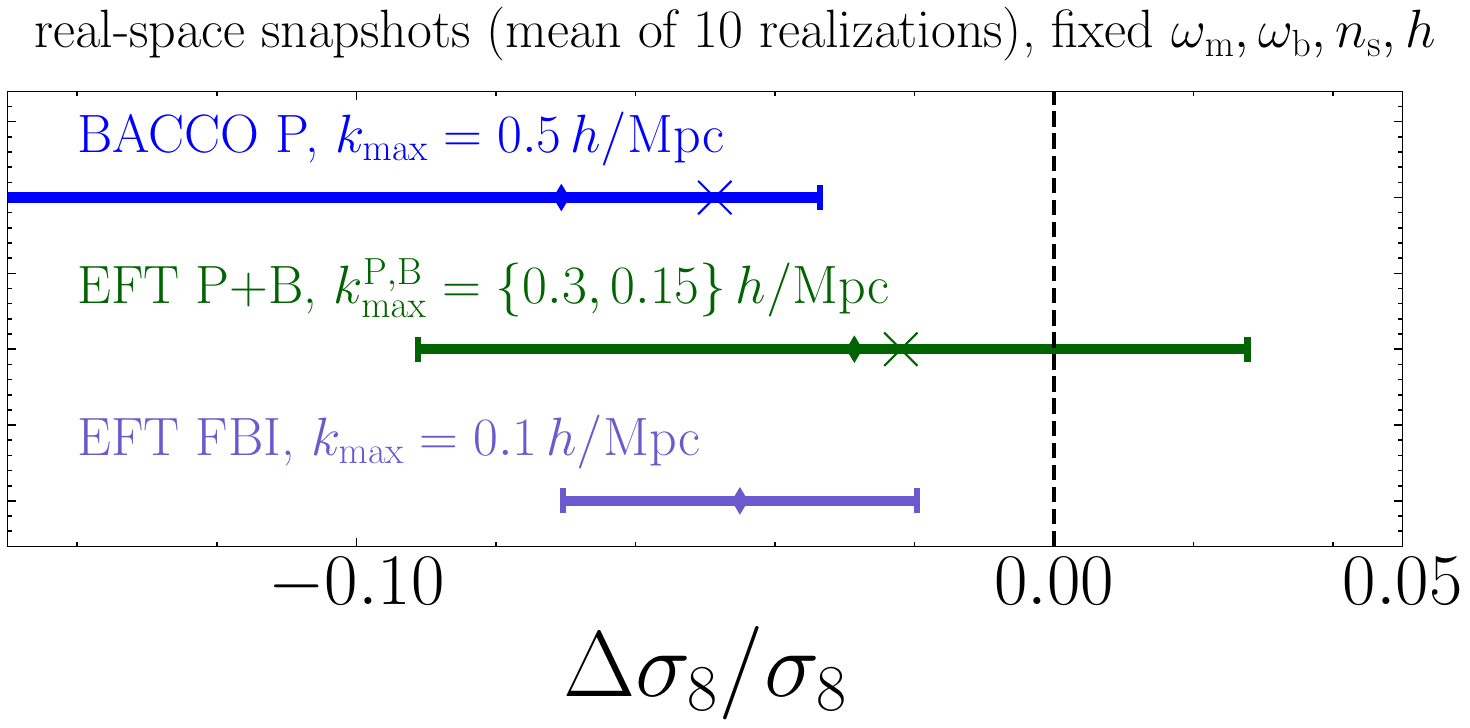}
    \caption{1D marginalized constraints on $\sigma_8$ for 
    analyses of real-space mocks (mean of 10 realizations, errors of 1 box), in a reduced parameter space with all cosmological parameters except $\sigma_8$ fixed at the fiducial values. Crosses indicate the maximum a posteriori (MAP) estimates, to illustrate projection effects.}
\label{fig:1D_summary_real}
\end{figure}

Figure~\ref{fig:1D_summary_rsd} shows the marginalized posteriors in 1D for $\Om$ and $\sigma_8$. 
The nominal distance between the mean and the ground truth (\textit{i.e.}\, bias) in $\Om$ is the smallest for the EFT P+B, \code{BACCO} and density split analyses, while for $\sigma_8$, the unmasking showed the lowest parameter bias for the EFT P+B, kNN RD, and SBI P+B analyses. After unmasking, further analysis by the \code{BACCO} team identified a large emulation uncertainty of the hexadecapole as potential source of this parameter bias (c.f. Sect.~\ref{subsec:bacco_inference}), which had been unnoticed in previous tests on smaller simulation volumes. We further discuss this post-unmasking reanalysis in Sect.~\ref{sec:discussion}. Focusing on the analyses with fixed $\theta_*$---the angular size of the sound horizon---we find similar error bars on $\Om$ from the density-split, EFT P+B and \code{BACCO} methods, implying that, for this particular parameter and mock data, the bulk of the information
comes from quasi-linear (BAO) scales. 
The tightest error bars on $\sigma_8$ are obtained with the kNN and \code{BACCO} methods.

Both density-split clustering and kNN statistics are modeled with emulators built on the \code{AbacusSummit} simulations, which fix $\theta_*$, while the SBI P+B analysis is built on the \code{Quijote} simulations which assume no such constraint. To ease the comparison, the EFT P+B and \code{BACCO} P teams ran additional analyses imposing the same prior. A comparison of the EFT P+B constraints in both parameter spaces (top and bottom part of \reffig{1D_summary_rsd}) indicates that this difference in cosmology priors has limited impact on the constraints shown here. The EFT P+B and SBI P+B results are quite similar in terms of $\sigma_8$ constraining power, despite the difference in survey volume. For $\Om$ however, the SBI P+B constraint is significantly wider than the EFT P+B result. We believe that the wide SBI P+B posterior on $\Om$ is a result of trimming the largest scales in the SBI P+B analysis, which down weights large scale modes that are important in the $\Omega_m$ recovery from the turnover of the galaxy power spectrum \citep[cf.][]{Ivanov:2019pdj,Philcox:2020xbv}.
\include{methods_summary_table_v3}

\subsection{Results on light-cone mock}
Unlike the redshift-space data discussed above, there exists only one realization for the light-cone mock. 
Hence scatter in recovered parameter values is expected due to cosmic variance of the summary statistics measurements. Since different analyses include (substantially) different scales, they are subject to different cosmic variance modes and may be scattered in different directions.

The effective volume of this mock catalog is similar to that of the BOSS DR12 LRG sample. 
For this mock, we have submissions from the EFT P+B, 
the void size function (VSF; Section~\ref{subsec:voids}), and the void-galaxy cross-correlation 
function (VGCF; Section~\ref{subsec:voids}) teams; the parameter constraints are shown in Fig.~\ref{fig:1D_summary_lc}. 
Both void methods and EFT P+B successfully recover the input values of $\Om$ within $95\%$ CL, and the EFT pipeline has additionally recovered the true input value of $\sigma_8$. We note that the true cosmological parameters are outside the informative prior (on $h$, $\omega_{\rm{b}}$ and $n_{\rm{s}}$) adopted by the VSF analysis. However, post-unmasking analyses indicate at most minor bias in $\Omega_{\rm{m}}$ due to the miscentered prior. The impact of prior width on the VSF constraining power is discussed in \refsec{void_choices}.
The void team has not submitted the $\sigma_8$ results, as its accurate inference from the VSF is not possible without an independent calibration of the tracer bias inside voids, e.g. taken from realistic mocks of the sample in question \citep[c.f.,][and Section~\ref{subsec:voids_modelling}]{Contarini2023BOSS}. For analyzing the VGCF, in this paper only the Alcock-Paczynski test is exploited for cosmological inference, which exclusively constrains parameters describing the background evolution.

\subsection{Results on real-space mock}
The real-space challenge boxes were analyzed with the EFT-based field-level Bayesian inference (EFT FBI) approach using the \code{LEFTfield} code (Section~\ref{subsec:EFT_fieldlevel_analysis_method}), EFT P+B, and the \code{BACCO} power spectrum emulator. As for the redshift-space analyses, the summary statistics measurements were averaged over the 10 realizations of the $\Lambda$CDM real-space galaxy catalog from a periodic box, and all \code{BACCO} and EFT P+B inferences were performed assuming a covariance of a $(2\,h^{-1}{\rm Gpc})^3$ box. For EFT FBI, we show the average of the 10 independent posteriors.

All three analysis teams fixed all cosmological parameters to the simulation true values in this challenge, except $\sigma_8$.
EFT P+B and BACCO had originally validated and unmasked their analyses of the real-space mocks in the full flat-$\Lambda$CDM parameter space. 
After unmasking, both teams kindly reran their analyses with all cosmological parameters except $\sigma_8$ set to their truth values to enable comparison with EFT FBI. This allows a direct comparison between $\sigma_8$ constraints obtained by the three statistics, as the EFT FBI team and their \code{LEFTfield} pipeline currently cannot perform a full inference over the $\Lambda$CDM cosmological parameter space. This simplified setup however still offers insights into the information content in (nonlinear) clustering beyond the 2-point function. 
The results are presented in Figure~\ref{fig:1D_summary_real}.
The constraints of the \code{BACCO} analysis are obtained with only the real-space power spectrum, and are thus heavily impacted by the $b_1$--$\sigma_8$ degeneracy present at leading order in the power spectrum. The degeneracy is broken only by non-linear contributions to the power spectrum.
The EFT P+B and FBI teams use extended data vectors to mitigate this $b_1$--$\sigma_8$ degeneracy.

All analysis teams recover the true value of $\sigma_8$ within 95\% CL of one single simulation box. 
At face value, the EFT FBI analysis achieves the tightest constraints on $\sigma_8$.
Compared with the EFT P+B analysis, the FBI analysis assumed a more restricted model for galaxy bias and galaxy stochasticity. The two EFT teams subsequently attempted to quantify the impact of their modeling differences in post-unmasking studies; we refer readers to Sec.~\ref{subsec:EFT_fieldlevel_analysis_method} and \refsec{discussion} for the results and discussions.

\section{Mock data}
\label{sec:mocks}

The challenge organizers present three different types of mock catalogs based on N-body simulations: real-space snapshot, redshift-space snapshot, and light cone---all using the HOD prescription. The mock catalogs are publicly available at this repository \github{https://github.com/ANSalcedo/Beyond2ptMock}.

\subsection{N-body simulations and halo catalogs}

\subsubsection{Flat \texorpdfstring{$\Lambda$}{L}CDM snapshot mocks}

After agreement on the range for the cosmological parameters described in Section \ref{sec:mask}, the organizers created the flat $\Lambda$CDM mock catalogs in $z = 1$.
Ten realizations have been created for each of real- and redshift-space mocks, each of which is a cubic simulation box with a comoving side length of $2 h^{-1}\,\mathrm{Gpc}$.
The $N$-body simulations have been run by using the cosmological $N$-body simulation code \code{GINKAKU} (Nishimichi, Tanaka \& Yoshikawa in prep.). 
This code employs the Tree Particle-Mesh (TreePM) method to compute the gravitational force in an expanding periodic box in comoving coordinates. The short-range tree force is implemented based on the Framework for Developing Particle Simulators \citep[\code{FDPS};][]{2016PASJ...68...54I,2018PASJ...70...70N}, a public library for general particle simulations. FDPS facilitates scalable computations on modern supercomputer systems,
ensuring optimized workload balance with efficient domain decomposition. The tree force is further accelerated by SIMD instructions implemented in the \code{Phantom-GRAPE} library \citep{Nitadori2006-ek,2012NewA...17...82T,2013NewA...19...74T}. Details of the long-range PM force can be found in \citet{Yoshikawa_2005} \citep[see also][for recent implementations]{2009PASJ...61.1319I,2012arXiv1211.4406I}.

The initial conditions are generated using second-order Lagrangian perturbation theory (2LPT; \citealt{scoccimarro98,crocce06}) at $z=49$.
Similarly to the existing $N$-body simulation ensembles like \code{AbacusSummit} or \code{Dark Quest}, the organizers have run the simulations without the effect of massive neutrino dynamics, and have incorporated massive neutrinos only in the matter transfer functions with the abundance $\omega_{\nu} = 0.000644$.
The transfer function has been computed at $z=0$ and rescaled to the starting redshift $z = 49$ using the linear growth factor with matter density $\Omega_\mathrm{m} = (\omega_{\mathrm{b}} + \omega_{\mathrm{cdm}} + \omega_{\nu}) h^{-2}$, but without the scale dependence induced by massive neutrinos.

For each realization, dark-matter halos are populated with mock galaxies as described in Sect.~\ref{sec:hod_galaxy_mocks}. We keep the choice of halo finder masked as differences in halo finder can contribute to model misspecification in the highly nonlinear regime \citep[e.g.,][]{Hahn2023a, Modi2023a}.
Neither real-space nor redshift-space mocks include the Alcock-Paczynski effect \citep{Alcock:1979mp}. This effect contributes to constraining the cosmological parameters in a more realistic galaxy spectroscopic survey setting and is taken into account in the light-cone mocks described next below.

\subsubsection{Flat \texorpdfstring{$\Lambda$}{L}CDM light-cone mocks}
The single-realization $\Lambda$CDM light-cone mock is based on one of the publicly available light cone halo catalogs from the \code{AbacusSummit} set of simulations \citep{Maksimova:2021ynf, 2021Hadzhiyska} generated with the \code{abacus} cosmological $N$-body code \citep{2019Garrison}. The specific \code{AbacusSummit} cosmology realization was chosen to be within the parameter range of all emulators participating in this challenge (c.f. Sect.~\ref{sec:mask}). The organizers shared that the light-cone mock is one of the \code{AbacusSummit} cosmologies with participants only after unmasking. 

Two teams, (\emph{EFT P\& B} and \emph{Cosmic Voids}), submitted results for the light-cone mock. Both analyses rely on analytic modeling prescriptions, eliminating concerns that the same halo mock catalog might inadvertently be used in training an emulator for this challenge.

\subsection{HOD galaxy mocks}
\label{sec:hod_galaxy_mocks}

To generate mock galaxy catalogs for our challenge, we take advantage of the HOD formalism \citep[e.g.][]{Berlind_2002, 2002MNRAS.329..246B,Zheng_et_al_2005, Gonzalez-Perez_et_al_2018, Salcedo_et_al_2022, Yuan_et_al_2023}.
In practice, we specify a parameterized form for the mean halo-mass-dependent occupation of galaxies and stochastically populate the dark matter halos with galaxies. The galaxy distribution within a host halo broadly traces the host's internal structure, but may differ in detail in a halo-mass-independent way. We may also in principle include some level of galaxy assembly bias, which refers to the possibility for the occupation of halos of a given mass to depend on properties other than mass \citep[e.g.][]{Hearin2016, McEwen_2018, Xu_et_al_2021, Salcedo_et_al_2022b, Wang_et_al_2022, Beltz-Mohrmann_et_al_2023, Contreras_et_al_2023, Zhongxu_et_al_2023}.
In this context we may implement galaxy assembly bias with respect to halo properties present in our simulation catalogs, or environmental properties of the simulation particle distribution. The light-cone mock may include redshift evolution of some of our HOD parameters. In our redshift-space mocks (including the light cone) we also add galaxy peculiar velocities that may include velocity bias \citep[e.g.][]{vdBosch_et_al_2005, Reid_et_al_2014, Guo_et_al_2016, Yuan_et_al_2018, Anbajagane_et_al_2022, Lange_et_al_2022, Beltz-Mohrmann_et_al_2023, Kwan_et_al_2023, Zhai_et_al_2023, Kwon_et_al_2024}.
The positions of mock galaxies are computed by modulating the real-space coordinates along the $z$-axis according to their peculiar velocities.
We emphasize that in constructing the mock galaxy catalogs in this challenge, we have remained broadly within the existing literature on the galaxy--halo connection and have by no means attempted to confound the challenge participants.

\section{Parameter Masking Implementation}
\label{sec:mask}
\subsection{Pre-Unmasking Information}
\label{sec:info}
Initially, the only information communicated to the analysis teams was the cosmology parameter space (flat $w$CDM) and that the galaxy assignment is based on a HOD technique, without specifying the detailed HOD implementation.

During initial test runs on these $w$CDM mocks, several teams realized that the (still masked) cosmologies were likely outside the parameter support of some emulator-based methods.

The analysis teams then agreed on common range for the cosmological parameters, which correspond to the intersection of the parameter spaces supported by the \code{AbacusSummit}, \code{Aemulus} and \code{BACCO} emulators, which includes a hard constraint on the angular scale of the sound horizon at decoupling $\theta_*$ imposed in the \code{AbacusSummit} simulation suite \citep{Maksimova:2021ynf}.

\begin{figure}
\centering \includegraphics[width=0.8\textwidth]{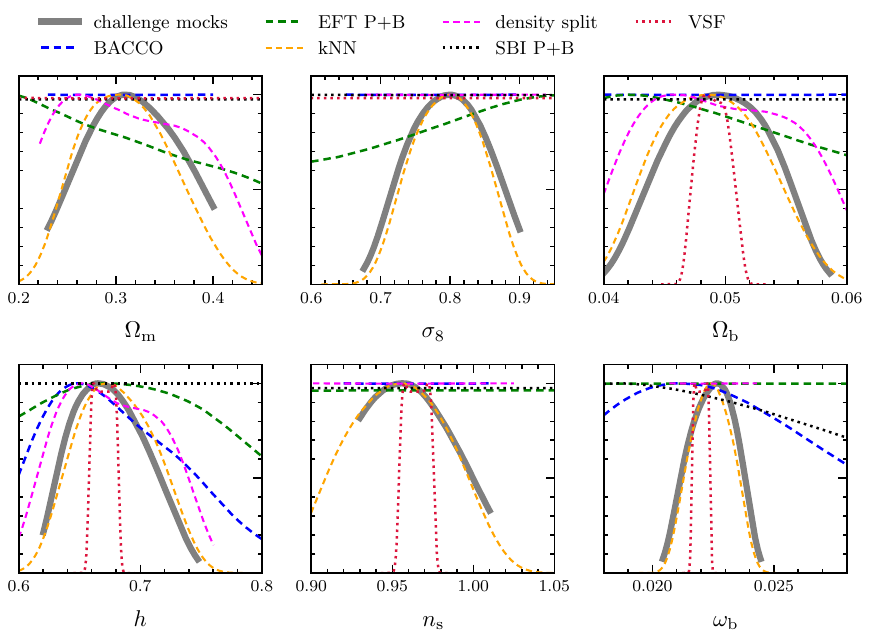}
    \caption{1D marginalized cosmology priors in flat $\Lambda\rm{CDM}$ for the challenge mocks (solid grey line) and those adopted by individual analysis teams. Analyses that fix $\theta_*$ are shown with dashed lines, analyses with free $\theta_*$  are shown as dotted lines. While there are only four/five independent parameters in flat $\Lambda\rm{CDM}$ with/without $\theta_*$ constraint, analysis teams specify their priors in different parameterizations. To facilitate comparison, we project priors in different parameters, noting that not all panels are independent.}
\label{fig:prior}
\end{figure}

A script implementing this parameter restriction was shared with all analysis teams\footnote{Available at \github{https://github.com/ANSalcedo/Beyond2ptMock/blob/main/parameters_theta_s_constrained.ipynb}, but not incorporated by any of the participating analyses.}, and the challenge organizers chose new sets of cosmological parameters to generate flat $\Lambda$CDM mock data within the specified parameter range. The cosmology prior for the mock catalogs, as well as the priors of individual analyses (documented in \refsec{methods}) are shown in \reffig{prior}.

Each analysis team documented their analysis choices and unmasking criteria, as well as potential caveats of their analysis (which could be revisited in the event of an unmasking surprise).

\emph{We show only parameter offsets to not unmask future participants and enable continued use of the mock datasets as a benchmark for testing novel analysis techniques.}
The organizers encourage future submissions from other analysis teams and commit to continuous unmasking.

\subsection{Unmasking Process}
There is only one parameter unmasking stage. Each analysis team chooses when they are ready to unmask and share their parameter chains with the challenge organizers, who produce the plots described in \refsec{results} and share values of the (marginalized) parameter shifts $\Delta \Omega_\mathrm{m}$ and $\Delta \sigma_8$ with the analysis team. Analysis teams agreed not to share parameter values with other teams but to only show plots with offset parameter values.

\subsection{Post-Unmasking Analyses}
In the event that an analysis team decides to adjust their analysis choice(s) after unmasking, they will document all post-unmasking tests in the paper and commit to showing the original unmasking result along with the post-unmasking update. 

In case some novel analyses may require expanded discussions and future work to reach the maturity required for precision cosmology analyses, post-unmasking analyses and discussions are encouraged for this paper.

\subsection{Accommodations}
In practice, the organizers unmasked submissions from all analysis teams but one in a joint video conference. To encourage submissions from analyses from emerging methods with less extensive validation, the unmasking plot shared in the unmasking video conference showed results from all submissions with anonymized labels (e.g., ``Team Blue" and ``Team Yellow"). Each analysis team was informed only of their own label in advance and the organizers did not share with participants the list of participating analyses. 
Organizers and participants had agreed on the option for analysis teams to withdraw from the challenge after unmasking to provide anonymity to unexpected results. Ultimately, no analysis team made use of the option to withdraw their submission.

The organizers communicated extensively with analysis teams to facilitate direct comparisons between analyses while preserving the parameter masking. 
Specifically, the organizers coordinated additional submissions from \code{BACCO} and EFT P+B with different cosmology priors: \textit{i)} $\Lambda$CDM without hard $\theta_*$ constraint (c.f.~\refsec{info}) to enable comparison with the SBI P+B analyses based on \code{Quijote} simulations without $\theta_*$ constraint (\reffig{1D_summary_rsd}), \textit{ii)} fixing all parameters except $\sigma_8$ to the true cosmology of the real-space mocks for comparison with EFT FBI results (\reffig{1D_summary_real}).
For the latter comparison, truth values were shared after unmasking of the \code{BACCO} and EFT P+B submissions for the full parameter space.
In addition, the organizers shared the normalized initial conditions of real-space mock \code{box1}
and truth values of all cosmological parameters except $\sigma_8$ with the EFT FBI team.

Feedback from participants suggests these accommodations enabled broader participation and further improved the participants' experience.

\section{Analysis Methods}
\label{sec:methods}
\begin{table}
\caption{Notation for parameter inference variables.}
\renewcommand{\arraystretch}{1.2}
\label{tab:variables}
\begin{center}
\begin{tabular}{l l}      
    	\hline
	Variable & Symbol\\
	\hline
     	data vector & $\data$ (or $\widehat{d}_i$)\\
    	cosmological parameters& $\cosPar\equiv\{\sigma_8,\Omega_{\rm m }, h,n_{\rm s},\cdots\}$\\
    	nuisance parameters& $\nuisPar\equiv\{b_1,b_2,\cdots\}$\\
    	parameters & $\Par = \cosPar\cup\nuisPar$ \\
    	model data vector & $\model(\Par)$\\
    	covariance & $\cov$ (or $\cov_{ij}$)\\
	covariance estimate & $\widehat{\cov}$\\
    	likelihood & $\lik[\data|\Par]$\\
    	posterior & $\post[\Par|\data]$\\
    	Gaussian distribution & $\gauss$, e.g. $b_1\sim \gauss(\mu,\sigma^2)$\\
    	uniform distribution (e.g., prior)& $\unif$, e.g., $h\sim \unif[0,1]$\\
	\hline
\renewcommand{\arraystretch}{1.0}
\end{tabular}
\end{center}
\end{table}

The following subsections detail the modeling, inference and parameter unmasking choices of the different analysis teams.

\paragraph{Notation} We adopt the notation in \reftab{variables} for common variables. For example, with this notation, the Gaussian data likelihood adopted in most inferences is given by
\be
-2\log\lik[\data|\Par] = (\model(\Par)-\data)^T\cov^{-1}(\model(\Par)-\data)+\text{constant}\,.
\ee
Throughout, we assume a flat $\Lambda$CDM cosmology, and use  $H(z)$ for the Hubble rate at redshift $z$ and $D_{\rm{A}}(z)$ for the angular diameter distance.

\paragraph{HOD models}
Analyses of galaxy clustering that extend into the highly nonlinear regime require a model for the galaxy--halo connection that relates galaxies to individual halos, for example HOD models (cf. \refsec{hod_galaxy_mocks} for extended references). This is in contrast to the EFT bias expansion relating the smoothed galaxy density field to the matter density on large scales.

Three of the participating analyses (SBI P+B \ref{subsec:SBI}, $k$NN statistics \ref{subsec:Abacus_kNN}, density-split clustering \ref{subsec:DSC}) follow the HOD approach, but with somewhat different implementations and model extensions. Specifically, analysis~\ref{subsec:SBI} uses one parameterization of the HOD model \citep[see][for details]{Hahn2023a} while analyses~\ref{subsec:Abacus_kNN} and \ref{subsec:DSC} use a somewhat different parameterization \citep[see][for details]{Yuan:2021hod}. For brevity, we refer to the two models as HOD1 and HOD2, respectively, in this overview. 

First, we briefly introduce the vanilla HOD formalism. Statistically, the HOD can be summarized as a probabilistic distribution $P(n_g|\boldsymbol{X}_h)$, where $n_g$ is the number of galaxies of the given halo, and $\boldsymbol{X}_h$ is some set of halo properties. Typically the galaxy population is divided into central and satellite populations, and assumes the central galaxy occupation follows a Bernoulli distribution whereas the satellites follow a Poisson distribution.

The vanilla HOD model has been extensively used to describe magnitude-limited galaxy samples \citep{Zheng2007}. It parameterizes the mean galaxy occupation as
\begin{align}
    \bar{n}_{\mathrm{cent}}(M) & = \frac{f_\mathrm{ic}}{2}\mathrm{erfc} \left[\frac{\log_{10}(M_{\mathrm{cut}}/M)}{\sqrt{2}\sigma}\right], \label{eq:zheng_hod_cent}\\
    \bar{n}_{\mathrm{sat}}(M) & = \left[\frac{M-M_{0}}{M_1}\right]^{\alpha},
    \label{eq:zheng_hod_sat}
\end{align}
where the five vanilla parameters characterizing the model are $M_{\mathrm{cut}}, M_1, \sigma, \alpha, M_0$, augmented by a central galaxy incompleteness parameter $f_{\mathrm{ic}}$, defined to be $0 < f_\mathrm{ic}\leq 1$. 
$M_{\mathrm{cut}}$ characterizes the minimum halo mass to host a central galaxy and $\sigma$ describes the steepness of the transition from 0 to $f_{\mathrm{ic}}$ in the number central galaxies.
$M_0$ gives the minimum halo mass to host a satellite galaxy, which is also commonly parameterized as $\kappa M_{\mathrm{cut}} =M_0 $, $M_1$ characterizes the typical halo mass that hosts one satellite galaxy and $\alpha$ is the power law index on the satellite galaxy occupation.

Both HOD models slightly deviate from this vanilla form. 
HOD1 fixes $f_\mathrm{ic} = 1$, effectively requiring all massive halos to host a central galaxy. HOD2 varies $f_\mathrm{ic}$ implicitly by rescaling the predicted number density to match the observed number density. 
HOD2 also adds a modulation term $\bar{n}_{\mathrm{cent}}(M)$ to the satellite occupation function to largely remove satellites from halos without centrals. 

The two models also differ in their ways of determining galaxy positions and velocities once the number of galaxies per halo is computed. 
In both models, the position and velocity of the central galaxy are set to be the same as those the halo center.
However, for satellite galaxies, HOD1 assigns the satellite galaxy positions within the virial radius of the halo following a Navarro-Frenk-White (NFW) profile \citep{Navarro:1996gj} while the velocities are solved from Jeans equations. 
In HOD2, the satellite galaxies are randomly assigned to halo particles with uniform weights, each satellite inheriting the position and velocity of its host particle. 

Both models also augment the vanilla HOD by adding additional flexibilities \citep[c.f.][for implementation details]{Hahn2023a,Yuan:2021hod}:
\begin{itemize}
    \item \textit{Velocity bias}: The galaxy velocities may not perfectly follow the dark-matter halo and particle velocities. HOD1 rescales galaxy velocities with parameter $\eta_\mathrm{cen}$ and $\eta_\mathrm{sat}$, which set the velocity dispersions of central and satellite galaxies relative to halo velocity dispersion: $\sigma_\mathrm{cen} = \eta_\mathrm{cen}\sigma_\mathrm{cen}$ and $\sigma_\mathrm{sat} = \eta_\mathrm{sat}\sigma_\mathrm{sat}$. In HOD2, we define velocity bias parameters $\alpha_\mathrm{vel, c}$ and $\alpha_\mathrm{vel, s}$ to modulate the peculiar velocities of the central and satellite galaxies with respect to the host halo center, respectively. 
    In this definition, $\alpha_\mathrm{vel, c} = 0$ and $\alpha_\mathrm{vel, s} = 1$ indicates no velocity bias.
    
    \item \textit{Galaxy assembly bias}: Galaxy occupation can also depend on secondary halo properties beyond halo mass, a phenomenon commonly referred to as galaxy assembly bias or galaxy secondary bias \citep[See][for a review]{2018Wechsler}. HOD1 adds the halo concentration as a secondary dependency in galaxy occupation via the Heaviside Assembly bias model, which is described in detail in \citet{Hearin2016}. HOD2 offers the option of using either the halo concentration or the halo environment in a $5h^{-1}$Mpc filter as the secondary dependency. Both models adopt two assembly bias parameters to modulate the central and satellite occupations separately. 
    
    \item \textit{Baryonic effects}: Both HOD implementations account for baryonic effects by modulating the radial distribution of satellite galaxies relative to the halo density profile. HOD1 includes a parameter $\eta_{\rm{conc}}$ which sets the ratio
    between the concentration of satellite and halo profile. HOD2 includes a parameter $s$ that modulates the radial satellite galaxy profile with $s = 0$ indicating no radial bias, $s > 0$ indicating a more extended (less concentrated) profile of satellites relative to the halo, and vice versa for negative $s$.
    
    \end{itemize}
\begin{table*}

\begin{center}
\setlength{\tabcolsep}{5pt} 
{\renewcommand{\arraystretch}{1.1}
\begin{tabular}{c|l|l}
\hline\hline
Ingredient & HOD1 \citep{Hahn2023a}& HOD2 \citep{Yuan:2021hod}\\ \hline
\multicolumn{3}{l}{1. Vanilla \citet{Zheng2007} HOD parameters 
\refeqs{zheng_hod_cent}{zheng_hod_sat}}\\
\hline
& satellite occupation cut off parameter: $M_0$ & satellite occupation cut off parameter: $ \kappa M_\mathrm{cut}$\\
& - & $\bar{n}_{\mathrm{sat}}(M)$ modulated by $\bar{n}_{\mathrm{cent}}(M)$\\
& satellites assigned via NFW profile & satellites assigned to halo particles. \\ 
& $f_{\rm{ic}}\equiv 1$ & $f_{\rm{ic}}$ determined by observed galaxy density\\
\hline

\multicolumn{3}{l}{2. Velocity bias parameters}  \\ \hline
$\alpha_\mathrm{vel, c}$ & - & central velocity bias defined relative to halo center \\ 
$\alpha_\mathrm{vel, s}$ & - & satellite velocity bias defined relative to particle \\
$\eta_\mathrm{cen}$ & central velocity bias defined as additional dispersion & - \\ 
$\eta_\mathrm{sat}$ & satellite velocity bias defined as additional dispersion & - \\ 
\hline

\multicolumn{3}{l}{3. Galaxy assembly bias parameters}  \\ \hline
$A_c$ & central occupation conditioned on concentration. & - \\ 
$A_s$ & satellite occupation conditioned on concentration. & - \\
$B_\mathrm{cent}$ & - & conditioned on environment or concentration. \\ 
$B_\mathrm{sat}$ & - & conditioned on environment or concentration.\\
\hline\multicolumn{3}{l}{4. Baryonic effects}  \\ \hline
$\eta_{\rm{conc}}$ &  concentration ratio of satellite and halo profile. & - \\
$s$ & - & modulation of radial satellite galaxy profile.\\
\hline\hline
\end{tabular}}
\end{center}
\caption{Summary of HOD parameterizations used by the SBI (HOD1), $k$NN and density split (HOD2) analysis teams.}
\label{Table:hod_summary}
\end{table*}
Note that the three HOD-based analyses all employ different priors for the HOD parameters, motivated by sensitivity studies specific to each statistic or based on conservative estimates, as detailed in their respective analysis sections.

\input{methods_BACCOemu.tex}
\input{methods_pt_pbj.tex}

\input{methods_EFT_fieldlevel.tex}
\input{methods_SBI.tex}
\input{methods_AbacuskNN.tex}
\input{methods_DensitySplit}
\input{methods_Voids.tex}

\section{Discussion}
\label{sec:discussion}
\begin{figure}
\centering \includegraphics[width=1.0\textwidth]{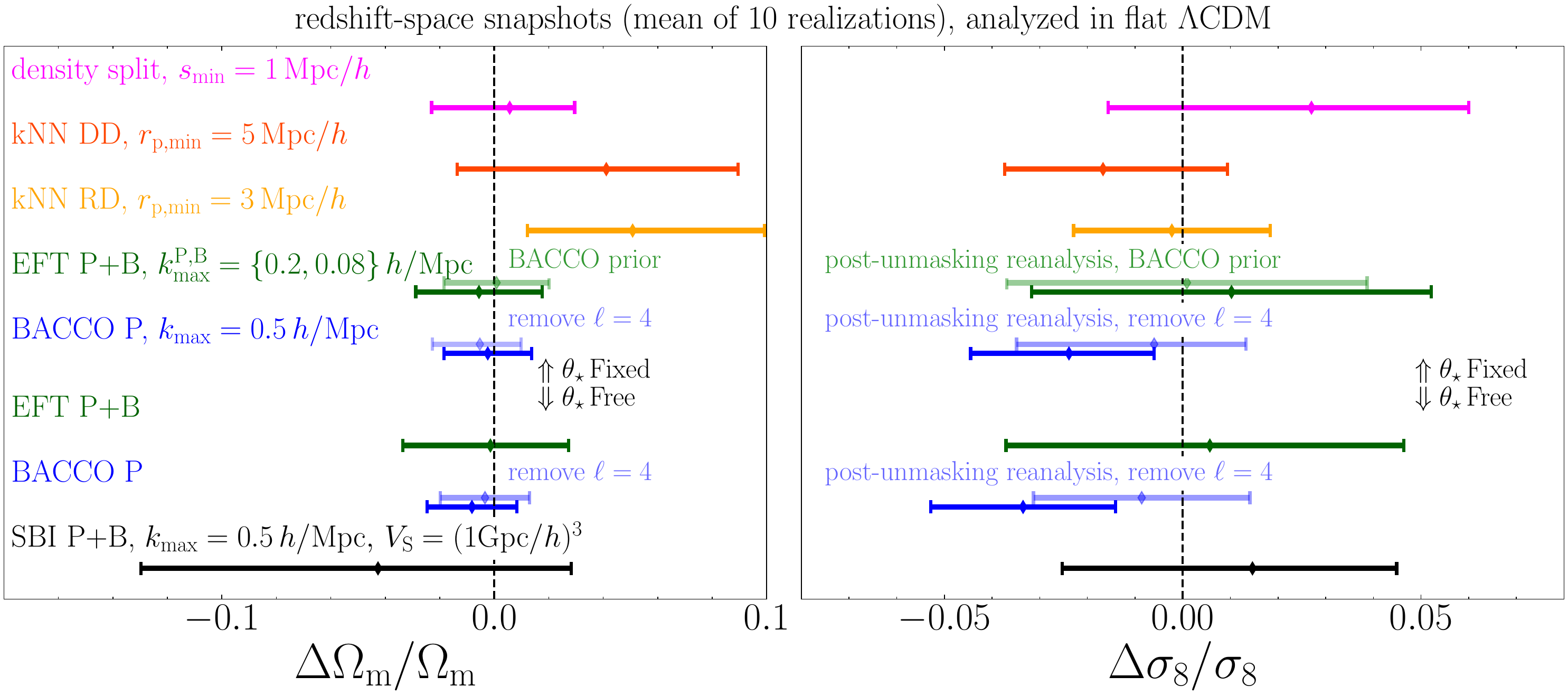}
    \caption{1D marginalized constraints on $\Omega_{\rm{m}}$ and $\sigma_8$ for analyses of redshift-space mocks, including post-unmasking reanalyses shown with light opacity.} 
\label{fig:1D_summary_rsd_post}
\end{figure}

After unmasking, several teams further investigate the accuracy and precision of their original submission. Several of the results of these post-unmasking reanalyses are shown in \reffigs{1D_summary_rsd_post}{1D_summary_real_post}. The discussions in this section underline the importance of accounting for systematics due to model and analysis choices. 
\subsection{Cosmology priors}
We show two post-unmasking analyses illustrate the impact of informative cosmology priors on parameter constraints. For the original unmasking submission, all analysis teams chose their own priors (which all differ from the priors from which the mock cosmologies are drawn), as illustrated in \reffig{prior}. 
While the posteriors on the target parameters ($\Omega_{\rm{m}},\,\sigma_8$) are substantially narrower than the priors in these parameters for all analyses, several analyses adopted informative priors for some of the other $\Lambda\mathrm{CDM}$ parameters.
The light green symbols in \reffig{1D_summary_rsd_post} show the impact of imposing the \code{BACCO} cosmology priors (\refeq{BACCO_priors_cosmo}) on the EFT P+B analysis, which originally used the widest cosmology priors among all participating analysis teams (\refeq{ptpb_priors_cosmo}). 
The approximately $15\%$ improvement in 1D-marginalized constraints is primarily caused by the informative prior on $n_{\rm{s}}$. 
On the other hand, the VSF analysis chose informative priors on $h$, $\omega_{\rm{b}}$ and $n_{\rm{s}}$ that excluded the light-cone mock cosmology at several $\sigma$.
Post-unmasking analyses indicate only a minor bias in $\Omega_{\rm{m}}$ due to the miscentered prior and a $23\%$ degradation in constraining power with the wider cosmology prior of the \code{BACCO} analysis. These examples highlight the importance of homogenizing cosmology priors for future detailed comparisons of different analysis methods.

\subsection{Post-unmasking \code{BACCO} analysis refinement}
The \code{BACCO} team identified a large emulation uncertainty of the hexadecapole previously unnoticed that became significant due to the increased simulation volume of the challenge mocks compared to their previous validation (c.f. Sect.~\ref{subsec:bacco_inference}). The light blue symbols show a post-unmasking reanalysis of monopole and quadrupole only, with all other analysis settings held fixed.

\subsection{Post-unmasking EFT analyses comparisons}
\begin{figure}
\centering \includegraphics[width=0.5\textwidth]{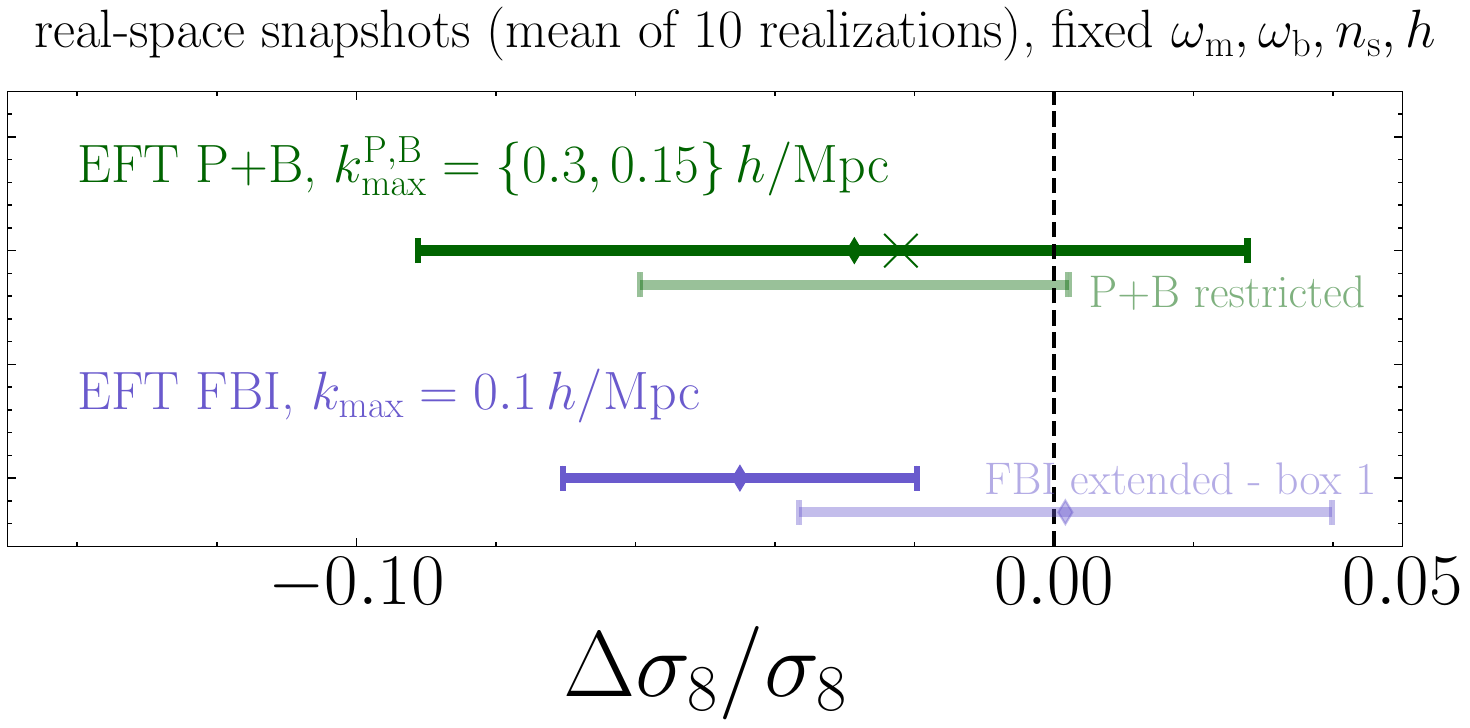}
    \caption{1D marginalized constraints on $\sigma_8$ for analyses of real-space mocks, including post-unmasking reanalyses shown with light opacity.The ``P+B restricted'' posterior mean is re-centered to the ``EFT P+B'' posterior mean in the EFT P+B baseline analysis. The ``FBI extended'' constraints is obtained from only one single realization (see \textit{post-unmasking studies} in Sec.~\ref{subsec:EFT_fieldlevel_analysis_method})}. 
\label{fig:1D_summary_real_post}
\end{figure}

The EFT P+B and EFT FBI teams adopted different modeling choices in their pre-unmasking analyses, which are partly motivated by the complexities of numerical implementation and computation.
To investigate the impact of these differences on their respective constraints, after unmasking, both teams have run new analyses with assumptions that aim to draw closer to the baseline analysis of the other team.
In \reffig{1D_summary_real_post}, we show results of these post-unmasking studies (light opacity), next to the original pre-unmasking results (full opacity).

To first understand the difference between the two EFT pre-unmasking analyses, note that the EFT FBI pre-unmasking analysis described the matter-galaxy connection with a second-order galaxy bias expansion and assumed that the stochastic contribution to the galaxy density field, $\varepsilon$, is Gaussian distributed with a white power spectrum $\sigma_{\rm Poisson}$, effectively neglecting the coupling between the stochastic and deterministic fields of the form $\delta_g\supset \delta_m \varepsilon$ \citep[e.g.][]{Desjacques:2016bnm}.
In contrast, the EFT P+B baseline analysis included (1) all second-order plus the most relevant third-order bias term, (2) the most general galaxy stochasticity model at the given order including a stochastic bispectrum and stochastic-deterministic cross bispectrum contributions (stemming from the $\delta_m \varepsilon$ term in the bias expansion), and (3) $\mathcal{O}(k^2)$ corrections to the stochasticity power spectrum.
    
To see how these differences between the EFT FBI and EFT P+B 
analysis affect their result, the EFT P+B team have run a ``P+B restricted'' analysis on the real-space mocks, at the same (pre-unmasking) scale cuts $\kmax=\{0.3,0.15\}\,\hMpc$, and fixing all cosmological parameters to their true values---except $\sigma_8$---and with $A_{\rm shot}=B_{\rm shot}=b_{\Gamma_3}=a_0=R_*^2 = 0$ (the higher derivative bias defined as in \citet{Chudaykin:2020hbf}). 
The first important observation is that EFT P+B restricted
analysis recovered $\sigma_8$ with a significant bias. 
This is because all components described above, such as non-Gaussian stochasticity in the bispectrum, third-order operators in the galaxy bias expansion, and stochastic-deterministic coupling are non-negligible at the scales involved in the EFT P+B analyses.\footnote{It is possible that these terms are less important on large scales, e.g. at $k\sim0.1\,\hMpc$ corresponding to $\kmax$ in the EFT FBI analyses.} 
In order to ease the comparison in the presence of this bias, the ``P+B restricted'' posterior mean in \reffig{1D_summary_real_post} is re-centered to the ``EFT P+B'' posterior mean in the EFT P+B baseline analysis. 
Thus, the ``P+B restricted'' results in \reffig{1D_summary_real_post} should only be considered in terms of the error bars. A direct comparison to the ``EFT FBI'' error bars shows that the nominal variances on $\sigma_8$ in both analyzes are relatively similar, suggesting that the difference between these two constraints likely comes from different analysis assumptions. 

Conversely, in their post-unmasking study, the EFT FBI team have run an ``FBI extended'' analysis that includes the full set of third-order bias operators in the galaxy bias expansion (i.e., three additional bias terms compared to the ``EFT P+B'' pre-unmasking analysis), as well as higher-derivative stochasticity (but still without mode coupling in the stochastic part), at their pre-unmasking scale cut $\kmax=0.1\,\hMpc$. The ``FBI extended'' posterior in \reffig{1D_summary_real_post} is completely consistent with the ground truth $\sigma_8$, with a noticeably increased width (in between those from the ``EFT P+B'' and ``P+B restricted'' analyses, pre- and post-unmasking).
    
The upshot of these post-unmasking analyses and comparisons by the two EFT teams is that we find a broad consistency between the EFT P+B and FBI results, but future work is required to extend this comparison to the full cosmology parameter space, and to redshift-space clustering.

\subsection{Towards correct cosmology constraints from the highly nonlinear regime}
\label{sec:discussion_HOD}
While there is undoubtedly significant signal-to-noise in galaxy clustering in the highly nonlinear regime, the conversion from signal-to-noise to cosmology parameter constraints crucially relies on parameterizations of the galaxy--halo connection in the nonlinear regime being sufficiently flexible to marginalize over all modeling uncertainties that could bias cosmological parameter estimation. Furthermore, models for clustering statistics in the highly nonlinear regime are typically evaluated using simulations and emulators. This type of model evaluation is subject to uncertainties from cosmic variance, emulation errors, and finite training sample size, which may further reduce constraining power.

Four participating analyses utilize scales within the highly nonlinear regime, of which three rely on HOD-based models. Here, we summarize pre-unmasking validation tests and post-unmasking reanalyses that illustrate the impact of highly nonlinear scales and discuss limitations to error quantification in the highly nonlinear regime. 

The \code{BACCO} P analyses with $k_{\rm{max}}=0.2\hMpc$ and $k_{\rm{max}}=0.5\hMpc$ (\reffig{BACCOconstraints}) indicate a substantial gain in constraining power from strongly nonlinear scales. The \code{BACCO} analysis relies on a ``hybrid'' Lagrangian galaxy bias expansion, which has been validated against HOD and SubHalo Abundance Matching extended (SHAMe, \citealt{ContrerasAnguloZennaro2020AB, ContrerasAnguloZennaro2020}) techniques. 

The SBI P+B analysis recovers the input cosmology to better than $1{-}\sigma$ with a scale cut $k_{\rm{max}}=0.5\hMpc$. While the SBI P+B team did not perform scale cut variations for the challenge mock catalogs, other analyses using the same model for galaxy samples with similar number density show substantial gains in constraining power from scales $0.25\hMpc < k \le 0.5\hMpc$ for the power spectrum \citep{Hahn2023b} and $0.3\hMpc < k \le 0.5\hMpc$ for the bispectrum \citep{Hahn:2023kky}.

The density-split clustering analysis includes measurements down to $1 \, h^{-1} \, \mathrm{Mpc}$. However, the gain in constraining power from $30 \, h^{-1} \, \mathrm{Mpc}$ to $1 \, h^{-1} \, \mathrm{Mpc}$ scales is limited (c.f. 
 \reffig{dsc_whisker}), as the density-split correlation functions are not sensitive to variations in the clustering below the smoothing kernel scale ($R_s = 10\,h^{-1}{\rm Mpc}$) and the small-scale measurement only contains information about the smoothed density PDF. Additionally, the emulator and training-set errors contribute significantly to the total covariance on small scales, further suppressing the information gain. Finally, previous studies \citep{Paillas:2023} showed that density split statistics can efficiently extract information from AP distortions, that were not included in the current redshift-space mocks.

The DD-$k$NN baseline analysis employs the scale cut $r_p > 5h^{-1}\rm{Mpc}$. Post-unmasking reanalyses with even more aggressive scale cuts shown in \reffig{knns_scales} do not pass robustness tests, indicating insufficient model flexibility or uncertainty modeling. Hence, further model refinements are required to quantify the cosmological constraining power of aggressive analyses in the 1-halo regime.

\subsubsection{Galaxy--halo connection models}
We reiterate that the organizers communicated to all participants that the challenge mock catalogs are based on the HOD formalism, and it is important to acknowledge that HODs, like any other model for the galaxy--halo connection, come with their own set of limitations due to assumptions about galaxy formation. While the specific HOD parameterization is not revealed, this information gives HOD-based modeling approaches an inherent advantage. Among the participating analyses, the modeling of density-split clustering, $k$NN statistics, and SBI P+B are explicitly based on HODs. We refer to \citet{CuestaLazaro2023:2309.16539} for parameter recovery validation of the HOD-based density-split statistics model on a SHAM-based galaxy mock, and \citet{Hahn2023a} for parameter recovery validation of the HOD-based SBI P+B model against different HOD models and different N-body halo catalogs. The $k$NN post-unmasking reanalyses with different HOD models (\reffig{knns_scales}) show that $k$NN-DD analyses with aggressive 1-halo scale cuts are currently subject to either error miscalibration or model misspecification, even within the HOD framework. Future research on more flexible parameterizations will be required to pass comprehensive parameter recovery tests \citep[see][for an early example of comprehensive non-HOD recovery tests]{Reddick2014} and to obtain well-calibrated error bars. Such increased modeling flexibility will likely be accompanied by a degradation in constraining power, especially when accounting for non-HOD galaxy--halo connection models \citep[see][for a review]{2018Wechsler}. Hence, the advance of novel analysis methods that exploit the highly nonlinear regime requires developing and validating more realistic mocks to establish a more accurate challenge framework that can encapsulate a broader set of models. 

Validation of the uncertainties reported by individual methods will ultimately require a suite of (ideally, parameter-masked) mock catalogs with variations in cosmology and galaxy--halo connection model. However, not just any suite will do, and the design of such a suite is essential for enabling Bayesian quantification of model performance and uncertainty calibration. Ideally, such a suite would be generated by drawing from posteriors of galaxy--halo connection model(s) that satisfy the following:
\begin{enumerate}
    \item The mock-generating model(s) should be consistent with observations within some measurement uncertainty.
    \item The mock-generating model(s) should be predicated upon different assumptions from the models in the analysis being tested.
\end{enumerate}
For this program to be successful, realism matching simulations to observations is essential. This is a challenging task because we do not yet have a suite of simulations available that fully matches all observational data. Still, paying attention to developing accurate and flexible mock sky surveys that have the right set of included systematics should be a high priority so that the field is not misled either by missing important systematics or trying to account for unphysical systematics. Furthermore, any Bayesian quantification of model sufficiency and uncertainty validation critically depends on the (hyper) parameter priors for the test suite, which should be determined in data space rather than through arbitrary parameterization choices.
This development will benefit from continued improvements in cosmological hydro-dynamical simulations \citep[for a review see e.g.][]{Crain_vdVoort_rev_2023} as well as from the increasing diversity of semi-empirical methods for generating realistic mock galaxy populations \citep[e.g.][]{Behroozi_et_al_2019, Hearin_et_al_2021, addgals, Hearin_et_al_2023, Kwon_et_al_2023}.

\section{Conclusion}
\label{sec:conclusion}
It is well-established that the galaxy density field contains valuable information beyond the power spectrum, and many ``novel'' statistics and analysis methods have emerged and matured over the last decade to exploit this information. This makes it timely to survey the state of these analyses. In this paper, we present a parameter-masked mock challenge for beyond-2pt galaxy clustering statistics. 
The challenge data set consists of mock catalogs created from N-body simulations with a flat $\Lambda$CDM cosmology and HOD galaxy--halo connection models, with parameter values known only to the organizers. While all parameter values are masked, analysis teams optimize scale cuts and determine other analysis choices (e.g., nuisance parameters and their priors) for each analysis method and submit one result per method for unmasking. Upon unmasking, the organizers share plots of the relative parameter biases and uncertainties of the target parameters ($\Omega_{\mathrm{m}}$, $\sigma_8$) but do not share information on other cosmological parameters, HOD parameterization, or HOD parameters. Post-unmasking analyses are encouraged to enable continued method development but need to be clearly labeled as such.
The main results of this mock challenge can be summarized in three themes:
\begin{itemize}
\item \textit{Design of pre-unmasking analysis strategies and validation studies.} The priors for the mock catalogs (e.g., $\sigma_8\sim \unif[0.68,0.9]$ and any HOD parameters; c.f. \refsec{mocks}) are significantly broader than implicit priors in (non-masked) data analyses from previous observations. The inability to iterate on \refsec{results} provided an incentive for the analysis teams to develop consistency checks and unmasking criteria, summarized in \refsec{methods}, even if this robustness comes at some cost in constraining power. These validation studies are an essential ingredient for future analyses to meet the accuracy requirements of next-generation data sets.
\item \textit{Constraints from parameter-masked mock challenge.} The unmasking results presented in \refsec{results} showcase the competitive constraining power of multiple beyond-2pt statistics and novel analysis methods in a parameter-masked mock challenge. This performance of multiple statistics, as well as the associated modeling and inference frameworks, lends credibility to obtaining accurate and precise cosmology constraints via these methods. Further, the consistency across different analysis approaches enables a level of cross-validation on real data that is impossible to achieve for a single method on its own.
The combination of multiple statistics and modeling,
in principle, will enable the most precise constraints \citep[e.g., P+B; see] [for other combinations of beyond-2pt and 2pt statistics]{kNN2021,Kreisch2022,2023JCAP...03..045H,2023arXiv231016116V,2023PhRvD.108d3521B,2023ApJ...951...70M,2024ApJ...961..208S}. Accurate joint covariances and consistent models of the galaxy--halo connection across different methods will however require further research and pipeline developments.

\item \textit{Post-unmasking method refinements.} The unmasking results assess the performance of an analysis based on previous method development and masked validation tests. After unmasking, several teams further investigate the accuracy and precision of their original submission (c.f.~\refsec{discussion}). With caution against excessive fine-tuning, post-unmasking reanalyses identify directions for model refinements and methodological improvements for future analyses. Additionally, post-unmasking comparisons between different analysis methods provide a starting point for future studies contrasting different approaches to understanding the source of cosmological information beyond the conventional, linear and quasi-linear galaxy 2pt analyses.
\end{itemize}
While most of these findings can in principle be obtained by each team on individually generated mock catalogs, an externally organized parameter-masked challenge provides a common benchmark and naturally separates the validation of analysis choices and post-unmasking refinements, enabling a clearer assessment of constraining power. Hence we invite future submissions from other analysis teams and offer to update summary results plots in the beyond-2pt challenge repository with new submissions \github{https://github.com/ANSalcedo/Beyond2ptMock}.

From a participant's perspective, the set-up of the challenge and unmasking procedure, as well as the exchanges with the organizers and other teams, allowed a beneficial constructive atmosphere and strongly encouraged scientific interactions. Overall, the positive work environment throughout the challenge encouraged the collaborative development of validation tests and contributed to a better understanding of the tools used by the various teams.

A single parameter-masked mock challenge offers no panacea: models and analysis methods for different statistics evolve (e.g., due to post-unmasking refinements motivated by this study), scale cuts and priors must be calibrated anew for different survey volumes and galaxy samples. Furthermore, this particular mock challenge was clearly labeled as consisting of HOD-based galaxies free of observational systematics. Therefore, future parameter-masked mock challenges with more realistic astrophysical and observational complications as well as with statistical precision mirroring the increasing survey volume and galaxy density of future surveys will be required to further method validation.
While this challenge provides a vital stress test and performance benchmark at one specific point in cosmology and HOD parameter space, the current setup leaves the uncertainty calibration and validation against model misspecification to the individual analysis teams, using individually generated mock catalogs with parameter and model variations. As discussed in \refsec{discussion_HOD}, developing a suite of mocks suitable for the validation of parameter constraints, including their uncertainties, from highly nonlinear scales at the accuracy of Stage-IV surveys will be a major research and computing project that is beyond the resources of individual groups. Hence it would be desirable for future community-wide challenges to provide suites of mock catalogs across parameter space and galaxy--halo connection models to facilitate a rigorous assessment of uncertainty quantification.

To conclude, we emphasize the maturity of multiple ``novel'' statistics and analysis methods that participated in this parameter-masked mock challenge. The individual constraining power of a particular statistic depends on the specific parameter space considered, and we caution against extrapolating relative constraining power in this challenge to other scenarios. The main strength of this emerging field is the complementarity of approaches, which will enable extensive cross-validations to yield reliable and competitive results.

\section*{Acknowledgments}
This mock challenge was initiated during the Aspen Center for Physics 
2022 Summer Program ``Large-Scale Structure Cosmology beyond 2-Point Statistics,'' co-organized by DJ, EK, Hiranya Peiris, and FS. We are grateful to Hiranya Peiris for co-organizing this workshop and input on the initial design of this challenge, and to the Aspen Center for Physics, supported by the National Science Foundation grant PHY-1607611.
We thank Steward Observatory, University of Arizona for hosting a second wokshop for participating analysis teams in spring 2023, with support from the David and Lucile Packard Foundation.

The light-cone mock galaxy catalog is based on the \code{AbacusSummit} simulation light cones and we thank the \code{AbacusSummit} team for making their data products publicly available. The $N$-body simulations and subsequent halo catalog creation for producing the $\Lambda$CDM snapshot mocks were carried out on Cray XC50 at Center for Computational Astrophysics, National Astronomical Observatory of Japan. We further acknowledge High Performance Computing (HPC) resources supported by the University of Arizona TRIF, UITS, and RDI and maintained by the UA Research Technologies department.

We are grateful to Boryana Hadzhiyska, Andrew Hearin, Johannes Lange, Ariel S\'{a}nchez, and Risa Wechsler for their valuable comments on the manuscript.
We further thank Camille Avestruz, Humna Awan, Andrew Hearin, Dragan Huterer, Nick Kokron, Martin Reinecke, Marko Simonovi\'{c}, Julia Stadler, Masahiro Takada, Kuan Wang, Risa Wechsler, and Martin White for helpful discussions. 

EK, YK and ANS were supported in part by David and Lucile Packard Foundation and a research fellowship from the Alfred P. Sloan foundation.
CH was supported by the AI Accelerator program of the Schmidt Futures Foundation.
NMN acknowledges support from the Leinweber Foundation. OHEP is a Junior Fellow of the Simons Society of Fellows.
The work of TA and SY was supported by the U.S. Department of Energy SLAC Contract DE-AC02-76SF00515. 
KA acknowledges the support from Fostering Joint International Research (B) under Contract No.~21KK0050. MPI is supported by STFC consolidated grant no. RA5496. AP acknowledges support from the Simons Foundation to the Center for Computational Astrophysics at the Flatiron Institute, as well as support from the European Research Council (ERC) under the European Union's Horizon programme (COSMOBEST ERC funded project, grant agreement 101078174). CCL is supported by the National Science Foundation under Cooperative Agreement PHY2019786 (The NSF AI Institute for Artificial Intelligence and Fundamental Interactions).

The EFT FBI analyses were conducted on the \code{COBRA} and \code{FREYA} HPC clusters at the Max Planck Computing and Data Facility.
\section*{Author contributions}
\input{contributions.tex}


\input{beyond2pt_apj.bbl}
\end{document}

%% file: methods_summary_table_v3.tex
\movetabledown=144pt
\begin{sidewaystable}
\centering
\caption{Overview of participating analyses and their analysis ingredients.\\
$^a$ Cosmology priors are listed for the baseline validation and original unmasking submission. Teams \code{BACCO} P and EFT P+B kindly reran their analyses with different cosmology priors (but otherwise identical analysis choices) to facilitate comparisons.
}
\begin{tabular}{lllllllll}
\hline
\hline
  Method & Section & Mock(s) & Gravity& Tracer & Model & Covariance & Cosmology & Team \\
  &  & analyzed & model & model & evaluation &  & prior &  \\
\hline
  \code{BACCO} P & \ref{subsec:bacco} & all & N-body & hybrid-EFT & emulator & analytic & \refeq{BACCO_priors_cosmo}$^a$ & M.~Pellejero, R.~Angulo\\
  & & & & & & + emulator cov.& & and M.~Zennaro \\[3pt]
  Density Split & \ref{subsec:DSC} & redshift space & N-body & HOD & emulator & 1500 mocks\ & \refeq{dsc_priors_cosmo} & E.~Paillas and C.~Cuesta-Lazaro \\
   & & & & & & + emulator cov. & & \\[3pt]
  EFT FBI & \ref{subsec:EFT_fieldlevel_analysis_method} & real space & perturbative & EFT & analytic & analytic & \refeq{EFT-FBI_priors_cosmo} & N.~M.~Nguyen and F.~Schmidt \\
  EFT P+B & \ref{subsec:PTPB} & all & perturbative & EFT & analytic & analytic & \refeq{ptpb_priors_cosmo}$^a$ & M.~Ivanov, O.~Philcox,\\
  & & & & & & & & G.~Cabass and K.~Akitsu \\[3pt]
  $k$NN & \ref{subsec:Abacus_kNN} & redshift space & N-body & HOD & emulator & jackknife & \refeq{knn_priors_cosmo} & S.~Yuan and T.~Abel \\
  & & & & & & + emulator cov. & & \\[3pt]
  SBI P+B & \ref{subsec:SBI} & redshift space & N-body & HOD & galaxy mocks & N/A & \refeq{sbipb_priors_cosmo} & C.~Modi and CH.~Hahn \\
  VGCF & \ref{subsec:voids} & light cone & perturbative & linear bias & analytic & jackknife & \reftab{void_priors} & N.~Hamaus, S.~Contarini,\\
  & &  & &  & & & &G.~Verza and A.~Pisani \\[3pt]
  VSF & \ref{subsec:voids} & light cone & perturbative & linear bias & analytic & analytic & \reftab{void_priors} & S.~Contarini, G.~Verza, \\
  & &  & &  & & & &N.~Hamaus and A.~Pisani\\
\hline
\end{tabular}
\label{table:methods}
\label{tab:methods}
\end{sidewaystable}

%% file: methods_BACCOemu.tex
\subsection{\texorpdfstring{\code{BACCO}}{BACCO} Hybrid emulator\footnote{Authors: Marcos Pellejero Iba\~nez, Raul E. Angulo, Matteo Zennaro.}}
\label{subsec:bacco}
In this section, we describe the results of the Beyond-2point challenge obtained by the \code{BACCO}-hybrid emulator approach. This emulator was presented in \citet{ZennaroAnguloPellejero2021} and \citet{PellejeroIbanez2023} and has been thoroughly tested on SubHalo Abundance Matching extended techniques (SHAMe, \citealt{ContrerasAnguloZennaro2020AB, ContrerasAnguloZennaro2020}), and survey-based HOD techniques \citep{Pezzotta_2024,Nicola_2024}. The model has been further extended to study Intrinsic Alignments in \citet{Maion_2023} and to generate field level predictions in \citet{PellejeroIbanez_2024}.

\subsubsection{Data \& Estimators}

\begin{figure*}	
\includegraphics[width=\textwidth]{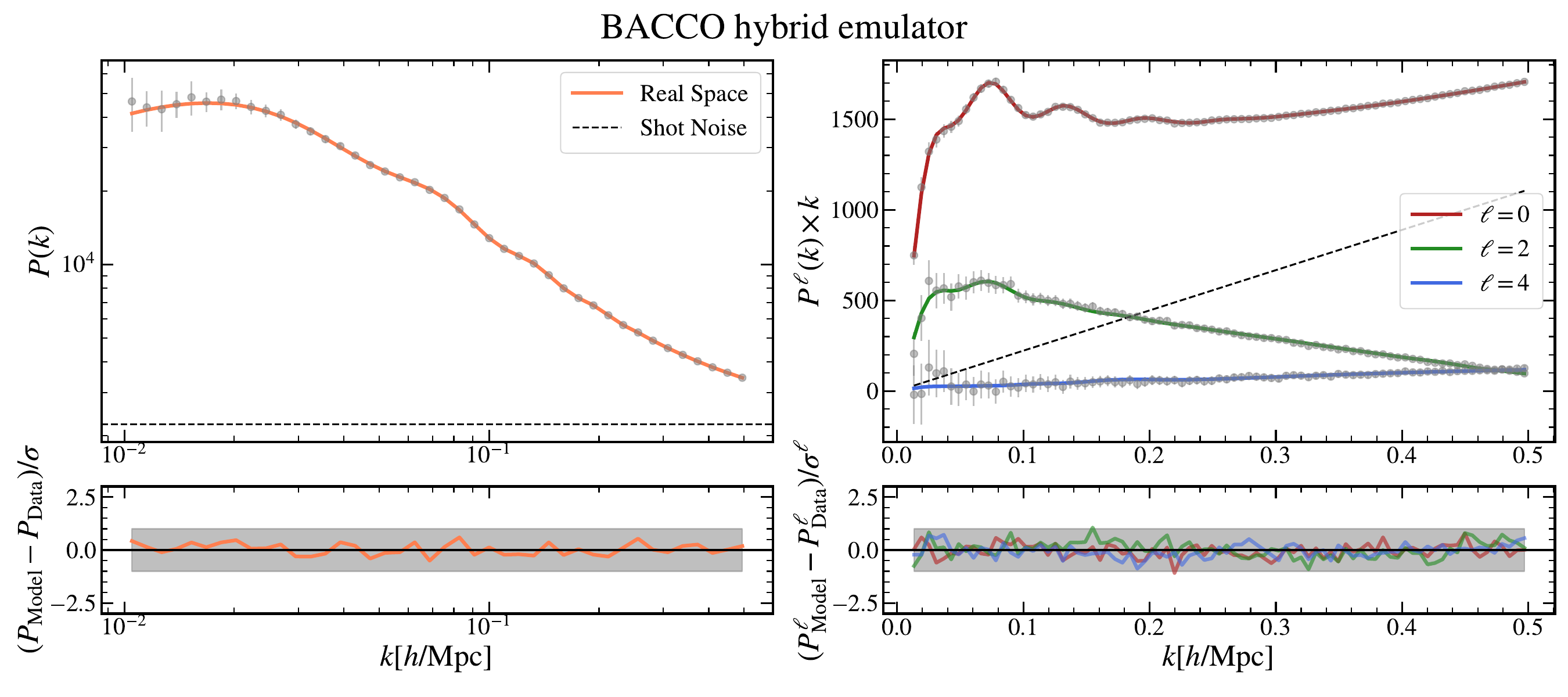}
    \caption{ Left panel: Power spectrum fit of the \code{BACCO} hybrid emulator model to the mean of the 10 $\Lambda$CDM mock boxes. The error bars correspond to the Gaussian approximation of the covariance matrix at the volume of one of the boxes. Right panel: Same as the left panel but for the 10 $\Lambda$CDM mock boxes in redshift space. We show the three multipoles used in this work.}
    \label{fig:BACCOmeasurements}
\end{figure*}

\begin{figure*}	
\includegraphics[width=0.49\textwidth]{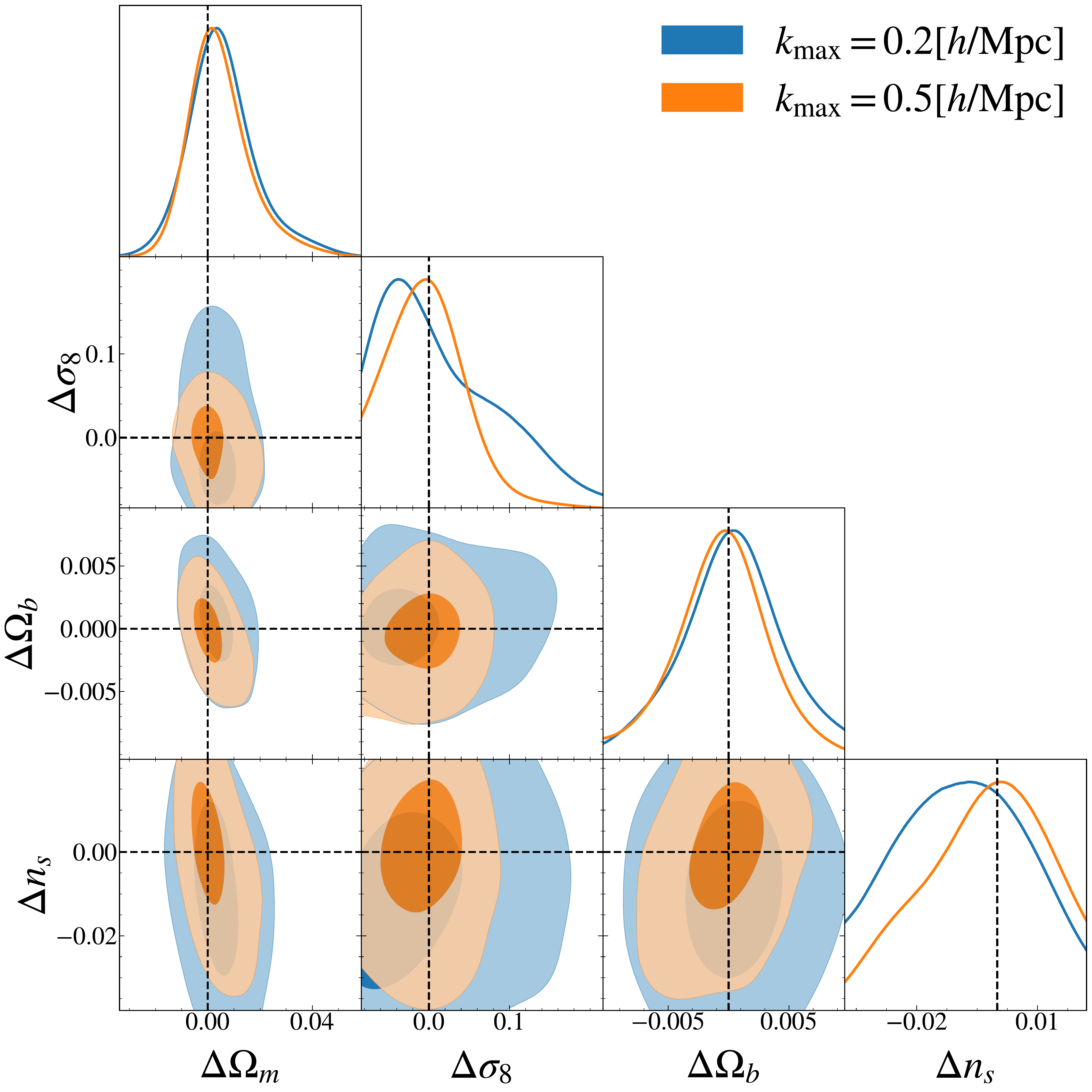}
\includegraphics[width=0.49\textwidth]{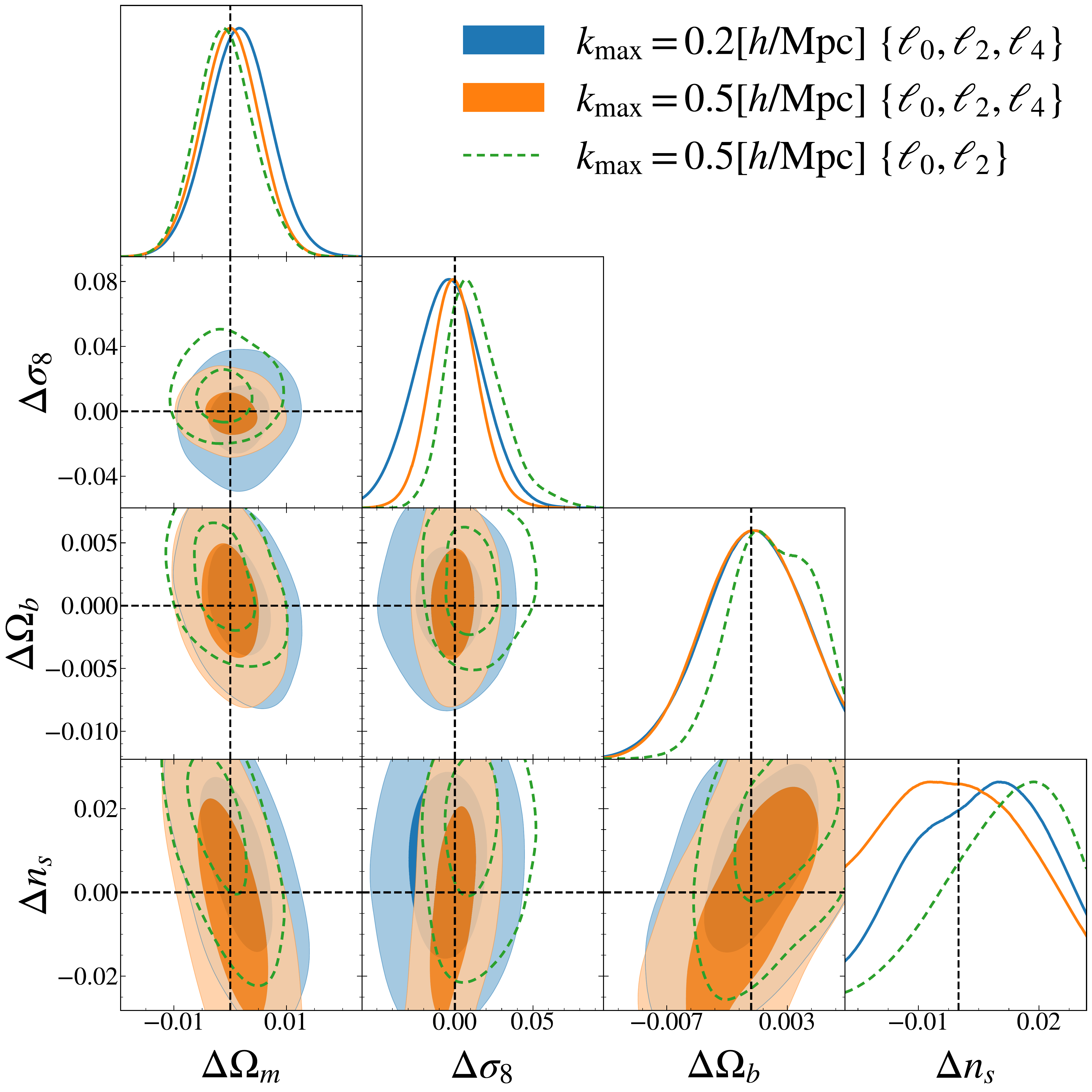}
    \caption{ Left panel: Posterior distribution inferred by the \code{BACCO} hybrid emulator model to the mean of the 10 $\Lambda$CDM mock boxes in real space as a function of $k_{\rm max}$. Right panel: Same as the left panel but for the 10 $\Lambda$CDM mock boxes in redshift space. We have displaced values by subtracting the mean at the $k_{\rm max}=0.5[h/$Mpc] result to keep the inferred cosmology masked.}
    \label{fig:BACCOconstraints}
\end{figure*}

This analysis focuses on the power spectrum $P(k)$ for the $\Lambda$CDM boxes in real space and the first non-zero multipoles ($\ell=0,2,4$) of the power spectrum $P_\ell(k)$ for the redshift space $\Lambda$CDM boxes, shown in Fig.~\ref{fig:BACCOmeasurements}. We measure these power spectra by mapping the galaxy positions into a mesh of 1024$^3$ cells and then perform a Fast Fourier transform as usual applying interlacing and ``triangular shape cloud'' as the deposit method.

\subsubsection{Model}

The modeling of the clustering statistics in the so-called ``hybrid'' approaches \citep{Modi_2020,PellejeroIbanez2022} contains two ingredients: i) a map from Lagrangian, $\pmb{q}$,  to Eulerian, $\pmb{x}$, space using the displacement field as measured on $N$-body simulations, $\pmb{\psi}(\pmb{q})$, and ii) a functional relation between the matter and galaxy density fields i.e. a bias model, $F(\delta_{\rm L}(\pmb{q}))$. 

\paragraph{$N-$body displacements}

Regarding the first ingredient, we can define completely the Eulerian overdensity field as:
\begin{equation}
\begin{split}
    1+\delta_{\rm tr}(\pmb{x})=\int \rm{d}^3q \, w(\pmb{q}) \, \delta_{\rm D}(\pmb{x}-\pmb{q}-\pmb{\psi}(\pmb{q}))\, .
    \label{eq:galmapp}
\end{split}
\end{equation}
For the \code{BACCO}-Hybrid model $\pmb{\psi}(\pmb{q})$ is measured in $N$-body simulations (as opposed to PT) by comparing the Lagrangian position of simulation particles with their position in a snapshot at any desired redshift.

\paragraph{Galaxy-halo connection}

We employ a 2nd-order Lagrangian bias model \citep{PhysRevD.78.083519,Desjacques2018}:

\begin{equation}
\begin{split}
    {\rm w}(\pmb{q}) = F(\delta_{\rm L}(\pmb{q})) = & 1 + b_1\delta_{\rm{L}}(\pmb{q}) + b_2 \left( \delta^2_{\rm{L}}(\pmb{q})-\langle\delta^2_{\rm{L}}(\pmb{q})\rangle \right) \\ & + b_s \left( s^2(\pmb{q}) - \langle s^2(\pmb{q})\rangle \right) + b_{\nabla} \nabla^2 \delta_{\rm{L}}(\pmb{q}) \; .
	\label{eq:model}
\end{split}
\end{equation}

\noindent Here $\delta_{\rm{L}}(\pmb{q})$ stands for the linear field and $s^2$ is the traceless part of the tidal field, $s^2=s_{ij}s^{ij}=(\partial_i\partial_j\phi(\pmb{q}) - 1/3\delta^{\rm{K}}_{ij}\delta_{\rm{L}}(\pmb{q}))^2$, with $\phi(\pmb{q})$ the linear gravitational potential.
The function $\mathrm{w}(\pmb{q})$ weighs the importance of different Lagrangian fields in representing the density of tracers at a given $\pmb{x}$. 

\paragraph{Redshift space}

We base our redshift-space modelling on our work in \citet{PellejeroIbanez2022}. In order to account for this effect, we substitute $\pmb{\psi}(\pmb{q})$ with $\pmb{\psi}^s(\pmb{q})$ accounting for the shifts in the line-of-sight direction due to the velocity field as follows,

\begin{equation}
\begin{split}
    \pmb{\psi}^s(\pmb{q}) = \pmb{\psi}(\pmb{q}) + \frac{\hat{\pmb{q}}_z \cdot \pmb{v}_{\rm{tr}}(\pmb{q})}{aH} \, \hat{\pmb{q}}_z \, ,
    \label{eq:galDispRSD}
\end{split}
\end{equation}
where we define the velocity $\pmb{v}_{\rm{tr}}$ based on the velocities of the $N$-body simulation. Concretely, $\pmb{v}_{\rm{tr}}(\pmb{x}) = \pmb{v}(\pmb{x})$, the velocity of the matter particle, if the tracer is outside of a halo and $\pmb{v}_{\rm{tr}}(\pmb{x}) = \pmb{v}_{\rm{halo}}(\pmb{x})$ if the tracer is inside of a halo. This distortion roughly accounts for the so-called ``Kaiser effect'' \citep{1987MNRAS.227....1K} but also incorporates additional contributions resulting from the nonlinearity of the halo velocity. We can model the effect of the intra-cluster velocities onto the redshift-space galaxy field by applying a convolution along the line-of-sight direction $\delta^s_{\rm{tr}}(\pmb{s}) \boldsymbol{\ast}_z \left[ (1-f_{\rm{sat}}) \delta_{\rm{D}}(s_z) +f_{\rm{sat}}\exp \left( {-\lambda_{\rm{FoG}} s_z} \right) \right]$,where $\boldsymbol{\ast}_z$ represents the convolution along the line of sight $z$.

\paragraph{Emulator}

The \code{BACCO} hybrid emulator employed a suite of high-resolution simulations (first introduced in \citealt{Angulo_2021}) together with cosmology rescaling \citep{AnguloWhite2010} to densely sample a target cosmological parameter space. Then, we compute the Eulerian fields weighted by their corresponding Lagrangian bias fields. We estimate their power and cross-power spectra and, finally, we use this data to train a neural network. As a result, this provides accurate and extremely fast predictions of the 2-point statistics, which makes it possible to use our model in cosmological data analyses, as presented in \citet{PellejeroIbanez2023}. At this point, we further include two noise terms to account for shot-noise to the power spectrum in the form of ${\rm Noise}=1/\bar{n}(\epsilon_1+\epsilon_2k^2)$. Note that in \citet{PellejeroIbanez2023} we tested the need to include $\mu$-dependent stochastic terms, finding them not required for the range of scales explored in this analysis. This does not imply the term is absent, rather, it indicates that the velocity from halos in the simulation, combined with the FoG free parameters, accounts for this dependency. We further tested if the inclusion of a $k^2\mu^2$ dependency in the noise changed our results, finding a negligible impact.

Another important aspect is how we address the binning discreteness effect caused by a finite fundamental $k$. To mitigate this effect, we apply a binning correction for each $k$ mode and each multipole. This is done by evaluating the theoretical power spectrum on the same mesh and using the resulting power spectrum divided by the non-binned one.

\subsubsection{Inference}
\label{subsec:bacco_inference}

\paragraph{Priors}

Our cosmological parameter priors are defined by the limits of the emulator. They are given by

\beq
    &&\Omega_{\rm m}\sim\unif[0.23, 0.4], \qquad \sigma_{8,\rm c}\sim\unif[0.65, 0.9], \qquad \Omega_{\rm b} \sim \unif[0.04, 0.06]\, ,
    \qquad n_s\sim\unif[0.92, 1.01] . 
\label{eq:BACCO_priors_cosmo}
\eeq
All beyond $\Lambda$CDM parameters are set to their default values. Specifically, neutrino mass is set to zero, and evolving Dark Energy parametrizations are not taken into account.
For bias parameters, we use uninformative priors, given by:
\beq
&& b_1\sim \unif[0,2]\,, \qquad b_2\sim\unif[-2,2]\,, 
\qquad b_{s}\sim \unif[-3,3] \,,
\qquad b_{\nabla}\sim \unif[-6,6]\,, \\\nonumber
&& \lambda_{\rm FoG} \sim \unif[0,1]\,, \qquad f_{\rm sat} \sim \unif[0,1]\,, \qquad \epsilon_{1} \sim \unif[0,2]\,, \qquad \epsilon_{2} \sim \unif[-4,4]\,.
\label{eq:BACCO_priors_bias}
\eeq

\paragraph{Covariance}

We make use of a Gaussian covariance. In real space, this is diagonal and proportional to the amplitude of $P(k)$ and shot noise. In redshift space, it is block diagonal in the multipoles with correlations given by the Wigner symbols. We are aware that this covariance is not accurate when pushing to scales as small as $k\approx 0.5 h/$Mpc since nonlinearities are not taken into account. However, for this challenge, the number density of mocks is such that the shot noise dominates the signal at scales of $k\approx 0.3 h/$Mpc. This shot noise contribution will dominate at scales where nonlinearities become relevant, making their contribution less important (see e.g. \citealt{Wadekar_2020,Blot_2019}). Nevertheless, an extended study on the impact of nonlinear and non-Gaussian covariance terms will be of interest for dense or highly-biases galaxy samples going forward.

\paragraph{Likelihood}

Due to the central limit theorem, it is safe to assume a Gaussian likelihood shape for the power spectrum in this case.

\paragraph{Sampling \& Validation}

We make use of the public code \code{MULTINEST}\footnote{Available at \href{https://github.com/farhanferoz/MultiNest}{Github.com/farhanferoz/MultiNest}} Bayesian inference tool for recovering credibility intervals \citep[see][for more details]{multinest1,multinest2,multinest3}. The best-fit values are also extracted from the maximum likelihood values of these chains. We set the number of live points to 1200 and the evidence tolerance to 0.08.

\subsubsection{Analysis Choices}

\paragraph{Scale cuts} 
We analyze real- and redshift-space power spectra down to nonlinear scales of $k_{\rm{max}}= 0.5h/$Mpc, motivated by our findings in previous works \citep{ZennaroAnguloPellejero2021, Zennaro_2022,PellejeroIbanez2022,PellejeroIbanez2023}. Specifically, in \citet{Zennaro_2022} we created thousands of SHAMe mocks with different physical parameters to validate the model and to compute physical priors on the bias parameters, including a mock that closely resembles the hydrodynamical simulation Illustris TNG galaxy clustering \citep{Contreras2023}.
These scale cuts pass all the parameters drift tests  described in the \emph{unmasking criteria} subsection and shown in Fig.~\ref{fig:BACCOconstraints}.

\paragraph{Unmasking criteria}

We follow the following main criteria before unmasking to assert the robustness of our measurements: 

\begin{itemize}
\item We verify that the reduced $\chi^2$ value of the best-fit value is $\approx$0.1 for the $\Lambda$CDM mocks sample mean. The value $\chi^2_{\rm red}=0.1$ is determined for the real space mocks, while the value $\chi^2_{\rm red}=0.12$ was determined for the redshift-space mocks.

\item We determine if any of the parameter posteriors intersect the priors' limits. This is remedied by extending the prior range for the nuisance parameters. In the case of cosmological parameters, however, our emulator provides the priors. For example, we discovered that the $\sigma_8$-$b_1$ degeneracy broadens the constraints and partially affects the real space emulator priors in the $2-3\sigma$ region. However, the $1-2\sigma$ region is well within our emulator's priors, so we believe these results to be reliable. 

\item We investigate the consistency of the inferred parameters with respect to various $k_{\rm max}$ values. In particular, we compute the evolution of the posterior distribution at $k_{\rm max}=\{0.1,0.2,0.3,0.4,0.5\}h/$Mpc and find no discernible change, as illustrated in Fig.~\ref{fig:BACCOconstraints}.

\item We test for redshift-space analysis whether adding or removing multipoles gives consistent cosmological parameter constraints. We find no tensions between results containing only monopole, monopole plus quadrupole, or monopole plus quadrupole and hexadecapole.

\item We include our theory error budget based on the emulator uncertainties. This is discussed in Appendix A of \citet{PellejeroIbanez2023}. For the test, we add these error estimates in quadrature to the covariance matrix diagonal elements. We find no strong dependence of the recovered cosmological marginalized values on this theory error budget. We note that the theory error should always be included, however, if this error is small compared to the data error, its effect becomes negligible.

\end{itemize}

\paragraph{Caveats}
We caution that several assumptions could potentially bias our constraints: First, the choice of a Gaussian covariance with no off-diagonal elements. Second, the possible underestimation of the emulator errors at the cosmology of interest. Third, the priors of the emulator. Even though we expect the true cosmologies to lie within these priors, strong degeneracies, such as the $\sigma_8$-$b_1$ direction in real space, might make our priors too informative. Fourth, possible model simplifications. To construct the \code{BACCO} emulator hybrid model, we make use of several approximations in the galaxy-to-matter and in the galaxy-velocity assignments. Specifically, the current galaxy velocity model lacks implementation of the mass-dependent Finger-of-God effect and strong velocity bias. These model misspecifications might bias our cosmological parameter estimations.    

\paragraph{Post-unmasking studies}

After unmasking, we found that our results for $\sigma_8$ exhibited a $1\sigma$ shift compared to the true values. This is not entirely unexpected, as cosmic variance can introduce shifts in the contours, and locating them within a $1-3\sigma$ range does not imply a failure of the model. Additionally, our assumption of a diagonal Gaussian covariance matrix exacerbates this issue by ignoring non-diagonal correlation terms that arise on small scales. 

However, to mitigate the impact of cosmic variance, parameter estimation was performed using the mean of 10 independent mocks. Naively, this should reduce the contour size by a factor of approximately $\sqrt{10}\sim 3$. Taking this into account, our contours appear to be at approximately $\sim 3\sigma$ from the true values. To confirm this, it would be necessary to reanalyze the data with reduced error bars corresponding to a volume 10 times larger than each individual mock. Unfortunately, all tests conducted on the \code{BACCO} hybrid model in redshift space thus far have been based on the assumption of a BOSS-like sample \citep{Pellejero2022,PellejeroIbanez2023} with an effective volume of $V_{\rm eff}=2.8 {\rm Gpc}^3\sim 0.9 {\rm Gpc}^3/h^3$ and our emulator's noise level remains too high for the galaxy sample and survey volume of this challenge. 

Nevertheless, we can investigate whether incorporating or removing theory errors impacts our predictions. As previously mentioned, we already conducted this test by including errors in quadrature. Based on \citet{PellejeroIbanez2023}, we are aware that typical uncertainties in the emulator account for $\sim 0.5\%$ in the monopole, $\sim 1\%$ in the quadrupole, and $\sim 10\%$ in the hexadecapole amplitudes. During this examination, we did not realize the low values of the hexadecapole, as depicted in Figure \ref{fig:BACCOconstraints}. Consequently, the theory error we included in the noisiest of our estimations was underestimated.

Therefore, we decided to present the measurement of cosmological parameters without the hexadecapole, as shown in dashed lines in Figure \ref{fig:BACCOconstraints}. The shift found is of around half a sigma and the slight decrease in constraining power deems our results unbiased with respect to the true values. It is important to reiterate that this is a post-unmasking finding, and further tests on the model are required to rule out potential failures when studying the HOD models employed in this work.      

%% file: methods_pt_pbj.tex
\subsection{{EFT P \& B: Analysis Method}\footnote{Authors: Mikhail M. Ivanov, Oliver Philcox, Kazuyuki Akitsu, Giovanni Cabass.}}
\label{subsec:PTPB}
In this section, we describe the analysis of the Beyond 2-point mocks with the Effective Field Theory of Large Scale Structure, focusing on the large-scale power spectrum and bispectrum. The
key theoretical underpinnings of this are described, e.g. in \citet{Baumann:2010tm,Carrasco:2012cv,Ivanov:2022mrd}\footnote{See footnote 6 of \citet{Nunes:2022bhn} for an extended set of references.}, with recent applications to galaxy surveys and other large-scale structure data shown in e.g.,~\citet{Ivanov:2019pdj,Ivanov:2019hqk,Philcox:2021kcw,DAmico:2019fhj,Chen:2021wdi}. 
We apply this methodology to all challenge catalogs. 

\paragraph{Information Sources}
We briefly discuss the various sources of information in the full shape of the galaxy power spectrum, which set the parameter degeneracy directions. More detail on the sources of information can be found in \citet{Ivanov:2019pdj}. From the shape of the galaxy power spectrum, one can determine $\omega_b$, $\omega_m \equiv \Omega_m h^2$, and $n_s$ regardless of the distance to the sample. These parameters then predict absolute scales such as that of matter-radiation-equality and the sound horizon ($k_{\rm eq}$ and $k_{\rm BAO}$ respectively).

In contrast, distance information is encoded in the angular scales $\theta_\parallel$ and $\theta_\perp$, for both BAO and the matter-radiation-equality. This constrains the Hubble parameter $h$, and, if multiple redshifts are available, $\Omega_m$  through the growth rate evolution. For periodic boxes, there is no distance information, since all quantities are measured in $\Mpch$ units, thus all distances are equivalent to $H^{-1}_0$.

From growth, we constrain $b_1\sigma_8(z_{\rm data})$ from the real-space power spectrum or $b_1^3\sigma^4_8(z_{\rm data})$ from the bispectrum. Loop corrections additionally yield $b^2_1\sigma^4_8(z_{\rm data})$ and infrared resummation directly measures $\sigma_8(z_{\rm data})$ \citep[see][]{Senatore:2014via,Baldauf:2015xfa,Blas:2015qsi,Blas:2016sfa,Ivanov:2018gjr,Vasudevan:2019ewf}, though these effects are comparatively small at the high redshifts of this challenge. In redshift space we instead measure $b_1\sigma_8(z_{\rm data})$ 
from the monopole and $f\sigma_8(z_{\rm data})$ from the quadrupole and hexadecapole power spectrum moments. Loops and the bispectrum monopole  help break degeneracies and give directly $f(z_{\rm data})$ and $\sigma_8(z_{\rm data})$.

\subsubsection{Data \& Estimators}
In all cases, we analyze the large-scale power spectrum and bispectrum extracted from the mock catalogs.
We use $P(k)$ and $B(k_1,k_2,k_3)$ in real space; in redshift space we consider the power spectrum monopole, quadrupole, and hexadecapole ($P_0$, $P_2$, and $P_4$), but restrict to the bispectrum monopole \citep[since higher moments do not add significant signal, see][] {Ivanov:2023qzb}.
For the redshift-space mocks, we supplement the data vector with the real-space power spectrum proxy, $Q_0$, which gives additional information on smaller scales without bias from fingers-of-God (\citealp{Ivanov:2021fbu}; see also \citealp{Scoccimarro:2004tg}, \citealp{DAmico:2021ymi}).

For the periodic box mocks, correlators are estimated using Fast Fourier Transforms, as usual \citep{Scoccimarro:2015bla}. For the light-cone mock catalog, we use window-free estimators \citep{Philcox:2020vbm,Philcox:2021ukg}, as implemented in the Spectra-Without-Windows code\footnote{Available at \href{https://github.com/oliverphilcox/Spectra-Without-Windows}{GitHub.com/oliverphilcox/Spectra-Without-Windows}}, making use of the mask and random file. In this case, we split the sample into two redshift bins of equal density, with effective redshifts $z_1 = 0.92$ and $z_2 = 1.17$. Knowledge of redshift evolution is necessary to break 
the geometric degeneracy.

\subsubsection{Model}
\paragraph{Perturbation theory}
We model the power spectrum and bispectrum with the Effective Field Theory of Large Scale Structure, as implemented in \code{CLASS-PT} \citep{Chudaykin:2020aoj}\footnote{Available at 
 \href{https://github.com/michalychforever/CLASS-PT}{GitHub.com/michalychforever/CLASS-PT}.} The power spectrum multipoles are modeled using the (infrared-resummed) one-loop theory \citep{Ivanov:2019pdj}, and we use tree-level theory for the bispectrum \citep{Ivanov:2021kcd}, noting that higher loops give limited gains \citep{Philcox:2022frc,DAmico:2022ukl}. 

\paragraph{Galaxy-matter connection}
We assume the following bias expansion up to third-order (renormalized) operators \citep{Ivanov:2019pdj}:
\beq
    \delta_g = b_1\delta + \frac{1}{2}b_2\delta^2 + b_{\mathcal{G}_2}\mathcal{G}_2+b_{\Gamma_3}\Gamma_3,
\eeq
where $\delta_g$ and $\delta$ are the galaxy and matter overdensities, $\mathcal{G}_2$ and $\Gamma_3$ are Galileon tidal operators. The model additionally includes stochastic contributions from shot-noise $P_{\rm shot},A_{\rm shot},B_{\rm shot},a_0,a_2$ 
(including $k^2$ scale dependence), and counterterms $\{c_0,c_2,c_4,\tilde c\}$, encapsulating small-scale physics such as halo formation and velocity effects. 
This makes no assumptions on the form of the galaxy-halo connection, except that it is statistically isotropic and homogeneous on large scales, and obeys 
Einstein's equivalence principle \citep{Desjacques:2016bnm}.

\subsubsection{Inference}
\paragraph{Priors}
We assume the following priors on cosmological parameters for all analyses:
\beq
    &&\omega_{\rm cdm}\sim\unif[-\infty,\infty], \qquad 10^9A_s\sim\unif[0.5,5], \qquad n_s\sim \unif[0.87,1.07],\qquad \omega_{\rm b}\sim\unif[0.01,0.035],
\label{eq:ptpb_priors_cosmo}
\eeq
with the angular size of the 
sound horizon $\theta_*$
fixed to the value known 
to all participants.
We fix the neutrino mass to zero and assume a flat Universe in all cases.
For bias parameters, we use weakly informative priors, given by~\citet{Philcox:2021kcw}:
\be
\begin{split}
& b_1\in \text{flat}[0,4]\,, \quad b_2\sim \mathcal{N}(0,1^2)\,, 
\quad b_{\mathcal{G}_2}\sim \mathcal{N}(0,1^2) \,,
\quad
b_{\Gamma_3}\sim \mathcal{N}\left(\frac{23}{42}(b_1-1),1^2\right),
\end{split}
\ee
where $\mathcal{N}(\mu,\sigma^2)$ indicates a Gaussian distribution with mean $\mu$ and variance $\sigma^2$. 
Note that these priors 
are consistent 
with the recent 
measurements \citep{Ivanov:2024hgq}, as well as earlier 
results \citep{Lazeyras:2017hxw,Abidi:2018eyd}.
Similarly, we use Gaussian priors for the counterterms $c_0,c_2,c_4,c_1,\tilde{c}$ and stochasticity parameters 
$a_0$,$a_2$,$P_{\rm shot}$, $A_{\rm shot}$, $B_{\rm shot}$:
\be
\begin{split}
& \frac{c_0}{[\text{Mpc}/h]^2} \sim \mathcal{N}(0,30^2)\,,\quad 
\frac{c_2}{[\text{Mpc}/h]^2} \sim \mathcal{N}(30,30^2)\,,\quad \frac{c_4}{[\text{Mpc}/h]^2} \sim \mathcal{N}(0,30^2)\,,\quad 
\frac{\tilde{c}}{[\text{Mpc}/h]^4} \sim \mathcal{N}(500,500^2)\,,\\
& \frac{c_1}{[\text{Mpc}/h]^2} \sim \mathcal{N}(0,51^2)\,,\quad P_{\rm shot} \sim \mathcal{N}(0,1^2)\,,\quad a_{0}
\sim \mathcal{N}(0,1^2)\,,\quad a_2\sim \mathcal{N}(0,1^2)\,,\\
& B_{\rm shot}\sim \mathcal{N}(1,1^2),\quad A_{\rm shot}\sim \mathcal{N}(0,1^2),
\end{split}
\label{eq:ptpb_priors_bias}
\ee
where we use the same convention\footnote{Note that \citet{Philcox:2021kcw,Philcox:2022frc} used 2 standard deviations for stochastic 
parameters of the high-z samples because their physical number density $\bar n$ 
was twice lower than the fiducial one, see the public likelihoods for more information.} as  \citet{Ivanov:2021kcd} \citep[see also][]{Philcox:2021kcw,Philcox:2022frc}, but with 
$\bar n\approx 5\times 10^{-4}\,h^3\,\mathrm{Mpc}^{-3}$ of the challenge boxes. 

\paragraph{Covariance}
We assume a Gaussian covariance matrix for the power spectrum and multipoles bispectrum monopole, defined explicitly in \citet{Chudaykin:2019ock,Ivanov:2021kcd,Ivanov:2023qzb}. This is exact in the linear regime (where modes are uncorrelated), and found to be highly accurate on our scales of interest, due to the high shot-noise, and limited cosmological information available at high-$k$ \citep{Wadekar:2020hax}. The covariance is diagonal in $k$, but includes correlations between different power spectrum multipoles, where necessary. We do not include cross-covariance between the power spectrum and bispectrum, since this is formally of higher order in EFT, and found to be unnecessary in \citet{Ivanov:2021kcd} for a similar choice of scale cuts.

The power spectrum covariance is computed using the specific realizations of the power spectrum multipoles (since these are measured at high significance). For the bispectrum covariance, we adopt an iterative procedure, first computing the covariance using a fiducial cosmology, then updating with the best-fit parameters from a likelihood analysis. This procedure is found to converge quickly.

\paragraph{Likelihood}
On large-scales, the data is well described by a Gaussian likelihood. This holds by the central limit theorem, and also follows from perturbation theory. 
The covariance can be computed precisely on large scales, and the number of bins is reasonable.

\paragraph{Sampling \& validation}
We sample the various parameters with \code{MontePython} \citep{Brinckmann:2018cvx}, analytically marginalizing over nuisance parameters that enter the model linearly \citep{Philcox:2020zyp}. This model has been validated on several suites of simulations, including MultiDark-Patchy, Las Damas, the Perturbation Theory Challenge, Nseries, and Outer Rim \citep{Ivanov:2023qzb,Ivanov:2019pdj,Philcox:2021kcw,Nishimichi:2020tvu,Philcox:2022frc,Ivanov:2021zmi,Ivanov:2021fbu,Chudaykin:2022nru,Ivanov:2023qzb}. We do not include a theoretical error covariance (due to our choice of scale cuts), but note that this can be, in principle, incorporated \citep{Baldauf:2016sjb,Chudaykin:2019ock,Chudaykin:2020hbf}. We consider our MCMC chains converged when they satisfy the Gelman-Rubin criterion $R-1<0.03$. 

\subsubsection{Analysis Choices}
\paragraph{Scale cuts}
The EFT P+B analyses adopt the following scale cuts (in $\hMpc$ units):
\begin{itemize}
    \item \textbf{real-space mock} (periodic box, $z=1$, $V=8$~(Gpc/$h$)$^3~\times 10$): $k^P\leq 0.3$, $k^B\leq 0.15$
      \item \textbf{redshift-space mock} (periodic box, $z=1$, $V=8$~(Gpc/$h$)$^3~\times 10$): $k^{P_\ell}\leq 0.20$, $0.20 < k^{Q_0}\leq 0.4$, $k^{B_0}\leq 0.08$.
     \item \textbf{light-cone mock} ($z=[0.8,1.3]$, $V=5.3$~(Gpc/$h$)$^3$ (for our fiducial cosmology $\Omega_m=0.31$, $h=0.676$)): $k^{P_\ell}\leq 0.25$, $0.25 < k^{Q_0}\leq 0.4$, $k^{B_0}\leq 0.08$
\end{itemize}

\begin{figure*}
    \centering
\includegraphics[width=0.99\textwidth]{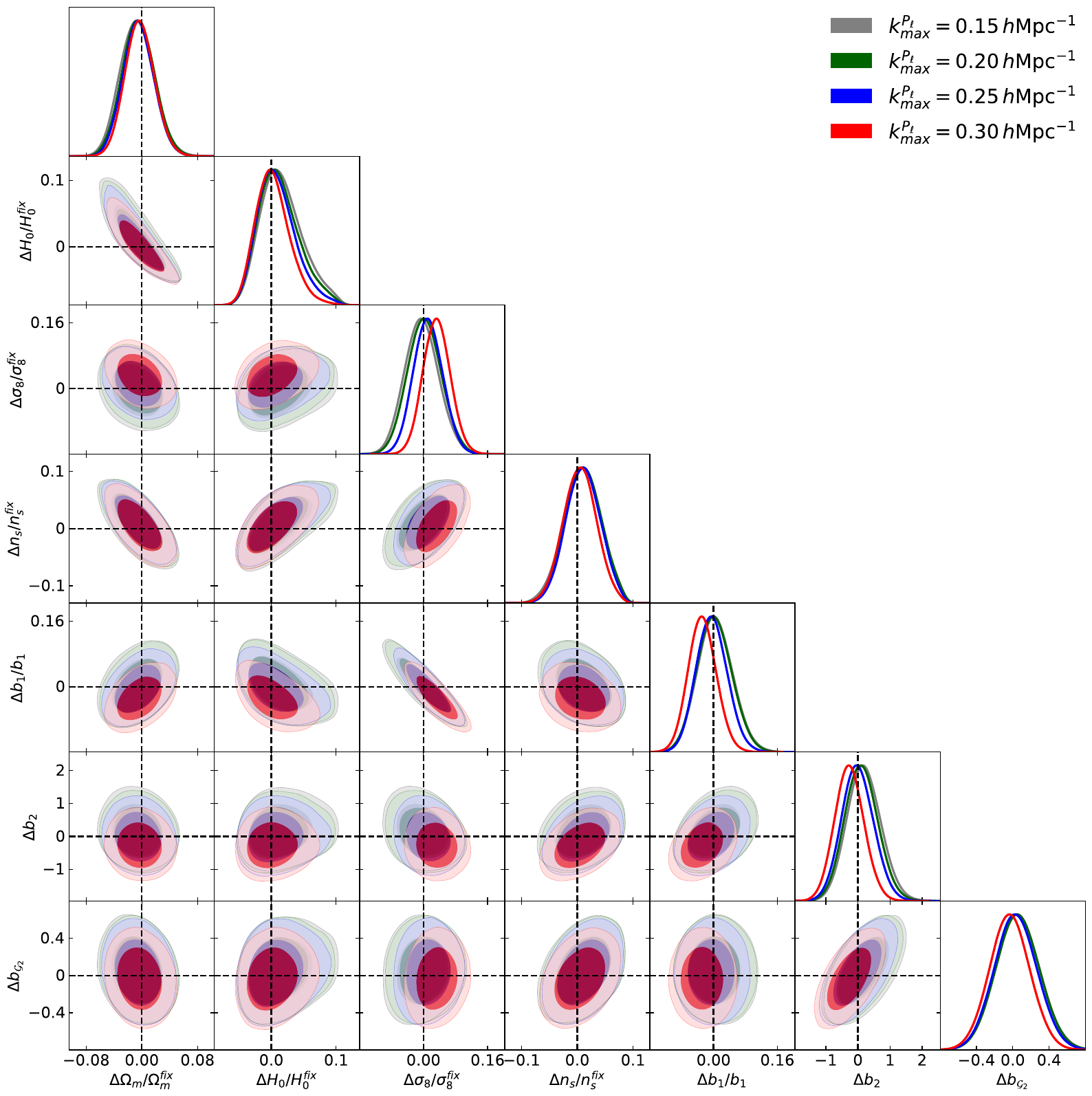}
    \caption{
    $\Lambda$CDM 
    redshift-space mock analysis: parameter constraints from EFT P+B for four choices of the power spectrum scale cut 
    $k_{\text{max}}^{P_\ell}$.
    All parameters are normalzied to 
    fixed values taken from the 
    best-fit at 
    $k_{\text{max}}^{P_\ell}=0.2\,h\text{Mpc}^{-1}$ 
    (our fiducial scale cut). Physical baryon density 
    $\omega_b$ 
    and some nuisance parameters are not displayed.
    \label{fig:mockJ}}
\end{figure*}

These are motivated by previous works and tests on large-volume simulations \citep{Ivanov:2019pdj,Philcox:2021kcw,Nishimichi:2020tvu,Ivanov:2021kcd,Ivanov:2021fbu} and validated for this challenge through parameter drift plots for all challenge mocks.
As an example, we show parameter posteriors for the redshift-space mock for four different
choices of $k_{\rm max}^{P_\ell}=0.15,0.2,0.25,0.3$ $h$Mpc$^{-1}$ in Figure~\ref{fig:mockJ}. 
The parameter drifts in this plot suggests that 
results at 
$k_{\rm max}^{P_\ell}=0.3$ $h$Mpc$^{-1}$ are biased, because one can clearly see an upward shift
of $\sigma_8$ by $1\sigma$, consistent
with the effects of two-loop
corrections previously observed \citep[e.g.]{Nishimichi:2020tvu,Chudaykin:2020hbf}.
The posteriors for the smaller 
scale cuts appear consistent with each other, and we select 
$k_{\rm max}^{P_\ell}=0.2$~$h$Mpc$^{-1}$ 
as a baseline choice in order to be conservative.
The same methodology is applied for the 
light-cone and 
real-space analyses, i.e. 
studying parameter drifts plus using 
estimates for the theoretical error based 
on the two-loop estimates (see 
also~\citep{Chudaykin:2020hbf,Ivanov:2021kcd} for studies of 
real-space clustering in EFT).
Using this methodology, all challenge mock catalogs were 
systematically analysed and tested 
(see consistency checks below). 
The particular analysis choices 
applied here were also
cross-validated on other mock catalogs such 
as Nseries, PT Challenge, and Las Damas. 
We choose a more aggressive scale cut for the light cones as their volume is small
compared to the periodic box simulations,
and hence the theory systematic error 
is smaller than cosmic variance 
down to smaller scales. In this case, 
the parameter drift plot is also less
conclusive because it is 
hard to disentangle parameters
shifts due to bias from those  
resulting from a statistical fluctuation
due to the addition of new data at a 
higher
$k_{\rm max}$.

We also note that the bispectrum is restricted to larger scales due to the tree-level modeling. The lower $k_{\rm max}$ for redshift-space snapshots (compared to real space ones) is also supported by 
expectations of the 
impact of velocity effects which become non-perturbative at larger scales (noting that the nonlinear scale is $k_{\rm NL}\sim \sigma_{\rm FoG}^{-1}$ in redshift space, as opposed to $R_{\rm halo}^{-1}$ in real space, if halo formation effects dominate), see~\citet{Ivanov:2021zmi,Ivanov:2021fbu} for 
detailed discussions. 

\paragraph{Consistency checks} We have run a number of consistency checks that include: variation of the bispectrum covariance matrix, stability w.r.t. scale cut changes, exclusion of $Q_0$, $B_0$, $P_4$ from the datavector, and the change of fiducial cosmology
in the case of the light-cone mock. 
We have not found any significant effect when perturbing 
our baseline choices within a reasonable range. 

As far as the choice of fiducial cosmology is concerned, we have explicitly tested that using $\Om=0.31$ (our baseline choice),
or  $\Om$ from the best-fit to the data
has negligible difference on the extracted 
cosmological parameters, even when 
$\Om$ is actually significantly different (by up to $\sim 30\%$) from its fiducial value 0.31. 
In this particular example of a 30\% difference (which constitutes $\simeq 10\sigma_{\Omega_m}$), the shift in the mean of $\Omega_m$
was found to be $0.1\sigma_{\Omega_m}$, while the shift in 
$\sigma_8$ was found to be $0.3\sigma_{\sigma_8}$.
This implies 
that the standard Alcock-Paczynski parametrization 
is quite accurate even for large deviations from 
the fiducial cosmological parameter values. 

\paragraph{Unmasking criteria}
We allow our results to be unmasked after visual confirmation of the relevant posteriors and their degeneracy directions, as well as simple tests based on the goodness of fit and analysis of variation with $k_{\rm max}$.  

For the real-space and redshift-space mocks we additionally analyzed the mean 
data vectors (from the 10 realizations) with the covariances 
that correspond to the full volume of 10 boxes. We found some non-negligible bias 
for $k_{\rm max}^{P_\ell}=0.25$~$h$Mpc$^{-1}$
for the redshift-space mock in this analysis,
which was another reason 
to stick to our 
baseline choice $k_{\rm max}^{P_\ell}=0.2$~$h$Mpc$^{-1}$. 
For the light-cone mock, 
however, we only have 
one realization for the volume which is approximately $40\%$ smaller than 
that of one snapshot. The two-loop systematic 
error is smaller than the 
cosmic variance in this case even at $k_{\rm max}^{P_\ell}=0.25$~$h$Mpc$^{-1}$, 
and therefore we could adopt   
this more optimistic 
scale cut in our analysis.

 \paragraph{Caveats}
Finally, we caution that failure modes of our analysis could include: overly optimistic scale-cuts (for example due to extremely large Fingers-of-God effects), cosmological or nuisance parameters lying outside the parameter ranges, or galaxy-halo connections violating our symmetry assumptions. The posteriors should be interpreted 
with caution, as for many cases (e.g. real-space analyses) their shape
is highly non-Gaussian. 
This may lead to parameter projection effects: i.e. 1d marginalized contours may be off due to 
marginalization over non-Gaussian posteriors, 
as it was discussed in detail e.g. in~\citet{Ivanov:2019pdj,Chudaykin:2020ghx,Philcox:2021kcw,Ivanov:2023qzb}.
Note that our choice of $k_{\rm max} = 0.3$~$h$Mpc$^{-1}$for the real-space analyses is 
somewhat aggressive (done in part in order to facilitate the sampling), so we expect (based on our tests on mock data) that the theory 
systematic error may be as large as $\sim 1\sigma$.

%% file: methods_EFT_fieldlevel.tex
\subsection{EFT FBI: Analysis Method \footnote{Authors: Nhat-Minh Nguyen, Fabian Schmidt.}}
\label{subsec:EFT_fieldlevel_analysis_method}

In this section, we review the Effective Field Theory Field-level Bayesian Inference (EFT FBI).
EFT FBI directly models and optimally extracts cosmological information from the three-dimensional galaxy density field within the EFT framework.
This requires an explicit sampling of all Fourier modes of the initial conditions---in our case, corresponding to $90^3\simeq730,000$ parameters---in addition to bias, stochastic, and cosmological parameters.
We refer readers to \citet{Schmidt:2018bkr} for the first EFT FBI groundwork and \citet{Nguyen:2024yth} for the most recent advance. We further note parallel efforts on cosmological field-level inference from galaxy clustering\footnote{We define the term ``field level'' in relation to the (uncompressed) three-dimensional density field of biased tracers or matter, thereby restricting the scope of the discussion and literature.}, within \citep{Baldauf:2015tla,Schmittfull:2018yuk,Cabass:2020jqo,Schmittfull:2020trd,Cabass:2023nyo} and outside of the EFT framework \citep{Seljak:2017rmr,Ramanah:2018eed,Andrews:2022nvv,Wang:2022itv,Bayer:2023rmj}, as well as the most recent advances within the community of machine learning for cosmology \citep[e.g][]{Cuesta-Lazaro:2023zuk,Doeser:2023yzv,Ho:2024whi,Saadeh:2024vuj}.

Our EFT FBI analysis uses \code{LEFTfield}, a Lagrangian, EFT-based forward modeling of cosmological density fields.
The code was first introduced in \citet{Schmidt:2020tao} with extensions and validations later shown in\citet{Kostic:2022vok,Stadler:2023hea,Tucci:2023bag,Nguyen:2024yth,Babic:2024wph}.

In this challenge, we analyze the real-space mocks. For the clustering of galaxies in their comoving rest frame, i.e. real space, both the matter $\delta$ and galaxy density field $\delta_g$ contain advection contributions that involve the Lagrangian displacement field. Since matter and galaxies co-move on large scales following Einstein's equivalence principle, the displacement is the same for galaxies and matter. This direct consequence of the equivalence principle therefore ensures that the advection contribution to $\delta_g$ can be uniquely predicted \citep{Desjacques:2016bnm}, hence the bias--$\sigma_8$ degeneracy breaking.

\subsubsection{Data and Estimators}
\label{sec:EFT_fieldlevel_data_vec}
Our data vector is the entire three-dimensional galaxy density field, namely $\data\equiv\delta_{\mathrm{g}}$, at $z=1.0$.
Similarly to \citet{Nguyen:2024yth}, we construct a filtered galaxy density field $\delta_{g,\Lambda}$ from the galaxies in each catalog by a nonuniform-to-uniform Fourier transform (NUFFT).
Here, $\Lambda$ is the cutoff scale of the sharp-$k$ filter, such that all Fourier modes above $\Lambda$ are set to zero.

\subsubsection{Model}
\label{sec:EFT_fieldlevel_modeling}

The \code{LEFTfield} forward model we employ in this challenge involves a \emph{Gravity model} evolving the initial conditions $\shat$ to the Eulerian matter density field and a \emph{Galaxy model} connecting the Eulerian matter density field to the model prediction for the data vector $\delta_{g,\Lambda}$.

We note that our pre-unmasking analyses in the challenge had concluded before the analysis in \citet{Nguyen:2024yth} started---where the latter analyzed different datasets than this challenge, consisting of mass-selected main halos.
Hence, while both works use the same \emph{Gravity model}, they differ in the \emph{Galaxy model}. Specifically, our baseline analyses here use a more restricted model compared to \citet{Nguyen:2024yth}.
We will highlight the differences below and revisit some of the main differences in our post-unmasking study.

\paragraph{Gravity model}
We predict gravitational evolution using the $n$-th order Lagrangian Perturbation Theory (LPT) framework \citep{Matsubara:2015ipa}, as described in Sec.~2 and App.~A-B of \citet{Schmidt:2020ovm}. The convergence of this $n-$LPT scheme was demonstrated by \citet{Schmidt:2020ovm} in their Sec.~6, up to $n=6$. In this challenge, we adopt $n=2$, i.e. the 2LPT model \citep{Buchert:1992ya,Bouchet:1994xp}.

\paragraph{Galaxy model}
We describe the matter-galaxy connection with the EFT bias expansion:
\beq
\delta_{g,\Lambda} = \sum_\op\,b_\op \op[\delta_\Lambda]
\label{eq:EFT_fieldlevel_bias_expansion}
\eeq
where $\op$ are operators constructed out of the filtered Eulerian matter density field $\delta_\Lambda$, together with their associated galaxy bias coefficients $b_\op$.
Similar to \citet{Nguyen:2024yth}, we expand \refeq{EFT_fieldlevel_bias_expansion} in the Eulerian basis. However, unlike \citet{Nguyen:2024yth}, here we do so only up to second-order operators:
\beq
\op\in\left[\delta,\delta^2-\<\delta^2\>,K^2-\<K^2\>,\nabla^2\delta\right],
\label{eq:op_secondorder}
\eeq
where
\beq
K^2\equiv(K_{ij}^2)=\left(\left[\frac{\partial_i\partial_j}{\nabla^2}-\frac{1}{3}\delta_{ij}\right]\delta\right)^2
\label{eq:K2_def}
\eeq
is the gravitational tidal field squared, and $\nabla^2\delta$ is the leading higher-derivative operator. Note that these coefficients are equivalent to the subset of coefficients $b_1, b_2, b_{\mathcal{G}_2}, c_0$ in the EFT P+B model (Sec.~\ref{subsec:PTPB}).
Compared to \refeq{op_secondorder}, \citet{Nguyen:2024yth} add the full set of four third-order bias operators, while the EFT P+B team adds a single third-order bias term ($\Gamma_3$) that is relevant for the one-loop power spectrum.
In our EFT-convergence test (see \reffig{EFT-FBI_kmaxconvergence_alpha_Stage1+2}) and post-unmasking study (see \reffig{1D_summary_real_post}), we additionally consider the full set of third-order bias terms and quantify the impact of on our $\sigma_8$ constraints [\reffig{1D_summary_real}].

\subsubsection{Inference}
\label{sec:EFT_fieldlevel_inference}

We sample and estimate the marginalized posterior of the amplitude rescaling parameter $\alpha=\sigma_8/\sigma_8^{\mathrm{fiducial}}$ given by
\begin{align}
\post\left(\alpha \Big| \data_\Lambda\right)
&= \int\mathcal{D}\shat\,\int d\sigma_\eps\,\int d\{b_\op\}\,
\post\left(\alpha, \{b_\op\}, \sigma_\eps, \shat \Big| \data_\Lambda\right) \nonumber\\
&\propto\int\mathcal{D}\shat\,\int d\sigma_\eps\,\int d\{b_\op\}\,
\lik\left(\data_\Lambda \Big| \alpha,\{b_\op\}, \sigma_\eps, \shat\right)\mathcal{P}(\alpha)\mathcal{P}(\{b_\op\})\mathcal{P}(\sigma_\eps)\mathcal{P}(\shat).
\label{eq:EFT_fieldlevel_joint_posterior}
\end{align}
where the parameters to be explicitly sampled and numerically marginalized over are $\nuisPar\equiv\{b_\op,\sigma_\eps,\shat\}$.
Here, $\sigma_\eps$ denotes the galaxy stochasticity, further described in the \emph{Covariance} section.

The normalized initial conditions $\shat$ (sometimes referred to as ``phases''), which are drawn from a unit Gaussian prior, are related to the linear density field via
\be
\delta^{(1)}(\vk,z)=\alpha\,\left[\frac{N_{\mathrm{grid}}^3}{L^3} \Plin(k,z)\right]^{1/2}\,\shat(\vk).
\label{eq:shat_deltalin}
\ee
In \refeq{shat_deltalin}, $\Plin(k,z)$ is the linear matter power spectrum in the fiducial cosmology, while $L,N_{\mathrm{grid}}$ are the length and grid size of the $\shat(\vk)$ grid, respectively. Note that the finite volume $L^3$ and finite cutoff $\Lambda$ imply that a finite grid is sufficient to explicitly represent every relevant Fourier mode of the initial conditions.

\paragraph{Likelihood}
Following \citet{Schmidt:2018bkr,Cabass:2019lqx,Schmidt:2020viy,Schmidt:2020tao,Kostic:2022vok,Nguyen:2024yth}, we adopt a Gaussian likelihood in Fourier space,
\begin{align*}
\ln\lik\left(\data_\Lambda \Big| \alpha, \shat,\{b_\op\}, \sigma_\eps\right) \,
=&
-\frac{1}{2}
\sum_{\vk \neq 0}^{\kmax}
\left[
\ln{2\pi\sigma_\eps^2}
+
\frac{1}{\sigma_\eps^2}
\Big\lvert
\data_\Lambda(\vk) - \delta_{g,\Lambda}[\alpha, \shat, \{b_\op\}](\vk)
\Big\rvert^2
\right]\\
&+ \text{const.} \,\, ,
\numberthis
\label{eq:EFT_fieldlevel_likelihood}
\end{align*}
where we have used the fact that the galaxy stochasticity is homogeneous and isotropic in the galaxy comoving rest frame, i.e. in real space, and hence has a diagonal covariance matrix in Fourier space $\cov=\sigma_\eps^2$.
Note that \refeq{EFT_fieldlevel_likelihood} introduces the analysis cutoff scale $\kmax$ where $\kmax\leq\Lambda$. Unlike \citet{Nguyen:2024yth}, here we choose a cubic sharp-$k$ filter to implement the analysis cutoff $\kmax$ in \refeq{EFT_fieldlevel_likelihood} and further set $\kmax=\Lambda/1.4$.
A value of $\kmax < \Lambda$ is chosen to reduce the magnitude of higher-derivative terms that are controlled by the length scale $\Lambda^{-1}$. We have neither performed a detailed study of the optimal choice of $\kmax/\Lambda$ here nor in \citet{Nguyen:2024yth}.

\refeq{EFT_fieldlevel_likelihood} does not capture a coupling of galaxy stochasticity to large-scale perturbations (``density-dependent noise''), or non-Gaussianity of the noise. While the former can be incorporated via a real-space formulation of the likelihood \citep{2020JCAP...07..051C}, the latter is technically more difficult to incorporate at the field level. Apart from the bias terms (see \emph{Galaxy model} above), this constitutes the second main model difference to the EFT P+B analysis.

\paragraph{Covariance}

Even when limiting to a diagonal covariance in Fourier space, galaxy stochasticity is not necessarily white noise on all scales. In fact, 
\beq
\label{eq:EFT_fieldlevel_epsilon_general}
\sigma_\eps=\sigma_\eps(k)=\sigma_{\epsilon,0}\left[1+\sigma_{\epsilon,k^2}k^2+\sigma_{\epsilon,k^4}k^4+\cdots\right],
\eeq
in full generality \citep{Desjacques:2016bnm}.
Though physically motivated, $\sigma_{\epsilon,k^2}$ and $\sigma_{\epsilon,k^4}$ add significant parameter degeneracies and correlations.
Therefore, unlike \citet{Nguyen:2024yth}, herein we consider only the scale-independent leading-order contribution\footnote{On sufficiently large scales, this $k$-dependent corrections should be minimal \citep{Schmittfull:2018yuk,Elsner:2019rql,Schmidt:2020viy,Schmidt:2020tao,Schmittfull:2020trd}.} in our baseline analyses to minimize sampling cost.\footnote{See our post-unmasking study and \citet{Nguyen:2024yth} for a more complete treatment of stochasticity.} \refeq{EFT_fieldlevel_epsilon_general} then reduces to
$\sigma_{\epsilon}(k)=\sigma_{\epsilon,0}$.
This still leaves open two possibilities: (1) fixing $\sigma_{\epsilon,0}=\sigma_{\eps_{\mathrm{Poisson}}}$ where the latter follows the expectation for Poisson shot noise, and (2) inferring $\sigma_{\epsilon,0}$ from the data itself. We choose option (1) in our baseline analyses, and consider the subleading correction $\propto k^2$, as employed in \citet{Nguyen:2024yth}, in a post-unmasking analysis described below.

\paragraph{Priors on cosmology and bias}
We adopt wide uniform priors for the amplitude rescaling parameter $\alpha$ and the bias parameters $b_\op$.
To summarize,
\begin{gather*}
\mathcal{P}(\alpha) = \mathcal{U}(0.5,1.5),\numberthis\label{eq:EFT-FBI_priors_cosmo} \\
\mathcal{P}(b_\delta) = \mathcal{U}(0.0,4.0), \quad \mathcal{P}(b_{\delta^2}) = \mathcal{P}(b_{K^2}) = \mathcal{U}(-4.0,4.0), \quad \mathcal{P}(b_{\nabla^2\delta}) = \mathcal{U}(-20.0,20.0), \quad \mathcal{P}(\sigma_\eps) = \delta_D(\sigma_{\eps_{\mathrm{Poisson}}}).\numberthis\label{eq:EFT-FBI_priors_bias}
\end{gather*}
\paragraph{Priors on initial conditions}

As mentioned above, the cosmological prior on the normalized initial conditions $\shat$, in the absence of primordial non-Gaussianity, is a unit Gaussian prior (free-IC). However, we will also consider a special analysis, on \code{box 1}, where we were given the ground-truth initial conditions used to initialize one of the N-body simulations underlying the challenge data set (fixed-IC). To summarize, we assume the following priors on $\shat$:
\beq
\mathcal{P}(\shat) =
\begin{cases}
	\delta_{D}(\shat - \shat_{\mathrm{true}}) & \text{(fixed-IC, \code{box 1})}, \\[3pt]
	\gauss(\shat; 0, \identity) & \text{(free-IC)},
\end{cases}
\label{eq:shat_prior}
\eeq
where, in a fixed-IC analysis, $\delta_{D}$ indicates the Dirac delta distribution fixing the initial conditions to the true initial conditions $\shat_{\rm true}$ of the simulation data while, in a free-IC analysis, $\gauss$ indicates the multivariate normal distribution on normalized initial conditions with zero mean and identity covariance.

\paragraph{Sampling}
We employ Hamiltonian Monte Carlo sampling to explore the initial conditions $\shat$ while adopting slice sampling to sample other parameters. We refer readers to Sec~3.2 of \citet{Kostic:2022vok} and references therein for the motivation behind this choice.

\begin{figure}[t!]
\centering
\includegraphics[width=\linewidth]{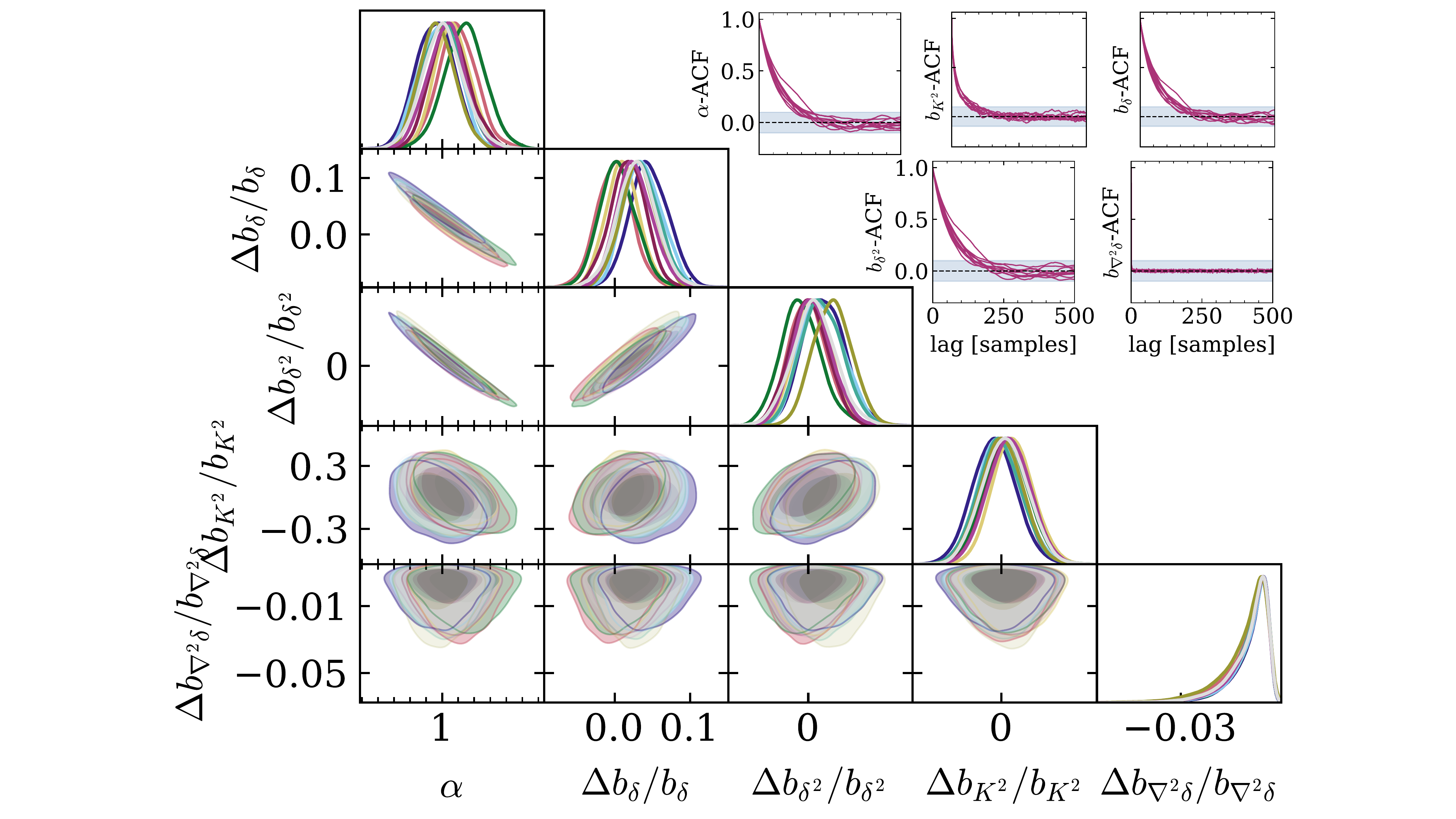}
\caption{Consistency between posteriors obtained in ten EFT FBI free-IC analyses on ten data realizations during stage 2. To avoid unmasking the values of bias parameters, we subtract and divide their values in each chain by the corresponding posterior means obtained from the analysis of \code{box 1}.}
\label{fig:EFT-FBI-Convergence+Correlation}
\end{figure}

\paragraph{MCMC convergence and autocorrelation}
Unconverged and correlated MCMC chains lead to biases and uncertainties in the estimation of moments of the parameter posteriors. Below, we focus on MCMC convergence diagnostics and autocorrelation estimates for stage-2 analyses (see \emph{Unmasking criteria} below), as their results can be directly compared to those obtained by other teams, namely EFT P+B and \code{BACCO} P.
\begin{itemize}[label=\textbullet]
\item MCMC convergence---We run two MCMC chains with distinct initial parameter values for each analysis. We first identify the warmup and equilibrium phases in each chain through the parameter drifts within individual chains and the Gelman-Rubin (G-R) statistics estimated from each and both chains.\footnote{We observe neither divergence nor slow mixing in MCMC chains across all pre-unmasking and post-unmasking analyses.} After removing the warmup phases, we find posteriors estimated from the two chains to be completely consistent.
\item MCMC autocorrelation---The univariate autocorrelation function (ACF) provides an estimate for the number of effective, i.e. independent, samples $n^{\mathrm{effective}}$ in the MCMC chain for the parameter of interest. In the upper right corner of \reffig{EFT-FBI-Convergence+Correlation}, we show the ACFs of the cosmological and bias parameters as a function of the lag between two samples in the MCMC chain.\footnote{We estimate the ACFs using the FFT method, as implemented in \code{GetDist}. See, e.g. \citet{NumericalRecipes}.} A null ACF indicates that two MCMC samples are independent draws from the posteriors.
\item MCMC Effective Sample Size (ESS)---We target a G-R value of $\hat{R}\sim0.01$ and an average number of $\sim255$ independent samples of $\alpha$.\footnote{We use \code{GetDist} \citep{GetDist} for these diagnostics and estimates. The package estimates the G-R statistics following the classical estimator in \citet{GelmanRubin} and the effective sample number following their Eq.~(22).} The MCMC sampling error on the means of $\alpha$ and $\sigma_8$ can be estimated from $\sigma_\alpha/n^{\mathrm{effective}}_\alpha$ where $n^{\mathrm{effective}}_\alpha$ is the ESS of $\alpha$. The ESS of ten stage-2 analyses range from 193 to 311, with a mean and median of 254.4 and 259.5, respectively. Their G-R statistics range from 0.003 (best) to 0.035 (worst), with a mean of 0.0115 and median of 0.0085.
\end{itemize}
%
%
%

\paragraph{Posterior consistency}
A unique feature of the EFT FBI (stage-2) analyses is that we individually analyze all ten data realizations in the real-space suite of mocks. \reffig{EFT-FBI-Convergence+Correlation} shows the posteriors obtained in all ten analyses, with their labels hidden to avoid revealing the cosmic variance in each realization. The main takeaway here is that all ten posteriors are consistent with each other within the 68\% confidence limit.

\subsubsection{Analysis choices}
\label{sec:EFT_fieldlevel_analysis_choices}

\begin{figure}[t!]
\centering
\includegraphics[width=0.8\linewidth]{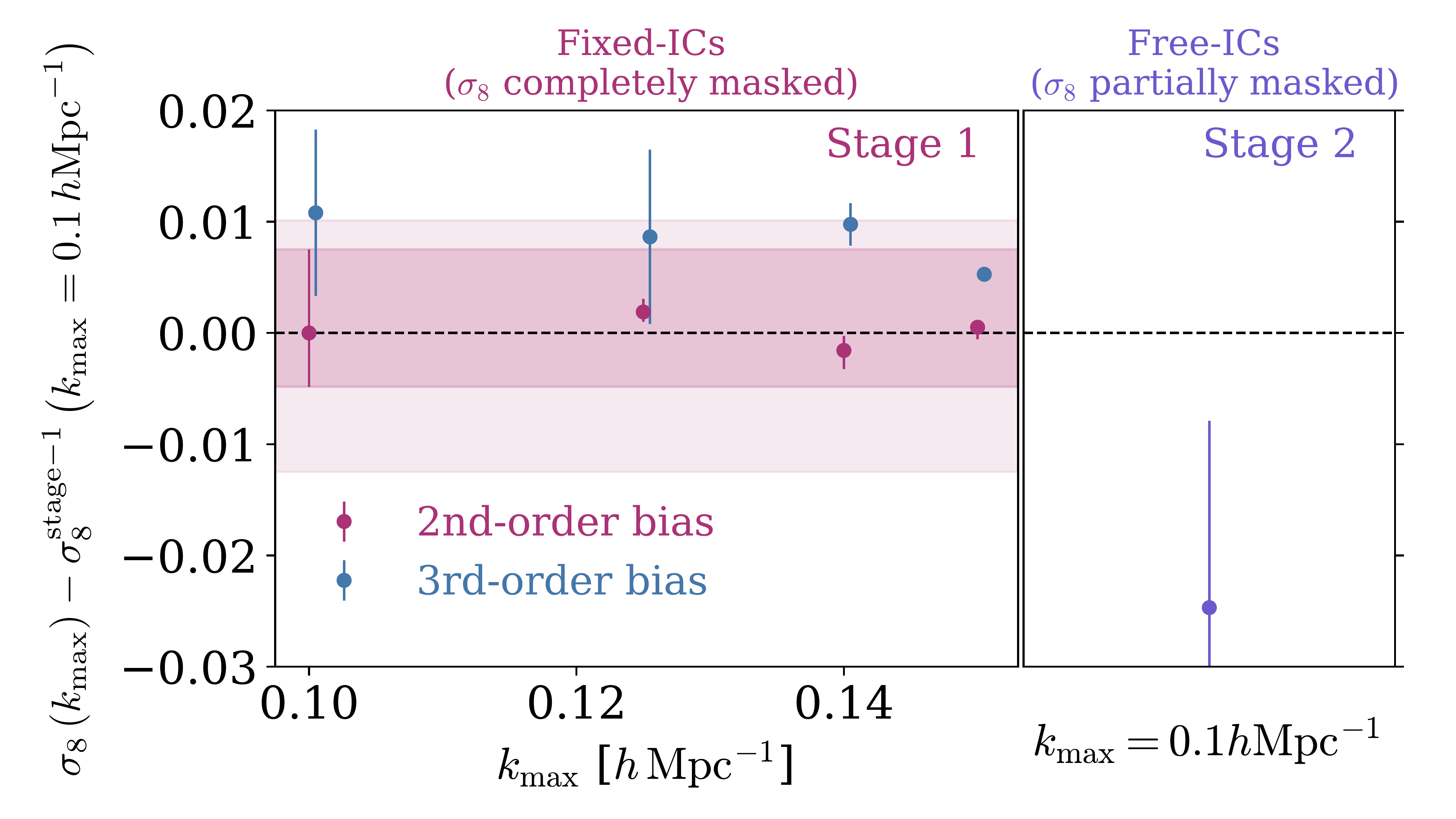}
\caption{Convergence of $\sigma_8(z=0)$ inferred in our stage-1 and stage-2 analyses, using the same data realization (\code{box 1}). In both panels, the error bars indicate 68\% lower and upper limits on $\alpha$. 
All $\sigma_8$ values are quoted relative to the stage-1 analysis using the fiducial second-order bias model at $k_{\rm max}=0.1$. We emphasize that the horizontal dashed line does \emph{not} indicate the ground truth $\sigma_8$. The horizontal bands indicate the 68\% and 95\% confidence intervals of the $\sigma_8$ posterior in that scenario.
Left (Stage 1): Convergence of fixed-IC constraints as a function of the analysis (EFT) cutoff scale $\kmax$ ($\Lambda$). All $\sigma_8$ posterior means are within 1-$\sigma$ of the corresponding value at $\kmax=0.1\,\hMpc$ for the respective bias model (same color). 
Magenta and blue points in the left panel indicate second-order and third-order bias model, respectively. The blue points are horizontally displaced to make error bars fully visible. Right (Stage 2): Free-IC constraint from \code{box 1}, again relative to the stage-1 result.}
\label{fig:EFT-FBI_kmaxconvergence_alpha_Stage1+2}
\end{figure}

\paragraph{Scale cuts}
We determine the fiducial scale cut $\kmax$ in our baseline analyses by choosing the maximum scale that maintains convergence of the amplitude rescaling parameter $\alpha$ inferred with different choices for the \emph{Galaxy model}, e.g. second- versus third-order galaxy bias expansion (\reffig{EFT-FBI_kmaxconvergence_alpha_Stage1+2}), Lagrangian versus Eulerian basis for the bias expansion.
Specifically, we define the relative parameter shift
\beq
\label{eq:relative_drift_alpha}
\Delta\alpha/\sigma_\alpha\equiv\frac{\<\alpha\>_A-\<\alpha\>_B}{\left(\sigma_{\alpha_A}^2+\sigma_{\alpha_B}^2\right)^{1/2}},
\eeq
where the indices $A,B$ label different analyses with their associated posterior means $\<\alpha\>$ and 68\% uncertainties $\sigma_{\alpha}$. We require \refeq{relative_drift_alpha} to be less than 2.0 for final submissions, i.e. the \code{LEFTfield} result shown in \reffig{1D_summary_real}.
The scale cut we end up with for our stage-1 and stage-2 submissions is $\kmax=0.1\,\hMpc$.
This then corresponds to the maximum scale up to which the second-order bias expansion is expected to be reliable. We anticipate that a third-order bias expansion will have higher reach in wavenumbers. For more details, see \citet{Nguyen:2024yth}.

\paragraph{Unmasking criteria}
We unmask the EFT FBI analysis in two stages:
\begin{enumerate}
    \item In stage 1, we performed the inference on \code{box 1}, with the initial conditions $\shat$ being fixed to $\shat_{\mathrm{true}}$ (fixed-IC) provided by the challenge organizers (only for \code{box 1}). During stage 1, we performed several EFT- and numerical-convergence tests because analyses are numerically cheap (at fixed initial conditions).
    After stage-1 submission, and before stage-2 analysis started, the organizers provided us a confirmation that stage-1 posterior mean of $\sigma_8$ is within 2-$\sigma$ of the true $\sigma_8$ (without specifying which side of the true $\sigma_8$ it actually is). For transparency, below in \reffig{EFT-FBI_kmaxconvergence_alpha_Stage1+2}, we also highlight the 2-$\sigma$ constraint of the stage-1 submission.
  \item In stage 2, \emph{without} any change in our analysis choices, we extended the analysis to all ten boxes, while also allowing $\shat$ to be explicitly sampled and marginalized over (free-IC). We emphasize that stage-2 analyses were performed \emph{without} the true initial conditions $\shat_{\mathrm{true}}$ of the simulation boxes.
\end{enumerate}    
In summary, the EFT FBI analyses and results are conducted and submitted in two stages, where $\sigma_8$ is fully masked, but $\shat$ is provided for \code{box 1} in stage 1.
The reason behind this choice is we initially agreed with the challenge organizers to conduct a fixed-IC analysis only.
After the submission of our fixed-IC analysis (stage 1), encouraged by the organizers, we decided to pursue a full free-IC analysis (stage 2), in coordination with the \code{BACCO} P and EFT P+B teams.

 \paragraph{Caveats}
In early tests or parallel studies, e.g. \citet{Nguyen:2020hxe,Nguyen:2024yth}, we have identified an issue of ``$\sigma_\eps$ collapse'' when trying to infer the galaxy stochasticity, parametrized by $\sigma_{\eps,0}$, from the data itself with wide uniform prior on $\sigma_{\eps,0}$: the inferred $\sigma_{\eps,0}$ (hence $\sigma_\eps$) tends to an unphysically low value.
This is a known issue of inference and sampling using a Gaussian likelihood with unknown covariance, especially on noisy data. That is, in this case, galaxy sample with a low number density.
Ongoing work is devoted to an extended noise model, and investigating physically-motivated parametrizations and priors for $\sigma_\eps$.

\paragraph{Post-unmasking studies}
Our baseline analyses opted for efficiency in model evaluation and sampling, specifically the ability to scale up our EFT FBI algorithm and individually analyze all the ten realizations of the real-space $\Lambda$CDM data suite during our stage-2 analysis. Therefore, we adopted the second-order bias expansion and assumed galaxy stochasticity to be white noise with Poisson amplitude (\refeq{EFT-FBI_priors_bias}).
Our primary motivation for a post-unmasking study is to facilitate a comparison between our $\sigma_8$ constraint and that reported by the EFT P+B team. To this end, we have reanalyzed \code{box 1} while extending the \emph{Galaxy model} in the \emph{Model} and the \emph{Covariance} in the \emph{Inference} as follows:
\begin{enumerate}
    \item \emph{Galaxy model}: going from second-order to third-order bias (cf. eq.~(A4) of \citet{Nguyen:2024yth});
    \item \emph{Covariance}: allowing for the galaxy stochasticity to be scale-dependent with unknown amplitudes, i.e. $\sigma_\eps(k)=\sigma_{\eps,0}\left[1+\sigma_{\eps,k^2}k^2\right]$, where $\sigma_{\eps,0}, \sigma_{\eps,k^2}$ are jointly inferred from the data.
\end{enumerate}
With these new choices, our \emph{Model} and \emph{Inference} correspond to the same adopted in \citet{Nguyen:2024yth}.
For the cubic bias parameters, we adopt the same uniform prior adopted for the quadractic bias parameters, i.e.
\beq
\mathcal{P}(b_{\delta^3})=\mathcal{P}(b_{K^3})=\mathcal{P}(b_{\delta K^2})=\mathcal{P}(b_{\op_{\mathrm{td}}})=\mathcal{U}(-4.,4.).
\label{eq:EFT-FBI_priors_cubic}
\eeq

For the additional stochastic parameters $\sigma_{\eps,0}$ and $\sigma_{\eps,k^2}$, we assume the following flat priors
\beq
\mathcal{P}(\sigma_{\eps,0}) = \mathcal{U}(0.9,1.1)\sigma_{\eps_{\mathrm{Poisson}}},\; \mathcal{P}(\sigma_{\eps,k^2}) = \mathcal{U}(-10.0,100.0).
\label{eq:EFT-FBI_priors_stochasticity}
\eeq

The result of this post-unmasking analysis is shown in \reffig{1D_summary_real_post} under the label ``FBI extended.'' The systematic bias in $\sigma_8$ is now completely relieved, while the increased parameter vector leads to a larger error bar compared to the pre-unmasking analysis.
Our constraints are further inline with the expectations from the EFT FBI constraints presented in \citet{Nguyen:2024yth}.
One technicality is that, here we implement the analysis cutoff $\kmax$ with a cubic sharp-$k$ filter, i.e. $\prod_{i=1}^3 \theta_H(|k_i|-\kmax)$, while \citet{Nguyen:2024yth} implemented the cutoff with a spherical sharp-$k$ filter, i.e. $\theta_H\left(|\vk|-\kmax\right)$. Therefore our analyses with $\kmax=0.1\,\hMpc$ here involves roughly the same number of modes as (hence should be compared to) the analyses with $\kmax=0.12\,\hMpc$ in \citet{Nguyen:2024yth}.
For more discussion and post-unmasking comparisons, we refer to \refsec{discussion}.

%% file: methods_SBI.tex
\subsection{SBI : P \& B method\footnote{Authors: Chirag Modi, ChangHoon Hahn.}}
\label{subsec:SBI}

In this section, we outline the analysis of the redshift-space mocks using simulation-based inference (SBI). 

SBI uses computationally simulated forward models of the observed data that are evaluated using 
parameter values spanning the prior~\citep{Cranmer2020}. 
It leverages the forward models to learn the likelihood distribution of any measurable data-statistics and
can, in principle, be applied to analyze many of the summary statistics presented in this paper.
Since the likelihood is learned from the forward models, it relaxes the assumptions on the 
likelihood (e.g., the emulator approach typically assumes a Gaussian likelihood).
Furthermore, the forward models can include observational effects to more robustly treat systematics\citep[e.g.][]{Yuan:2022ibz}. SBI has been applied to large-scale structure analysis \citep{Alsing2019, Jeffrey2021, Hahn2023a, Hahn2023b, Tucci:2023bag} 
and our analysis here will most closely follow \citet{Hahn2023b}.

\subsubsection{Data \& Estimators}
We focus on analyzing the power spectrum multipoles $P_\ell(k)$ for ($\ell=0,2,4)$ and bispectrum monopole 
$B_0(k_1, k_2, k_3)$ using the mean measurement from the 10 $\Lambda$CDM redshift-space boxes.
Power spectrum multipoles are measured with Fast Fourier Transforms using Nbodykit \citep{nbodykit} on a 512$^3$ mesh,
while bispectrum is measured using \code{pySpectrum} code \citep{pyspectrum} on a 760$^3$ mesh. 
In addition, we supplement our data vector with the galaxy number density, $\bar{n}$\footnote{
We find that explicitly including number density does not add any significant constraining power, but helps make the analysis more stable when the number density varies widely over different cosmologies of the latin hypercube for the same HOD parameters.}.
We use both the statistics between $k_{\mathrm{min}}=0.009 h/$Mpc and $k_{\mathrm{max}}=0.5 h/$Mpc. This results in the power-spectrum data vector of size 79$\times$ 3 (for three multipoles) and bispectrum data-vector of size 1898. 

\subsubsection{Model}
For SBI, we need to accurately model galaxy clustering statistics over a range of cosmology and galaxy formation parameters. 
To this end, we use \code{Quijote} $N$-Body simulations suite \citep{Villaescusa-Navarro:2019bje}, specifically 
the high resolution $\Lambda$-CDM Latin-hypercube (LH) which consists of 2000 simulations varying over 
5 cosmological parameters- $\Omega_m,\, \Omega_b,\, h,\, n_s$ and $\sigma_8$.
Each simulation evolves 1024$^3$ dark matter particles with TreePM Gadget-III code in a volume of 1${\rm Gpc}/h$.
For each simulation, there are two sets of dark matter halos identified with {\sc Rockstar} \citep{Behroozi:2011ju} 
and Friends-of-friends (FoF) halo finder, respectively.
Recent work has shown that SBI models trained on one set can lead to biased results on another \citep{Modi2023a}, 
hence we will use both halo catalogs to evaluate robustness of our analysis. 
However, the final results of this challenge will be derived using halos from {\sc Rockstar}, which more accurately 
identifies the position and velocity of halos.

Next, we populate these dark matter halos with galaxies. We use an extended 9-parameter HOD model that includes assembly, concentration and velocity biases, summarized in Table~\ref{Table:hod_summary}.
Five of these parameters are the same as the standard Zheng07 HOD \citep{Zheng2007} model- 
$\log M_{\mathrm{min}}$, $\sigma_{\log M}$, $\log M_0$, $\log M_1$, $\alpha$. 
These are supplemented with `assembly' bias parameters ($A_c$, $A_s$) which modify the number of centrals 
and satellites based on the halo concentration\footnote{FoF halos do not measure concentration accurately and 
hence we use an analytic relation to estimate this value from halo mass \citep{Dutton2014}. In this sense it is not strictly assembly bias, but a different parameterization with respect to halo mass.},
`concentration' bias parameter ($\eta_{\rm conc}$) which allows the concentration of satellite galaxies to deviate from NFW profile, 
and `velocity' dispersion parameters ($\eta_c,\, \eta_s$)  which rescale the central and satellites velocities over the halo velocity.
For further details on implementation of this decorated HOD, we refer the reader to \citep{Hahn2023a, Hearin2016}.

\subsubsection{Inference}

\paragraph{Methodology}
SBI requires a training dataset of $(\vec{\theta},\, \vec{x})$ pairs where $\vec{\theta}$ are the model parameters of interest (here, cosmology and HOD parameters) and $\vec{x}$ are the corresponding summary statistic (here, power spectrum multipoles, bispectrum and number density).

$\vec{x}$ is generated using the simulations and estimators described above. 
The dataset of $\{(\vec{\theta},\, \vec{x})\}$, therefore, corresponds to samples drawn from the joint 
distribution  $p(\vec{\theta}, \vec{x})$. 
More importantly, $\{(\vec{\theta},\, \vec{x})\}$ can be used to infer the posterior $p(\vec{\theta}|\vec{x})$.

While different methods can be used to infer $p(\vec{\theta}|\vec{x})$ using 
$\{(\vec{\theta},\, \vec{x})\}$, we use neural density estimators, which have been used to accurately estimate
even high dimensional distributions with limited training data. 
More specifically, we train a conditional neural density estimator, $q_{\phi}$, with parameters $\phi$ to 
approximate $p(\vec{\theta}|\vec{x}) \approx q_{\phi}(\vec{\theta}|\vec{x})$.
We do this by minimizing the Kullback-Leibler divergence between $p(\vec{\theta}|\vec{x})$ and 
$q_{\phi}(\vec{\theta}|\vec{x})$, estimated by the log-probability of $q_\phi$ evaluated over the training 
dataset.

Once $q_{\phi}(\vec{\theta}|\vec{x})$ is successfully trained, we can infer the posterior for 
observations, $\vec{x'}$, by simply querying the trained $q_{\phi^*}$ to generate samples from the posterior:
$\vec{\theta'} \sim q_{\phi^*}(\vec{\theta}|\vec{x'})$

\paragraph{Training data and priors}
To generate the training dataset for SBI, we use the 
{\sc Quijote} LH described in the previous section. 
For each simulation in LH, we sample 10 different HOD parameter values over a prior range and generate 10 galaxy catalogs. 
Thus in total, we have 20,000 galaxy catalogs. We use catalogs for 1500 of these simulations (i.e. 15,000 galaxy catalogs) for training, 200 for validation, and held out 300 for testing. 
The prior range of cosmological parameters is set by the bounds of {\sc Quijote} LH.
\beq
    \Omega_{\rm m}\sim\unif[0.1, 0.5], \,\, \sigma_{8}\sim\unif[0.6, 1.0], \,\, \Omega_{\rm b} \sim \unif[0.03, 0.07], \,\, n_s\sim\unif[0.8, 1.2], \,\, h\sim\unif[0.5, 0.9] 
\label{eq:sbipb_priors_cosmo}
\eeq
For HOD parameters, we shift the central values of prior on $\log M_{\mathrm{min}},\, \log M_0$ and $\log M_1$ for different cosmologies so as to match the given number density on average. 
This allows us to more efficiently sample the parameter space by focusing only on the regions that will 
generate simulations loosely consistent with the data. 
These central values are estimated as follows: we assume a fiducial value of $\alpha=0.7$ and satellite fraction of 0.2, then for every cosmology we set $\log M_{\mathrm{min}}^{\theta} = \log M_0^{\theta} = M_h$, the halo-mass above which the number of halos is the same as the number of centrals, and then estimate $\log M_1^{\theta}$ to match the satellites. The resulting priors are as follows:
$$
\log M_{\mathrm{min}}\sim\unif[\log M_{\mathrm{min}}^{\theta}\pm0.15], 
    \quad \log M_0\sim\unif[\log M_0^{\theta}\pm0.2],  \quad \log M_1\sim\unif[\log M_1^{\theta}\pm0.3]
$$
For remaining HOD parameters, we use the following priors for all cosmologies:
\beq
    && \quad \alpha\sim\unif[0.4, 1.0],  \quad \quad  \sigma_{\log M}\sim\unif[0.3, 0.5], \quad \quad  A_c, A_s\sim \mathcal{N}(0, 0.2)\, {\rm over}\, [-1, 1], \nonumber\\
   && \quad  \eta_{\rm conc}\sim\unif[0.2, 2.0],  \quad \eta_{c}\sim\unif[0., 0.7],  \quad \eta_s\sim\unif[0.2, 2.0] \nonumber
\eeq
We estimate the power spectrum multipoles and bispectrum for each of the 20,000 simulated galaxy catalogs.

\paragraph{Posterior inference}
We use \code{sbi}\footnote{Available at \href{https://github.com/mackelab/sbi}{Github.com/mackelab/sbi}} package to train masked auto-regressive flows as conditional neural density estimators and learn the posterior, $q_{\phi}(\vec{\theta}|\vec{x}) \sim p(\vec{\theta}|\vec{x})$.
To minimize stochasticity in training, we use \code{Weights-and-Biases}\footnote{Available at \href{https://wandb.ai/site}{wandb.ai}} package and train 200 networks for each data-statistic by varying hyperparameters such as the width and the number of layers, learning rate and batch size. 
After training, we collect 10 neural density estimators with best validation loss and use them as an ensemble {\em i.e.} we construct a mixture distribution with uniform weighting to approximate the posterior. 

\subsubsection{Analysis Choices}
\paragraph{Scale cuts}
The {\sc Quijote} simulations are only $(1 {\rm Gpc}/h)^3$ in volume, while the data-challenge 
simulations are $(2 {\rm Gpc}/h)^3$ in volume.
Thus in addition to the typical small-scale cut, we also impose a large scale cut on the power-spectrum 
and bispectrum measurements.
We use only the scales between $k_{\mathrm{min}}=0.009 h/$Mpc and $k_{\mathrm{max}}=0.5 h/$Mpc.
We point out that the smaller volume of our training simulations than the mock data results in 
our learnt-likelihood, and hence the posterior, being over-dispersed due to larger cosmic variance. This is further exacerbated for $\Omega_m$ since we miss the informative large scale modes due to larger $k_{\mathrm{min}}$ than the fundamental wavenumber of the challenge mocks, as discussed in \refsec{resultsredshift}.
Thus we expect our results to be more conservative than if the simulations had the same volume. We also note that in the future, such limitations on the volume of the training simulations can be overcome using hybrid simulation-based inference techniques to combine EFT analysis on the large scale analysis with SBI analysis only on the small scales \citep{hysbi}.

\paragraph{Validation}
\begin{figure}
\centering
\includegraphics[width=\linewidth]{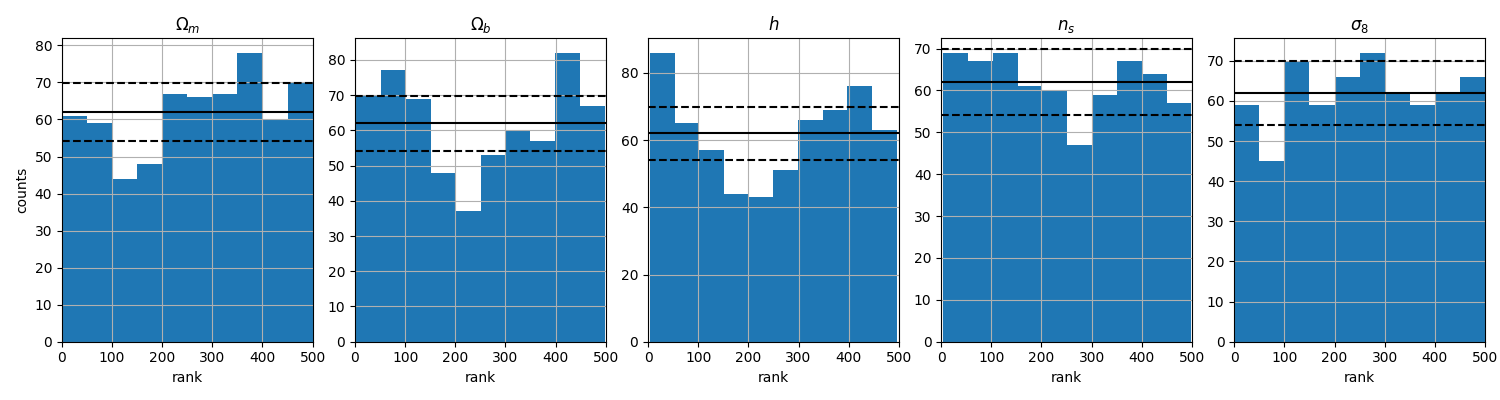}
\caption{Rank histograms of the five cosmology parameters for the trained ensemble evaluated using the held out test dataset. The solid horizontal black line is the  mean number of counts expected for a uniform distribution and the dashed lines indicate the expected Poisson scatter. Broadly, all the histograms are consistent with uniform distribution.}
\label{fig:SBI_rank}
\end{figure}
We use our trained ensemble to predict the cosmological parameters over the held-out test dataset to do coverage tests as described in \citet{Talts2018} and \citet{Hahn2023b}.
We verify that all the rank histograms are uniformly distributed within the rank scatter, which suggests (but does not guarantee) that the posterior marginal are accurate. These histograms are shown for the five cosmology parameters in \reffig{SBI_rank}.
To test for model-misspecification, we repeat the exercise with {\sc Quijote}-FoF latin hypercube and find consistent results. 

\paragraph{Caveats}
The primary source of error with SBI is model mis-specification \citep{Modi2023a, Cannon2022}.
Previous analysis \citep{Hahn2023b, Modi2023a} suggest that inference with power-spectrum up to $k_{\mathrm{max}}=0.5 h/$Mpc should be robust to these modeling differences. 
However, more powerful statistics such as bispectrum and combined power-spectrum and bispectrum analysis can exacerbate 
model mis-specification, both due to increased sensitivity and smaller error bars. 
This parameter-masked challenge provides a unique opportunity to investigate this aspect.

%% file: methods_AbacuskNN.tex
\subsection{\textit{k}NNs with \texorpdfstring{\code{AbacusSummit}}{AbacusSummit} emulator\footnote{\texorpdfstring{Author: Sihan Yuan}{Author: Sihan Yuan}}}
\label{subsec:Abacus_kNN}
In this section, we summarize the methodology of analyzing the 2D $k$-th nearest neighbor statistics ($k$NN) with an extended HOD emulator based on the \code{AbacusSummit} simulation suite. The detailed methodology is described in a dedicated supporting paper \citep{knnpaper}. 

\subsubsection{Data \& Estimators}
\begin{figure*}
    \centering
    \hspace*{-0.3cm}
    \includegraphics[width = 0.5\textwidth]{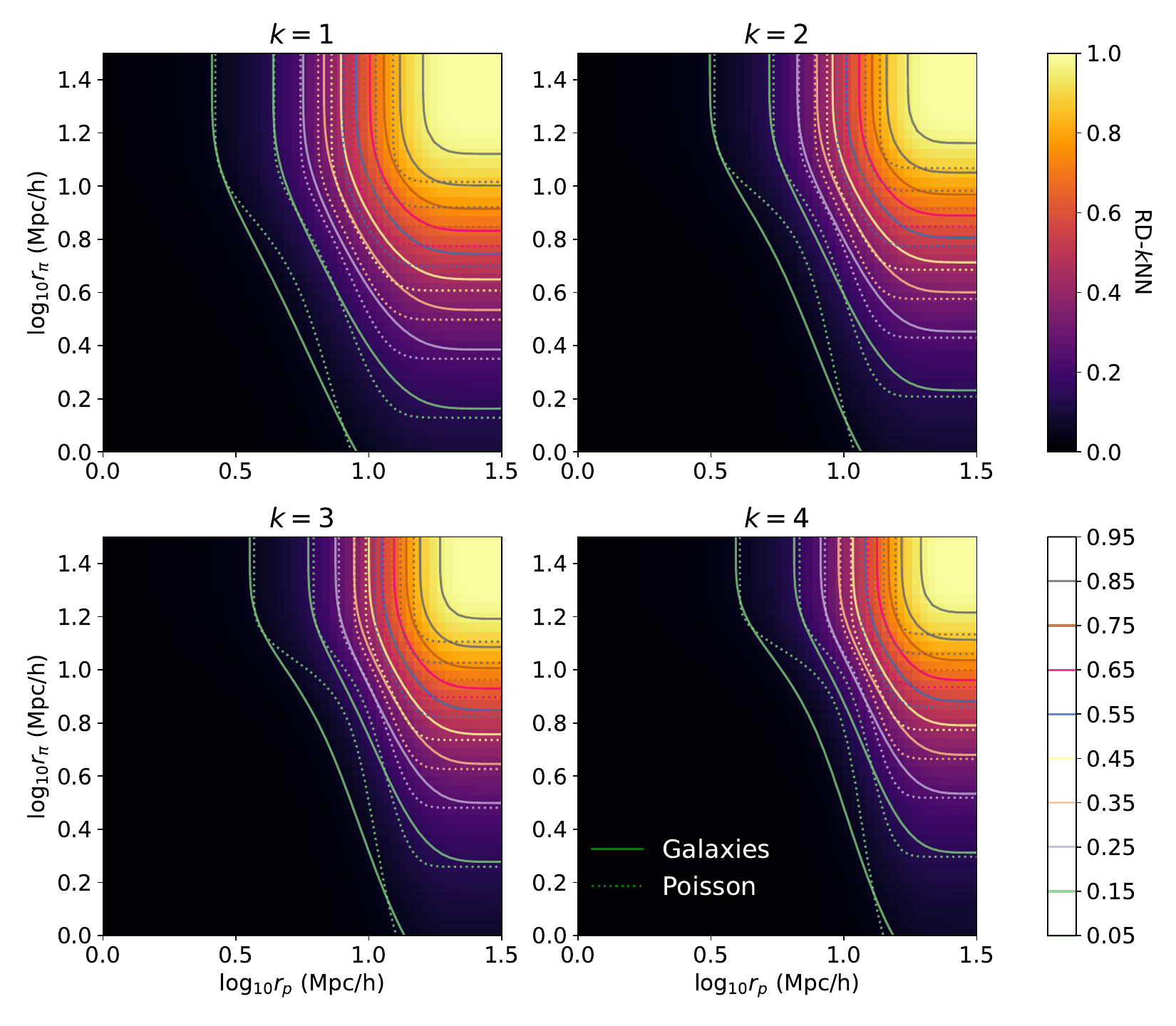}
    \includegraphics[width =  0.5\textwidth]{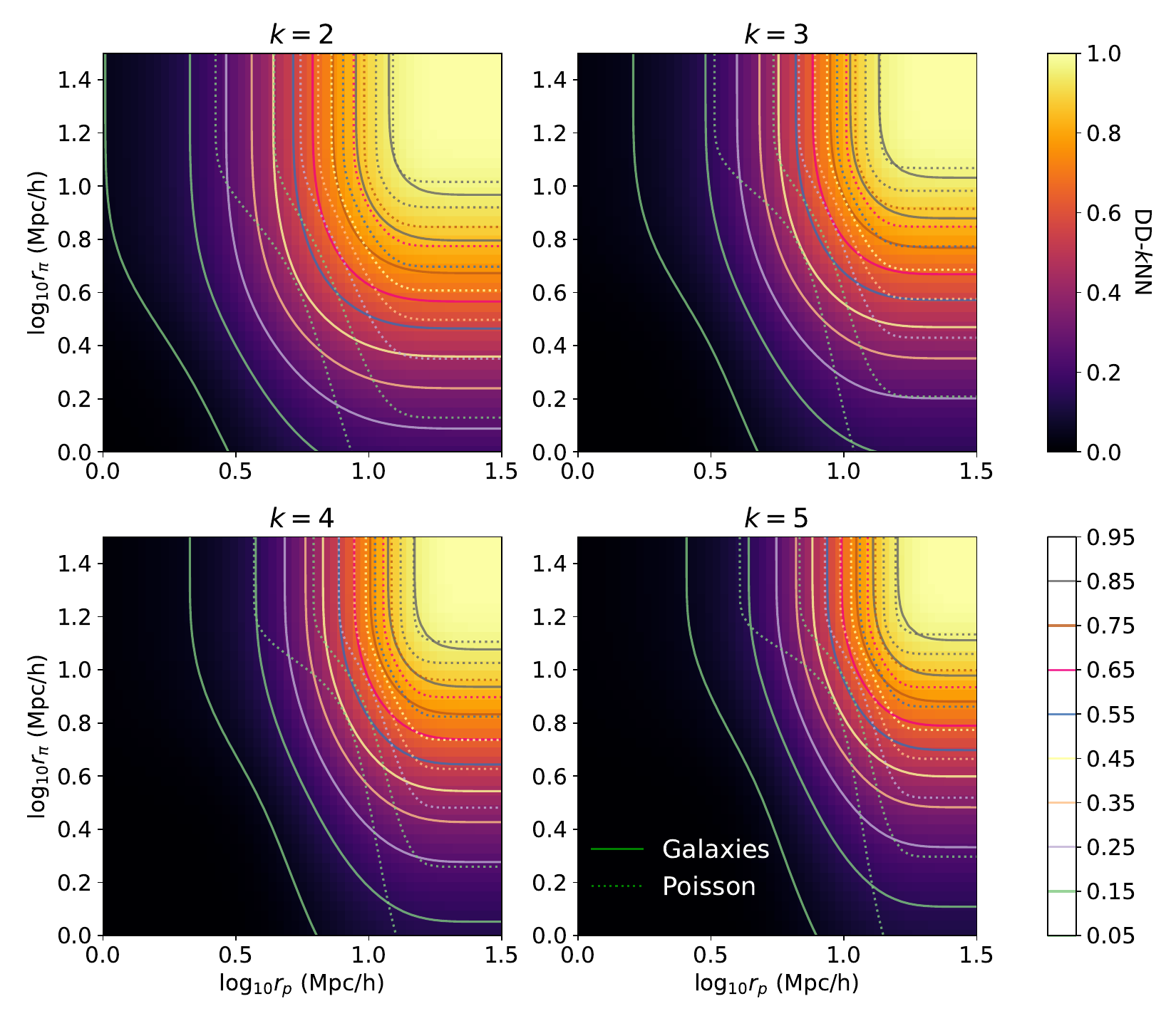}
    \caption{Visualizations of the RD-$k$NN statistics (left) and DD-$k$NN statistics (right), calculated on the redshift-space mock, averaged over 10 realizations. The dotted lines show the contours of a un-clustered Poisson random sample. We only showcase the first 4 $k$-orders for brevity. Note that we have defined DD-$k$NN$=1$, thus $k = 2$ is the first meaningful order.}
    \label{fig:knns_target_RD}
\end{figure*}

For this analysis, we use two different 2D $k$NN statistics: (1) the random-data $k$NN statistics (RD-$k$NN); (2) the data-data $k$NN statistics (DD-$k$NN). For a detailed description of these statistics, we refer the readers to \citet{Yuan:2023ddknn} and \citet{knnpaper}. Conceptually, the $k$NN statistics captures the probability distribution of the separation between random query points (or in the case of DD-$k$NN, the query points are galaxies) and their $k$-th nearest galaxies. Thus, in principle, $k$NNs are parametrized as a function of $k$ and the separation, which we further decompose in 2D to a transverse component $r_p$ and a line-of-sight (LOS) component $r_\pi$. Figure~\ref{fig:knns_target_RD} visualizes the RD and DD $k$NN statistics up to the first four orders, where each panel shows how the cumulative distribution $k$NN-CDF depends on the separation ($r_p, r_\pi$). While the heatmap and the solid contours showcases the $k$NN measurement on the provided redshift-space mocks, the dashed contours visualize the measurement on an unclustered Poisson random sample of the same size for comparison. The difference between the solid and dashed contours represent the informative features in the statistics.

The exact compression we use to construct our data vectors for this analysis are as follows.
We use $k = 1,2,3,..., 9$, and for each $k$ we use 8 logarithmic bins along the $r_p$ direction between $0.63 \, h^{-1}$Mpc and $63 \, h^{-1}$Mpc, and 5 logarithmic bins along the $r_\pi$ direction between $0.5 \, h^{-1}$Mpc and $32 \, h^{-1}$Mpc. We also remove bins where the CDF is less than 0.05 or greater than 0.95 to increase the overall signal-to-noise of our statistics. As a result, the final data vector is summarized with 114 bins in the RD case and 144 bins in the DD case. Our $r_p$ and $r_\pi$ binning choices are designed to expose our analysis to the nonlinear and quasi-nonlinear scales. We are not sensitive to large-scale features such as the BAO by design. Our $k$ choices are fairly arbitrary and are mostly chosen to confine us to nonlinear scales and computational efficiency. We reserve the discussion of optimal $k$ choices for a future paper. 

\subsubsection{Model}

To forward model the galaxy $k$NNs, we employ a neural network based emulator that learns the summary statistics as a function of cosmology and extended HOD parameters. In this sub-section, we introduce the relevant model details and emulation techniques. 

\paragraph{Parameterization} For the galaxy--halo connection, we employ the \code{AbacusHOD} package for efficiency and beyond vanilla HOD extensions \citep{Yuan:2021hod}. The model notably assigns satellites to halo particles and includes velocity bias, two types of galaxy assembly bias, and baryonic modulations. The model parameters were summarized as model HOD2 in Table~\ref{Table:hod_summary}. Here we expand on the relevant model extensions.

    Velocity bias is parametrized with two parameters $\alpha_\mathrm{vel, c}$ and $\alpha_\mathrm{vel, s}$. The central velocity bias parameter $\alpha_\mathrm{vel, c}$ modulates the peculiar velocity of the central galaxy relative to the halo center along the LoS. $\alpha_\mathrm{vel, c} = 0$ corresponds to no central velocity bias, i.e. centrals perfectly track the velocities of halo centers. We also define $\alpha_\mathrm{vel, c}$ as non-negative.
    
     The satellite velocity bias parameter $\alpha_\mathrm{vel, s}$ modulates how the satellite galaxy peculiar velocity deviates from that of the local dark matter particle. 
    $\alpha_\mathrm{vel, s} = 1$ indicates no satellite velocity bias, i.e. satellites perfectly track the velocity of their underlying particles. Detailed descriptions of our implementation can be found in \citet{Yuan:2021hod}.

    The assembly bias parameters can be defined against the halo concentration or the local over-densities over a $r_\mathrm{env} = 5h^{-1}$Mpc tophat filter. $B_\mathrm{cent} = 0, B_\mathrm{sat} = 0$ indicate no galaxy assembly bias. 
    In this analysis, we consider assembly bias against both halo concentration and local environment when we compare different models. 

 To summarize, the extended HOD model employed in the $k$NN analysis consists of 10 parameters: (1) 5 vanilla HOD parameters $M_{\mathrm{cut}}$, $M_1$, $\sigma$, $\alpha$, $\kappa$; (2) an incompleteness parameter $f_\mathrm{ic}$; (3) velocity bias parameters $\alpha_\mathrm{vel, c}$ and $\alpha_\mathrm{vel, s}$; (4) galaxy assembly bias parameters $B_\mathrm{cent}$ or $B_\mathrm{sat}$; (5) baryonic modulation parameter $s$.

\paragraph{Simulation details} This analysis employs the \code{AbacusSummit} simulation suite \citep[][]{2021Maksimova}, a set of large, high-accuracy cosmological N-body simulations using the \code{Abacus} N-body code \citep{2019Garrison}, designed to meet and exceed the Cosmological Simulation Requirements of the Dark Energy Spectroscopic Instrument (DESI) survey \citep{2013DESI}. \code{AbacusSummit} consists of over 150 simulations, containing approximately 60 trillion particles at 97 different cosmologies. The \code{AbacusSummit} suite also uses a new specialized spherical-overdensity based halo finder known as {\sc CompaSO} \citep{2021Hadzhiyska}.
For this analysis, we use the ``base'' configuration boxes for forward modeling, each of which contains $6912^3$ particles within a $(2h^{-1}$Gpc$)^3$ volume, corresponding to a particle mass of $2.1 \times 10^9 h^{-1}M_\odot$. \footnote{For more details, see \url{https://abacussummit.readthedocs.io/en/latest/abacussummit.html}}

To model cosmology dependencies, we use the 85 emulator boxes. The cosmology parameter basis used for our emulator includes 8 parameters spanning the $w$CDM+$N\mathrm{eff}$+running parameter space: the baryon density $\omega_b = \Omega_b h^2$, the cold dark matter density $\omega_\mathrm{cdm} = \Omega_\mathrm{cdm}h^2$, the amplitude of structure $\sigma_8$, the spectral tilt $n_s$, running of the spectral tilt $\alpha_s$, the density of massless relics $N_\mathrm{eff}$, and dark energy equation of state parameters $w_0$ and $w_a$ ($w(a) = w_0+(1-a)w_a$). 
The different cosmologies are indexed by \code{cXXX}, where \code{XXX} ranges from 000 to 181. The details of each cosmology are described on the \code{AbacusSummit} website \footnote{\url{https://abacussummit.readthedocs.io/en/latest/cosmologies.html}}.

\paragraph{Emulator details} We build a neural-network-based emulator to interpolate between the finite set of cosmologies. Specifically, we follow the approach of \citet{2022bYuan}, where we take advantage of the high efficiency of the \code{AbacusHOD} code and run MCMC chains in the HOD parameter space against the target data vector at each cosmology. We stop the MCMC chains after 20,000 evaluations (limited by computational resources) in each box and select samples whose likelihood is greater than $\log L > -9000$ (to ensure large training sample around the maximum likelihood region) as the training set for the subsequent emulator model. This approach constrains the emulator training to a compact region in the cosmology+HOD parameter space, improving the emulator precision. We also find consistent behavior when selecting our training set with different likelihood cutoffs. Similar approaches were adopted in \citet{2022bYuan} and \citet{knnpaper}. For the emulator model, we adopt a fully connected neural network of 5 layers and 500 nodes per layer with Randomised Leaky Rectified Linear Units (RReLU) activation. We train the network following a mini-batch routine with the Adam optimiser and a mean squared loss function, where we use the diagonal terms of the covariance matrix as bin weights. The performance of the emulator is characterized on 9 hold-out cosmologies (\code{c001-004} and \code{c171-175}). We refer the readers to \citet{knnpaper} for details. 

\subsubsection{Inference}
\paragraph{Likelihood Function}
For this analysis, we assume a Gaussian likelihood function, see \citet{knnpaper} for validation of the Gaussian likelihood assumptions.
To sample the likelihood function, we employ the nested sampling package \code{dynesty} \citep{2018Speagle, 2019Speagle}. 

\paragraph{Covariance}
We compute jackknife covariances for 2D-$k$NNs using the 10 realizations provided. We divide each realization into 125 cubic chunks, each of size ($400h^{-1}$Mpc)$^3$. We visualize the resulting correlation matrix in Figure~A1 and A2 of \citet{knnpaper}. Both covariances are invertible. The RD-$k$NN covariance matrix has a condition number of 1$\times 10^{7}$ and the DD-$k$NN covariance matrix has a condition number of 4$\times 10^{7}$.
To account for emulation errors, we compute a separate emulator covariance matrix from the hold-out cosmologies and add that onto the data jackknife covariance matrix. 

\paragraph{Priors} This analysis employs cosmology priors based on the $\Lambda$CDM parameter range covered by the \code{AbacusSummit} simulations,
\beq \label{eq:knn_priors_cosmo}
    &&\omega_\mathrm{cdm}\in[0.099,0.140], \qquad \sigma_8\in[0.68,0.94], \qquad n_s\in[0.90, 1.02],\nonumber\\
    &&\omega_{\rm b}\in[0.021,0.024]\,,
\eeq
with the prior distribution specified by the \code{AbacusSummit} parameter envelope, $N_{\rm ur} = 2.0328$ and $\alpha_s = 0$.  For the HOD parameters, we adopt the following flat priors: 
\beq \label{eq:knn_priors_hod}
    &&\log_{10} M_{\rm cut}\sim\unif[12.0, 14.5], \qquad \log_{10} M_1\sim\unif[13.0, 15.0], \qquad \alpha\sim \unif[0.5, 1.5],\nonumber\\
    &&\alpha_{\rm{vel, c}}\sim\unif[0.0, 1.0],\qquad \alpha_{\rm {vel,s}}\sim\unif[0.2, 1.8], \qquad \log_{10} \sigma\sim \unif[-3.5, 1.5],\nonumber\\
    &&\kappa\sim \unif[0.0, 2.0], \qquad B_{\rm cent}\sim \unif[-1.0, 1.0], \qquad B_{\rm sat}\sim\unif[-1.0, 1.0]\,.
\eeq

\subsubsection{Analysis choices}
\label{subsec:Abacus_kNN_validation}
\paragraph{Scale cut validation}
For both $k$NN configurations, the first cut we apply is removing bins where the CDF value measured on the target mock is either less than 0.05 or greater than 0.95. This removes the noisiest bins and Gaussianizes the likelihood function. 

We apply an additional scale cut to the DD-$k$NN statistics in $r_p > 5h^{-1}$Mpc. We refer the readers to \citet{knnpaper} for justification of this cut. Essentially, the DD-$k$NN statistic is highly sensitive to small-scale modeling, and we find our models to not be flexible enough to accurately predict the DD-$k$NN statistic at $r_p < 5h^{-1}$Mpc given the precision of a ($2\, h^{-1}$Gpc)$^{3}$ volume. We do not apply any scale cut to the RD-$k$NN statistic as it is less sensitive to small-scale modeling. Unfortunately, our scale cut at $r_p < 5h^{-1}$Mpc comes at significant cost to our constraining power on cosmology. This highlights an important avenue of future work. We discuss this point further in Sect.~\ref{sec:discussion}. 

\paragraph{Consistency Checks} 

Our supporting paper \citep{knnpaper} details how we test and compare different HOD models invoking different extensions. We leverage both goodness-of-fit metrics and Bayesian evidence. We also conduct cross-validation tests where we fit one subset of the data vectors and predict a different set of data vectors. 

Specifically, we set RD-$k$NN and DD-$k$NN as separate data vectors as they summarize different information of the density field. RD-$k$NN captures the density information and DD-$k$NN captures the clustering information. We also include the standard 2PCF as a third statistics that share some information with DD-$k$NN but also captures clustering out to larger scales. We start with three different HOD models that include some combinations of baryonic effect treatment and two different types of galaxy assembly bias in addition to the vanilla HOD. We fit all three different HOD models to one of the three statistics and predict one of the other two statistics. 

\begin{figure*}
    \centering
    \hspace*{-0.5cm}
    \includegraphics[width = 3.5in]{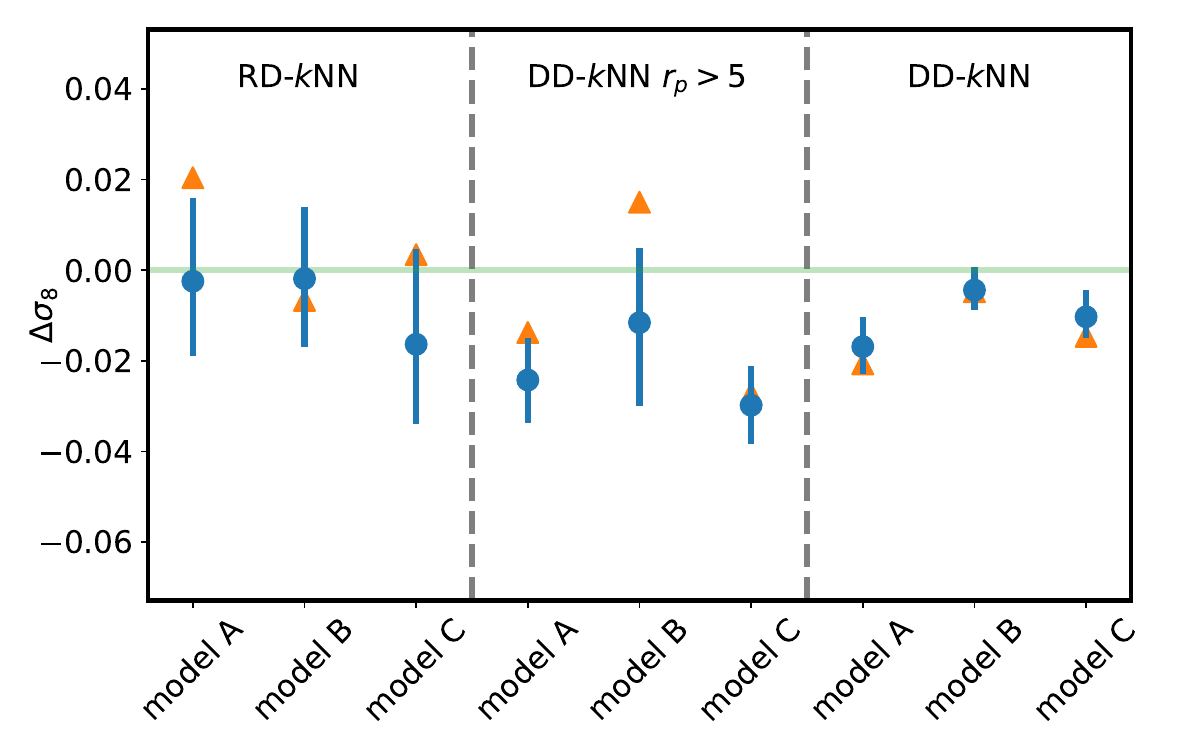}
    \vspace{-0.5cm}
    \caption{(Plot adapted from Figure~9 of \citealt{knnpaper}) The constraints of $k$NN statistics on $\sigma_8$ for different HOD models and different scale cuts. Model A, B, C refer to the three HOD candidate HOD models. The blue error bars show the $1\sigma$ marginalized constraints. The orange triangles show the maximum likelihood points. The green line represents the truth. The left, middle, right blocks represent the RD-$k$NN constraints, DD-$k$NN constraints with scale cut $r_p > 5h^{-1}$Mpc, and DD-$k$NN constraints without minimum scale cuts ($r_\mathrm{min} = 0.67h^{-1}$Mpc), respectively. Our pre-unmasking consistency tests favored model B, and we see post-unmasking that model B indeed results in the least biased constraints. Comparing the middle block to the right block shows that the inclusion of small scales drastically tightens the constraints on $\sigma_8$. Unfortunately, our models were not flexible enough to self-consistently model $r_p < 5h^{-1}$Mpc. This comparison motivates for more robust modeling of small scales. }
    \label{fig:knns_scales}
\end{figure*}

Our tests clearly favor one HOD model, which results in good fits and no tension between different subsets of the data vectors. We show the constraints of the three models on $\sigma_8$ in Figure~\ref{fig:knns_scales}, where we see that the favored model B indeed achieves the least biased constraints post-unmasking. We do not reveal the exact details of the favored HOD model to keep the challenge for future participants masked. Note that we submitted only the cosmology constraints for the favored model B for the official unmasking.

\paragraph{Unmasking criteria}
Our unmasked criteria consist of the series of tests that we have described so far. To summarize, we first test the Gaussianity of our summary statistics. As a result, we removed $k$NN bins where the CDF value is less than 0.05 or greater than 0.95. Then we construct our covariance matrix and make sure the matrices are invertible and have reasonable condition numbers. 

The second set of tests were associated to the emulators. Specifically, we ensure that mean emulator error computed on the hold-out tests is sub-dominant compared to the statistical error computed from covariances. In principle, because we are adding the emulator error to the final covariance matrix, the emulator error should not bias our results. However, because the emulator error is intrinsically a function of model parameter values and our treatment of emulator errors (only looking at mean error) is approximate, we would like to make sure potential issues with the emulator error do not dominate our final error budget and lead to significant biases. 

Finally, we conduct a suite of consistency tests outlined in the ``consistency checks'' section as the final part of our unmasking criteria. We explored different scale cuts and different HOD models until we found a configuration (model B and $r_p>5h^{-1}$Mpc) that yields acceptable goodness-of-fits and self-consistent results in our cross-validation tests. These validation tests are critical in increasing our confidence in our results. Again, all these steps were described in detail in \citet{knnpaper}.

The $68\%$  marginalized cosmology constraints with our final results using our favored model are shown in Figure~\ref{fig:2D_summary} and Figure~\ref{fig:1D_summary_rsd}. The RD-$k$NN fit achieves a $\chi^2$/d.o.f. = 0.44 whereas the DD-$k$NN fit achieves a $\chi^2$/d.o.f. = 1.23. 

 \paragraph{Caveats}
 The main limitation is that we had to employ aggressive scale cuts because our extended HOD models were not flexible enough to reliably model scales smaller than 5$h^{-1}$Mpc. In \citet{knnpaper}, we argue that this is specifically due to inability to model the small-scale velocities correctly. Thus, limitations with our galaxy--halo connection modeling is our main area of concern and improvement. 

 Another issue that we were aware of before unmasking is the 
 potential mis-characterization of emulator errors since we only accounted for the mean errors calculated on a limited set of hold-out tests. The issue with emulator errors could be mitigated with either running coverage tests or by following a likelihood-free inference approach. Both require a large amount of training data, which requires additional simulations. 

%% file: methods_DensitySplit.tex
\subsection{Density-split clustering\footnote{Authors: Enrique Paillas, Carolina Cuesta-Lazaro.}}
\label{subsec:DSC}

In this section, we describe the methodology for analyzing the density-split clustering statistics (DSC), based on an emulator trained on the \code{AbacusSummit} simulations, which can learn about the dependence of these statistics on cosmology and HOD parameters. More details about the implementation of this emulator can be found in \citet{CuestaLazaro2023:2309.16539}

\subsubsection{Data \& Estimators}

\begin{figure}
    \centering
    \includegraphics[width=0.5\textwidth]{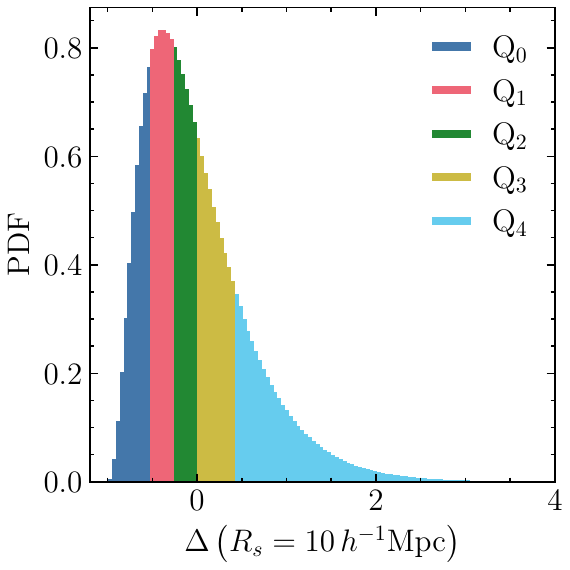}
    \caption{Probability distribution function of the galaxy overdensity measured at random query points from one of the $\Lambda$CDM simulations, where the galaxy density field has been smoothed with a Gaussian filter of radius $R_s = 10\,h^{-1}{\rm Mpc}$. The points are classified into five density quintiles, shown by the different colours.}
    \label{fig:dsc_density_pdf}
\end{figure}

Density-split clustering \citep{Paillas:2021, Paillas:2023} is a method that characterizes the environmental dependence of galaxy clustering, exploiting the sensitivity of each environment to cosmology. We focus our analysis on measuring the DSC statistics on the 10 $\Lambda$CDM mocks in redshift space. 
Here we briefly summarize the algorithm to measure the DSC data vector:

We begin by painting the galaxy overdensity field to a rectangular grid of $5\,h^{-1}{\rm Mpc}$ resolution. We smooth the field with a Gaussian kernel of width $R_s = 10\,h^{-1}{\rm Mpc}$, and we then sample it at $N_{\rm query}$ random query points, where $N_{\rm query}$ is equal to five times the number of galaxies. These query points are then split into five quintiles, according to the value of the overdensity at their locations. Figure~\ref{fig:dsc_density_pdf} shows an example of the measured overdensity distribution and its splitting into quintiles, as measured from one of the $\Lambda$CDM simulations.

\begin{figure*}
    \centering
    \begin{tabular}{cc}
      \includegraphics[width=0.47\textwidth]{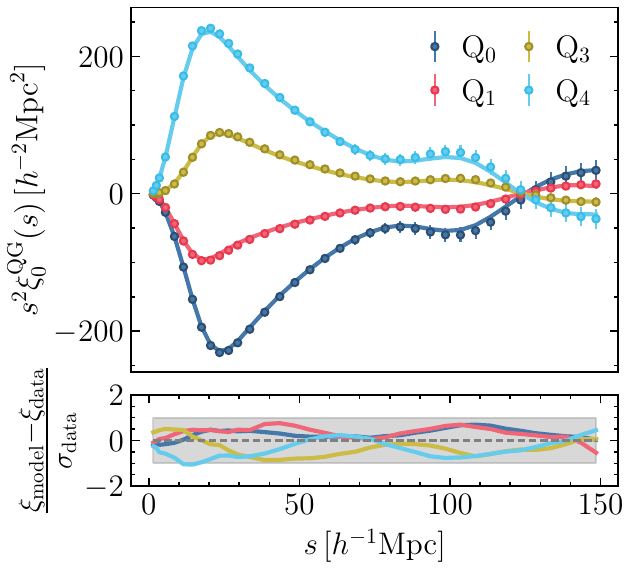}  & \includegraphics[width=0.47\textwidth]
      {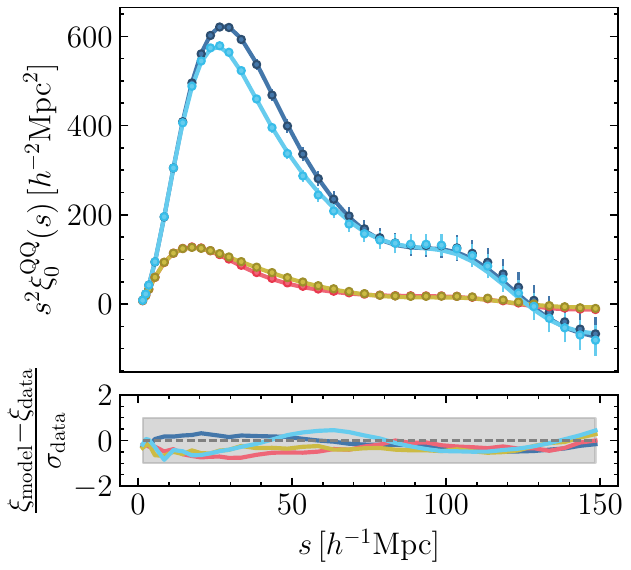} \\
      \includegraphics[width=0.47\textwidth]{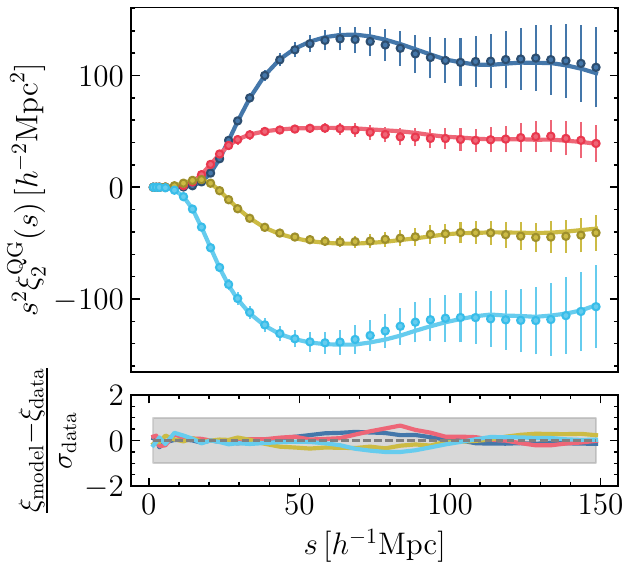}  & \includegraphics[width=0.47\textwidth]
      {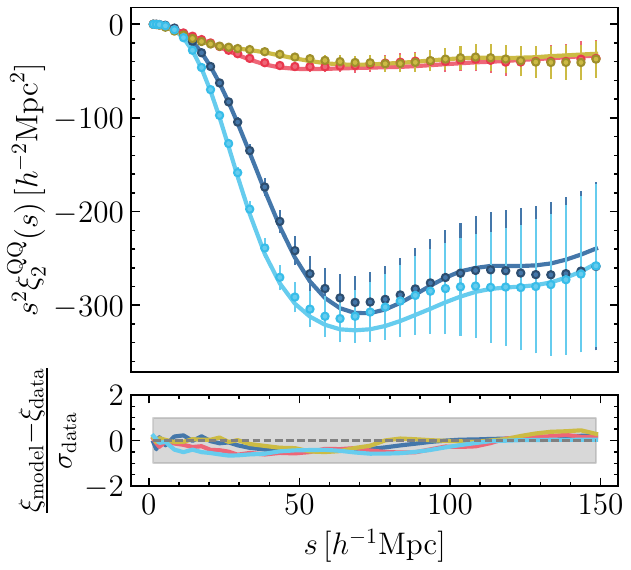}
    \end{tabular}
    \caption{A visualization of the density-split clustering data vectors, along with the best-fit model from our emulator. Markers and solid lines show the data vectors and the emulator predictions, respectively. Left: multipoles of the quintile-galaxy cross-correlation functions. Right: multipoles of the quintile autocorrelation functions. Upper and lower panels show the monopole and quadrupole moments, respectively. We also display the difference between the model and the data, in units of the data error. Each color correspond to a different density quintile, as exemplified in Fig.~\ref{fig:dsc_density_pdf}.}
    \label{fig:dsc_multipoles}
\end{figure*}

We proceed to measure the cross-correlation function between each quintile and the redshift-space galaxy field, as well as the autocorrelation function of the quintiles themselves. Both of these are measured in bins of $s$ and $\mu$, using 241 $\mu$ bins from -1 to 1, and radial bins of scale-dependent widths: $1\, h^{-1}{\rm Mpc}$ width for $s \in [0, 4]\,h^{-1}{\rm Mpc}$, $3\, h^{-1}{\rm Mpc}$ width for $s \in [4, 30]\,h^{-1}{\rm Mpc}$, and $5\, h^{-1}{\rm Mpc}$ width for $s \in [30, 150]\,h^{-1}{\rm Mpc}$. The correlation functions are then decomposed into multipoles. Here, we restrict the analysis to the monopole and quadrupole moments. Figure~\ref{fig:dsc_multipoles} shows the measured multipoles, averaged over the 10 simulations. 

Finally, the data vector is constructed as a concatenation of the monopole and quadrupole moments of the quintile-galaxy cross-correlation functions, $\xi_{0+2}^{\rm QG}(s)$, and the quintile autocorrelation functions , $\xi_{0+2}^{\rm QQ}(s)$,
\begin{equation}
\data_{\rm{DSC}} =\left(\xi_{0}^{\rm Q=(0,1,3,4)G}(s),\xi_{2}^{\rm Q=(0,1,3,4)G}(s),\xi_{0}^{\rm QQ=(00,11,33,44)},\xi_{2}^{\rm QQ=(00,11,33,44)}(s)\right)\,,
\end{equation}
using the same radial binning as describe above. When doing so, we discard the middle quintile, ${\rm Q}_2$, before concatenating the data vector. The reason for this is that the information from all five quintiles is redundant, since the sum of all cross-correlation functions averages to zero by construction. Therefore, all the information available from this method is already contained in the remaining four quintiles.

\begin{figure}
    \centering
    \includegraphics[width=\textwidth]{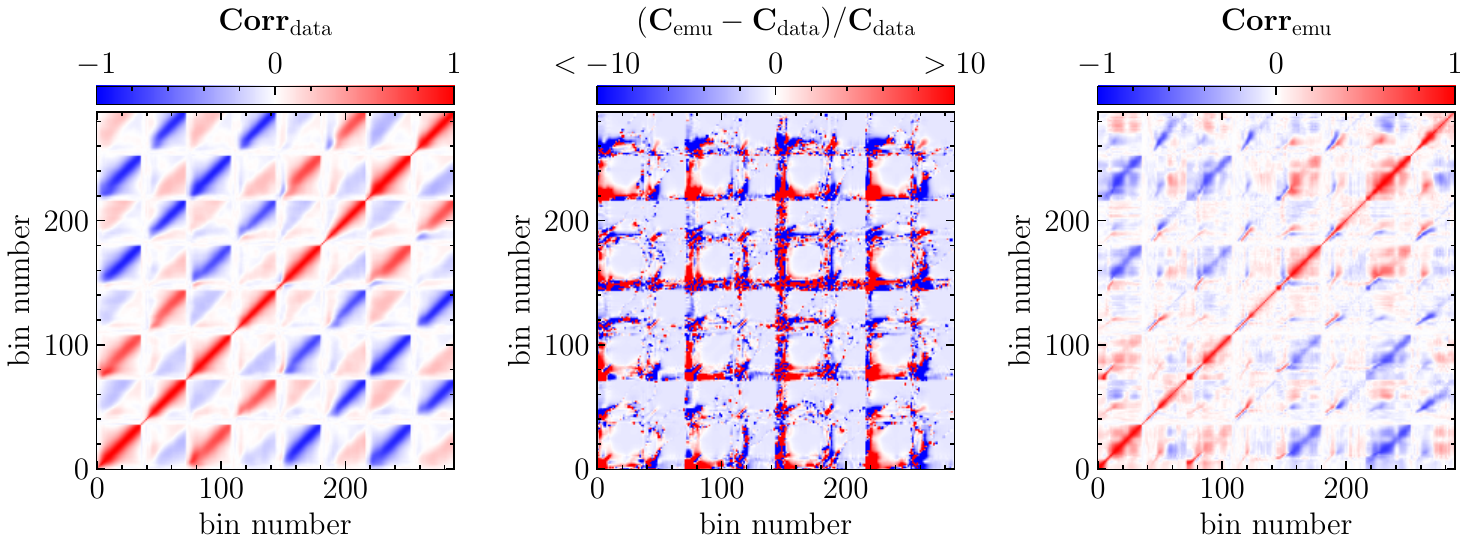}
    \caption{Error sources in the density-split clustering analysis. Left: correlation matrix associated to the sample variance of the data vector. Right: correlation matrix associated to the imperfect emulation of the multipoles. Middle: relative change between the emulator and data covariance. The error associated to the finite size of the training sample, ${\bf C}_{\rm train}$, is the same as the data error in this case, so we do not display it.}
    \label{fig:dsc_covariance}
\end{figure}

\subsubsection{Model}

We model the DSC multipoles using a neural-network based emulator that learns about the dependence of these multipoles on cosmology and HOD parameters. The emulator is trained on the \code{AbacusSummit} suite of simulations, using an extended-HOD parametrization, as described in detail in Sect.~\ref{subsec:Abacus_kNN}.

\paragraph{Parametrization and simulations} We start from the dark matter halo catalogues from the base \code{AbacusSummit} simulation boxes at 85 different cosmologies, which span an 8-dimensional $w_0w_a$CDM parameter space, including the physical cold dark matter density $\omega_{\rm cdm}$, the physical baryon density $\omega_{b}$, the RMS of density fluctuations $\sigma_8$, the spectral index $n_s$, the running of the spectral index $\alpha_{\rm s}$, the effective number of massless relics $N_{\rm eff}$, and the dark energy equation of state parameters, $w_0$ and $w_a$. However, when we perform the cosmological inference, we restrict ourselves to a base-$\Lambda$CDM parameter space, using the priors on cosmological parameters defined in \refeq{dsc_priors_cosmo}.

We generate a Latin hypercube with 100,000 samples of an extended-HOD parameter space, including velocity bias and environment-based secondary bias, using the prior range defined in \refeq{dsc_priors_hod}. Each cosmology is assigned 1,000 HOD parameters from the Latin hypercube, and dark matter haloes are populated with HOD galaxies from those parameters, using the \code{AbacusHOD} code \citep{Yuan:2021hod}. When the resulting number density of an HOD catalogue is larger than the average number density of the $\Lambda$CDM simulations ($n_{\rm gal} \approx 4.5\times 10^{-4}\,(h/{\rm Mpc})^{-3}$), we invoke an incompleteness parameter $f_{\rm ic}$ and subsample the catalogue to match the target number density. If the number density is lower than the target, we simply keep it as is. Galaxy positions are converted to redshift space by perturbing their real-space positions with their peculiar velocities along the line of sight (taken to be one of the axes of the simulation box).




\paragraph{Neural-network emulator}
We split the collection of HOD catalogues into training, validation and test sets. We construct two neural-network emulators, one for each correlation type (quintile-galaxy cross-correlation and quintile autocorrelation). The inputs to the neural network are the cosmological and HOD parameters, whereas the outputs are the concatenated monopole and quadrupole for the four quintiles.

The networks are fully connected, using SiLU activation functions, a mean absolute error loss function, and an AdamW optimizer. To avoid overfitting due to the limited size of our training set, we also introduce dropout. To improve the performance of the model and reduce training time, we decrease the learning rate by a factor of $10$ every $5$ epochs over which the validation loss does not improve, until reaching a minimum learning rate of $10^{-6}$. We use the \code{optuna} framework\footnote{Available at \href{https://github.com/optuna/optuna}{Github.com/optuna/optuna}} to optimize the hyperparameters of the neural network, aiming at minimizing the validation loss. More details related to the architecture and its optimisation can be found in our public repository\footnote{Available at \href{https://github.com/florpi/sunbird}{Github.com/florpi/sunbird}}.

\subsubsection{Inference}

\paragraph{Likelihood} We fit our emulator to the mean measurement of the 10 realizations of the $\Lambda$CDM redshift-space mocks. We assume a Gaussian data likelihood with covariance $\cov_{\rm{tot}}$, which incorporates the total error budget of our analysis. \footnote{The validity of the Gaussian likelihood assumption for density-split clustering is explored in Appendix~A of \cite{CuestaLazaro2023:2309.16539}.}
This covariance includes contributions from sample variance in the data vector ($\cov_{\rm data}$), sample variance in the simulations used for training ($\cov_{\rm train}$) and the intrinsic emulator error ($\cov_{\rm emu}$):

\begin{equation}
   \cov_{\rm tot} = \cov_{\rm data} + \cov_{\rm train} + \cov_{\rm emu}\, .
\end{equation}
\paragraph{Covariance} To calculate $\cov_{\rm data}$, we measure the DSC multipoles from 1500 realizations of the small \code{AbacusSummit} simulations, which are $500\, \Mpch$ on a side and use the baseline \code{AbacusSummit} cosmology. We choose a combination of HOD parameters that produce multipoles that minimize the $\chi^2$ with respect to the data vector measured from the $\Lambda$CDM simulations. We then divide the covariance by a factor of 64 to compensate for the volume difference between the small \code{AbacusSummit} boxes and the $2000\,\Mpch$ $\Lambda$CDM boxes.
We note that here, we have assumed that the noise in the data vector is that associated to a single box of the $\Lambda$CDM simulations. To account for the finite volume of the simulations used for training, we repeat the same procedure to calculate $\cov_{\rm train}$, so effectively $\cov_{\rm train} = \cov_{\rm data}$ for this analysis. Finally, to estimate $\cov_{\rm emu}$, which accounts for the error due to an imperfect emulation of the multipoles, we compare the emulator predictions $\model(\Par^{\rm test})$ against the multipoles measured from our test set of simulations, $\data^{\rm test}$, and compute the covariance of the absolute emulator error
$\Delta^{{\rm test}}= \model(\Par^{\rm test})-\data^{\rm test}$ 
as
\begin{equation} \label{eq:cov_emu}
    {\bf C}_{\rm emu} = \frac{1}{n_{\rm test - 1}}\sum_{k=1}^{n_{\rm test}}\left(\Delta^{{\rm test}_k} - \overline{\Delta^{{\rm test}}} \right) 
   \left(\Delta^{{\rm test}_k} - \overline{\Delta^{{\rm test}}} \right) ^\top\, ,
\end{equation}
where the overline denotes an average across test samples.

Figure~\ref{fig:dsc_covariance} shows the correlation matrices associated to $\cov_{\rm data}$ and $\cov_{\rm emu}$, highlighting how the different quintiles, multipoles and correlation types complement each other. We also display the ratio between the emulator and data covariance, where we can see that, while the data error is the dominant source of error at large scales, the emulator error becomes larger for small separation bins.

\paragraph{Sampler and priors}
We sample the posterior distribution using the \code{Dynesty} nested sampler \citep{dynesty}. We assume the following flat priors on cosmological parameters:
\beq \label{eq:dsc_priors_cosmo}
    &&\omega_{\rm cdm}\sim\unif[0.1032,0.140], \qquad \sigma_8\sim\unif[0.678, 0.938], \qquad n_s\sim \unif[0.9012, 1.025],\nonumber\\
    &&\omega_{\rm b}\sim\unif[0.0207,0.0243]
\eeq
We fix those parameters beyond the base-$\Lambda$CDM model to the following default values: $w_0 = -1$, $w_a = 0$, $N_{\rm ur} = 2.0328$, and $\alpha_s = 0$. For the HOD parameters, we adopt the following flat priors: 
\beq \label{eq:dsc_priors_hod}
    &&\log_{10} M_{\rm cut}\sim\unif[12.0, 13.5], \qquad \log_{10} M_1\sim\unif[12.5, 15.0], \qquad \alpha\sim \unif[0.3, 1.5],\nonumber\\
    &&\alpha_{v, {\rm cen}}\sim\unif[0.0, 1.0],\qquad \alpha_{v,{\rm sat}}\sim\unif[0.0, 2.0], \qquad \log_{10} \sigma\sim \unif[-7.0, 0.0],\nonumber\\
    &&\kappa\sim \unif[0.0, 1.5], \qquad B_{\rm cen}\sim \unif[-0.5, 0.5], \qquad B_{\rm sat}\sim\unif[-1.0, 1.0]
\eeq

\subsubsection{Analysis choices}

\begin{figure}
    \centering
    \includegraphics[width=0.8\textwidth]{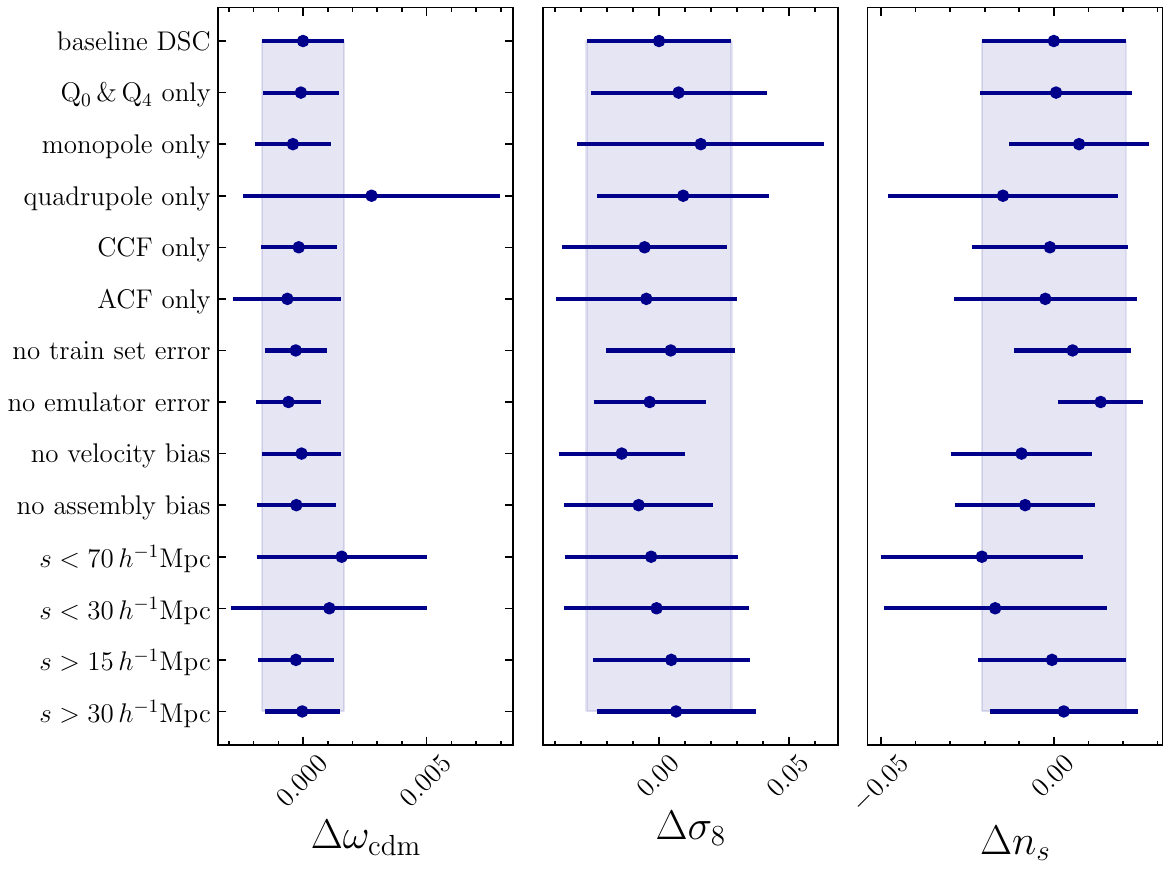}
    \caption{Masked constraints from density-split clustering, assuming different configurations for the analysis. Blue dots show the mean of the marginalized posterior for each parameter (relative to the mean of the baseline constraints), and the error bars show the associated 68\% confidence interval. The blue-shadowed region shows the constraints for the baseline configuration adopted in the paper.}
    \label{fig:dsc_whisker}
\end{figure}

To validate the baseline setup of our analysis, we run our inference pipeline adopting different choices of scales, summary statistics, HOD models and treatment of errors. The results are summarized in Fig.~\ref{fig:dsc_whisker}. Our baseline configuration, shown at the top of each panel, uses scales between $1.0\, \Mpch < s < 151\, \Mpch$, including the monopole and quadrupole of the quintile-galaxy CCF and the quintile ACF, using quintiles ${\rm Q}_{0}$, ${\rm Q}_{1}$, ${\rm Q}_{3}$, and ${\rm Q}_{0}$. The extended HOD model includes both velocity and assembly bias, and the emulator systematic error, as well as the variance associated to our finite training sample, are included in the likelihood calculation. We see that adjusting these choices one by one produces changes in the recovered parameter constraints, but overall, all such variations are consistent within $1\sigma$ of the values inferred from our fiducial configuration, highlighting the robustness of the recovered constraints against the tuning of these settings. Below, we comment on each of these aspects, based on the results from Fig.~\ref{fig:dsc_whisker}.

\paragraph{Scale cut validation} We observe that the recovered constraints do not change significantly when excluding the very small scale information at $s < 30 \Mpch$. This is a direct consequence of the inclusion of the systematic emulator error in our analysis, which starts to dominate the error budget on those scales. We also see that the inclusion of large-scale information adds a significant amount of constraining power. Moreover, we have run independent recovery tests on our test set from the \code{AbacusSummit} simulations, which further confirm that our fiducial range $1\, \Mpch < s < 151\, \Mpch$ seems robust at recovering unbiased cosmological constraints for the volume of the $\Lambda$CDM simulations. Based on this, we choose to adopt this scale range for the baseline analysis. 

\paragraph{Choice of data vector} Concerning the DSC statistics, the quintile-galaxy CCF amounts for most of the information, with the quintile ACF only adding a small contribution to the budget. In terms of the multipoles, the constraints are mostly driven by the monopole. We also see that the quintiles at the extrema, ${\rm Q}_0$ and ${\rm Q}_4$, carry most of the information. This is in line with previous findings based on Fisher forecasts for DSC statistics \citep{Paillas:2023}. 

\paragraph{Choice of HOD model} The choice of the HOD model has little impact on $\omega_{\rm cdm}$, but becomes noticeable for $\sigma_8$ and $n_s$. More specifically, not including central and satellite velocity bias in the model shifts the mean of the posterior for $\sigma_8$ and $n_s$ to smaller values, although well within the $1 \sigma$ region of the baseline setup. A similar trend is seen when excluding assembly bias. Given the findings of previous studies, which suggest that velocity and assembly bias are important for accurately fitting the small-scale clustering from observational data and simulations \citep{Guo_et_al_2015,Yuan:2021hod,2022bYuan}, and to ensure our emulator is robust against the potential presence of such effects in the $\Lambda$CDM simulations fitted in this study, we choose to include them in the default analysis.

\paragraph{Choice of error budget} We see that the exclusion of the emulator systematic error has an important effect in the analysis. Removing this error has a dramatic increase on the precision at which we recover the spectral tilt $n_s$, and a smaller effect is observed for $\omega_{\rm cdm}$ and $\sigma_8$. A similar trend is spotted for the exclusion of the error associated to the training sample, although in a lesser extent. Based on separate tests with the \code{AbacusSummit} simulations, we have concluded that this increase of constraining power comes at the expense of potential biases in the recovered cosmological parameters \citep{CuestaLazaro2023:2309.16539}. Therefore, to ensure the robustness of our inference analysis, we include both error contributions in the likelihood calculation. 

\paragraph{Choice of likelihood}
The choice of a  Gaussian likelihood in our analysis is motivated by previous tests we have carried out with the \code{Quijote} and \code{AbacusSummit} simulations, which show that the likelihood associated to the DSC data vector is indeed close to Gaussian for the scales and the volume of interest \citep[see e.g., Fig.~C1 from]{Paillas:2023}.

\paragraph{Unmasking criteria}
Having explored the different choices of settings in our inference pipeline, we have found that the resulting cosmological constraints are robust against such choices. Moreover, we have verified that our best-fit model provides a good $\chi^2$ to the measured data vector. Therefore, we allow our results to be unmasked, adopting the baseline configuration for our analysis described in the previous section. 

\paragraph{Caveats}
Possible caveats that could lead to unexpected results in our analysis after unmasking include:
\begin{itemize}
    \itemsep0em 
    \item[-] HOD parameters lying outside the range of our priors.
    \item[-] An HOD model including effects that were not implemented in our emulator, such as satellite profile flexibilities to account for baryonic effects.
    \item[-] An overly-conservative error budget in our likelihood, mainly related to the systematic error coming from the emulator predictions. We have chosen to add the emulator error calculated across the full prior range of our parameters, which could lead to an over-estimation of the error bars at small scales, and consequently, to an unnecessary degradation of the precision of our constraints.
    \item[-] Previous analyses have shown that DSC is particularly sensitive to Alcock-Paczynski distortions \citep{Paillas:2021}, which can increase the relative gain of information with respect to standard two-point measurements. Here, we have worked directly with the true galaxy positions in redshift space, ignoring the Alcock-Paczynski effect, which could potetially degrade the relative improvement in parameter constraints found in previous studies.
    \item[-] Due to computational constraints, we have not explored changing the number of quantiles that the density PDF is split into, nor the width of the kernel that is used to smooth the density field. We followed choices made in previous analyses that were calibrated for the number density and redshift of BOSS CMASS. These choices could be sub-optimal for the galaxy sample used in this study, leading to weaker constraints than what could be achieved when varying those settings.
\end{itemize}

%% file: methods_Voids.tex
\subsection{Cosmic Voids\footnote{Authors: Sofia Contarini, Giovanni Verza, Nico Hamaus, Alice Pisani.}}
\label{subsec:voids}
Cosmic voids, the extended under-dense regions of the cosmic web, are becoming an increasingly important probe in the galaxy-clustering and cosmology communities~\citep{Pisani2019,Moresco2022}. While originating in the 3D distribution of matter, voids can also be identified in tracers of the latter, such as galaxies. 
In this work we focus on two summary statistics: the void size function (VSF) and the void-galaxy cross-correlation function (VGCF). The VSF specifies the number of voids as a function of their size and is sometimes referred to as void abundance. The VGCF describes the spatial clustering between the centers of voids and their surrounding galaxies in the form of a cross-correlation function. In the case where all spatial separations are expressed in units of individual void sizes, the VGCF is sometimes simply denoted as a void \emph{stack}. Both techniques have already been applied to SDSS BOSS data~\citep{Dawson2013} to derive constraints on cosmological parameters~\citep[e.g.,][]{Hamaus2016,Hamaus2020,Contarini2023BOSS,Contarini2024}. In this challenge we consider both the VSF and the VGCF measured in the mock light cone.

\subsubsection{Data vector \& estimators}
We rely on the popular void finding algorithm \code{VIDE}\footnote{Available at \href{https://bitbucket.org/cosmicvoids/vide_public/wiki/Home}{Bitbucket.org/cosmicvoids/vide\_public/wiki/Home}}~\citep{Sutter2014}, which is based on \code{ZOBOV}~\citep{Neyrinck2008} to estimate a tracer density field via Voronoi tessellation and makes use of the watershed technique~\citep{Platen2007} to identify local depressions therein. \code{VIDE} has been used extensively for void analyses in both simulations~\citep[e.g.,][]{Hamaus2014a,Chan2014,Pollina2017,Schuster2019,Contarini2019,Verza2019,Kreisch2022} and observational data~\citep[e.g.,][]{Sutter2012b,Hamaus2016,Pollina2019,Fang2019,Contarini2023BOSS,Contarini2024}.
In order to run on light-cone data, \code{VIDE} requires a mask to specify the (mock) survey footprint. For this we generate a healpix mask of resolution $n_\mathrm{side}=128$ from the angular distribution of mock-galaxy coordinates provided in the catalog. This resolution is sufficient to capture the simple rectangular geometry with the given ranges in RA and Dec. The final data product of \code{VIDE} consists of various catalogs containing different void properties, such as their sky coordinates, redshifts, effective radii, shape and density information. Because \code{VIDE} identifies voids in comoving space, angles and redshifts are transformed to comoving coordinates assuming a fiducial cosmology. Therefore, void properties defined in comoving space, such as effective radii, volumes and densities, depend on this assumption. The same fiducial cosmology is used to transform the comoving coordinates of voids back to sky coordinates, which are therefore less affected by model assumptions. Here we focus our analysis on the \lcdm\ light-cone mock catalog, where \code{VIDE} identifies more than $14,000$ voids.

\paragraph{Void size function (VSF)}
The data vector for the VSF is obtained by measuring the number of voids found by \code{VIDE} in logarithmic bins of radius. In our analysis we split the void sample in three equispaced bins of redshift, i.e. $0.812 < z \le 0.968$, $0.968 < z \le 1.124$ and $1.124 < z \le 1.280$.
The division into multiple bins allows us to capture the redshift evolution. While different binning options have been tested (finding consistent results), the subdivision in three bins provides a good balance in terms of number of bins and void statistics within each bin.

We present in Fig.~\ref{fig:cleaned_VSF} (first plot) the number of identified voids as a function of their effective radius $R$, for the three aforementioned redshift bins. $R$ is defined as the radius of a sphere whose volume equals that of the void, which is non-spherical in general. To align the measured VSF with the theoretical model and remove spurious voids, subsequent cleaning is performed. The cleaning procedure is described below.
The VSF data vector is given by
\begin{equation}
\data_{\rm{VSF}} = \left(\widehat{N}( \tilde{R}_{i=1,\ldots, N_r},z_0),\widehat{N}( \tilde{R}_{i=1,\ldots, N_r},z_1),\widehat{N}( \tilde{R}_{i=1,\ldots, N_r},z_2)\right)\,,
\end{equation}
where $ \tilde{R}_{i=1,\ldots, N_r}$ denotes the $N_r$ bins in cleaned void size, and $z_{0,1,2}$ the three tomographic bins described above.

\paragraph{Void-galaxy cross-correlation function (VGCF)}
The data vector of the VGCF is constructed by counting the number of galaxies around each void center as a function of their separation in directions along and perpendicular to the line of sight, $s_\parallel$ and $s_\perp$, measured in units of the effective radius $R$. A stack over a sample of voids then yields a cross-correlation function between void centers and galaxies~\citep{Hamaus2015}. We use the Landy-Szalay estimator~\citep{Landy1993} to obtain a measurement of the VGCF data vector via
\begin{equation}
        \widehat{d}_{{\mathrm{VGCF},i}} \equiv \widehat{\xi}_{\void\gal,i}(s_\perp,s_\parallel) \;,
        \label{VGCF_estimator}
\end{equation}
where we use $15$ linearly spaced bins in each direction, ranging from the void center to about twice the effective void radius. 
The randoms are constructed by sampling from the smoothed redshift distributions of galaxies and voids identified in the mock data, and uniformly across the sky, with an oversampling factor of $10$ (we have checked that a factor of $50$ yields indistinguishable results). In order to impose the identical sky footprint as in the mock data, we apply the same healpix mask to the randoms.

\subsubsection{Model}\label{subsec:voids_modelling}

\paragraph{VSF Parameterization}
The VSF model, developed by \citet{Sheth2004} and further modified in \citet{Jennings2013} to account for the volume conservation in the transition to nonlinearity, relies on the excursion-set theory. The so-called \textit{Vdn} (\textit{volume conserving}) model is defined as: 

\begin{equation}
\frac{\diff n}{\diff \ln R} = \frac{f_{\ln \sigma_{\rm R}}(\sigma_{\rm R})}{V(R)} \, \frac{\diff \ln \sigma_{\rm R}^{-1}}{\diff \ln R_\mathrm{L}} \biggr \rvert_{R_\mathrm{L} = R_\mathrm{L}(R)}\, ,
\end{equation}
where $\sigma_{\rm R}$ represents the root mean square variance of linear matter perturbations on a scale $R_\mathrm{L}$ and where, to compute the number density of voids, we rely on the multiplicity function: 
\begin{equation}\label{eq:multiplicity}
\begin{gathered}
f_{\ln \sigma_{\rm R}}(\sigma_{\rm R}) = 2 \sum_{j=1}^{\infty} \, \exp{\bigg(-\frac{(j \pi x)^2}{2}\bigg)} \, j \pi x^2 \, \sin{\left( j \pi D \right)}\, ,\\
\text{with} \quad D \equiv \frac{|\delta_\mathrm{v}^\mathrm{L}|}{\delta_\mathrm{c}^\mathrm{L} + |\delta_\mathrm{v}^\mathrm{L}|}\, \quad 
\text{and} \quad x \equiv \frac{D}{|\delta_\mathrm{v}^\mathrm{L}|} \sigma_{\rm R} \, ,
\end{gathered}
\end{equation}
Considering linear theory, the function $f_{\ln \sigma_{\rm R}}(\sigma_{\rm R})$ represents the volume fraction of the Universe occupied by cosmic voids, with radii in the range $(R,R+{\rm d}R)$. The additional cosmological dependence of this function lies in the density contrasts defining the formation of dark matter halos and cosmic voids, $\delta_\mathrm{c}^\mathrm{L}$ and $\delta_\mathrm{v}^\mathrm{L}$, respectively.
To predict the number of voids identified from observations of biased tracers, a number of effects have to be considered: 
\begin{enumerate}[label=(\roman*)]
    \item the link of $\delta_\mathrm{v}^\mathrm{L}$ to its nonlinear counterpart $\delta_\mathrm{c}^\mathrm{NL}$ (corresponding to the same threshold but considering nonlinear theory) in the biased tracer field;
    \item the dynamic distortions caused by the peculiar velocities of the tracers; 
    \item the impact of geometric distortions. 
\end{enumerate}

Firstly, to account for the link of $\delta_\mathrm{v}^\mathrm{L}$ to its nonlinear counterpart in the biased tracer field, the sample of voids analyzed has to be prepared to align with the definition of voids given by the VSF theory. For this purpose, we apply to the void catalog built with \code{VIDE} the cleaning algorithm developed by \citet{Ronconi2017}, publicly available in the libraries \code{CosmoBolognaLib}\footnote{Available at \href{https://gitlab.com/federicomarulli/CosmoBolognaLib}{Gitlab.com/federicomarulli/CosmoBolognaLib}} \citep{CBL}. The output of this procedure is a catalog of non-overlapping spheres with cleaned void radius $\tilde{R}$, embedding a fixed density contrast in the tracer density field, $\delta_\mathrm{v, tr}^\mathrm{NL}$. We chose to fix the latter to $\delta_\mathrm{v, tr}^\mathrm{NL}=-0.7$.While different values of this threshold can be used, the selected value leads to a good compromise. Indeed, as discussed for example in \citet{Contarini2022}, a more negative threshold implies a decrease in the effective void radius, which in turn reduces the void sample since only the deepest regions can fulfill the requirement. Moreover it provides voids with fewer tracers, therefore increasing the uncertainty on the radius itself. On the other hand, a less negative threshold provides a void sample composed of larger voids (more likely to overlap and to be excluded from the cleaned sample) and including a higher number of shallower voids, enhancing the likeliness of sample contamination by spurious underdensities \citep[see][]{Neyrinck2008,Cousinou2019}.

We then use a bias relation to convert the density contrast in the tracer density field, $\delta_\mathrm{v, tr}^\mathrm{NL}$, to the corresponding one in the matter distribution, $\delta_\mathrm{v}^\mathrm{NL}$. Galaxy bias in voids cannot be properly described by the large-scale effective galaxy bias
\citep{Contarini2019,Verza2022}. To model it, we therefore rely on a linear function of the effective bias~\citep{Contarini2019,Contarini2021,Contarini2022,Contarini2023BOSS,Contarini2024}:

\begin{equation}\label{eq:thr_conversion}
\begin{gathered}
\delta_\mathrm{v}^\mathrm{NL} = \frac{\delta_\mathrm{v,tr}^\mathrm{NL}}{\mathcal{F}(b_\mathrm{eff}, \sigma_8)} \,, \text{ with} \\
\mathcal{F}(b_\mathrm{eff} \, \sigma_8) = C_\mathrm{slope} \, b_\mathrm{eff} \, \sigma_8 + C_\mathrm{offset} \, .
\end{gathered}
\end{equation}

Here $b_\mathrm{eff}$ represents the galaxy effective bias and, in this analysis, the combined quantity $b_\mathrm{eff} \, \sigma_8$ is derived from the galaxy 2pt correlation function and inserted in the void size function model, marginalizing over it. In particular, we used the \citet{Landy1993} estimator to measure the 2D redshift-space 2pt correlation function and we selected its first three non-null multipole moments. Then we modeled the 2pt correlation function multipoles relying on the prescriptions of \citet{Taruya:2010mx} \citep[see Appendix A of][for further details]{Contarini2023BOSS}.
Secondly, tracer peculiar velocities cause an enlargement along the observer's line of sight, further increasing the mean void radius  ~\citep{Pisani2015b,Hamaus2020,Verza2023}. The above parametrization is used to also correct for this enlargement~\citep{Contarini2022}, allowing us to statistically align voids observed in the redshift-space galaxy distribution to theoretical voids, identified in real space with unbiased tracers.
In \refeq{thr_conversion}, $C_{\rm slope}$ and $C_{\rm offset}$ are redshift-independent coefficients of the linear function $\mathcal{F}$ and can be calibrated by using simulations or, as in this case, marginalized over. Adopting the conservative choice of using wide uniform priors on these parameters, we are able to constrain only the parameter $\Omega_{\rm m}$. We underline, however, that exploiting the information derived from mock catalogs to limit the priors of $C_{\rm slope}$ and $C_{\rm offset}$ would allow us to provide cosmological constraints on $\sigma_8$ as well~\citep{Contarini2023BOSS}.

Thirdly, we account for the Alcock-Paczyński (AP) effect, i.e. geometric distortions due to mismatch between the fiducial and the true cosmology.
Following the prescriptions in \citet{SanchezA2017,Hamaus2020,Correa2021a},
we rescale the observed void radius $R^*$ to $R=q_\parallel^{1/3}q_\perp^{2/3} R^*$, where $q_\parallel$ and $q_\perp$ are defined via
\begin{equation}
\begin{gathered}
r_\parallel = \frac{H^*(z)}{H(z)} \, r^*_\parallel \equiv q_\parallel \, r^*_\parallel \,, \\
r_\perp = \frac{D_{\rm A}(z)}{D^*_{\rm A}(z)} \, r^*_\perp \equiv q_\perp \, r^*_\perp \, ,
\end{gathered}
\end{equation}
and the asterisk symbol indicates quantities evaluated in the fiducial cosmology. Here, $r^*_\parallel$ and $r^*_\perp$ are the observed comoving distances between two points at redshift $z$, projected along the parallel and perpendicular line-of-sight directions, respectively. $H(z)$ is the Hubble parameter and $D_{\rm A}(z)$ is the comoving angular-diameter distance.

\begin{figure*}
    \centering
    \includegraphics[height = 2.15in, width = 2.38in, trim = 10 0 0 5]{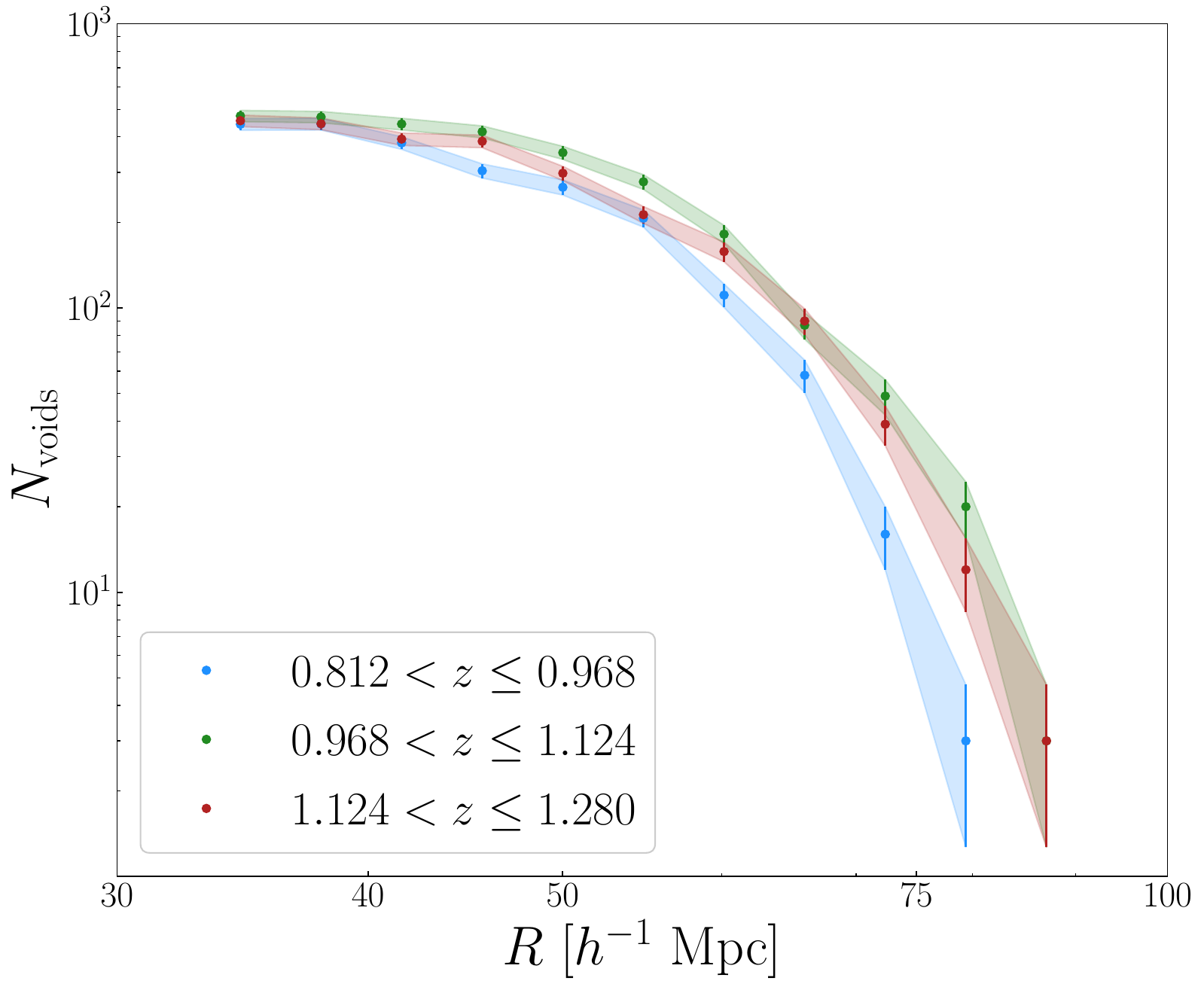}
    \includegraphics[height = 2.3in, width = 4.68in, trim = 10 10 0 0]{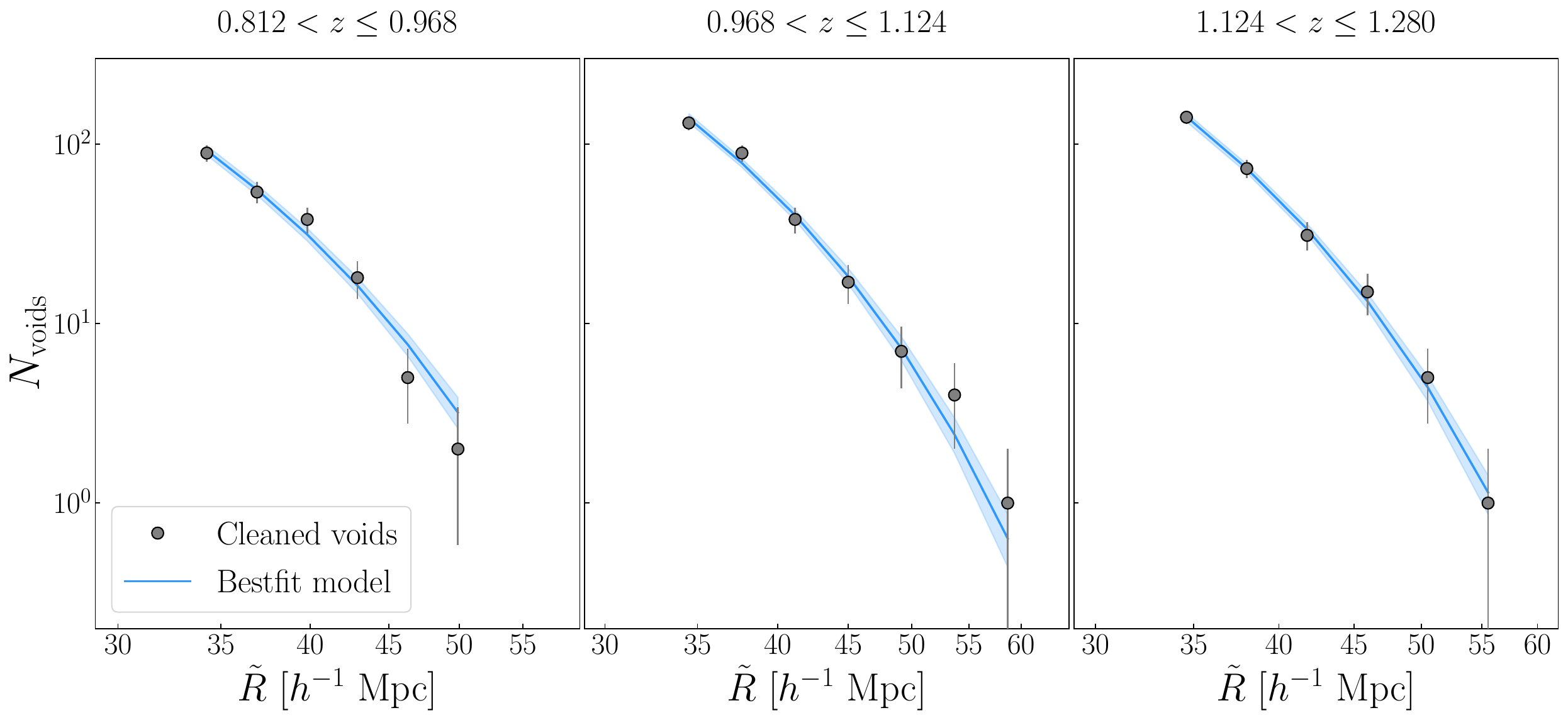}
    \caption{The first plot shows the number of voids as a function of their effective radius $R$ (pre-cleaned VSF), identified with \code{VIDE} in three equispaced redshift bins (see legend). The second, third and fourth plots show the number counts of voids used for the VSF analysis, after the cleaning procedure, as a function of the cleaned radius $\tilde{R}$ (for details on the procedure, see section \ref{subsec:voids_modelling}),
    measured in three equispaced redshift bins. We represent the data with round markers and Poisson error bars. We show in blue the best-fit model derived with the Bayesian analysis described in section \ref{subsec:voids_inference}. The continuous line indicates the median of the model, while the shaded area covers from the $16$-th to the $84$-th percentile.}
    \label{fig:cleaned_VSF}
\end{figure*}

The final VSF model depends on cosmology through the quantities $\sigma_{\rm R}$, $q_\parallel$ and $q_\perp$, i.e. those parameters determining the amplitude of the matter power spectrum and cosmological distances. Additionally, the model depends on the nuisance parameters $C_{\rm slope}$ and $C_{\rm offset}$, as well as on the effective bias of tracers. The latter is estimated from their two-point correlation function and marginalized over. Hence, the model space that we consider for the VSF is spanned by 7 parameters: $\Omega_{\rm m}, \sigma_8, h, \Omega_{\rm b} h^2, n_{\rm s}, C_{\rm slope}, C_{\rm offset}$. In addition, we assume one massive neutrino species of minimal mass. As already mentioned, in this analysis we will provide constraints on $\Omega_{\rm m}$ only, because no calibration of the parameters $C_{\rm slope}$ and $C_{\rm offset}$ is available with the cosmic tracers used in this work. Figure~\ref{fig:cleaned_VSF} presents measurements of the pre-cleaning VSF, and of the cleaned VSF including best-fit models in the considered redshift bins.

\paragraph{VGCF Parameterization}
We rely on three fundamental assumptions to model the VGCF. Firstly, we make use of the cosmological principle, implying that stacked voids are spherically symmetric and hence statistically isotropic in real space. This means that they can be used as \emph{standard spheres} to measure distance ratios via the AP effect. Secondly, we assume that the peculiar velocity field $\mathbf{u}_\void$ around the void center follows linear dynamics according to the linearized continuity equation~\citep{Peebles1980},
\begin{equation}
        \mathbf{u}_\void(\mathbf{r}) = -f(z)\frac{H(z)}{1+z}\,\frac{\mathbf{r}}{r^3}\int_0^r\delta(r')\,r'^{\,2}\,\mathrm{d}r'\;,
        \label{eq:VGCF_velocity}
\end{equation}
where $f(z)$ is the linear growth rate of the matter-density contrast $\delta$. Peculiar velocities cause voids to appear anisotropic in redshift space, because the Doppler effect contributes an additional component to the cosmological redshift along the~$\mathbf{r}_\parallel$ direction. In order to successfully exploit voids as standard spheres, these redshift-space distortions have to be accurately modeled. For this purpose, the validity of \refeq{VGCF_velocity} has been thoroughly tested in simulations and was found to be extremely accurate in void environments~\citep{Hamaus2014b,Schuster2023}, opening up the opportunity to use the AP test with voids as a precision probe of cosmology. However, in this mock challenge we do not have access to the full matter distribution, so our third assumption relies on a linear bias relation between the galaxy- and matter-density contrasts, such that $\xi_{\void\gal}(r) = b\delta(r)$. This relation has been verified in various simulation studies~\citep{Sutter2014a,Pollina2017, Pollina2019,Contarini2019,Ronconi2019,Schuster2023}, finding that the proportionality constant $b$ asymptotes to the linear large-scale tracer bias $b_1$ for large voids, while smaller voids tend to yield higher, but scale-independent values.

\begin{figure*}
    \centering
    \includegraphics[width = 4in, trim = 0 10 0 0]{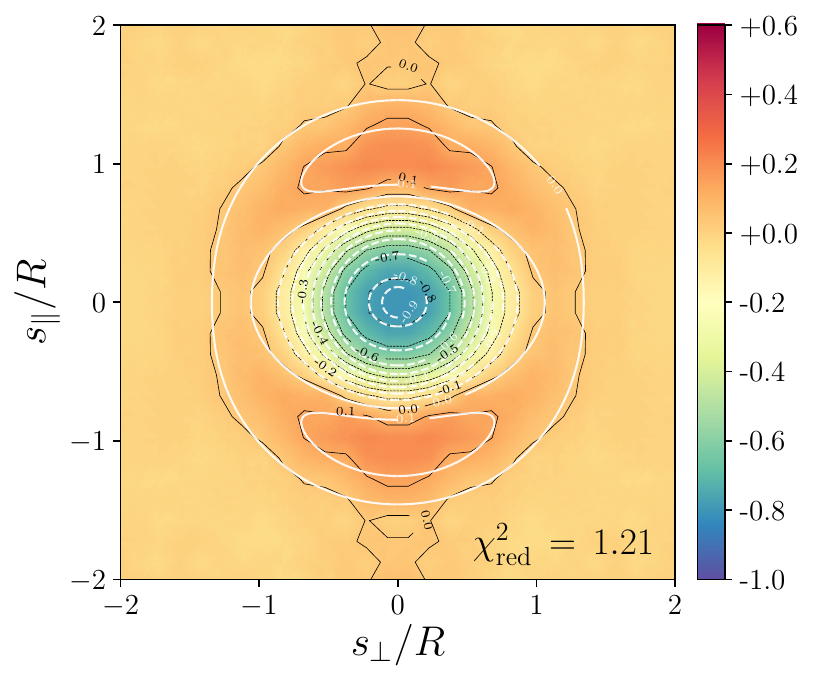}
    \caption{Void-galaxy cross-correlation function in separations along and perpendicular to the line of sight, extracted from the entire \lcdm\ light cone (color scale with black contour lines). White contours show the best-fit model with a reduced chi-square value shown on the bottom right.}
    \label{fig:VIDE_VGCF}
\end{figure*}

With these ingredients we can model the coordinate transformation between the real-space vector $\mathbf{r}$ and redshift-space vector $\mathbf{s}$ for the separation of void centers and galaxies, 
\begin{equation}
        \mathbf{s} = \mathbf{r} + \frac{1+z}{H(z)}\,\mathbf{u}_\parallel = \mathbf{r} - \frac{f(z)}{b(z)}\,\frac{\mathbf{r}_\parallel}{r^3}\int_0^r\xi_{\void\gal}(r')\,r'^{\,2} \mathrm{d}r'\;.
        \label{eq:VGCF_coordinates}
\end{equation}
This requires the VGCF in real space, $\xi_{\void\gal}(r)$, which is not directly observable. We can, however, observe the line-of-sight projected VGCF, $\tilde{\xi}_{\void\gal}(s_\perp) = \int\xi_{\void\gal}(\mathbf{s})\,\mathrm{d}s_\parallel$, which is insensitive to redshift-space distortions, as it only depends on separations $s_\perp$ on the plane of the sky. From it we obtain the real-space VGCF by deprojection via the inverse Abel transform~\citep{Pisani2014,Hawken2017},
\begin{equation}
        \xi_{\void\gal}(r) = -\frac{1}{\pi}\int_r^\infty\frac{\mathrm{d}\tilde{\xi}_{\void\gal}(s_\perp)}{\mathrm{d}s_\perp}\left(s_\perp^2-r^2\right)^{-1/2}\mathrm{d}s_\perp\;.
        \label{eq:VGCF_deprojection}
\end{equation}
\refeqs{VGCF_velocity}{VGCF_deprojection} fully specify the redshift-space VGCF at linear order in perturbation theory and a closed-form expression for $\xi_{\void\gal}(s_\perp,s_\parallel)$ can be derived with the Jacobian of \refeq{VGCF_coordinates}~\citep{Cai2016,Hamaus2017,Hamaus2020}. In order to account for systematic effects, such as inaccuracies in the deprojection procedure and selection effects in the void sample due to sparse sampling of tracers with nonlinear redshift-space distortions (see caveats below), we augment this model with two additional nuisance parameters $\mathcal{M}$ (for monopole-like) and $\mathcal{Q}$ (for quadrupole-like), and adopt the semi-empirical expression derived in \citet{Hamaus2022},
\begin{equation}
        \xi_{\void\gal}(s_\perp,s_\parallel) = \mathcal{M}\left\{\xi_{\void\gal}(r) + \frac{f}{b}\xibar_{\void\gal}(r) + 2\mathcal{Q}\,\frac{f}{b}\mu^2\left[\xi_{\void\gal}(r)-\xibar_{\void\gal}(r)\right]\right\}\;,
        \label{eq:VGCF_model}
\end{equation}
where $\mu=r_\parallel/r$ and $\xibar_{\void\gal}(r) = 3r^{-3}\!\int_0^r\xi_{\void\gal}(r')\,r'^2\,\mathrm{d}r'$. We make use of \refeq{VGCF_coordinates} to map the coordinates from observed redshift space to real space, where the model is evaluated, and include the AP parameters to account for geometric distortions,
\begin{equation}
        r_\perp = q_\perp s_\perp\;,\qquad r_\parallel = q_\parallel s_\parallel\left[1-\frac{1}{3}\frac{f}{b}\mathcal{M}\,\xibar_{\void\gal}(r)\right]^{-1}\!\;,\label{VGCF_model_coordinates}
\end{equation}
which can be solved via iteration upon setting an initial value of $r=s$~\citep{Hamaus2020}. Because we measure void-centric distances in units of the effective void radius $R$, which scales as $q_\parallel^{1/3}q_\perp^{2/3}$ with the AP parameters, only ratios of $q_\perp$ and $q_\parallel$ appear in Eq.~(\ref{VGCF_model_coordinates}). Hence, the final model space for the VGCF is spanned by four parameters; $f/b$, $q_\perp/q_\parallel$, $\mathcal{M}$, and $\mathcal{Q}$. To summarize, our modeling of the VGCF involves the following steps:
\begin{enumerate}[label=(\roman*)]
    \item estimate the real-space VGCF via deprojection of its line-of-sight projected counterpart using \refeq{VGCF_deprojection};
    \item account for dynamic (redshift-space) distortions along the line of sight using \refeq{VGCF_coordinates};
    \item account for geometric (AP) distortions using Eq.~(\ref{VGCF_model_coordinates});
    \item account for systematic effects arising via the deprojection procedure, or via selection effects due to sparse sampling and nonlinear RSD, by including nuisance parameters for the monopole and quadrupole terms in \refeq{VGCF_model}.
\end{enumerate}
 Figure~\ref{fig:VIDE_VGCF} depicts a representation of our data vector, the VGCF in separations along and perpendicular to the line of sight, along with the best-fit model based on \refeq{VGCF_model}.

\subsubsection{Inference}\label{subsec:voids_inference}

\paragraph{VSF Covariance}
We assume the off-diagonal terms of the VSF covariance matrix to be negligible on the basis of previous works, such as \citet{Bayer2021,Kreisch2022,Contarini2023BOSS,Pelliciari2022}. For its diagonal elements, we assume Poisson statistics.

\paragraph{VSF Likelihood Function}
Since the VSF is based on the number counts of cosmic voids, we assume the likelihood to follow the Poisson statistic \citep{sahlen2016,Thiele2023}:

\begin{equation}
\lik[\data_{\rm{VSF}}|\Par] =\prod_{i}\frac{m_{{\rm{VSF}},i}(\Par)^{\widehat{d}_{{\rm{VSF}},i}}\,
    \exp{\left[ -m_{{\rm{VSF}},i}(\Par)\right]}}{\widehat{d}_{{\rm{VSF}},i}!}\,,
\end{equation}
with model parameters $\Par$ listed in Table~\ref{tab:void_priors}.
We use the functions implemented in the \code{CosmoBolognaLib} to sample the likelihood.

\paragraph{VSF Priors}
We impose wide uniform priors for $\Omega_{\rm m}$ and $\sigma_8$, i.e. $\unif[0.05,0.9]$ and $\unif[0.2,2]$, respectively. We impose instead restricted priors on the remaining cosmological parameters, using the uncertainty provided by Planck2018~\citep{Planck2018} multiplied by a factor of three. This latter prior choice is aimed at reducing the parameter-space volume on those parameters that would be weakly constrained by the VSF. This allows us to increase the speed of our MCMC without introducing biased results.
Then we marginalize over the nuisance parameters  $C_{\rm slope}$ and $C_{\rm offset}$, assigning uniform priors as well.
For a summary of all our parameters and priors, see Table~\ref{tab:void_priors}.

\begin{table}
\centering
    \begin{minipage}{.25\linewidth}
      \centering
      VSF \\
    \vspace{-0.25cm}
    \begin{tabular}{lcc}
        \hline
        \hline
        Parameter & Prior         \\
        \hline
        $\Omega_\mathrm{m}$  & $\unif[0.05, 0.9]$ \\
        $\sigma_8$           & $\unif[0.2, 2]$  \\
        $h$                        &    $\unif[0.657,0.683]$ \\
        $\Omega_\mathrm{b} \ h^2$  &    $\unif[0.0216,0.0224]$ \\
        $n_\mathrm{s}$             &    $\unif[0.9535,0.9763]$ \\
        $C_\mathrm{slope}$   & $\unif[-50, 10]$ \\
        $C_\mathrm{offset}$  & $\unif[-10, 50]$  \\
        \hline        
    \end{tabular}
    \end{minipage}%
    \hspace{20pt}
    \begin{minipage}{.25\linewidth}
    \vspace{-1pt}
      \centering
        VGCF \\
        \vspace{-0.25cm}
    \begin{tabular}{lcc}
        \hline
        \hline
        Parameter & Prior         \\
        \hline
        $\Omega_\mathrm{m}$  & $\unif[0, 1]$ \\
        $f/b$                  & $\unif[-10,10]$ \\
        $q_\perp/q_\parallel$  & $\unif[-10,10]$ \\
        $\mathcal{M}$  & $\unif[-10,10]$ \\
        $\mathcal{Q}$  & $\unif[-10,10]$ \\
        \hline  
        \\~\\
    \end{tabular}
    \end{minipage} 
    \caption{Parameters and prior bounds for the VSF (left) and VGCF (right) analyses. }
    \label{tab:void_priors}
\end{table}

\paragraph{VGCF Covariance}
We estimate the covariance matrix $\cov$ of the VGCF by means of jackknife resampling the selected void sample, which is spatially non-overlapping. To this end we apply Eq.~(\ref{VGCF_estimator}) after removal of one void at a time and calculate the covariance over all jackknife samples.


\paragraph{VGCF Likelihood Function}
We adopt a Gaussian likelihood for the VGCF, which has previously been validated on the QPM~\citep{White2014} and PATCHY~\citep{Kitaura2016a} mocks in \citet{Hamaus2017,Hamaus2020}.
We apply the Hartlap correction~\citep{Hartlap2007} to estimate the inverse covariance matrix and maximize the likelihood with respect to the parameter vector $\Par=(f/b,q_\perp/q_\parallel,\mathcal{M},\mathcal{Q})$. To sample the posterior probability distribution of these parameters, we use the affine-invariant MCMC ensemble sampler \code{emcee}~\citep{Foreman-Mackey2019}. Two parameters of our model explicitly depend on cosmology. One via the growth rate, which can be approximated as $f(z)\simeq\Omega_\mathrm{m}^{0.6}(z)$~\citep{Lahav1991}, the other one via the AP parameter $q_\perp/q_\parallel\propto D_\mathrm{A}(z)H(z)$. The latter is particularly well constrained by the AP effect from voids and is therefore a sensitive probe of the expansion history~\citep{Lavaux2012,Hamaus2015}. In a flat \lcdm\ cosmology (neglecting the presence of neutrinos and radiation),
the product $D_\mathrm{A}(z)H(z)$ is fully determined by the parameter $\Omega_\mathrm{m}$. We obtain a posterior on the latter by sampling from a Gaussian likelihood containing measurements of $q_\perp/q_\parallel$ and their uncertainty within each redshift bin.

\paragraph{VGCF Priors}
We impose identical uniform priors for each of our model parameters, $\Par\sim\unif[-10,10]$. These prior boundaries are wide enough to be uninformative, and we have checked that our results are insensitive to this particular choice. For our cosmological parameter of interest in the \lcdm\ light-cone mock, we further impose $\Omega_\mathrm{m}\sim\unif[0,1]$.

\subsubsection{Analysis choices}
\label{sec:void_choices}
\paragraph{VSF scale cut validation}
We select voids above the cleaned radius $\tilde{R}_\mathrm{min}=33 \ h^{-1} \ \mathrm{Mpc}$, for all the considered redshift bins, i.e. $0.812 < z \le 0.968$, $0.968 < z \le 1.124$, $1.124 < z \le 1.280$. 
This selection is applied to avoid those spatial scales affected by a loss of void counts and depends on the resolution of the catalog. 
We measure the void counts as a function of the cleaned radius fixing the value of $\tilde{R}_\mathrm{min}$, and the number of resulting non-empty radius bins are $[6,7,6]$, respectively. This is shown in Fig.~\ref{fig:cleaned_VSF}, where we represent the measured void abundances and the VSF best-fit model derived by the MCMC analysis. The agreement between data and theory is such that all measurements are reproduced by the model within $68\%$ uncertainty.

\begin{figure*}
    \centering
    \includegraphics[height = 2.5in]{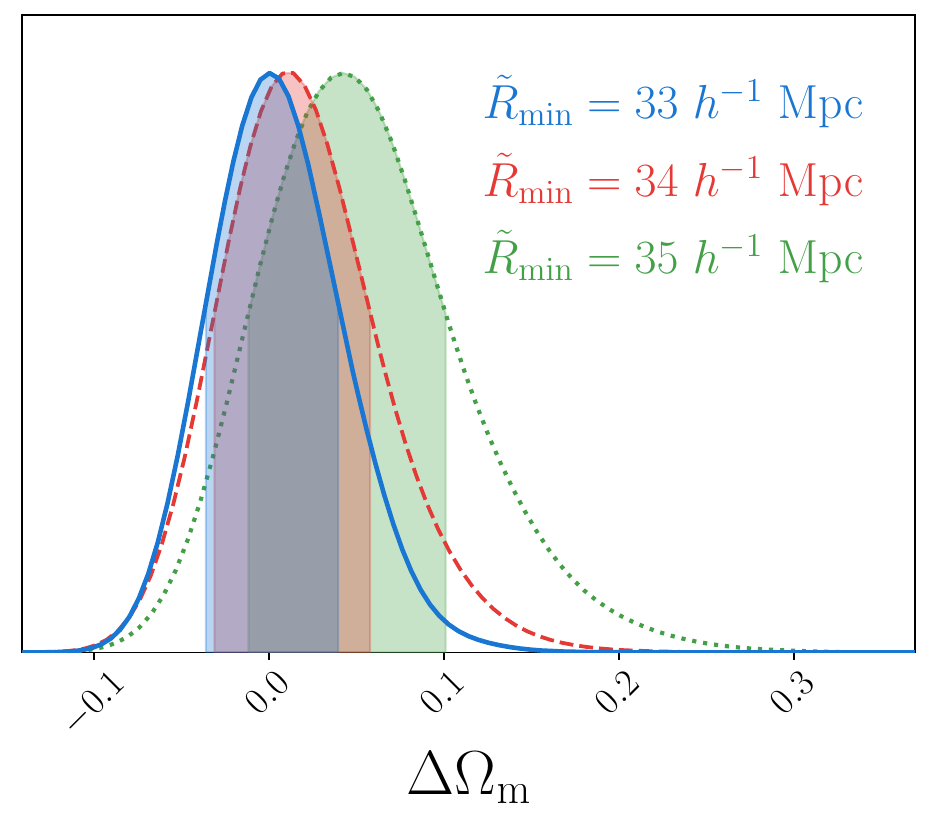}
    \includegraphics[height = 2.5in, trim = 0 5 0 0]{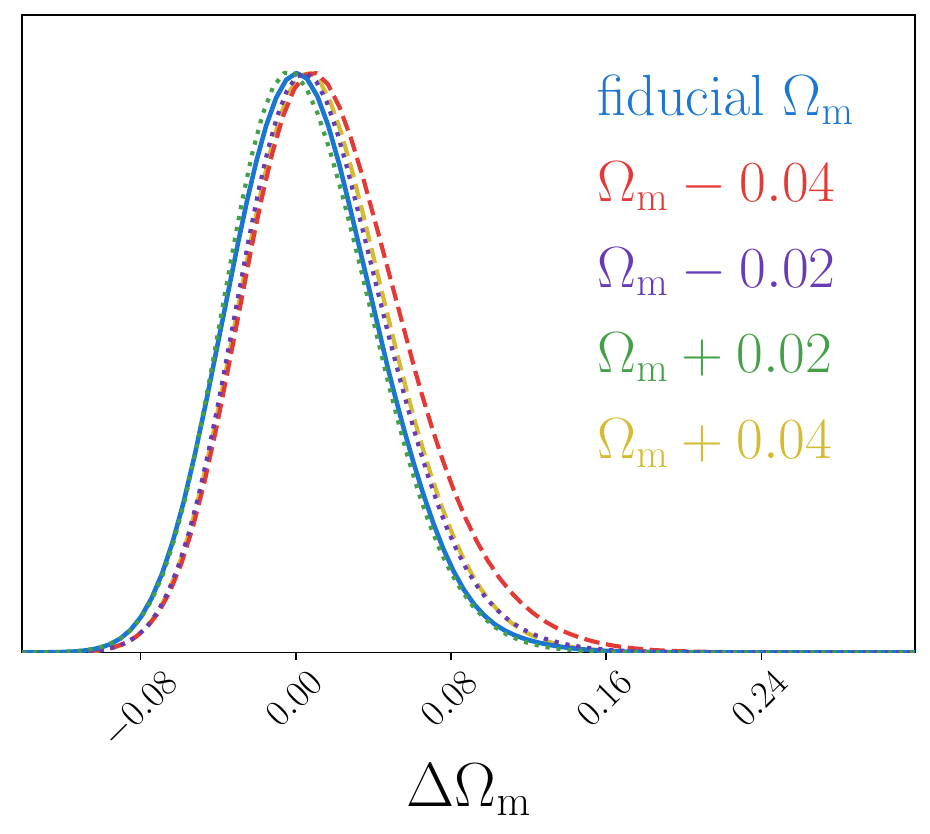}
    \caption{Posterior distribution for the value of $\Omega_{\rm m}$ based on the modeling of the VSF in the $\Lambda$CDM light cone. The mean of the reference posterior (in blue in each panel) has been subtracted to mask the results. We show in the left panel the impact of different cuts on void radius, indicating with shaded regions the $68\%$ confidence interval around the maximum of the posterior distribution. The right panel illustrates the effect of varying the fiducial $\Omega_\mathrm{m}$ value in our analysis pipeline.}
    \label{fig:VSF_posterior}
\end{figure*}

\paragraph{VSF consistency checks} 
We performed a number of tests to check the stability of the results with the choice of the minimum cleaned void radius, i.e. $\tilde{R}_{\rm min}$. As expected, when increasing the value of $\tilde{R}_{\rm min}$ (and thus reducing the number of voids analyzed), the constraining power is reduced. This is shown in the left panel of Fig.~\ref{fig:VSF_posterior}, where we report the posterior distribution of $\Omega_{\rm m}$ derived by imposing $\tilde{R}_{\rm min} = [33, 34, 35] \ h^{-1} \ \mathrm{Mpc}$. Over a certain scale, that is above the region where the void counts incompleteness is stronger, the cosmological constraints on $\Omega_{\rm m}$ are all consistent within $68\%$ of uncertainty. However, we note a moderate dependency of the best-fit value of $\Omega_{\rm m}$ with the choice of $\tilde{R}_{\rm min}$, which could impact the robustness of our results (see caveats below).

We also assessed the impact of assuming different fiducial cosmologies, varying in particular the value of $\Omega_{\rm m}$. The results of this test are reported in the right panel of Fig.~\ref{fig:VSF_posterior}. We find constraints consistent within $68\%$ of uncertainty in all the analyzed cases, i.e. in a range of $\pm 0.04$ from our fiducial value of $\Omega_{\rm m}$. Moreover, we tested different void radius and redshift binning choices, finding consistent results despite the number of cleaned voids being low, thus affected by statistical noise. Finally, we performed consistency checks on the impact of galaxy effective bias modeling and on the bias relation in voids from \refeq{thr_conversion}), finding a negligible impact on the posterior distribution of $\Omega_{\rm m}$.

\paragraph{VGCF scale cut validation}
In contrast to most standard probes of large-scale structure, the VGCF analysis is not restricted to the largest scales. In fact, linear theory accurately describes the VGCF on all scales, even arbitrarily close to the void center. This is because the dynamics inside voids remain linear, in accordance with \refeq{VGCF_velocity}, which has been tested in simulations for a wide range of void sizes down to only a few Mpc in effective radius~\citep{Schuster2023}.

\begin{figure*}
    \centering
    \includegraphics[width = 2.5in, trim = 0 10 0 0]{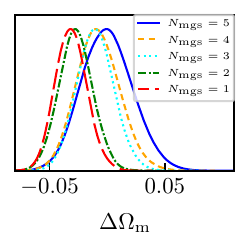}
    \includegraphics[width = 2.5in, trim = 0 10 0 0]{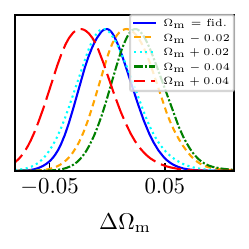}
    \caption{Posterior distributions for the value of $\Omega_\mathrm{m}$ based on the AP test with the VGCF in the \lcdm\ light cone. The mean of the first distribution (solid blue line in each panel) has been subtracted to mask the results. The left panel shows the influence of different cuts on void effective radius in units of the mean galaxy separation (mgs), while the right panel illustrates the impact of variations in the assumed fiducial values for $\Omega_\mathrm{m}$ in our analysis pipeline.}
    \label{fig:VGCF_posterior}
\end{figure*}

However, both the void identification method and estimators for the VGCF are affected by sparse sampling of tracers~\citep{Sutter2014a,Cousinou2019,Schuster2023}. A characteristic scale of a galaxy distribution is its mean galaxy separation (mgs), below which discreteness effects from sparse sampling become important. For example, in this regime random Poisson noise can create or disrupt void detections and counts-in-shell estimators for the VGCF can return biased results due to low or no particle-count statistics. We therefore restrict our VGCF analysis to voids whose effective radius is larger than a multiple of the mean galaxy separation. This can be implemented as a redshift-dependent cut with
\begin{equation}
        R > N_\mathrm{mgs}\left(\frac{4\pi}{3}\bar{n}(z)\right)^{-1/3}\;,
        \label{eq:VGCF_cut}
\end{equation}
where $N_\mathrm{mgs}$ is a tuning parameter and $\bar{n}(z)$ is the mean mock-galaxy density as a function of redshift. The choice of $N_\mathrm{mgs}$ is a trade-off between maximizing the void sample size (lower $N_\mathrm{mgs}$) and minimizing the number of spurious voids (higher $N_\mathrm{mgs}$). In addition, we restrict our sample to the largest $50\%$ of all voids passing this cut. We repeated our analysis for a range of values with $N_\mathrm{mgs}\in[1,5]$ and find consistent results with decreasing uncertainty in our posterior constraints down to a value of $N_\mathrm{mgs}=3$. Below that the posteriors start shifting and stop shrinking (see the left panel of Fig.~\ref{fig:VGCF_posterior}), indicating the onset of stochastic bias or noise in the void sample. However, we select $N_\mathrm{mgs}=5$ as our more conservative default, which results in a minimum void radius of $R\simeq50.6h^{-1}\mathrm{Mpc}$ and corresponds to the data vector shown in Fig.~\ref{fig:VIDE_VGCF}.

Furthermore, we have the option to restrict or subdivide the provided redshift range. The choice of redshift binning is a compromise between detecting the redshift evolution of the VGCF on the one hand, and maintaining sufficient statistics for its estimation on the other. For example, in a $w$CDM cosmology, measuring the AP parameter in multiple redshift bins is necessary to break the degeneracy between $w$ and $\Omega_\mathrm{m}$ entering $D_\mathrm{A}(z)H(z)$. However, in \lcdm\ a single redshift bin is sufficient to determine $\Omega_\mathrm{m}$. Hence, for the \lcdm\ light-cone mock we use the entire redshift range available without binning to estimate the VGCF with reduced statistical noise.

\paragraph{VGCF consistency checks} 
A strategy to gain confidence in our analysis is to investigate its dependence on the assumed fiducial cosmology. We began using a flat \lcdm\ cosmology with Planck2015~\citep{Planck2015} parameters, both in the void finder and for the construction of the VGCF data vector. The best-fit cosmology obtained from our posterior is then used to update our fiducial cosmology and this procedure is repeated until convergence. In practice, only one such iteration was necessary in updating our void catalog to obtain a stable result within our error margins. The right panel of Fig.~\ref{fig:VGCF_posterior} shows the impact of varying our fiducial value of $\Omega_\mathrm{m}$ when constructing the VGCF data vector and running our analysis pipeline. It is evident that the posterior distribution moves diametrically to changes in the fiducial value of $\Omega_\mathrm{m}$. This allows us to identify a fiducial value which agrees most closely with its corresponding posterior mean. We select this posterior as our final result.

\paragraph{Unmasking criteria}
We calculate the reduced $\chi^2$ statistic for the best-fit model of the VGCF with parameters $\Par^*$ as
\begin{equation}
        \chi^2_\mathrm{red} = -\frac{2}{N_\mathrm{dof}}\log\lik(\data_{\rm{VGCF}}|\Par^*) \;,
        \label{VGCF_chi2}
\end{equation}
with $N_\mathrm{dof} = N_\mathrm{bin} - N_\mathrm{par}$ degrees of freedom, where $N_\mathrm{bin}$ is the number of bins in the data and $N_\mathrm{par}$ the number of model parameters. A value close to unity indicates a satisfactory model fit, but does not guarantee unbiased parameter constraints. However, combined with the iteration strategy over fiducial cosmologies described above, it can be used to indicate convergence to an optimal result.

\paragraph{Caveats}
This section describes failure modes for our analysis that could impact the final result for one or both the considered void statistics. These can entail small shifts for the posterior, or complete failure of predicted values. 

\begin{itemize}

\item Void catalogs can be contaminated by spurious voids due to sparse sampling~\citep[e.g.,][]{Neyrinck2008,Sutter2014a}. 
Recent work has shown that small voids are more likely to be spurious~\citep{Pisani2015b,Cousinou2019,Schuster2023}, therefore our analysis cuts based on void size should reduce the probability to include spurious voids in both the VSF and VGCF measurements.
An additional purpose of this cut in the VSF analysis is to exclude spatial scales that are affected by incompleteness, i.e. loss of void counts \citep{Pisani2015a}. This occurs due to the sparsity of tracers and is currently not modeled by the VSF theory.
The posteriors of both analyses mildly depend on the choice of these cuts, we therefore caution that they can have an impact on our cosmological constraints. This additional source of error is so far neither quantified, nor included as a systematic error in our analysis.

\item Despite the geometric distortion correction applied, our final results exhibit a mild dependence on the fiducial cosmology assumed. This indicates residual geometric effects that may play a role, but currently this dependence is not quantified as a potential systematic error.

\item Based on previous studies \citep{Pelliciari2022,Contarini2023BOSS} the VSF analysis assumes the off-diagonal terms of the covariance matrix to be null. However, we acknowledge the possibility that this assumption could have a minor influence on our results. Also, the assumed Poisson errors associated with the VSF are generally smaller than the true uncertainty related to this measure, so the errors on cosmological parameters may be underestimated.
The jackknife approach to determine the covariance of the VGCF can be limited by the finite sample size of voids, which typically leads to an overestimation of uncertainties~\citep{Hamaus2022}. Moreover, a super-sample covariance is not included in our analysis, but we expect it to have a negligible impact on the final cosmological constraints~(for the VSF, see \citet{Bayer2022}). 
\item Possible systematic errors related to the void finding and cleaning algorithms can also have an impact. A void sample identified via sparse tracers in redshift space is subject to various selection effects~\citep[e.g.,][]{Correa2021a,Correa2021b}. While we can limit the impact of sparse sampling via \refeq{VGCF_cut}, residual selection biases may remain present in our sample. A full exploration of their correlation with various void properties and their mitigation is beyond the scope of this work, however.

\item The impact of HOD models on void statistics has barely been investigated in the literature so far. Because these models are usually tuned to reproduce the two-point statistics of galaxies, their performance on void statistics is largely unknown and potentially less accurate. For example, a particular choice of HOD model may produce a strong Finger-of-God effect, which results in elongated features along the line of sight. We do observe some elongation beyond the void boundaries in the VGCF shown in Fig.~\ref{fig:VIDE_VGCF}. While this may be statistically insignificant, Fingers of God could impact the inner shape of the VGCF resulting in overly flattened contours. In turn, this would bias the AP parameter. Furthermore, the HOD scheme may affect the VSF modeling via the bias-in-void parametrization in \refeq{thr_conversion}.

\item The deprojection based on \refeq{VGCF_deprojection} applied to the VGCF is exact for noiseless data, but can amplify statistical fluctuations in noisy data by large amounts~\citep{Pisani2014,Hamaus2020}. The survey characteristics, like tracer density, redshift range, and sky coverage provided in the challenge mocks are more restricted than those available in BOSS, or expected by Euclid, for example. This limits the total number of bins (e.g., in redshift and effective radius) we can divide our void sample by to a single one. We therefore expect stronger systematic errors due to deprojection effects in this analysis.

\end{itemize}

\noindent Finally, we emphasize that the current analysis of the presented void statistics does not make
use of any external simulation suites that can be used for (nuisance-) parameter calibration, template fitting, and the training of emulators. Such tools are helpful to significantly reduce parameter biases and uncertainties, but require a substantial amount of pre-processing and computing resources. We leave such more extensive approaches to future work.
\paragraph{Post-unmasking studies} After unmasking, we realized that the informative prior on $h,\, n_{\rm s},\, \omega_{\rm b}$ adopted by the baseline VSF analysis excluded the true cosmology of the light-cone mock. In subsequent analyses, we repeated the VSF analysis with (a) the informative cosmology prior re-centered to include the true cosmology and (b) the broad cosmology prior of the \code{BACCO} analysis (\refeq{BACCO_priors_cosmo}). The central value of $\Omega_{\rm{m}}$ is robust to both prior variations, and the wide prior broadens the marginalized $1$-$\sigma$ uncertainty on $\Omega_{\rm{m}}$ from $12.7\%$ to $15.6\%$.
\paragraph{Lessons learned}
One of the most relevant lessons learned in this challenge concerns the importance of gaining confidence for the treatment of small voids. In particular, performing a masked analysis reveals that when we are uncertain of the outcome, it is tempting to be overly conservative with scale cuts. Our first tests based on less conservative cuts resulted in best-fit cosmologies with a significantly different mean galaxy number density as compared to the one provided by the challenge. Although we were aware that this mismatch could have had multiple origins apart from the background cosmology (e.g. the HOD prescription), we still opted for very cautious scale cuts due to this observation.

In hindsight, however, it is reassuring to realize that the initial results based on larger void samples that extend down to smaller effective radii actually lie closer to the truth with smaller error bars (see the left panels of Figs.~\ref{fig:VSF_posterior} and~\ref{fig:VGCF_posterior}). This suggests we could have been more ambitious with including smaller voids in our analyses, but also that our level of confidence in them was too low for doing so. For the future, it demonstrates the importance for increasing our understanding with regards to the systematics of small voids, so that their constraining power can be fully exploited.

%% file: contributions.tex
All authors contributed to the interpretation of results and reviewed the manuscript. Author contributions to individual analyses and challenge organization are listed below. \\

Elisabeth Krause -- \Challenge: challenge conceptualization and organization, workshop organization, design of mock catalogs, coordination with analysis teams. Pre-unmasking review of analysis sections, general writing and end-to-end editing.\\[3pt]
Yosuke Kobayashi -- \Challenge: design of mock catalogs, N-body simulations, creation of mock galaxy catalogs, communication with analysis teams.\\[3pt]
Andr\'{e}s N. Salcedo -- \Challenge: design of mock catalogs, creation of mock galaxy catalogs, communication with analysis teams, writing and editing of general sections.\\[3pt]
Mikhail Ivanov -- \Challenge: customized EFT P+B analysis runs for comparisons with other methods, writing and editing of general sections. EFT P+B \Analysis: design of validation tests, analysis runs, interpretation of results, and writing of analysis section.\\[3pt]

Tom Abel -- $k$NN \Analysis: 
design of validation tests, interpretation of results, post-unmasking studies. \\[3pt]
Kazuyuki Akitsu -- EFT P+B \Analysis: analysis runs, interpretation of results, and writing of analysis section.\\
Raul Angulo -- \code{BACCO} P \Analysis: design of validation tests, interpretation of results, post-unmasking studies.\\[3pt]
Giovanni Cabass -- EFT P+B \Analysis: design of validation tests and interpretation of results.\\[3pt]
Sofia Contarini -- Void \Analysis: development of analysis pipeline, design of validation tests, analysis runs, interpretation of results, and writing of analysis section.\\[3pt]
Carolina Cuesta-Lazaro -- Density-Split \Analysis: design of validation tests, analysis runs, interpretation of results, and writing of analysis section.\\[3pt]
ChangHoon Hahn -- SBI P+B \Analysis: design of validation tests, analysis runs, interpretation of results, and writing of analysis section.\\[3pt]
Nico Hamaus -- Void \Analysis: development of analysis pipeline, design of validation tests, analysis runs, interpretation of results, writing of analysis section, and analysis team coordination.\\[3pt]
Donghui Jeong -- \Challenge: challenge conceptualization and workshop organization.\\[3pt]
Chirag Modi -- SBI P+B \Analysis: design of validation tests, analysis runs, interpretation of results, and writing of analysis section.\\[3pt]
Nhat-Minh Nguyen -- \Challenge: General writing and end-to-end editing. EFT FBI \Analysis: development of analysis pipeline, design of validation tests, analysis runs, interpretation of results, post-unmasking studies, and writing of analysis section. \\[3pt]
Takahiro Nishimishi -- \Challenge: development of N-body simulation code \code{GINKAKU}, optimization of accuracy parameters for the use in the challenge, writing of simulation section.\\[3pt]
Enrique Paillas -- Density-Split \Analysis: design of validation tests, analysis runs, interpretation of results, and writing of analysis section.\\[3pt]
Marcos Pellejero Iba\~nez -- \code{BACCO} P \Analysis: analysis team coordination, design of validation tests, analysis runs, interpretation of results, post-unmasking studies, and writing of analysis section.\\[3pt]
Oliver H.~E. Philcox -- EFT P+B \Analysis: design of validation tests, analysis runs, interpretation of results, and writing of analysis section.\\[3pt]
Alice Pisani -- Void \Analysis: design of validation tests, analyses interpretation, interpretation of results, writing of analysis section, and analysis team coordination.\\[3pt]
Fabian Schmidt -- \Challenge: challenge conceptualization and coordination, workshop organization. EFT FBI \Analysis: development of analysis pipeline, design of validation test, interpretation of results, post-unmasking studies.\\[3pt]
Satoshi Tanaka -- \Challenge: development and tuning of N-body simulation code \code{GINKAKU}.\\[3pt]
Giovanni Verza -- Void \Analysis: design of validation tests, analysis runs, interpretation of results, and writing of analysis section.\\[3pt]
Sihan Yuan -- \Challenge: writing of HOD overview. $k$NN \Analysis: design of validation tests, analysis runs, interpretation of results, and writing of analysis section.\\[3pt]
Matteo Zennaro -- \code{BACCO} P \Analysis: design of validation tests, interpretation of results, post-unmasking studies.

%% file: beyond2pt_apj.bbl
\begin{thebibliography}{}
\expandafter\ifx\csname natexlab\endcsname\relax\def\natexlab#1{#1}\fi
\providecommand{\url}[1]{\href{#1}{#1}}
\providecommand{\dodoi}[1]{doi:~\href{http://doi.org/#1}{\nolinkurl{#1}}}
\providecommand{\doeprint}[1]{\href{http://ascl.net/#1}{\nolinkurl{http://ascl.net/#1}}}
\providecommand{\doarXiv}[1]{\href{https://arxiv.org/abs/#1}{\nolinkurl{https://arxiv.org/abs/#1}}}

\bibitem[{Abidi \& Baldauf(2018)}]{Abidi:2018eyd}
Abidi, M.~M., \& Baldauf, T. 2018, JCAP, 07, 029,
  \dodoi{10.1088/1475-7516/2018/07/029}

\bibitem[{Aghamousa {et~al.}(2016)}]{DESI:2016fyo}
Aghamousa, A., {et~al.} 2016.
\newblock \doarXiv{1611.00036}

\bibitem[{{Alam} {et~al.}(2021){Alam}, {Aubert}, {Avila}, {Balland},
  {Bautista}, {Bershady}, {Bizyaev}, {Blanton}, {Bolton}, {Bovy}, {Brinkmann},
  {Brownstein}, {Burtin}, {Chabanier}, {Chapman}, {Choi}, {Chuang}, {Comparat},
  {Cousinou}, {Cuceu}, {Dawson}, {de la Torre}, {de Mattia}, {Agathe}, {des
  Bourboux}, {Escoffier}, {Etourneau}, {Farr}, {Font-Ribera}, {Frinchaboy},
  {Fromenteau}, {Gil-Mar{\'\i}n}, {Le Goff}, {Gonzalez-Morales},
  {Gonzalez-Perez}, {Grabowski}, {Guy}, {Hawken}, {Hou}, {Kong}, {Parker},
  {Klaene}, {Kneib}, {Lin}, {Long}, {Lyke}, {de la Macorra}, {Martini},
  {Masters}, {Mohammad}, {Moon}, {Mueller}, {Mu{\~n}oz-Guti{\'e}rrez}, {Myers},
  {Nadathur}, {Neveux}, {Newman}, {Noterdaeme}, {Oravetz}, {Oravetz},
  {Palanque-Delabrouille}, {Pan}, {Paviot}, {Percival}, {P{\'e}rez-R{\`a}fols},
  {Petitjean}, {Pieri}, {Prakash}, {Raichoor}, {Ravoux}, {Rezaie}, {Rich},
  {Ross}, {Rossi}, {Ruggeri}, {Ruhlmann-Kleider}, {S{\'a}nchez}, {S{\'a}nchez},
  {S{\'a}nchez-Gallego}, {Sayres}, {Schneider}, {Seo}, {Shafieloo}, {Slosar},
  {Smith}, {Stermer}, {Tamone}, {Tinker}, {Tojeiro}, {Vargas-Maga{\~n}a},
  {Variu}, {Wang}, {Weaver}, {Weijmans}, {Y{\`e}che}, {Zarrouk}, {Zhao},
  {Zhao}, \& {Zheng}}]{2021PhRvD.103h3533A}
{Alam}, S., {Aubert}, M., {Avila}, S., {et~al.} 2021, \prd, 103, 083533,
  \dodoi{10.1103/PhysRevD.103.083533}

\bibitem[{Alcock \& Paczynski(1979)}]{Alcock:1979mp}
Alcock, C., \& Paczynski, B. 1979, Nature, 281, 358, \dodoi{10.1038/281358a0}

\bibitem[{{Alsing} {et~al.}(2019){Alsing}, {Charnock}, {Feeney}, \&
  {Wandelt}}]{Alsing2019}
{Alsing}, J., {Charnock}, T., {Feeney}, S., \& {Wandelt}, B. 2019, \mnras, 488,
  4440, \dodoi{10.1093/mnras/stz1960}

\bibitem[{{Anbajagane} {et~al.}(2022){Anbajagane}, {Aung}, {Evrard}, {Farahi},
  {Nagai}, {Barnes}, {Cui}, {Dolag}, {McCarthy}, {Rasia}, \&
  {Yepes}}]{Anbajagane_et_al_2022}
{Anbajagane}, D., {Aung}, H., {Evrard}, A.~E., {et~al.} 2022, \mnras, 510,
  2980, \dodoi{10.1093/mnras/stab3587}

\bibitem[{Andrews {et~al.}(2022)Andrews, Jasche, Lavaux, \&
  Schmidt}]{Andrews:2022nvv}
Andrews, A., Jasche, J., Lavaux, G., \& Schmidt, F. 2022.
\newblock \doarXiv{2203.08838}

\bibitem[{{Angulo} \& {White}(2010)}]{AnguloWhite2010}
{Angulo}, R.~E., \& {White}, S.~D.~M. 2010, \mnras, 405, 143,
  \dodoi{10.1111/j.1365-2966.2010.16459.x}

\bibitem[{{Angulo} {et~al.}(2021){Angulo}, {Zennaro}, {Contreras}, {Aric{\`o}},
  {Pellejero-Iba{\~n}ez}, \& {St{\"u}cker}}]{Angulo_2021}
{Angulo}, R.~E., {Zennaro}, M., {Contreras}, S., {et~al.} 2021, \mnras, 507,
  5869, \dodoi{10.1093/mnras/stab2018}

\bibitem[{Babi\'c {et~al.}(2024)Babi\'c, Schmidt, \& Tucci}]{Babic:2024wph}
Babi\'c, I., Schmidt, F., \& Tucci, B. 2024.
\newblock \doarXiv{2407.01524}

\bibitem[{Balaguera-Antol\'\i{}nez
  {et~al.}(2023)}]{Balaguera-Antolinez:2022xko}
Balaguera-Antol\'\i{}nez, A., {et~al.} 2023, Astron. Astrophys., 673, A130,
  \dodoi{10.5281/zenodo.7330776}

\bibitem[{Baldauf {et~al.}(2015)Baldauf, Mirbabayi, Simonovi\'c, \&
  Zaldarriaga}]{Baldauf:2015xfa}
Baldauf, T., Mirbabayi, M., Simonovi\'c, M., \& Zaldarriaga, M. 2015, Phys.
  Rev. D, 92, 043514, \dodoi{10.1103/PhysRevD.92.043514}

\bibitem[{Baldauf {et~al.}(2016{\natexlab{a}})Baldauf, Mirbabayi, Simonovi\'c,
  \& Zaldarriaga}]{Baldauf:2016sjb}
---. 2016{\natexlab{a}}.
\newblock \doarXiv{1602.00674}

\bibitem[{Baldauf {et~al.}(2016{\natexlab{b}})Baldauf, Schaan, \&
  Zaldarriaga}]{Baldauf:2015tla}
Baldauf, T., Schaan, E., \& Zaldarriaga, M. 2016{\natexlab{b}}, JCAP, 03, 017,
  \dodoi{10.1088/1475-7516/2016/03/017}

\bibitem[{{Banerjee} \& {Abel}(2021)}]{kNN2021}
{Banerjee}, A., \& {Abel}, T. 2021, \mnras, 500, 5479,
  \dodoi{10.1093/mnras/staa3604}

\bibitem[{Baumann {et~al.}(2012)Baumann, Nicolis, Senatore, \&
  Zaldarriaga}]{Baumann:2010tm}
Baumann, D., Nicolis, A., Senatore, L., \& Zaldarriaga, M. 2012, JCAP, 07, 051,
  \dodoi{10.1088/1475-7516/2012/07/051}

\bibitem[{{Bayer} {et~al.}(2022){Bayer}, {Liu}, {Terasawa}, {Barreira},
  {Zhong}, \& {Feng}}]{Bayer2022}
{Bayer}, A.~E., {Liu}, J., {Terasawa}, R., {et~al.} 2022, arXiv e-prints,
  arXiv:2210.15647.
\newblock \doarXiv{2210.15647}

\bibitem[{{Bayer} {et~al.}(2023){Bayer}, {Liu}, {Terasawa}, {Barreira},
  {Zhong}, \& {Feng}}]{2023PhRvD.108d3521B}
---. 2023, \prd, 108, 043521, \dodoi{10.1103/PhysRevD.108.043521}

\bibitem[{Bayer {et~al.}(2023)Bayer, Seljak, \& Modi}]{Bayer:2023rmj}
Bayer, A.~E., Seljak, U., \& Modi, C. 2023, in {40th International Conference
  on Machine Learning}.
\newblock \doarXiv{2307.09504}

\bibitem[{{Bayer} {et~al.}(2021){Bayer}, {Villaescusa-Navarro}, {Massara},
  {Liu}, {Spergel}, {Verde}, {Wandelt}, {Viel}, \& {Ho}}]{Bayer2021}
{Bayer}, A.~E., {Villaescusa-Navarro}, F., {Massara}, E., {et~al.} 2021, \apj,
  919, 24, \dodoi{10.3847/1538-4357/ac0e91}

\bibitem[{{Behroozi} {et~al.}(2019){Behroozi}, {Wechsler}, {Hearin}, \&
  {Conroy}}]{Behroozi_et_al_2019}
{Behroozi}, P., {Wechsler}, R.~H., {Hearin}, A.~P., \& {Conroy}, C. 2019,
  \mnras, 488, 3143, \dodoi{10.1093/mnras/stz1182}

\bibitem[{Behroozi {et~al.}(2013)Behroozi, Wechsler, \& Wu}]{Behroozi:2011ju}
Behroozi, P.~S., Wechsler, R.~H., \& Wu, H.-Y. 2013, Astrophys. J., 762, 109,
  \dodoi{10.1088/0004-637X/762/2/109}

\bibitem[{{Beltz-Mohrmann} {et~al.}(2023){Beltz-Mohrmann}, {Szewciw},
  {Berlind}, \& {Sinha}}]{Beltz-Mohrmann_et_al_2023}
{Beltz-Mohrmann}, G.~D., {Szewciw}, A.~O., {Berlind}, A.~A., \& {Sinha}, M.
  2023, \apj, 948, 100, \dodoi{10.3847/1538-4357/acc576}

\bibitem[{{Berlind} \& {Weinberg}(2002)}]{Berlind_2002}
{Berlind}, A.~A., \& {Weinberg}, D.~H. 2002, \apj, 575, 587,
  \dodoi{10.1086/341469}

\bibitem[{Blas {et~al.}(2016{\natexlab{a}})Blas, Garny, Ivanov, \&
  Sibiryakov}]{Blas:2015qsi}
Blas, D., Garny, M., Ivanov, M.~M., \& Sibiryakov, S. 2016{\natexlab{a}}, JCAP,
  07, 052, \dodoi{10.1088/1475-7516/2016/07/052}

\bibitem[{Blas {et~al.}(2016{\natexlab{b}})Blas, Garny, Ivanov, \&
  Sibiryakov}]{Blas:2016sfa}
---. 2016{\natexlab{b}}, JCAP, 07, 028, \dodoi{10.1088/1475-7516/2016/07/028}

\bibitem[{{Blot} {et~al.}(2019){Blot}, {Crocce}, {Sefusatti}, {Lippich},
  {S{\'a}nchez}, {Colavincenzo}, {Monaco}, {Alvarez}, {Agrawal}, {Avila},
  {Balaguera-Antol{\'\i}nez}, {Bond}, {Codis}, {Dalla Vecchia}, {Dorta},
  {Fosalba}, {Izard}, {Kitaura}, {Pellejero-Ibanez}, {Stein}, {Vakili}, \&
  {Yepes}}]{Blot_2019}
{Blot}, L., {Crocce}, M., {Sefusatti}, E., {et~al.} 2019, \mnras, 485, 2806,
  \dodoi{10.1093/mnras/stz507}

\bibitem[{Bouchet {et~al.}(1995)Bouchet, Colombi, Hivon, \&
  Juszkiewicz}]{Bouchet:1994xp}
Bouchet, F.~R., Colombi, S., Hivon, E., \& Juszkiewicz, R. 1995, Astron.
  Astrophys., 296, 575.
\newblock \doarXiv{astro-ph/9406013}

\bibitem[{Brieden {et~al.}(2021)Brieden, Gil-Mar\'\i{}n, \&
  Verde}]{Brieden:2021edu}
Brieden, S., Gil-Mar\'\i{}n, H., \& Verde, L. 2021, JCAP, 12, 054,
  \dodoi{10.1088/1475-7516/2021/12/054}

\bibitem[{Brieden {et~al.}(2022)Brieden, Gil-Mar\'\i{}n, \&
  Verde}]{Brieden:2022ieb}
---. 2022, JCAP, 06, 005, \dodoi{10.1088/1475-7516/2022/06/005}

\bibitem[{Brinckmann \& Lesgourgues(2019)}]{Brinckmann:2018cvx}
Brinckmann, T., \& Lesgourgues, J. 2019, Phys. Dark Univ., 24, 100260,
  \dodoi{10.1016/j.dark.2018.100260}

\bibitem[{Brooks \& Gelman(1998)}]{GelmanRubin}
Brooks, S.~P., \& Gelman, A. 1998, Journal of Computational and Graphical
  Statistics, 7, 434, \dodoi{10.1080/10618600.1998.10474787}

\bibitem[{Buchert(1992)}]{Buchert:1992ya}
Buchert, T. 1992, Mon. Not. Roy. Astron. Soc., 254, 729

\bibitem[{{Bullock} {et~al.}(2002){Bullock}, {Wechsler}, \&
  {Somerville}}]{2002MNRAS.329..246B}
{Bullock}, J.~S., {Wechsler}, R.~H., \& {Somerville}, R.~S. 2002, \mnras, 329,
  246, \dodoi{10.1046/j.1365-8711.2002.04959.x}

\bibitem[{Cabass(2021)}]{Cabass:2020jqo}
Cabass, G. 2021, JCAP, 01, 067, \dodoi{10.1088/1475-7516/2021/01/067}

\bibitem[{Cabass \& Schmidt(2020)}]{Cabass:2019lqx}
Cabass, G., \& Schmidt, F. 2020, JCAP, 04, 042,
  \dodoi{10.1088/1475-7516/2020/04/042}

\bibitem[{{Cabass} \& {Schmidt}(2020)}]{2020JCAP...07..051C}
{Cabass}, G., \& {Schmidt}, F. 2020, \jcap, 2020, 051,
  \dodoi{10.1088/1475-7516/2020/07/051}

\bibitem[{Cabass {et~al.}(2024)Cabass, Simonovi\'c, \&
  Zaldarriaga}]{Cabass:2023nyo}
Cabass, G., Simonovi\'c, M., \& Zaldarriaga, M. 2024, Phys. Rev. D, 109,
  043526, \dodoi{10.1103/PhysRevD.109.043526}

\bibitem[{{Cai} {et~al.}(2016){Cai}, {Taylor}, {Peacock}, \&
  {Padilla}}]{Cai2016}
{Cai}, Y.-C., {Taylor}, A., {Peacock}, J.~A., \& {Padilla}, N. 2016, \mnras,
  462, 2465, \dodoi{10.1093/mnras/stw1809}

\bibitem[{{Cannon} {et~al.}(2022){Cannon}, {Ward}, \& {Schmon}}]{Cannon2022}
{Cannon}, P., {Ward}, D., \& {Schmon}, S.~M. 2022, arXiv e-prints,
  arXiv:2209.01845, \dodoi{10.48550/arXiv.2209.01845}

\bibitem[{Carrasco {et~al.}(2012)Carrasco, Hertzberg, \&
  Senatore}]{Carrasco:2012cv}
Carrasco, J. J.~M., Hertzberg, M.~P., \& Senatore, L. 2012, JHEP, 09, 082,
  \dodoi{10.1007/JHEP09(2012)082}

\bibitem[{{Chan} {et~al.}(2014){Chan}, {Hamaus}, \& {Desjacques}}]{Chan2014}
{Chan}, K.~C., {Hamaus}, N., \& {Desjacques}, V. 2014, \prd, 90, 103521,
  \dodoi{10.1103/PhysRevD.90.103521}

\bibitem[{Chen {et~al.}(2021)Chen, Vlah, Castorina, \& White}]{Chen:2020zjt}
Chen, S.-F., Vlah, Z., Castorina, E., \& White, M. 2021, JCAP, 03, 100,
  \dodoi{10.1088/1475-7516/2021/03/100}

\bibitem[{Chen {et~al.}(2020)Chen, Vlah, \& White}]{Chen:2020fxs}
Chen, S.-F., Vlah, Z., \& White, M. 2020, JCAP, 07, 062,
  \dodoi{10.1088/1475-7516/2020/07/062}

\bibitem[{Chen {et~al.}(2022)Chen, Vlah, \& White}]{Chen:2021wdi}
---. 2022, JCAP, 02, 008, \dodoi{10.1088/1475-7516/2022/02/008}

\bibitem[{{Cheng} {et~al.}(2024){Cheng}, {Marques}, {Grand{\'o}n}, {Thiele},
  {Shirasaki}, {M{\'e}nard}, \& {Liu}}]{2024arXiv240416085C}
{Cheng}, S., {Marques}, G.~A., {Grand{\'o}n}, D., {et~al.} 2024, arXiv
  e-prints, arXiv:2404.16085, \dodoi{10.48550/arXiv.2404.16085}

\bibitem[{Chudaykin {et~al.}(2021{\natexlab{a}})Chudaykin, Dolgikh, \&
  Ivanov}]{Chudaykin:2020ghx}
Chudaykin, A., Dolgikh, K., \& Ivanov, M.~M. 2021{\natexlab{a}}, Phys. Rev. D,
  103, 023507, \dodoi{10.1103/PhysRevD.103.023507}

\bibitem[{Chudaykin \& Ivanov(2019)}]{Chudaykin:2019ock}
Chudaykin, A., \& Ivanov, M.~M. 2019, JCAP, 11, 034,
  \dodoi{10.1088/1475-7516/2019/11/034}

\bibitem[{{Chudaykin} \& {Ivanov}(2023)}]{Chudaykin:2022nru}
{Chudaykin}, A., \& {Ivanov}, M.~M. 2023, \prd, 107, 043518,
  \dodoi{10.1103/PhysRevD.107.043518}

\bibitem[{Chudaykin {et~al.}(2020)Chudaykin, Ivanov, Philcox, \&
  Simonovi\'c}]{Chudaykin:2020aoj}
Chudaykin, A., Ivanov, M.~M., Philcox, O. H.~E., \& Simonovi\'c, M. 2020, Phys.
  Rev. D, 102, 063533, \dodoi{10.1103/PhysRevD.102.063533}

\bibitem[{Chudaykin {et~al.}(2021{\natexlab{b}})Chudaykin, Ivanov, \&
  Simonovi\'c}]{Chudaykin:2020hbf}
Chudaykin, A., Ivanov, M.~M., \& Simonovi\'c, M. 2021{\natexlab{b}}, Phys. Rev.
  D, 103, 043525, \dodoi{10.1103/PhysRevD.103.043525}

\bibitem[{{Contarini} {et~al.}(2021){Contarini}, {Marulli}, {Moscardini},
  {Veropalumbo}, {Giocoli}, \& {Baldi}}]{Contarini2021}
{Contarini}, S., {Marulli}, F., {Moscardini}, L., {et~al.} 2021, \mnras, 504,
  5021, \dodoi{10.1093/mnras/stab1112}

\bibitem[{{Contarini} {et~al.}(2023){Contarini}, {Pisani}, {Hamaus}, {Marulli},
  {Moscardini}, \& {Baldi}}]{Contarini2023BOSS}
{Contarini}, S., {Pisani}, A., {Hamaus}, N., {et~al.} 2023, \apj, 953, 46,
  \dodoi{10.3847/1538-4357/acde54}

\bibitem[{{Contarini} {et~al.}(2024){Contarini}, {Pisani}, {Hamaus}, {Marulli},
  {Moscardini}, \& {Baldi}}]{Contarini2024}
---. 2024, \aap, 682, A20, \dodoi{10.1051/0004-6361/202347572}

\bibitem[{{Contarini} {et~al.}(2019){Contarini}, {Ronconi}, {Marulli},
  {Moscardini}, {Veropalumbo}, \& {Baldi}}]{Contarini2019}
{Contarini}, S., {Ronconi}, T., {Marulli}, F., {et~al.} 2019, \mnras, 488,
  3526, \dodoi{10.1093/mnras/stz1989}

\bibitem[{{Contarini} {et~al.}(2022){Contarini}, {Verza}, {Pisani}, {Hamaus},
  {Sahl{\'e}n}, {Carbone}, {Dusini}, {Marulli}, {Moscardini}, {Renzi},
  {Sirignano}, {Stanco}, {Aubert}, {Bonici}, {Castignani}, {Courtois},
  {Escoffier}, {Guinet}, {Kovacs}, {Lavaux}, {Massara}, {Nadathur}, {Pollina},
  {Ronconi}, {Ruppin}, {Sakr}, {Veropalumbo}, {Wandelt}, {Amara}, {Auricchio},
  {Baldi}, {Bonino}, {Branchini}, {Brescia}, {Brinchmann}, {Camera},
  {Capobianco}, {Carretero}, {Castellano}, {Cavuoti}, {Cledassou}, {Congedo},
  {Conselice}, {Conversi}, {Copin}, {Corcione}, {Courbin}, {Cropper}, {Da
  Silva}, {Degaudenzi}, {Dubath}, {Duncan}, {Dupac}, {Ealet}, {Farrens},
  {Ferriol}, {Fosalba}, {Frailis}, {Franceschi}, {Garilli}, {Gillard},
  {Gillis}, {Giocoli}, {Grazian}, {Grupp}, {Guzzo}, {Haugan}, {Holmes},
  {Hormuth}, {Jahnke}, {K{\"u}mmel}, {Kermiche}, {Kiessling}, {Kilbinger},
  {Kunz}, {Kurki-Suonio}, {Laureijs}, {Ligori}, {Lilje}, {Lloro}, {Maiorano},
  {Mansutti}, {Marggraf}, {Markovic}, {Massey}, {Melchior}, {Meneghetti},
  {Meylan}, {Moresco}, {Munari}, {Niemi}, {Padilla}, {Paltani}, {Pasian},
  {Pedersen}, {Percival}, {Pettorino}, {Pires}, {Polenta}, {Poncet}, {Popa},
  {Pozzetti}, {Raison}, {Rhodes}, {Rossetti}, {Saglia}, {Sartoris},
  {Schneider}, {Secroun}, {Seidel}, {Sirri}, {Surace}, {Tallada-Cresp{\'\i}},
  {Taylor}, {Tereno}, {Toledo-Moreo}, {Torradeflot}, {Valentijn}, {Valenziano},
  {Wang}, {Weller}, {Zamorani}, {Zoubian}, {Andreon}, {Maino}, \&
  {Mei}}]{Contarini2022}
{Contarini}, S., {Verza}, G., {Pisani}, A., {et~al.} 2022, \aap, 667, A162,
  \dodoi{10.1051/0004-6361/202244095}

\bibitem[{{Contreras} {et~al.}(2021{\natexlab{a}}){Contreras}, {Angulo}, \&
  {Zennaro}}]{ContrerasAnguloZennaro2020AB}
{Contreras}, S., {Angulo}, R.~E., \& {Zennaro}, M. 2021{\natexlab{a}}, \mnras,
  504, 5205, \dodoi{10.1093/mnras/stab1170}

\bibitem[{{Contreras} {et~al.}(2021{\natexlab{b}}){Contreras}, {Angulo}, \&
  {Zennaro}}]{ContrerasAnguloZennaro2020}
---. 2021{\natexlab{b}}, \mnras, 508, 175, \dodoi{10.1093/mnras/stab2560}

\bibitem[{{Contreras} {et~al.}(2023{\natexlab{a}}){Contreras},
  {Chaves-Montero}, \& {Angulo}}]{Contreras2023}
{Contreras}, S., {Chaves-Montero}, J., \& {Angulo}, R.~E. 2023{\natexlab{a}},
  \mnras, 525, 3149, \dodoi{10.1093/mnras/stad2434}

\bibitem[{{Contreras} {et~al.}(2023{\natexlab{b}}){Contreras}, {Angulo},
  {Springel}, {White}, {Hadzhiyska}, {Hernquist}, {Pakmor}, {Kannan},
  {Hern{\'a}ndez-Aguayo}, {Barrera}, {Ferlito}, {Delgado}, {Bose}, \&
  {Frenk}}]{Contreras_et_al_2023}
{Contreras}, S., {Angulo}, R.~E., {Springel}, V., {et~al.} 2023{\natexlab{b}},
  \mnras, 524, 2489, \dodoi{10.1093/mnras/stac3699}

\bibitem[{{Correa} {et~al.}(2022){Correa}, {Paz}, {Padilla}, {S{\'a}nchez},
  {Ruiz}, \& {Angulo}}]{Correa2021b}
{Correa}, C.~M., {Paz}, D.~J., {Padilla}, N.~D., {et~al.} 2022, \mnras, 509,
  1871, \dodoi{10.1093/mnras/stab3070}

\bibitem[{{Correa} {et~al.}(2021){Correa}, {Paz}, {S{\'a}nchez}, {Ruiz},
  {Padilla}, \& {Angulo}}]{Correa2021a}
{Correa}, C.~M., {Paz}, D.~J., {S{\'a}nchez}, A.~G., {et~al.} 2021, \mnras,
  500, 911, \dodoi{10.1093/mnras/staa3252}

\bibitem[{{Cousinou} {et~al.}(2019){Cousinou}, {Pisani}, {Tilquin}, {Hamaus},
  {Hawken}, \& {Escoffier}}]{Cousinou2019}
{Cousinou}, M.~C., {Pisani}, A., {Tilquin}, A., {et~al.} 2019, Astronomy and
  Computing, 27, 53, \dodoi{10.1016/j.ascom.2019.03.001}

\bibitem[{{Crain} \& {van de Voort}(2023)}]{Crain_vdVoort_rev_2023}
{Crain}, R.~A., \& {van de Voort}, F. 2023, \araa, 61, 473,
  \dodoi{10.1146/annurev-astro-041923-043618}

\bibitem[{{Cranmer} {et~al.}(2020){Cranmer}, {Brehmer}, \&
  {Louppe}}]{Cranmer2020}
{Cranmer}, K., {Brehmer}, J., \& {Louppe}, G. 2020, Proceedings of the National
  Academy of Science, 117, 30055, \dodoi{10.1073/pnas.1912789117}

\bibitem[{{Crocce} {et~al.}(2006){Crocce}, {Pueblas}, \&
  {Scoccimarro}}]{crocce06}
{Crocce}, M., {Pueblas}, S., \& {Scoccimarro}, R. 2006, Mon. Not. Roy. Astron.
  Soc., 373, 369, \dodoi{10.1111/j.1365-2966.2006.11040.x}

\bibitem[{Crocce \& Scoccimarro(2008)}]{Crocce:2007dt}
Crocce, M., \& Scoccimarro, R. 2008, Phys. Rev. D, 77, 023533,
  \dodoi{10.1103/PhysRevD.77.023533}

\bibitem[{Cuesta-Lazaro \& Mishra-Sharma(2023)}]{Cuesta-Lazaro:2023zuk}
Cuesta-Lazaro, C., \& Mishra-Sharma, S. 2023.
\newblock \doarXiv{2311.17141}

\bibitem[{Cuesta-Lazaro {et~al.}(2023)Cuesta-Lazaro, Paillas, Yuan, Cai,
  Nadathur, Percival, Beutler, de~Mattia, Eisenstein, Forero-Sanchez, Padilla,
  Pinon, Ruhlmann-Kleider, Sánchez, Valogiannis, \&
  Zarrouk}]{CuestaLazaro2023:2309.16539}
Cuesta-Lazaro, C., Paillas, E., Yuan, S., {et~al.} 2023.
\newblock \doarXiv{2309.16539}

\bibitem[{D'Amico {et~al.}(2022{\natexlab{a}})D'Amico, Donath, Lewandowski,
  Senatore, \& Zhang}]{DAmico:2022osl}
D'Amico, G., Donath, Y., Lewandowski, M., Senatore, L., \& Zhang, P.
  2022{\natexlab{a}}.
\newblock \doarXiv{2206.08327}

\bibitem[{D'Amico {et~al.}(2022{\natexlab{b}})D'Amico, Donath, Lewandowski,
  Senatore, \& Zhang}]{DAmico:2022ukl}
---. 2022{\natexlab{b}}.
\newblock \doarXiv{2211.17130}

\bibitem[{D'Amico {et~al.}(2020)D'Amico, Gleyzes, Kokron, Markovic, Senatore,
  Zhang, Beutler, \& Gil-Mar\'\i{}n}]{DAmico:2019fhj}
D'Amico, G., Gleyzes, J., Kokron, N., {et~al.} 2020, JCAP, 05, 005,
  \dodoi{10.1088/1475-7516/2020/05/005}

\bibitem[{{D'Amico} {et~al.}(2024){D'Amico}, {Senatore}, {Zhang}, \&
  {Nishimichi}}]{DAmico:2021ymi}
{D'Amico}, G., {Senatore}, L., {Zhang}, P., \& {Nishimichi}, T. 2024, \jcap,
  2024, 037, \dodoi{10.1088/1475-7516/2024/01/037}

\bibitem[{{Dawson} {et~al.}(2013){Dawson}, {Schlegel}, {Ahn}, {Anderson},
  {Aubourg}, {Bailey}, {Barkhouser}, {Bautista}, {Beifiori}, {Berlind},
  {Bhardwaj}, {Bizyaev}, {Blake}, \& et~al.}]{Dawson2013}
{Dawson}, K.~S., {Schlegel}, D.~J., {Ahn}, C.~P., {et~al.} 2013, \aj, 145, 10,
  \dodoi{10.1088/0004-6256/145/1/10}

\bibitem[{{DeRose} {et~al.}(2019){DeRose}, {Wechsler}, {Becker}, {Busha},
  {Rykoff}, {MacCrann}, {Erickson}, {Evrard}, {Kravtsov}, {Gruen}, {Allam},
  {Avila}, {Bridle}, {Brooks}, {Buckley-Geer}, {Carnero Rosell}, {Carrasco
  Kind}, {Carretero}, {Castander}, {Cawthon}, {Crocce}, {da Costa}, {Davis},
  {De Vicente}, {Dietrich}, {Doel}, {Drlica-Wagner}, {Fosalba}, {Frieman},
  {Garcia-Bellido}, {Gutierrez}, {Hartley}, {Hollowood}, {Hoyle}, {James},
  {Krause}, {Kuehn}, {Kuropatkin}, {Lima}, {Maia}, {Menanteau}, {Miller},
  {Miquel}, {Ogando}, {Plazas Malag{\'o}n}, {Romer}, {Sanchez}, {Schindler},
  {Serrano}, {Sevilla-Noarbe}, {Smith}, {Suchyta}, {Swanson}, {Tarle}, \&
  {Vikram}}]{Buzzard}
{DeRose}, J., {Wechsler}, R.~H., {Becker}, M.~R., {et~al.} 2019, arXiv
  e-prints, arXiv:1901.02401, \dodoi{10.48550/arXiv.1901.02401}

\bibitem[{{DeRose} {et~al.}(2023){DeRose}, {Kokron}, {Banerjee}, {Chen},
  {White}, {Wechsler}, {Storey-Fisher}, {Tinker}, \& {Zhai}}]{aemulusnu}
{DeRose}, J., {Kokron}, N., {Banerjee}, A., {et~al.} 2023, \jcap, 2023, 054,
  \dodoi{10.1088/1475-7516/2023/07/054}

\bibitem[{{DESI Collaboration} {et~al.}(2024){DESI Collaboration}, {Adame},
  {Aguilar}, {Ahlen}, {Alam}, {Alexander}, {Alvarez}, {Alves}, {Anand},
  {Andrade}, {Armengaud}, {Avila}, {Aviles}, {Awan}, {Bahr-Kalus}, {Bailey},
  {Baltay}, {Bault}, {Behera}, {BenZvi}, {Bera}, {Beutler}, {Bianchi}, {Blake},
  {Blum}, {Brieden}, {Brodzeller}, {Brooks}, {Buckley-Geer}, {Burtin},
  {Calderon}, {Canning}, {Carnero Rosell}, {Cereskaite}, {Cervantes-Cota},
  {Chabanier}, {Chaussidon}, {Chaves-Montero}, {Chen}, {Chen}, {Claybaugh},
  {Cole}, {Cuceu}, {Davis}, {Dawson}, {de la Macorra}, {de Mattia}, {Deiosso},
  {Dey}, {Dey}, {Ding}, {Doel}, {Edelstein}, {Eftekharzadeh}, {Eisenstein},
  {Elliott}, {Fagrelius}, {Fanning}, {Ferraro}, {Ereza}, {Findlay}, {Flaugher},
  {Font-Ribera}, {Forero-S{\'a}nchez}, {Forero-Romero}, {Frenk},
  {Garcia-Quintero}, {Gazta{\~n}aga}, {Gil-Mar{\'\i}n}, {Gontcho},
  {Gonzalez-Morales}, {Gonzalez-Perez}, {Gordon}, {Green}, {Gruen}, {Gsponer},
  {Gutierrez}, {Guy}, {Hadzhiyska}, {Hahn}, {Hanif}, {Herrera-Alcantar},
  {Honscheid}, {Howlett}, {Huterer}, {Ir{\v{s}}i{\v{c}}}, {Ishak}, {Juneau},
  {Kara{\c{c}}ayl{\i}}, {Kehoe}, {Kent}, {Kirkby}, {Kremin}, {Krolewski},
  {Lai}, {Lan}, {Landriau}, {Lang}, {Lasker}, {Le Goff}, {Le Guillou},
  {Leauthaud}, {Levi}, {Li}, {Linder}, {Lodha}, {Magneville}, {Manera},
  {Margala}, {Martini}, {Maus}, {McDonald}, {Medina-Varela}, {Meisner},
  {Mena-Fern{\'a}ndez}, {Miquel}, {Moon}, {Moore}, {Moustakas}, {Mudur},
  {Mueller}, {Mu{\~n}oz-Guti{\'e}rrez}, {Myers}, {Nadathur}, {Napolitano},
  {Neveux}, {Newman}, {Nguyen}, {Nie}, {Niz}, {Noriega}, {Padmanabhan},
  {Paillas}, {Palanque-Delabrouille}, {Pan}, {Penmetsa}, {Percival}, {Pieri},
  {Pinon}, {Poppett}, {Porredon}, {Prada}, {P{\'e}rez-Fern{\'a}ndez},
  {P{\'e}rez-R{\`a}fols}, {Rabinowitz}, {Raichoor}, {Ram{\'\i}rez-P{\'e}rez},
  {Ramirez-Solano}, {Rashkovetskyi}, {Rezaie}, {Rich}, {Rocher}, {Rockosi},
  {Roe}, {Rosado-Marin}, {Ross}, {Rossi}, {Ruggeri}, {Ruhlmann-Kleider},
  {Samushia}, {Sanchez}, {Saulder}, {Schlafly}, {Schlegel}, {Schubnell}, {Seo},
  {Shafieloo}, {Sharples}, {Silber}, {Slosar}, {Smith}, {Sprayberry}, {Tan},
  {Tarl{\'e}}, {Taylor}, {Trusov}, {Ure{\~n}a-L{\'o}pez}, {Vaisakh}, {Valcin},
  {Valdes}, {Vargas-Maga{\~n}a}, {Verde}, {Walther}, {Wang}, {Wang}, {Weaver},
  {Weaverdyck}, {Wechsler}, {Weinberg}, {White}, {Yu}, {Yu}, {Yuan},
  {Y{\`e}che}, {Zaborowski}, {Zarrouk}, {Zhang}, {Zhao}, {Zhao}, {Zhou},
  {Zhuang}, \& {Zou}}]{2024arXiv240403002D}
{DESI Collaboration}, {Adame}, A.~G., {Aguilar}, J., {et~al.} 2024, arXiv
  e-prints, arXiv:2404.03002, \dodoi{10.48550/arXiv.2404.03002}

\bibitem[{{Desjacques} {et~al.}(2018){Desjacques}, {Jeong}, \&
  {Schmidt}}]{Desjacques2018}
{Desjacques}, V., {Jeong}, D., \& {Schmidt}, F. 2018, \physrep, 733, 1,
  \dodoi{10.1016/j.physrep.2017.12.002}

\bibitem[{Desjacques {et~al.}(2018)Desjacques, Jeong, \&
  Schmidt}]{Desjacques:2016bnm}
Desjacques, V., Jeong, D., \& Schmidt, F. 2018, Phys. Rept., 733, 1,
  \dodoi{10.1016/j.physrep.2017.12.002}

\bibitem[{Doeser {et~al.}(2023)Doeser, Jamieson, Stopyra, Lavaux, Leclercq, \&
  Jasche}]{Doeser:2023yzv}
Doeser, L., Jamieson, D., Stopyra, S., {et~al.} 2023.
\newblock \doarXiv{2312.09271}

\bibitem[{{Dutton} \& {Macci{\`o}}(2014)}]{Dutton2014}
{Dutton}, A.~A., \& {Macci{\`o}}, A.~V. 2014, \mnras, 441, 3359,
  \dodoi{10.1093/mnras/stu742}

\bibitem[{Elsner {et~al.}(2020)Elsner, Schmidt, Jasche, Lavaux, \&
  Nguyen}]{Elsner:2019rql}
Elsner, F., Schmidt, F., Jasche, J., Lavaux, G., \& Nguyen, N.-M. 2020, JCAP,
  01, 029, \dodoi{10.1088/1475-7516/2020/01/029}

\bibitem[{{Euclid Collaboration} {et~al.}(2023){Euclid Collaboration},
  {Pezzotta}, {Moretti}, {Zennaro}, {Moradinezhad Dizgah}, {Crocce},
  {Sefusatti}, {Ferrero}, {Pardede}, {Eggemeier}, {Barreira}, {Angulo},
  {Marinucci}, {Camacho Quevedo}, {de la Torre}, {Alkhanishvili}, {Biagetti},
  {Breton}, {Castorina}, {D'Amico}, {Desjacques}, {Guidi}, {K{\"a}rcher},
  {Oddo}, {Pellejero Ibanez}, {Porciani}, {Pugno}, {Salvalaggio}, {Sarpa},
  {Veropalumbo}, {Vlah}, {Amara}, {Andreon}, {Auricchio}, {Baldi}, {Bardelli},
  {Bender}, {Bodendorf}, {Bonino}, {Branchini}, {Brescia}, {Brinchmann},
  {Camera}, {Capobianco}, {Carbone}, {Cardone}, {Carretero}, {Casas},
  {Castander}, {Castellano}, {Cavuoti}, {Cimatti}, {Congedo}, {Conselice},
  {Conversi}, {Copin}, {Corcione}, {Courbin}, {Courtois}, {Da Silva},
  {Degaudenzi}, {Di Giorgio}, {Dinis}, {Dupac}, {Dusini}, {Ealet}, {Farina},
  {Farrens}, {Fosalba}, {Frailis}, {Franceschi}, {Galeotta}, {Gillis},
  {Giocoli}, {Granett}, {Grazian}, {Grupp}, {Guzzo}, {Haugan}, {Hormuth},
  {Hornstrup}, {Jahnke}, {Joachimi}, {Keih{\"a}nen}, {Kermiche}, {Kiessling},
  {Kilbinger}, {Kitching}, {Kubik}, {Kunz}, {Kurki-Suonio}, {Ligori}, {Lilje},
  {Lindholm}, {Lloro}, {Maiorano}, {Mansutti}, {Marggraf}, {Markovic},
  {Martinet}, {Marulli}, {Massey}, {Medinaceli}, {Mellier}, {Meneghetti},
  {Merlin}, {Meylan}, {Moresco}, {Moscardini}, {Munari}, {Niemi}, {Padilla},
  {Paltani}, {Pasian}, {Pedersen}, {Percival}, {Pettorino}, {Pires}, {Polenta},
  {Pollack}, {Poncet}, {Popa}, {Pozzetti}, {Raison}, {Renzi}, {Rhodes},
  {Riccio}, {Romelli}, {Roncarelli}, {Rossetti}, {Saglia}, {Sapone},
  {Sartoris}, {Schneider}, {Schrabback}, {Secroun}, {Seidel}, {Seiffert},
  {Serrano}, {Sirignano}, {Sirri}, {Stanco}, {Surace}, {Tallada-Cresp{\'\i}},
  {Taylor}, {Tereno}, {Toledo-Moreo}, {Torradeflot}, {Tutusaus}, {Valentijn},
  {Valenziano}, {Vassallo}, {Wang}, {Weller}, {Zamorani}, {Zoubian}, {Zucca},
  {Biviano}, {Bozzo}, {Burigana}, {Colodro-Conde}, {Di Ferdinando}, {Mainetti},
  {Martinelli}, {Mauri}, {Sakr}, {Scottez}, {Tenti}, {Viel}, {Wiesmann},
  {Akrami}, {Allevato}, {Anselmi}, {Baccigalupi}, {Ballardini}, {Bernardeau},
  {Blanchard}, {Borgani}, {Bruton}, {Cabanac}, {Cappi}, {Carvalho},
  {Castignani}, {Castro}, {Ca\textbackslash \{n\}as-Herrera}, {Chambers},
  {Contarini}, {Cooray}, {Coupon}, {Davini}, {De Lucia}, {Desprez}, {Di
  Domizio}, {Dole}, {D{\'\i}az-S{\'a}nchez}, {Escartin Vigo}, {Escoffier},
  {Ferreira}, {Finelli}, {Gabarra}, {Ganga}, {Garc{\'\i}a-Bellido},
  {Giacomini}, {Gozaliasl}, {Hall}, {Ili{\'c}}, {Joudaki}, {Kajava}, {Kansal},
  {Kirkpatrick}, {Legrand}, {Loureiro}, {Macias-Perez}, {Magliocchetti},
  {Mannucci}, {Maoli}, {Martins}, {Matthew}, {Maurin}, {Metcalf}, {Migliaccio},
  {Monaco}, {Morgante}, {Nadathur}, {Walton}, {Patrizii}, {Popa}, {Potter},
  {Pourtsidou}, {P{\"o}ntinen}, {Risso}, {Rocci}, {S{\'a}nchez}, {Sahl{\'e}n},
  {Schneider}, {Sereno}, {Simon}, {Spurio Mancini}, {Steinwagner}, {Testera},
  {Teyssier}, {Toft}, {Tosi}, {Troja}, {Tucci}, {Valiviita}, {Vergani},
  {Verza}, \& {Vielzeuf}}]{Pezzotta_2024}
{Euclid Collaboration}, {Pezzotta}, A., {Moretti}, C., {et~al.} 2023, arXiv
  e-prints, arXiv:2312.00679, \dodoi{10.48550/arXiv.2312.00679}

\bibitem[{{Fang} {et~al.}(2019){Fang}, {Hamaus}, {Jain}, {Pandey}, {Pollina},
  {S{\'a}nchez}, {Kov{\'a}cs}, {Chang}, {Carretero}, {Castander}, {Choi},
  {Crocce}, {DeRose}, {Fosalba}, {Gatti}, {Gazta{\~n}aga}, et~al., \& {DES
  Collaboration}}]{Fang2019}
{Fang}, Y., {Hamaus}, N., {Jain}, B., {et~al.} 2019, \mnras, 490, 3573,
  \dodoi{10.1093/mnras/stz2805}

\bibitem[{Feldman {et~al.}(2001)Feldman, Frieman, Fry, \&
  Scoccimarro}]{Feldman:2000vk}
Feldman, H.~A., Frieman, J.~A., Fry, J.~N., \& Scoccimarro, R. 2001, Phys. Rev.
  Lett., 86, 1434, \dodoi{10.1103/PhysRevLett.86.1434}

\bibitem[{Feroz \& Hobson(2008)}]{multinest1}
Feroz, F., \& Hobson, M.~P. 2008, \mnras, 384, 449,
  \dodoi{10.1111/j.1365-2966.2007.12353.x}

\bibitem[{{Feroz} {et~al.}(2009){Feroz}, {Hobson}, \& {Bridges}}]{multinest2}
{Feroz}, F., {Hobson}, M.~P., \& {Bridges}, M. 2009, \mnras, 398, 1601,
  \dodoi{10.1111/j.1365-2966.2009.14548.x}

\bibitem[{{Feroz} {et~al.}(2019){Feroz}, {Hobson}, {Cameron}, \&
  {Pettitt}}]{multinest3}
{Feroz}, F., {Hobson}, M.~P., {Cameron}, E., \& {Pettitt}, A.~N. 2019, The Open
  Journal of Astrophysics, 2, 10, \dodoi{10.21105/astro.1306.2144}

\bibitem[{{Foreman-Mackey} {et~al.}(2019){Foreman-Mackey}, {Farr}, {Sinha},
  {Archibald}, {Hogg}, {Sanders}, {Zuntz}, {Williams}, {Nelson}, {de
  Val-Borro}, {Erhardt}, {Pashchenko}, \& {Pla}}]{Foreman-Mackey2019}
{Foreman-Mackey}, D., {Farr}, W., {Sinha}, M., {et~al.} 2019, The Journal of
  Open Source Software, 4, 1864, \dodoi{10.21105/joss.01864}

\bibitem[{{Garrison} {et~al.}(2019){Garrison}, {Eisenstein}, \&
  {Pinto}}]{2019Garrison}
{Garrison}, L.~H., {Eisenstein}, D.~J., \& {Pinto}, P.~A. 2019, \mnras, 485,
  3370, \dodoi{10.1093/mnras/stz634}

\bibitem[{{Gatti} {et~al.}(2022){Gatti}, {Jain}, {Chang}, {Raveri},
  {Z{\"u}rcher}, {Secco}, {Whiteway}, {Jeffrey}, {Doux}, {Kacprzak}, {Bacon},
  {Fosalba}, {Alarcon}, {Amon}, {Bechtol}, {Becker}, {Bernstein}, {Blazek},
  {Campos}, {Choi}, {Davis}, {Derose}, {Dodelson}, {Elsner}, {Elvin-Poole},
  {Everett}, {Ferte}, {Gruen}, {Harrison}, {Huterer}, {Jarvis}, {Krause},
  {Leget}, {Lemos}, {Maccrann}, {Mccullough}, {Muir}, {Myles}, {Navarro},
  {Pandey}, {Prat}, {Rollins}, {Roodman}, {Sanchez}, {Sheldon}, {Shin},
  {Troxel}, {Tutusaus}, {Yin}, {Aguena}, {Allam}, {Andrade-Oliveira}, {Annis},
  {Bertin}, {Brooks}, {Burke}, {Carnero Rosell}, {Carrasco Kind}, {Carretero},
  {Cawthon}, {Costanzi}, {da Costa}, {Pereira}, {De Vicente}, {Desai}, {Diehl},
  {Dietrich}, {Doel}, {Drlica-Wagner}, {Eckert}, {Evrard}, {Ferrero},
  {Garc{\'\i}a-Bellido}, {Gaztanaga}, {Giannantonio}, {Gruendl}, {Gschwend},
  {Gutierrez}, {Hinton}, {Hollowood}, {Honscheid}, {James}, {Kuehn},
  {Kuropatkin}, {Lahav}, {Lidman}, {Maia}, {Marshall}, {Melchior}, {Menanteau},
  {Miquel}, {Morgan}, {Palmese}, {Paz-Chinch{\'o}n}, {Pieres}, {Plazas
  Malag{\'o}n}, {Reil}, {Rodriguez-Monroyv}, {Romer}, {Sanchez}, {Schubnell},
  {Serrano}, {Sevilla-Noarbe}, {Smith}, {Soares-Santos}, {Suchyta}, {Tarle},
  {Thomas}, {To}, {Varga}, \& {DES Collaboration}}]{2022PhRvD.106h3509G}
{Gatti}, M., {Jain}, B., {Chang}, C., {et~al.} 2022, \prd, 106, 083509,
  \dodoi{10.1103/PhysRevD.106.083509}

\bibitem[{Gaztanaga {et~al.}(2009)Gaztanaga, Cabre, Castander, Crocce, \&
  Fosalba}]{Gaztanaga:2008sq}
Gaztanaga, E., Cabre, A., Castander, F., Crocce, M., \& Fosalba, P. 2009, Mon.
  Not. Roy. Astron. Soc., 399, 801, \dodoi{10.1111/j.1365-2966.2009.15313.x}

\bibitem[{Gil-Mar\'\i{}n {et~al.}(2017)Gil-Mar\'\i{}n, Percival, Verde,
  Brownstein, Chuang, Kitaura, Rodr\'\i{}guez-Torres, \&
  Olmstead}]{Gil-Marin:2016wya}
Gil-Mar\'\i{}n, H., Percival, W.~J., Verde, L., {et~al.} 2017, Mon. Not. Roy.
  Astron. Soc., 465, 1757, \dodoi{10.1093/mnras/stw2679}

\bibitem[{{Gonzalez-Perez} {et~al.}(2018){Gonzalez-Perez}, {Comparat},
  {Norberg}, {Baugh}, {Contreras}, {Lacey}, {McCullagh}, {Orsi}, {Helly}, \&
  {Humphries}}]{Gonzalez-Perez_et_al_2018}
{Gonzalez-Perez}, V., {Comparat}, J., {Norberg}, P., {et~al.} 2018, \mnras,
  474, 4024, \dodoi{10.1093/mnras/stx2807}

\bibitem[{{Guo} {et~al.}(2015){Guo}, {Zheng}, {Zehavi}, {Dawson}, {Skibba},
  {Tinker}, {Weinberg}, {White}, \& {Schneider}}]{Guo_et_al_2015}
{Guo}, H., {Zheng}, Z., {Zehavi}, I., {et~al.} 2015, \mnras, 446, 578,
  \dodoi{10.1093/mnras/stu2120}

\bibitem[{{Guo} {et~al.}(2016){Guo}, {Zheng}, {Behroozi}, {Zehavi}, {Chuang},
  {Comparat}, {Favole}, {Gottloeber}, {Klypin}, {Prada},
  {Rodr{\'\i}guez-Torres}, {Weinberg}, \& {Yepes}}]{Guo_et_al_2016}
{Guo}, H., {Zheng}, Z., {Behroozi}, P.~S., {et~al.} 2016, \mnras, 459, 3040,
  \dodoi{10.1093/mnras/stw845}

\bibitem[{{Hadzhiyska} {et~al.}(2021){Hadzhiyska}, {Eisenstein}, {Bose},
  {Garrison}, \& {Maksimova}}]{2021Hadzhiyska}
{Hadzhiyska}, B., {Eisenstein}, D., {Bose}, S., {Garrison}, L.~H., \&
  {Maksimova}, N. 2021, arXiv e-prints, arXiv:2110.11408.
\newblock \doarXiv{2110.11408}

\bibitem[{{Hahn}(2020)}]{pyspectrum}
{Hahn}, C. 2020, {pySpectrum: Power spectrum and bispectrum calculator},
  Astrophysics Source Code Library, record ascl:2009.014.
\newblock \doeprint{2009.014}

\bibitem[{Hahn {et~al.}(2019)Hahn, Beutler, Sinha, Berlind, Ho, \&
  Hogg}]{Hahn:2018zja}
Hahn, C., Beutler, F., Sinha, M., {et~al.} 2019, Mon. Not. Roy. Astron. Soc.,
  485, 2956, \dodoi{10.1093/mnras/stz558}

\bibitem[{{Hahn} {et~al.}(2022){Hahn}, {Eickenberg}, {Ho}, {Hou}, {Lemos},
  {Massara}, {Modi}, {Moradinezhad Dizgah}, {R{\'e}galdo-Saint Blancard}, \&
  {Abidi}}]{Hahn2023b}
{Hahn}, C., {Eickenberg}, M., {Ho}, S., {et~al.} 2022, arXiv e-prints,
  arXiv:2211.00723, \dodoi{10.48550/arXiv.2211.00723}

\bibitem[{Hahn {et~al.}(2023)Hahn, Eickenberg, Ho, Hou, Lemos, Massara, Modi,
  Moradinezhad~Dizgah, Parker, \& Blancard}]{Hahn:2023kky}
Hahn, C., Eickenberg, M., Ho, S., {et~al.} 2023.
\newblock \doarXiv{2310.15243}

\bibitem[{{Hahn} {et~al.}(2023){Hahn}, {Eickenberg}, {Ho}, {Hou}, {Lemos},
  {Massara}, {Modi}, {Moradinezhad Dizgah}, {R{\'e}galdo-Saint Blancard}, \&
  {Abidi}}]{Hahn2023a}
{Hahn}, C., {Eickenberg}, M., {Ho}, S., {et~al.} 2023, \jcap, 2023, 010,
  \dodoi{10.1088/1475-7516/2023/04/010}

\bibitem[{{Hamaus} {et~al.}(2017){Hamaus}, {Cousinou}, {Pisani}, {Aubert},
  {Escoffier}, \& {Weller}}]{Hamaus2017}
{Hamaus}, N., {Cousinou}, M.-C., {Pisani}, A., {et~al.} 2017, \jcap, 7, 014,
  \dodoi{10.1088/1475-7516/2017/07/014}

\bibitem[{{Hamaus} {et~al.}(2020){Hamaus}, {Pisani}, {Choi}, {Lavaux},
  {Wandelt}, \& {Weller}}]{Hamaus2020}
{Hamaus}, N., {Pisani}, A., {Choi}, J.-A., {et~al.} 2020, \jcap, 2020, 023,
  \dodoi{10.1088/1475-7516/2020/12/023}

\bibitem[{{Hamaus} {et~al.}(2016){Hamaus}, {Pisani}, {Sutter}, {Lavaux},
  {Escoffier}, {Wandelt}, \& {Weller}}]{Hamaus2016}
{Hamaus}, N., {Pisani}, A., {Sutter}, P.~M., {et~al.} 2016, Physical Review
  Letters, 117, 091302, \dodoi{10.1103/PhysRevLett.117.091302}

\bibitem[{{Hamaus} {et~al.}(2015){Hamaus}, {Sutter}, {Lavaux}, \&
  {Wandelt}}]{Hamaus2015}
{Hamaus}, N., {Sutter}, P.~M., {Lavaux}, G., \& {Wandelt}, B.~D. 2015, \jcap,
  11, 036, \dodoi{10.1088/1475-7516/2015/11/036}

\bibitem[{{Hamaus} {et~al.}(2014{\natexlab{a}}){Hamaus}, {Sutter}, \&
  {Wandelt}}]{Hamaus2014b}
{Hamaus}, N., {Sutter}, P.~M., \& {Wandelt}, B.~D. 2014{\natexlab{a}}, Physical
  Review Letters, 112, 251302, \dodoi{10.1103/PhysRevLett.112.251302}

\bibitem[{{Hamaus} {et~al.}(2014{\natexlab{b}}){Hamaus}, {Wandelt}, {Sutter},
  {Lavaux}, \& {Warren}}]{Hamaus2014a}
{Hamaus}, N., {Wandelt}, B.~D., {Sutter}, P.~M., {Lavaux}, G., \& {Warren},
  M.~S. 2014{\natexlab{b}}, Physical Review Letters, 112, 041304,
  \dodoi{10.1103/PhysRevLett.112.041304}

\bibitem[{{Hamaus} {et~al.}(2022){Hamaus}, {Aubert}, {Pisani}, {Contarini},
  {Verza}, {Cousinou}, {Escoffier}, {Hawken}, {Lavaux}, {Pollina}, {Wandelt},
  {Weller}, {Bonici}, {Carbone}, {Guzzo}, {Kovacs}, {Marulli}, {Massara},
  {Moscardini}, {Ntelis}, {Percival}, {Radinovi{\'c}}, {Sahl{\'e}n}, {Sakr},
  {S{\'a}nchez}, {Winther}, {Auricchio}, {Awan}, {Bender}, {Bodendorf},
  {Bonino}, {Branchini}, {Brescia}, {Brinchmann}, {Capobianco}, {Carretero},
  {Castander}, {Castellano}, {Cavuoti}, {Cimatti}, {Cledassou}, {Congedo},
  {Conversi}, {Copin}, {Corcione}, {Cropper}, {Da Silva}, {Degaudenzi},
  {Douspis}, {Dubath}, {Duncan}, {Dupac}, {Dusini}, {Ealet}, {Ferriol},
  {Fosalba}, {Frailis}, {Franceschi}, {Franzetti}, {Fumana}, {Garilli},
  {Gillis}, {Giocoli}, {Grazian}, {Grupp}, {Haugan}, {Holmes}, {Hormuth},
  {Jahnke}, {Kermiche}, {Kiessling}, {Kilbinger}, {Kitching}, {K{\"u}mmel},
  {Kunz}, {Kurki-Suonio}, {Ligori}, {Lilje}, {Lloro}, {Maiorano}, {Marggraf},
  {Markovic}, {Massey}, {Maurogordato}, {Melchior}, {Meneghetti}, {Meylan},
  {Moresco}, {Munari}, {Niemi}, {Padilla}, {Paltani}, {Pasian}, {Pedersen},
  {Pettorino}, {Pires}, {Poncet}, {Popa}, {Pozzetti}, {Rebolo}, {Rhodes},
  {Rix}, {Roncarelli}, {Rossetti}, {Saglia}, {Schneider}, {Secroun}, {Seidel},
  {Serrano}, {Sirignano}, {Sirri}, {Starck}, {Tallada-Cresp{\'\i}},
  {Tavagnacco}, {Taylor}, {Tereno}, {Toledo-Moreo}, {Torradeflot}, {Valentijn},
  {Valenziano}, {Wang}, {Welikala}, {Zamorani}, {Zoubian}, {Andreon}, {Baldi},
  {Camera}, {Mei}, {Neissner}, \& {Romelli}}]{Hamaus2022}
{Hamaus}, N., {Aubert}, M., {Pisani}, A., {et~al.} 2022, \aap, 658, A20,
  \dodoi{10.1051/0004-6361/202142073}

\bibitem[{{Hand} {et~al.}(2018){Hand}, {Feng}, {Beutler}, {Li}, {Modi},
  {Seljak}, \& {Slepian}}]{nbodykit}
{Hand}, N., {Feng}, Y., {Beutler}, F., {et~al.} 2018, \aj, 156, 160,
  \dodoi{10.3847/1538-3881/aadae0}

\bibitem[{{Harnois-D{\'e}raps} {et~al.}(2021){Harnois-D{\'e}raps}, {Martinet},
  {Castro}, {Dolag}, {Giblin}, {Heymans}, {Hildebrandt}, \&
  {Xia}}]{2021MNRAS.506.1623H}
{Harnois-D{\'e}raps}, J., {Martinet}, N., {Castro}, T., {et~al.} 2021, \mnras,
  506, 1623, \dodoi{10.1093/mnras/stab1623}

\bibitem[{{Hartlap} {et~al.}(2007){Hartlap}, {Simon}, \&
  {Schneider}}]{Hartlap2007}
{Hartlap}, J., {Simon}, P., \& {Schneider}, P. 2007, \aap, 464, 399,
  \dodoi{10.1051/0004-6361:20066170}

\bibitem[{{Hawken} {et~al.}(2017){Hawken}, {Granett}, {Iovino}, {Guzzo},
  {Peacock}, {de la Torre}, {Garilli}, {Bolzonella}, {Scodeggio}, {Abbas},
  {Adami}, \& et~al.}]{Hawken2017}
{Hawken}, A.~J., {Granett}, B.~R., {Iovino}, A., {et~al.} 2017, \aap, 607, A54,
  \dodoi{10.1051/0004-6361/201629678}

\bibitem[{{Hearin} {et~al.}(2023){Hearin}, {Chaves-Montero}, {Alarcon},
  {Becker}, \& {Benson}}]{Hearin_et_al_2023}
{Hearin}, A.~P., {Chaves-Montero}, J., {Alarcon}, A., {Becker}, M.~R., \&
  {Benson}, A. 2023, \mnras, 521, 1741, \dodoi{10.1093/mnras/stad456}

\bibitem[{{Hearin} {et~al.}(2021){Hearin}, {Chaves-Montero}, {Becker}, \&
  {Alarcon}}]{Hearin_et_al_2021}
{Hearin}, A.~P., {Chaves-Montero}, J., {Becker}, M.~R., \& {Alarcon}, A. 2021,
  The Open Journal of Astrophysics, 4, 7, \dodoi{10.21105/astro.2105.05859}

\bibitem[{{Hearin} {et~al.}(2016){Hearin}, {Zentner}, {van den Bosch},
  {Campbell}, \& {Tollerud}}]{Hearin2016}
{Hearin}, A.~P., {Zentner}, A.~R., {van den Bosch}, F.~C., {Campbell}, D., \&
  {Tollerud}, E. 2016, \mnras, 460, 2552, \dodoi{10.1093/mnras/stw840}

\bibitem[{Heitmann {et~al.}(2019)}]{Heitmann:2019ytn}
Heitmann, K., {et~al.} 2019, Astrophys. J. Suppl., 245, 16,
  \dodoi{10.3847/1538-4365/ab4da1}

\bibitem[{{Heydenreich} {et~al.}(2022){Heydenreich}, {Br{\"u}ck}, {Burger},
  {Harnois-D{\'e}raps}, {Unruh}, {Castro}, {Dolag}, \&
  {Martinet}}]{2022A&A...667A.125H}
{Heydenreich}, S., {Br{\"u}ck}, B., {Burger}, P., {et~al.} 2022, \aap, 667,
  A125, \dodoi{10.1051/0004-6361/202243868}

\bibitem[{Ho {et~al.}(2024)}]{Ho:2024whi}
Ho, M., {et~al.} 2024.
\newblock \doarXiv{2402.05137}

\bibitem[{{Hou} {et~al.}(2023){Hou}, {Moradinezhad Dizgah}, {Hahn}, \&
  {Massara}}]{2023JCAP...03..045H}
{Hou}, J., {Moradinezhad Dizgah}, A., {Hahn}, C., \& {Massara}, E. 2023, \jcap,
  2023, 045, \dodoi{10.1088/1475-7516/2023/03/045}

\bibitem[{{Hou} {et~al.}(2024){Hou}, {Moradinezhad Dizgah}, {Hahn},
  {Eickenberg}, {Ho}, {Lemos}, {Massara}, {Modi}, {Parker}, \&
  {R{\'e}galdo-Saint Blancard}}]{2024arXiv240115074H}
{Hou}, J., {Moradinezhad Dizgah}, A., {Hahn}, C., {et~al.} 2024, arXiv
  e-prints, arXiv:2401.15074, \dodoi{10.48550/arXiv.2401.15074}

\bibitem[{{Ishiyama} {et~al.}(2009){Ishiyama}, {Fukushige}, \&
  {Makino}}]{2009PASJ...61.1319I}
{Ishiyama}, T., {Fukushige}, T., \& {Makino}, J. 2009, \pasj, 61, 1319,
  \dodoi{10.1093/pasj/61.6.1319}

\bibitem[{{Ishiyama} {et~al.}(2012){Ishiyama}, {Nitadori}, \&
  {Makino}}]{2012arXiv1211.4406I}
{Ishiyama}, T., {Nitadori}, K., \& {Makino}, J. 2012, arXiv e-prints,
  arXiv:1211.4406.
\newblock \doarXiv{1211.4406}

\bibitem[{{Ishiyama} {et~al.}(2021){Ishiyama}, {Prada}, {Klypin}, {Sinha},
  {Metcalf}, {Jullo}, {Altieri}, {Cora}, {Croton}, {de la Torre},
  {Mill{\'a}n-Calero}, {Oogi}, {Ruedas}, \& {Vega-Mart{\'\i}nez}}]{Uuchu}
{Ishiyama}, T., {Prada}, F., {Klypin}, A.~A., {et~al.} 2021, \mnras, 506, 4210,
  \dodoi{10.1093/mnras/stab1755}

\bibitem[{Ivanov(2021)}]{Ivanov:2021zmi}
Ivanov, M.~M. 2021, Phys. Rev. D, 104, 103514,
  \dodoi{10.1103/PhysRevD.104.103514}

\bibitem[{Ivanov(2023)}]{Ivanov:2022mrd}
---. 2023, {Effective Field Theory for Large-Scale Structure},
  \dodoi{10.1007/978-981-19-3079-9_5-1}

\bibitem[{Ivanov {et~al.}(2024)Ivanov, Cuesta-Lazaro, Mishra-Sharma, Obuljen,
  \& Toomey}]{Ivanov:2024hgq}
Ivanov, M.~M., Cuesta-Lazaro, C., Mishra-Sharma, S., Obuljen, A., \& Toomey,
  M.~W. 2024.
\newblock \doarXiv{2402.13310}

\bibitem[{Ivanov {et~al.}(2023)Ivanov, Philcox, Cabass, Nishimichi,
  Simonovi\'c, \& Zaldarriaga}]{Ivanov:2023qzb}
Ivanov, M.~M., Philcox, O. H.~E., Cabass, G., {et~al.} 2023, Phys. Rev. D, 107,
  083515, \dodoi{10.1103/PhysRevD.107.083515}

\bibitem[{Ivanov {et~al.}(2022{\natexlab{a}})Ivanov, Philcox, Nishimichi,
  Simonovi\'c, Takada, \& Zaldarriaga}]{Ivanov:2021kcd}
Ivanov, M.~M., Philcox, O. H.~E., Nishimichi, T., {et~al.} 2022{\natexlab{a}},
  Phys. Rev. D, 105, 063512, \dodoi{10.1103/PhysRevD.105.063512}

\bibitem[{Ivanov {et~al.}(2022{\natexlab{b}})Ivanov, Philcox, Simonovi\'c,
  Zaldarriaga, Nischimichi, \& Takada}]{Ivanov:2021fbu}
Ivanov, M.~M., Philcox, O. H.~E., Simonovi\'c, M., {et~al.} 2022{\natexlab{b}},
  Phys. Rev. D, 105, 043531, \dodoi{10.1103/PhysRevD.105.043531}

\bibitem[{Ivanov \& Sibiryakov(2018)}]{Ivanov:2018gjr}
Ivanov, M.~M., \& Sibiryakov, S. 2018, JCAP, 07, 053,
  \dodoi{10.1088/1475-7516/2018/07/053}

\bibitem[{Ivanov {et~al.}(2020{\natexlab{a}})Ivanov, Simonovi\'c, \&
  Zaldarriaga}]{Ivanov:2019pdj}
Ivanov, M.~M., Simonovi\'c, M., \& Zaldarriaga, M. 2020{\natexlab{a}}, JCAP,
  05, 042, \dodoi{10.1088/1475-7516/2020/05/042}

\bibitem[{Ivanov {et~al.}(2020{\natexlab{b}})Ivanov, Simonovi\'c, \&
  Zaldarriaga}]{Ivanov:2019hqk}
---. 2020{\natexlab{b}}, Phys. Rev. D, 101, 083504,
  \dodoi{10.1103/PhysRevD.101.083504}

\bibitem[{{Iwasawa} {et~al.}(2016){Iwasawa}, {Tanikawa}, {Hosono}, {Nitadori},
  {Muranushi}, \& {Makino}}]{2016PASJ...68...54I}
{Iwasawa}, M., {Tanikawa}, A., {Hosono}, N., {et~al.} 2016, \pasj, 68, 54,
  \dodoi{10.1093/pasj/psw053}

\bibitem[{{Jeffrey} {et~al.}(2021){Jeffrey}, {Alsing}, \&
  {Lanusse}}]{Jeffrey2021}
{Jeffrey}, N., {Alsing}, J., \& {Lanusse}, F. 2021, \mnras, 501, 954,
  \dodoi{10.1093/mnras/staa3594}

\bibitem[{{Jennings} {et~al.}(2013){Jennings}, {Li}, \& {Hu}}]{Jennings2013}
{Jennings}, E., {Li}, Y., \& {Hu}, W. 2013, \mnras, 434, 2167,
  \dodoi{10.1093/mnras/stt1169}

\bibitem[{{Kaiser}(1987)}]{1987MNRAS.227....1K}
{Kaiser}, N. 1987, \mnras, 227, 1, \dodoi{10.1093/mnras/227.1.1}

\bibitem[{{Kitaura} {et~al.}(2016){Kitaura}, {Rodr{\'\i}guez-Torres}, {Chuang},
  {Zhao}, {Prada}, {Gil-Mar{\'\i}n}, {Guo}, {Yepes}, {Klypin}, {Sc{\'o}ccola},
  {Tinker}, {McBride}, {Reid}, {S{\'a}nchez}, {Salazar-Albornoz}, {Grieb},
  {Vargas-Magana}, {Cuesta}, {Neyrinck}, {Beutler}, {Comparat}, {Percival}, \&
  {Ross}}]{Kitaura2016a}
{Kitaura}, F.-S., {Rodr{\'\i}guez-Torres}, S., {Chuang}, C.-H., {et~al.} 2016,
  \mnras, 456, 4156, \dodoi{10.1093/mnras/stv2826}

\bibitem[{Kobayashi {et~al.}(2022)Kobayashi, Nishimichi, Takada, \&
  Miyatake}]{Kobayashi:2021oud}
Kobayashi, Y., Nishimichi, T., Takada, M., \& Miyatake, H. 2022, Phys. Rev. D,
  105, 083517, \dodoi{10.1103/PhysRevD.105.083517}

\bibitem[{Kobayashi {et~al.}(2020)Kobayashi, Nishimichi, Takada, Takahashi, \&
  Osato}]{Kobayashi:2020zsw}
Kobayashi, Y., Nishimichi, T., Takada, M., Takahashi, R., \& Osato, K. 2020,
  Phys. Rev. D, 102, 063504, \dodoi{10.1103/PhysRevD.102.063504}

\bibitem[{{Kosti{\'c}} {et~al.}(2023){Kosti{\'c}}, {Nguyen}, {Schmidt}, \&
  {Reinecke}}]{Kostic:2022vok}
{Kosti{\'c}}, A., {Nguyen}, N.-M., {Schmidt}, F., \& {Reinecke}, M. 2023,
  \jcap, 2023, 063, \dodoi{10.1088/1475-7516/2023/07/063}

\bibitem[{{Kreisch} {et~al.}(2022){Kreisch}, {Pisani}, {Villaescusa-Navarro},
  {Spergel}, {Wandelt}, {Hamaus}, \& {Bayer}}]{Kreisch2022}
{Kreisch}, C.~D., {Pisani}, A., {Villaescusa-Navarro}, F., {et~al.} 2022, \apj,
  935, 100, \dodoi{10.3847/1538-4357/ac7d4b}

\bibitem[{{Kwan} {et~al.}(2023){Kwan}, {Saito}, {Leauthaud}, {Heitmann},
  {Habib}, {Frontiere}, {Guo}, {Huang}, {Pope}, \&
  {Rodrigu{\'e}z-Torres}}]{Kwan_et_al_2023}
{Kwan}, J., {Saito}, S., {Leauthaud}, A., {et~al.} 2023, \apj, 952, 80,
  \dodoi{10.3847/1538-4357/acd92f}

\bibitem[{{Kwon} \& {Hahn}(2024)}]{Kwon_et_al_2024}
{Kwon}, K.~J., \& {Hahn}, C. 2024, arXiv e-prints, arXiv:2401.12318,
  \dodoi{10.48550/arXiv.2401.12318}

\bibitem[{{Kwon} {et~al.}(2023){Kwon}, {Hahn}, \& {Alsing}}]{Kwon_et_al_2023}
{Kwon}, K.~J., {Hahn}, C., \& {Alsing}, J. 2023, \apjs, 265, 23,
  \dodoi{10.3847/1538-4365/acba14}

\bibitem[{{Lahav} {et~al.}(1991){Lahav}, {Lilje}, {Primack}, \&
  {Rees}}]{Lahav1991}
{Lahav}, O., {Lilje}, P.~B., {Primack}, J.~R., \& {Rees}, M.~J. 1991, \mnras,
  251, 128, \dodoi{10.1093/mnras/251.1.128}

\bibitem[{{Landy} \& {Szalay}(1993)}]{Landy1993}
{Landy}, S.~D., \& {Szalay}, A.~S. 1993, \apj, 412, 64, \dodoi{10.1086/172900}

\bibitem[{{Lange} {et~al.}(2022){Lange}, {Hearin}, {Leauthaud}, {van den
  Bosch}, {Guo}, \& {DeRose}}]{Lange_et_al_2022}
{Lange}, J.~U., {Hearin}, A.~P., {Leauthaud}, A., {et~al.} 2022, \mnras, 509,
  1779, \dodoi{10.1093/mnras/stab3111}

\bibitem[{Lange {et~al.}(2023)Lange, Hearin, Leauthaud, van~den Bosch, Xhakaj,
  Guo, Wechsler, \& DeRose}]{Lange:2023khv}
Lange, J.~U., Hearin, A.~P., Leauthaud, A., {et~al.} 2023,
  \dodoi{10.1093/mnras/stad473}

\bibitem[{{Laureijs} {et~al.}(2011){Laureijs}, {Amiaux}, {Arduini},
  {Augu{\`e}res}, {Brinchmann}, {Cole}, {Cropper}, {Dabin}, {Duvet}, {Ealet},
  {Garilli}, {Gondoin}, {Guzzo}, {Hoar}, {Hoekstra}, {Holmes}, {Kitching},
  {Maciaszek}, {Mellier}, {Pasian}, {Percival}, {Rhodes}, {Saavedra Criado},
  {Sauvage}, {Scaramella}, {Valenziano}, {Warren}, {Bender}, {Castander},
  {Cimatti}, {Le F{\`e}vre}, {Kurki-Suonio}, {Levi}, {Lilje}, {Meylan},
  {Nichol}, {Pedersen}, {Popa}, {Rebolo Lopez}, {Rix}, {Rottgering},
  {Zeilinger}, {Grupp}, {Hudelot}, {Massey}, {Meneghetti}, {Miller}, {Paltani},
  {Paulin-Henriksson}, {Pires}, {Saxton}, {Schrabback}, {Seidel}, {Walsh},
  {Aghanim}, {Amendola}, {Bartlett}, {Baccigalupi}, {Beaulieu}, {Benabed},
  {Cuby}, {Elbaz}, {Fosalba}, {Gavazzi}, {Helmi}, {Hook}, {Irwin}, {Kneib},
  {Kunz}, {Mannucci}, {Moscardini}, {Tao}, {Teyssier}, {Weller}, {Zamorani},
  {Zapatero Osorio}, {Boulade}, {Foumond}, {Di Giorgio}, {Guttridge}, {James},
  {Kemp}, {Martignac}, {Spencer}, {Walton}, {Bl{\"u}mchen}, {Bonoli},
  {Bortoletto}, {Cerna}, {Corcione}, {Fabron}, {Jahnke}, {Ligori}, {Madrid},
  {Martin}, {Morgante}, {Pamplona}, {Prieto}, {Riva}, {Toledo}, {Trifoglio},
  {Zerbi}, {Abdalla}, {Douspis}, {Grenet}, {Borgani}, {Bouwens}, {Courbin},
  {Delouis}, {Dubath}, {Fontana}, {Frailis}, {Grazian}, {Koppenh{\"o}fer},
  {Mansutti}, {Melchior}, {Mignoli}, {Mohr}, {Neissner}, {Noddle}, {Poncet},
  {Scodeggio}, {Serrano}, {Shane}, {Starck}, {Surace}, {Taylor},
  {Verdoes-Kleijn}, {Vuerli}, {Williams}, {Zacchei}, {Altieri}, {Escudero
  Sanz}, {Kohley}, {Oosterbroek}, {Astier}, {Bacon}, {Bardelli}, {Baugh},
  {Bellagamba}, {Benoist}, {Bianchi}, {Biviano}, {Branchini}, {Carbone},
  {Cardone}, {Clements}, {Colombi}, {Conselice}, {Cresci}, {Deacon}, {Dunlop},
  {Fedeli}, {Fontanot}, {Franzetti}, {Giocoli}, {Garcia-Bellido}, {Gow},
  {Heavens}, {Hewett}, {Heymans}, {Holland}, {Huang}, {Ilbert}, {Joachimi},
  {Jennins}, {Kerins}, {Kiessling}, {Kirk}, {Kotak}, {Krause}, {Lahav}, {van
  Leeuwen}, {Lesgourgues}, {Lombardi}, {Magliocchetti}, {Maguire}, {Majerotto},
  {Maoli}, {Marulli}, {Maurogordato}, {McCracken}, {McLure}, {Melchiorri},
  {Merson}, {Moresco}, {Nonino}, {Norberg}, {Peacock}, {Pello}, {Penny},
  {Pettorino}, {Di Porto}, {Pozzetti}, {Quercellini}, {Radovich}, {Rassat},
  {Roche}, {Ronayette}, {Rossetti}, {Sartoris}, {Schneider}, {Semboloni},
  {Serjeant}, {Simpson}, {Skordis}, {Smadja}, {Smartt}, {Spano}, {Spiro},
  {Sullivan}, {Tilquin}, {Trotta}, {Verde}, {Wang}, {Williger}, {Zhao},
  {Zoubian}, \& {Zucca}}]{2011arXiv1110.3193L}
{Laureijs}, R., {Amiaux}, J., {Arduini}, S., {et~al.} 2011, arXiv e-prints,
  arXiv:1110.3193, \dodoi{10.48550/arXiv.1110.3193}

\bibitem[{{Lavaux} \& {Wandelt}(2012)}]{Lavaux2012}
{Lavaux}, G., \& {Wandelt}, B.~D. 2012, \apj, 754, 109,
  \dodoi{10.1088/0004-637X/754/2/109}

\bibitem[{Lazeyras \& Schmidt(2018)}]{Lazeyras:2017hxw}
Lazeyras, T., \& Schmidt, F. 2018, JCAP, 09, 008,
  \dodoi{10.1088/1475-7516/2018/09/008}

\bibitem[{{Levi} {et~al.}(2013){Levi}, {Bebek}, {Beers}, {Blum}, {Cahn},
  {Eisenstein}, {Flaugher}, {Honscheid}, {Kron}, {Lahav}, {McDonald}, {Roe},
  {Schlegel}, \& {representing the DESI collaboration}}]{2013DESI}
{Levi}, M., {Bebek}, C., {Beers}, T., {et~al.} 2013, arXiv e-prints,
  arXiv:1308.0847.
\newblock \doarXiv{1308.0847}

\bibitem[{Lewis(2019)}]{GetDist}
Lewis, A. 2019.
\newblock \doarXiv{1910.13970}

\bibitem[{{Maion} {et~al.}(2023){Maion}, {Angulo}, {Bakx}, {Chisari}, {Kurita},
  \& {Pellejero-Ib{\'a}{\~n}ez}}]{Maion_2023}
{Maion}, F., {Angulo}, R.~E., {Bakx}, T., {et~al.} 2023, arXiv e-prints,
  arXiv:2307.13754, \dodoi{10.48550/arXiv.2307.13754}

\bibitem[{Maksimova {et~al.}(2021)Maksimova, Garrison, Eisenstein, Hadzhiyska,
  Bose, \& Satterthwaite}]{Maksimova:2021ynf}
Maksimova, N.~A., Garrison, L.~H., Eisenstein, D.~J., {et~al.} 2021, Mon. Not.
  Roy. Astron. Soc., 508, 4017, \dodoi{10.1093/mnras/stab2484}

\bibitem[{{Maksimova} {et~al.}(2021){Maksimova}, {Garrison}, {Eisenstein},
  {Hadzhiyska}, {Bose}, \& {Satterthwaite}}]{2021Maksimova}
{Maksimova}, N.~A., {Garrison}, L.~H., {Eisenstein}, D.~J., {et~al.} 2021,
  \mnras, \dodoi{10.1093/mnras/stab2484}

\bibitem[{Marinoni {et~al.}(2008)Marinoni, Guzzo, Cappi, Le~Fevre, Mazure,
  Meneux, \& Pollo}]{Marinoni:2008wx}
Marinoni, C., Guzzo, L., Cappi, A., {et~al.} 2008, Astron. Astrophys., 487, 7,
  \dodoi{10.1051/0004-6361:20078891}

\bibitem[{{Marulli} {et~al.}(2016){Marulli}, {Veropalumbo}, \& {Moresco}}]{CBL}
{Marulli}, F., {Veropalumbo}, A., \& {Moresco}, M. 2016, Astronomy and
  Computing, 14, 35, \dodoi{10.1016/j.ascom.2016.01.005}

\bibitem[{{Massara} {et~al.}(2023){Massara}, {Villaescusa-Navarro}, {Hahn},
  {Abidi}, {Eickenberg}, {Ho}, {Lemos}, {Dizgah}, \&
  {Blancard}}]{2023ApJ...951...70M}
{Massara}, E., {Villaescusa-Navarro}, F., {Hahn}, C., {et~al.} 2023, \apj, 951,
  70, \dodoi{10.3847/1538-4357/acd44d}

\bibitem[{Matsubara(2008)}]{PhysRevD.78.083519}
Matsubara, T. 2008, Phys. Rev. D, 78, 083519,
  \dodoi{10.1103/PhysRevD.78.083519}

\bibitem[{Matsubara(2015)}]{Matsubara:2015ipa}
---. 2015, Phys. Rev. D, 92, 023534, \dodoi{10.1103/PhysRevD.92.023534}

\bibitem[{{McEwen} \& {Weinberg}(2018)}]{McEwen_2018}
{McEwen}, J.~E., \& {Weinberg}, D.~H. 2018, \mnras, 477, 4348,
  \dodoi{10.1093/mnras/sty882}

\bibitem[{Modi {et~al.}(2020)Modi, Chen, \& White}]{Modi_2020}
Modi, C., Chen, S.-F., \& White, M. 2020, Monthly Notices of the Royal
  Astronomical Society, 492, 5754, \dodoi{10.1093/mnras/staa251}

\bibitem[{{Modi} {et~al.}(2023){Modi}, {Pandey}, {Ho}, {Hahn}, {R'egaldo-Saint
  Blancard}, \& {Wandelt}}]{Modi2023a}
{Modi}, C., {Pandey}, S., {Ho}, M., {et~al.} 2023, arXiv e-prints,
  arXiv:2309.15071, \dodoi{10.48550/arXiv.2309.15071}

\bibitem[{{Modi} \& {Philcox}(2023)}]{hysbi}
{Modi}, C., \& {Philcox}, O. H.~E. 2023, arXiv e-prints, arXiv:2309.10270,
  \dodoi{10.48550/arXiv.2309.10270}

\bibitem[{{Moresco} {et~al.}(2022){Moresco}, {Amati}, {Amendola}, {Birrer},
  {Blakeslee}, {Cantiello}, {Cimatti}, {Darling}, {Della Valle}, {Fishbach},
  {Grillo}, {Hamaus}, {Holz}, {Izzo}, {Jimenez}, {Lusso}, {Meneghetti},
  {Piedipalumbo}, {Pisani}, {Pourtsidou}, {Pozzetti}, {Quartin}, {Risaliti},
  {Rosati}, \& {Verde}}]{Moresco2022}
{Moresco}, M., {Amati}, L., {Amendola}, L., {et~al.} 2022, Living Reviews in
  Relativity, 25, 6, \dodoi{10.1007/s41114-022-00040-z}

\bibitem[{{Namekata} {et~al.}(2018){Namekata}, {Iwasawa}, {Nitadori},
  {Tanikawa}, {Muranushi}, {Wang}, {Hosono}, {Nomura}, \&
  {Makino}}]{2018PASJ...70...70N}
{Namekata}, D., {Iwasawa}, M., {Nitadori}, K., {et~al.} 2018, \pasj, 70, 70,
  \dodoi{10.1093/pasj/psy062}

\bibitem[{Navarro {et~al.}(1997)Navarro, Frenk, \& White}]{Navarro:1996gj}
Navarro, J.~F., Frenk, C.~S., \& White, S. D.~M. 1997, Astrophys. J., 490, 493,
  \dodoi{10.1086/304888}

\bibitem[{{Neyrinck}(2008)}]{Neyrinck2008}
{Neyrinck}, M.~C. 2008, \mnras, 386, 2101,
  \dodoi{10.1111/j.1365-2966.2008.13180.x}

\bibitem[{Nguyen {et~al.}(2021)Nguyen, Schmidt, Lavaux, \&
  Jasche}]{Nguyen:2020hxe}
Nguyen, N.-M., Schmidt, F., Lavaux, G., \& Jasche, J. 2021, JCAP, 03, 058,
  \dodoi{10.1088/1475-7516/2021/03/058}

\bibitem[{Nguyen {et~al.}(2024)Nguyen, Schmidt, Tucci, Reinecke, \&
  Kosti\'c}]{Nguyen:2024yth}
Nguyen, N.-M., Schmidt, F., Tucci, B., Reinecke, M., \& Kosti\'c, A. 2024.
\newblock \doarXiv{2403.03220}

\bibitem[{{Nicola} {et~al.}(2024){Nicola}, {Hadzhiyska}, {Findlay},
  {Garc{\'\i}a-Garc{\'\i}a}, {Alonso}, {Slosar}, {Guo}, {Kokron}, {Angulo},
  {Aviles}, {Blazek}, {Dunkley}, {Jain}, {Pellejero}, {Sullivan}, {Walter},
  {Zennaro}, \& {LSST Dark Energy Science Collaboration}}]{Nicola_2024}
{Nicola}, A., {Hadzhiyska}, B., {Findlay}, N., {et~al.} 2024, \jcap, 2024, 015,
  \dodoi{10.1088/1475-7516/2024/02/015}

\bibitem[{Nishimichi {et~al.}(2020)Nishimichi, D'Amico, Ivanov, Senatore,
  Simonovi\'c, Takada, Zaldarriaga, \& Zhang}]{Nishimichi:2020tvu}
Nishimichi, T., D'Amico, G., Ivanov, M.~M., {et~al.} 2020, Phys. Rev. D, 102,
  123541, \dodoi{10.1103/PhysRevD.102.123541}

\bibitem[{Nishimichi {et~al.}(2019)}]{Nishimichi:2018etk}
Nishimichi, T., {et~al.} 2019, Astrophys. J., 884, 29,
  \dodoi{10.3847/1538-4357/ab3719}

\bibitem[{Nitadori {et~al.}(2006)Nitadori, Makino, \& Hut}]{Nitadori2006-ek}
Nitadori, K., Makino, J., \& Hut, P. 2006, New Astron., 12, 169,
  \dodoi{10.1016/j.newast.2006.07.007}

\bibitem[{Nunes {et~al.}(2022)Nunes, Vagnozzi, Kumar, Di~Valentino, \&
  Mena}]{Nunes:2022bhn}
Nunes, R.~C., Vagnozzi, S., Kumar, S., Di~Valentino, E., \& Mena, O. 2022,
  Phys. Rev. D, 105, 123506, \dodoi{10.1103/PhysRevD.105.123506}

\bibitem[{{Paillas} {et~al.}(2021){Paillas}, {Cai}, {Padilla}, \&
  {S{\'a}nchez}}]{Paillas:2021}
{Paillas}, E., {Cai}, Y.-C., {Padilla}, N., \& {S{\'a}nchez}, A.~G. 2021,
  \mnras, 505, 5731, \dodoi{10.1093/mnras/stab1654}

\bibitem[{Paillas {et~al.}(2023)}]{Paillas:2023cpk}
Paillas, E., {et~al.} 2023.
\newblock \doarXiv{2309.16541}

\bibitem[{{Paillas} {et~al.}(2023){Paillas}, {Cuesta-Lazaro}, {Zarrouk}, {Cai},
  {Percival}, {Nadathur}, {Pinon}, {de Mattia}, \& {Beutler}}]{Paillas:2023}
{Paillas}, E., {Cuesta-Lazaro}, C., {Zarrouk}, P., {et~al.} 2023, \mnras, 522,
  606, \dodoi{10.1093/mnras/stad1017}

\bibitem[{{Peebles}(1980)}]{Peebles1980}
{Peebles}, P.~J.~E. 1980, {The large-scale structure of the universe}
  (Princeton University Press, Princeton U.S.A.)

\bibitem[{{Pellejero-Iba{\~n}ez} {et~al.}(2024){Pellejero-Iba{\~n}ez},
  {Angulo}, {Jamieson}, \& {Li}}]{PellejeroIbanez_2024}
{Pellejero-Iba{\~n}ez}, M., {Angulo}, R.~E., {Jamieson}, D., \& {Li}, Y. 2024,
  \mnras, \dodoi{10.1093/mnras/stae489}

\bibitem[{{Pellejero Iba{\~n}ez} {et~al.}(2023){Pellejero Iba{\~n}ez},
  {Angulo}, {Zennaro}, {St{\"u}cker}, {Contreras}, {Aric{\`o}}, \&
  {Maion}}]{PellejeroIbanez2023}
{Pellejero Iba{\~n}ez}, M., {Angulo}, R.~E., {Zennaro}, M., {et~al.} 2023,
  \mnras, 520, 3725, \dodoi{10.1093/mnras/stad368}

\bibitem[{{Pellejero Iba{\~n}ez} {et~al.}(2022){Pellejero Iba{\~n}ez},
  {St{\"u}cker}, {Angulo}, {Zennaro}, {Contreras}, \&
  {Aric{\`o}}}]{PellejeroIbanez2022}
{Pellejero Iba{\~n}ez}, M., {St{\"u}cker}, J., {Angulo}, R.~E., {et~al.} 2022,
  \mnras, 514, 3993, \dodoi{10.1093/mnras/stac1602}

\bibitem[{{Pellejero-Ibanez} {et~al.}(2022){Pellejero-Ibanez}, {Angulo},
  {Zennaro}, {Stuecker}, {Contreras}, {Arico}, \& {Maion}}]{Pellejero2022}
{Pellejero-Ibanez}, M., {Angulo}, R.~E., {Zennaro}, M., {et~al.} 2022, arXiv
  e-prints, arXiv:2207.06437, \dodoi{10.48550/arXiv.2207.06437}

\bibitem[{{Pelliciari} {et~al.}(2022){Pelliciari}, {Contarini}, {Marulli},
  {Moscardini}, {Giocoli}, {Lesci}, \& {Dolag}}]{Pelliciari2022}
{Pelliciari}, D., {Contarini}, S., {Marulli}, F., {et~al.} 2022, arXiv
  e-prints, arXiv:2210.07248, \dodoi{10.48550/arXiv.2210.07248}

\bibitem[{{Petri} {et~al.}(2015){Petri}, {Liu}, {Haiman}, {May}, {Hui}, \&
  {Kratochvil}}]{2015PhRvD..91j3511P}
{Petri}, A., {Liu}, J., {Haiman}, Z., {et~al.} 2015, \prd, 91, 103511,
  \dodoi{10.1103/PhysRevD.91.103511}

\bibitem[{Philcox(2021{\natexlab{a}})}]{Philcox:2020vbm}
Philcox, O. H.~E. 2021{\natexlab{a}}, Phys. Rev. D, 103, 103504,
  \dodoi{10.1103/PhysRevD.103.103504}

\bibitem[{Philcox(2021{\natexlab{b}})}]{Philcox:2021ukg}
---. 2021{\natexlab{b}}, Phys. Rev. D, 104, 123529,
  \dodoi{10.1103/PhysRevD.104.123529}

\bibitem[{Philcox \& Ivanov(2022)}]{Philcox:2021kcw}
Philcox, O. H.~E., \& Ivanov, M.~M. 2022, Phys. Rev. D, 105, 043517,
  \dodoi{10.1103/PhysRevD.105.043517}

\bibitem[{Philcox {et~al.}(2022)Philcox, Ivanov, Cabass, Simonovi\'c,
  Zaldarriaga, \& Nishimichi}]{Philcox:2022frc}
Philcox, O. H.~E., Ivanov, M.~M., Cabass, G., {et~al.} 2022, Phys. Rev. D, 106,
  043530, \dodoi{10.1103/PhysRevD.106.043530}

\bibitem[{Philcox {et~al.}(2021{\natexlab{a}})Philcox, Ivanov, Zaldarriaga,
  Simonovic, \& Schmittfull}]{Philcox:2020zyp}
Philcox, O. H.~E., Ivanov, M.~M., Zaldarriaga, M., Simonovic, M., \&
  Schmittfull, M. 2021{\natexlab{a}}, Phys. Rev. D, 103, 043508,
  \dodoi{10.1103/PhysRevD.103.043508}

\bibitem[{Philcox {et~al.}(2021{\natexlab{b}})Philcox, Sherwin, Farren, \&
  Baxter}]{Philcox:2020xbv}
Philcox, O. H.~E., Sherwin, B.~D., Farren, G.~S., \& Baxter, E.~J.
  2021{\natexlab{b}}, Phys. Rev. D, 103, 023538,
  \dodoi{10.1103/PhysRevD.103.023538}

\bibitem[{{Pisani} {et~al.}(2014){Pisani}, {Lavaux}, {Sutter}, \&
  {Wandelt}}]{Pisani2014}
{Pisani}, A., {Lavaux}, G., {Sutter}, P.~M., \& {Wandelt}, B.~D. 2014, \mnras,
  443, 3238, \dodoi{10.1093/mnras/stu1399}

\bibitem[{{Pisani} {et~al.}(2015{\natexlab{a}}){Pisani}, {Sutter}, {Hamaus},
  {Alizadeh}, {Biswas}, {Wandelt}, \& {Hirata}}]{Pisani2015a}
{Pisani}, A., {Sutter}, P.~M., {Hamaus}, N., {et~al.} 2015{\natexlab{a}}, \prd,
  92, 083531, \dodoi{10.1103/PhysRevD.92.083531}

\bibitem[{{Pisani} {et~al.}(2015{\natexlab{b}}){Pisani}, {Sutter}, \&
  {Wandelt}}]{Pisani2015b}
{Pisani}, A., {Sutter}, P.~M., \& {Wandelt}, B.~D. 2015{\natexlab{b}}, ArXiv
  e-prints.
\newblock \doarXiv{1506.07982}

\bibitem[{{Pisani} {et~al.}(2019){Pisani}, {Massara}, {Spergel}, {Alonso},
  {Baker}, {Cai}, {Cautun}, {Davies}, {Demchenko}, {Dor{\'e}}, {Goulding},
  {Habouzit}, {Hamaus}, {Hawken}, {Hirata}, {Ho}, {Jain}, {Kreisch}, {Marulli},
  {Padilla}, {Pollina}, {Sahl{\'e}n}, {Sheth}, {Somerville}, {Szapudi}, {van de
  Weygaert}, {Villaescusa-Navarro}, {Wandelt}, \& {Wang}}]{Pisani2019}
{Pisani}, A., {Massara}, E., {Spergel}, D.~N., {et~al.} 2019,
  \textnormal{BAAS}, 51, 40.
\newblock \doarXiv{1903.05161}

\bibitem[{{Planck Collaboration} {et~al.}(2016){Planck Collaboration}, {Ade},
  {Aghanim}, {Arnaud}, {Ashdown}, {Aumont}, {Baccigalupi}, {Banday},
  {Barreiro}, {Bartlett}, {Bartolo}, {Battaner}, {Battye}, {Benabed},
  {Beno{\^\i}t}, {Benoit-L{\'e}vy}, {Bernard}, {Bersanelli}, {Bielewicz},
  {Bock}, {Bonaldi}, {Bonavera}, {Bond}, {Borrill}, {Bouchet}, {Boulanger},
  {Bucher}, {Burigana}, {Butler}, {Calabrese}, {Cardoso}, {Catalano},
  {Challinor}, {Chamballu}, {Chary}, {Chiang}, {Chluba}, {Christensen},
  {Church}, {Clements}, {Colombi}, {Colombo}, {Combet}, {Coulais}, {Crill},
  {Curto}, {Cuttaia}, {Danese}, {Davies}, {Davis}, {de Bernardis}, {de Rosa},
  {de Zotti}, {Delabrouille}, {D{\'e}sert}, {Di Valentino}, {Dickinson},
  {Diego}, {Dolag}, {Dole}, {Donzelli}, {Dor{\'e}}, {Douspis}, {Ducout},
  {Dunkley}, {Dupac}, {Efstathiou}, {Elsner}, {En{\ss}lin}, {Eriksen},
  {Farhang}, {Fergusson}, {Finelli}, {Forni}, {Frailis}, {Fraisse},
  {Franceschi}, {Frejsel}, {Galeotta}, {Galli}, {Ganga}, {Gauthier}, {Gerbino},
  {Ghosh}, {Giard}, {Giraud-H{\'e}raud}, {Giusarma}, {Gjerl{\o}w},
  {Gonz{\'a}lez-Nuevo}, {G{\'o}rski}, {Gratton}, {Gregorio}, {Gruppuso},
  {Gudmundsson}, {Hamann}, {Hansen}, {Hanson}, {Harrison}, {Helou},
  {Henrot-Versill{\'e}}, {Hern{\'a}ndez-Monteagudo}, {Herranz}, {Hildebrandt},
  {Hivon}, {Hobson}, {Holmes}, {Hornstrup}, {Hovest}, {Huang}, {Huffenberger},
  {Hurier}, {Jaffe}, {Jaffe}, {Jones}, {Juvela}, {Keih{\"a}nen}, {Keskitalo},
  {Kisner}, {Kneissl}, {Knoche}, {Knox}, {Kunz}, {Kurki-Suonio}, {Lagache},
  {L{\"a}hteenm{\"a}ki}, {Lamarre}, {Lasenby}, {Lattanzi}, {Lawrence}, {Leahy},
  {Leonardi}, {Lesgourgues}, {Levrier}, {Lewis}, {Liguori}, {Lilje},
  {Linden-V{\o}rnle}, {L{\'o}pez-Caniego}, {Lubin}, {Mac{\'\i}as-P{\'e}rez},
  {Maggio}, {Maino}, {Mandolesi}, {Mangilli}, {Marchini}, {Maris}, {Martin},
  {Martinelli}, {Mart{\'\i}nez-Gonz{\'a}lez}, {Masi}, {Matarrese}, {McGehee},
  {Meinhold}, {Melchiorri}, {Melin}, {Mendes}, {Mennella}, {Migliaccio},
  {Millea}, {Mitra}, {Miville-Desch{\^e}nes}, {Moneti}, {Montier}, {Morgante},
  {Mortlock}, {Moss}, {Munshi}, {Murphy}, {Naselsky}, {Nati}, {Natoli},
  {Netterfield}, {N{\o}rgaard-Nielsen}, {Noviello}, {Novikov}, {Novikov},
  {Oxborrow}, {Paci}, {Pagano}, {Pajot}, {Paladini}, {Paoletti}, {Partridge},
  {Pasian}, {Patanchon}, {Pearson}, {Perdereau}, {Perotto}, {Perrotta},
  {Pettorino}, {Piacentini}, {Piat}, {Pierpaoli}, {Pietrobon}, {Plaszczynski},
  {Pointecouteau}, {Polenta}, {Popa}, {Pratt}, {Pr{\'e}zeau}, {Prunet},
  {Puget}, {Rachen}, {Reach}, {Rebolo}, {Reinecke}, {Remazeilles}, {Renault},
  {Renzi}, {Ristorcelli}, {Rocha}, {Rosset}, {Rossetti}, {Roudier},
  {Rouill{\'e} d'Orfeuil}, {Rowan-Robinson}, {Rubi{\~n}o-Mart{\'\i}n},
  {Rusholme}, {Said}, {Salvatelli}, {Salvati}, {Sandri}, {Santos},
  {Savelainen}, {Savini}, {Scott}, {Seiffert}, {Serra}, {Shellard}, {Spencer},
  {Spinelli}, {Stolyarov}, {Stompor}, {Sudiwala}, {Sunyaev}, {Sutton},
  {Suur-Uski}, {Sygnet}, {Tauber}, {Terenzi}, {Toffolatti}, {Tomasi},
  {Tristram}, {Trombetti}, {Tucci}, {Tuovinen}, {T{\"u}rler}, {Umana},
  {Valenziano}, {Valiviita}, {Van Tent}, {Vielva}, {Villa}, {Wade}, {Wandelt},
  {Wehus}, {White}, {White}, {Wilkinson}, {Yvon}, {Zacchei}, \&
  {Zonca}}]{Planck2015}
{Planck Collaboration}, {Ade}, P.~A.~R., {Aghanim}, N., {et~al.} 2016, \aap,
  594, A13, \dodoi{10.1051/0004-6361/201525830}

\bibitem[{{Planck Collaboration} {et~al.}(2020){Planck Collaboration},
  {Aghanim}, {Akrami}, {Ashdown}, {Aumont}, {Baccigalupi}, {Ballardini},
  {Banday}, {Barreiro}, {Bartolo}, {Basak}, {Battye}, {Benabed}, {Bernard},
  {Bersanelli}, {Bielewicz}, {Bock}, {Bond}, {Borrill}, {Bouchet}, {Boulanger},
  {Bucher}, {Burigana}, {Butler}, {Calabrese}, {Cardoso}, {Carron},
  {Challinor}, {Chiang}, {Chluba}, {Colombo}, {Combet}, {Contreras}, {Crill},
  {Cuttaia}, {de Bernardis}, {de Zotti}, {Delabrouille}, {Delouis}, {Di
  Valentino}, {Diego}, {Dor{\'e}}, {Douspis}, {Ducout}, {Dupac}, {Dusini},
  {Efstathiou}, {Elsner}, {En{\ss}lin}, {Eriksen}, {Fantaye}, {Farhang},
  {Fergusson}, {Fernandez-Cobos}, {Finelli}, {Forastieri}, {Frailis},
  {Fraisse}, {Franceschi}, {Frolov}, {Galeotta}, {Galli}, {Ganga},
  {G{\'e}nova-Santos}, {Gerbino}, {Ghosh}, {Gonz{\'a}lez-Nuevo}, {G{\'o}rski},
  {Gratton}, {Gruppuso}, {Gudmundsson}, {Hamann}, {Handley}, {Hansen},
  {Herranz}, {Hildebrandt}, {Hivon}, {Huang}, {Jaffe}, {Jones}, {Karakci},
  {Keih{\"a}nen}, {Keskitalo}, {Kiiveri}, {Kim}, {Kisner}, {Knox},
  {Krachmalnicoff}, {Kunz}, {Kurki-Suonio}, {Lagache}, {Lamarre}, {Lasenby},
  {Lattanzi}, {Lawrence}, {Le Jeune}, {Lemos}, {Lesgourgues}, {Levrier},
  {Lewis}, {Liguori}, {Lilje}, {Lilley}, {Lindholm}, {L{\'o}pez-Caniego},
  {Lubin}, {Ma}, {Mac{\'\i}as-P{\'e}rez}, {Maggio}, {Maino}, {Mandolesi},
  {Mangilli}, {Marcos-Caballero}, {Maris}, {Martin}, {Martinelli},
  {Mart{\'\i}nez-Gonz{\'a}lez}, {Matarrese}, {Mauri}, {McEwen}, {Meinhold},
  {Melchiorri}, {Mennella}, {Migliaccio}, {Millea}, {Mitra},
  {Miville-Desch{\^e}nes}, {Molinari}, {Montier}, {Morgante}, {Moss}, {Natoli},
  {N{\o}rgaard-Nielsen}, {Pagano}, {Paoletti}, {Partridge}, {Patanchon},
  {Peiris}, {Perrotta}, {Pettorino}, {Piacentini}, {Polastri}, {Polenta},
  {Puget}, {Rachen}, {Reinecke}, {Remazeilles}, {Renzi}, {Rocha}, {Rosset},
  {Roudier}, {Rubi{\~n}o-Mart{\'\i}n}, {Ruiz-Granados}, {Salvati}, {Sandri},
  {Savelainen}, {Scott}, {Shellard}, {Sirignano}, {Sirri}, {Spencer},
  {Sunyaev}, {Suur-Uski}, {Tauber}, {Tavagnacco}, {Tenti}, {Toffolatti},
  {Tomasi}, {Trombetti}, {Valenziano}, {Valiviita}, {Van Tent}, {Vibert},
  {Vielva}, {Villa}, {Vittorio}, {Wandelt}, {Wehus}, {White}, {White},
  {Zacchei}, \& {Zonca}}]{Planck2018}
{Planck Collaboration}, {Aghanim}, N., {Akrami}, Y., {et~al.} 2020, \aap, 641,
  A6, \dodoi{10.1051/0004-6361/201833910}

\bibitem[{{Platen} {et~al.}(2007){Platen}, {van de Weygaert}, \&
  {Jones}}]{Platen2007}
{Platen}, E., {van de Weygaert}, R., \& {Jones}, B. J.~T. 2007, \mnras, 380,
  551, \dodoi{10.1111/j.1365-2966.2007.12125.x}

\bibitem[{{Pollina} {et~al.}(2017){Pollina}, {Hamaus}, {Dolag}, {Weller},
  {Baldi}, \& {Moscardini}}]{Pollina2017}
{Pollina}, G., {Hamaus}, N., {Dolag}, K., {et~al.} 2017, \mnras, 469, 787,
  \dodoi{10.1093/mnras/stx785}

\bibitem[{{Pollina} {et~al.}(2019){Pollina}, {Hamaus}, {Paech}, {Dolag},
  {Weller}, {S{\'a}nchez}, {Rykoff}, {Jain}, {Abbott}, {Allam}, {Avila},
  {Bernstein}, {Bertin}, et~al., \& {DES Collaboration}}]{Pollina2019}
{Pollina}, G., {Hamaus}, N., {Paech}, K., {et~al.} 2019, \mnras, 487, 2836,
  \dodoi{10.1093/mnras/stz1470}

\bibitem[{{Press} {et~al.}(1986){Press}, {Flannery}, \&
  {Teukolsky}}]{NumericalRecipes}
{Press}, W.~H., {Flannery}, B.~P., \& {Teukolsky}, S.~A. 1986, {Numerical
  recipes. The art of scientific computing}

\bibitem[{Ramanah {et~al.}(2019)Ramanah, Lavaux, Jasche, \&
  Wandelt}]{Ramanah:2018eed}
Ramanah, D.~K., Lavaux, G., Jasche, J., \& Wandelt, B.~D. 2019, Astron.
  Astrophys., 621, A69, \dodoi{10.1051/0004-6361/201834117}

\bibitem[{{Reddick} {et~al.}(2014){Reddick}, {Tinker}, {Wechsler}, \&
  {Lu}}]{Reddick2014}
{Reddick}, R.~M., {Tinker}, J.~L., {Wechsler}, R.~H., \& {Lu}, Y. 2014, \apj,
  783, 118, \dodoi{10.1088/0004-637X/783/2/118}

\bibitem[{{R{\'e}galdo-Saint Blancard} {et~al.}(2023){R{\'e}galdo-Saint
  Blancard}, {Hahn}, {Ho}, {Hou}, {Lemos}, {Massara}, {Modi}, {Moradinezhad
  Dizgah}, {Parker}, {Yao}, \& {Eickenberg}}]{Blancard:2023iab}
{R{\'e}galdo-Saint Blancard}, B., {Hahn}, C., {Ho}, S., {et~al.} 2023, arXiv
  e-prints, arXiv:2310.15250, \dodoi{10.48550/arXiv.2310.15250}

\bibitem[{{Reid} {et~al.}(2014){Reid}, {Seo}, {Leauthaud}, {Tinker}, \&
  {White}}]{Reid_et_al_2014}
{Reid}, B.~A., {Seo}, H.-J., {Leauthaud}, A., {Tinker}, J.~L., \& {White}, M.
  2014, \mnras, 444, 476, \dodoi{10.1093/mnras/stu1391}

\bibitem[{Rimes \& Hamilton(2005)}]{Rimes:2005xs}
Rimes, C.~D., \& Hamilton, A. J.~S. 2005, Mon. Not. Roy. Astron. Soc., 360,
  L82, \dodoi{10.1111/j.1745-3933.2005.00051.x}

\bibitem[{{Ronconi} {et~al.}(2019){Ronconi}, {Contarini}, {Marulli}, {Baldi},
  \& {Moscardini}}]{Ronconi2019}
{Ronconi}, T., {Contarini}, S., {Marulli}, F., {Baldi}, M., \& {Moscardini}, L.
  2019, \mnras, 488, 5075, \dodoi{10.1093/mnras/stz2115}

\bibitem[{{Ronconi} \& {Marulli}(2017)}]{Ronconi2017}
{Ronconi}, T., \& {Marulli}, F. 2017, \aap, 607, A24,
  \dodoi{10.1051/0004-6361/201730852}

\bibitem[{Saadeh {et~al.}(2024)Saadeh, Koyama, \&
  Morice-Atkinson}]{Saadeh:2024vuj}
Saadeh, D., Koyama, K., \& Morice-Atkinson, X. 2024.
\newblock \doarXiv{2406.03374}

\bibitem[{{Sahl{\'e}n} {et~al.}(2016){Sahl{\'e}n}, {Zubeld{\'\i}a}, \&
  {Silk}}]{sahlen2016}
{Sahl{\'e}n}, M., {Zubeld{\'\i}a}, {\'I}., \& {Silk}, J. 2016, \apjl, 820, L7,
  \dodoi{10.3847/2041-8205/820/1/L7}

\bibitem[{{Salcedo} {et~al.}(2022{\natexlab{a}}){Salcedo}, {Weinberg}, {Wu}, \&
  {Wibking}}]{Salcedo_et_al_2022}
{Salcedo}, A.~N., {Weinberg}, D.~H., {Wu}, H.-Y., \& {Wibking}, B.~D.
  2022{\natexlab{a}}, \mnras, 510, 5376, \dodoi{10.1093/mnras/stab3793}

\bibitem[{{Salcedo} {et~al.}(2022{\natexlab{b}}){Salcedo}, {Zu}, {Zhang},
  {Wang}, {Yang}, {Wu}, {Jing}, {Mo}, \& {Weinberg}}]{Salcedo_et_al_2022b}
{Salcedo}, A.~N., {Zu}, Y., {Zhang}, Y., {et~al.} 2022{\natexlab{b}}, Science
  China Physics, Mechanics, and Astronomy, 65, 109811,
  \dodoi{10.1007/s11433-022-1955-7}

\bibitem[{{S{\'a}nchez} {et~al.}(2017){S{\'a}nchez}, {Scoccimarro}, {Crocce},
  {Grieb}, {Salazar-Albornoz}, {Dalla Vecchia}, {Lippich}, {Beutler},
  {Brownstein}, {Chuang}, {Eisenstein}, {Kitaura}, {Olmstead}, {Percival},
  {Prada}, {Rodr{\'{\i}}guez-Torres}, {Ross}, {Samushia}, {Seo}, {Tinker},
  {Tojeiro}, {Vargas-Maga{\~n}a}, {Wang}, \& {Zhao}}]{SanchezA2017}
{S{\'a}nchez}, A.~G., {Scoccimarro}, R., {Crocce}, M., {et~al.} 2017, \mnras,
  464, 1640, \dodoi{10.1093/mnras/stw2443}

\bibitem[{Schaye {et~al.}(2023)}]{Schaye:2023jqv}
Schaye, J., {et~al.} 2023, Mon. Not. Roy. Astron. Soc., 526, 4978,
  \dodoi{10.1093/mnras/stad2419}

\bibitem[{Schmidt(2021{\natexlab{a}})}]{Schmidt:2020tao}
Schmidt, F. 2021{\natexlab{a}}, JCAP, 04, 032,
  \dodoi{10.1088/1475-7516/2021/04/032}

\bibitem[{Schmidt(2021{\natexlab{b}})}]{Schmidt:2020ovm}
---. 2021{\natexlab{b}}, JCAP, 04, 033, \dodoi{10.1088/1475-7516/2021/04/033}

\bibitem[{Schmidt {et~al.}(2020)Schmidt, Cabass, Jasche, \&
  Lavaux}]{Schmidt:2020viy}
Schmidt, F., Cabass, G., Jasche, J., \& Lavaux, G. 2020, JCAP, 11, 008,
  \dodoi{10.1088/1475-7516/2020/11/008}

\bibitem[{Schmidt {et~al.}(2019)Schmidt, Elsner, Jasche, Nguyen, \&
  Lavaux}]{Schmidt:2018bkr}
Schmidt, F., Elsner, F., Jasche, J., Nguyen, N.~M., \& Lavaux, G. 2019, JCAP,
  01, 042, \dodoi{10.1088/1475-7516/2019/01/042}

\bibitem[{Schmittfull {et~al.}(2019)Schmittfull, Simonovi\'c, Assassi, \&
  Zaldarriaga}]{Schmittfull:2018yuk}
Schmittfull, M., Simonovi\'c, M., Assassi, V., \& Zaldarriaga, M. 2019, Phys.
  Rev. D, 100, 043514, \dodoi{10.1103/PhysRevD.100.043514}

\bibitem[{Schmittfull {et~al.}(2021)Schmittfull, Simonovi\'c, Ivanov, Philcox,
  \& Zaldarriaga}]{Schmittfull:2020trd}
Schmittfull, M., Simonovi\'c, M., Ivanov, M.~M., Philcox, O. H.~E., \&
  Zaldarriaga, M. 2021, JCAP, 05, 059, \dodoi{10.1088/1475-7516/2021/05/059}

\bibitem[{{Schuster} {et~al.}(2023){Schuster}, {Hamaus}, {Dolag}, \&
  {Weller}}]{Schuster2023}
{Schuster}, N., {Hamaus}, N., {Dolag}, K., \& {Weller}, J. 2023, \jcap, 2023,
  031, \dodoi{10.1088/1475-7516/2023/05/031}

\bibitem[{{Schuster} {et~al.}(2019){Schuster}, {Hamaus}, {Pisani}, {Carbone},
  {Kreisch}, {Pollina}, \& {Weller}}]{Schuster2019}
{Schuster}, N., {Hamaus}, N., {Pisani}, A., {et~al.} 2019, \jcap, 2019, 055,
  \dodoi{10.1088/1475-7516/2019/12/055}

\bibitem[{{Scoccimarro}(1998)}]{scoccimarro98}
{Scoccimarro}, R. 1998, Mon. Not. Roy. Astron. Soc., 299, 1097,
  \dodoi{10.1046/j.1365-8711.1998.01845.x}

\bibitem[{Scoccimarro(2004)}]{Scoccimarro:2004tg}
Scoccimarro, R. 2004, Phys. Rev. D, 70, 083007,
  \dodoi{10.1103/PhysRevD.70.083007}

\bibitem[{Scoccimarro(2015)}]{Scoccimarro:2015bla}
---. 2015, Phys. Rev. D, 92, 083532, \dodoi{10.1103/PhysRevD.92.083532}

\bibitem[{Seljak {et~al.}(2017)Seljak, Aslanyan, Feng, \&
  Modi}]{Seljak:2017rmr}
Seljak, U., Aslanyan, G., Feng, Y., \& Modi, C. 2017, JCAP, 12, 009,
  \dodoi{10.1088/1475-7516/2017/12/009}

\bibitem[{Senatore \& Zaldarriaga(2015)}]{Senatore:2014via}
Senatore, L., \& Zaldarriaga, M. 2015, JCAP, 02, 013,
  \dodoi{10.1088/1475-7516/2015/02/013}

\bibitem[{{Sheth} \& {van de Weygaert}(2004)}]{Sheth2004}
{Sheth}, R.~K., \& {van de Weygaert}, R. 2004, \mnras, 350, 517,
  \dodoi{10.1111/j.1365-2966.2004.07661.x}

\bibitem[{{Speagle} \& {Barbary}(2018)}]{2018Speagle}
{Speagle}, J., \& {Barbary}, K. 2018, {dynesty: Dynamic Nested Sampling
  package}, Astrophysics Source Code Library.
\newblock \doeprint{1809.013}

\bibitem[{{Speagle}(2020{\natexlab{a}})}]{2019Speagle}
{Speagle}, J.~S. 2020{\natexlab{a}}, \mnras, 493, 3132,
  \dodoi{10.1093/mnras/staa278}

\bibitem[{{Speagle}(2020{\natexlab{b}})}]{dynesty}
---. 2020{\natexlab{b}}, \mnras, 493, 3132, \dodoi{10.1093/mnras/staa278}

\bibitem[{{Stadler} {et~al.}(2023){Stadler}, {Schmidt}, \&
  {Reinecke}}]{Stadler:2023hea}
{Stadler}, J., {Schmidt}, F., \& {Reinecke}, M. 2023, \jcap, 2023, 069,
  \dodoi{10.1088/1475-7516/2023/10/069}

\bibitem[{Stevens {et~al.}(2023)Stevens, Sinha, Rohl, Sammons, Hadzhiyska,
  Hern\'andez-Aguayo, \& Hernquist}]{Stevens:2023ozg}
Stevens, A. R.~H., Sinha, M., Rohl, A., {et~al.} 2023.
\newblock \doarXiv{2312.04137}

\bibitem[{{Storey-Fisher} {et~al.}(2024){Storey-Fisher}, {Tinker}, {Zhai},
  {DeRose}, {Wechsler}, \& {Banerjee}}]{2024ApJ...961..208S}
{Storey-Fisher}, K., {Tinker}, J.~L., {Zhai}, Z., {et~al.} 2024, \apj, 961,
  208, \dodoi{10.3847/1538-4357/ad0ce8}

\bibitem[{{Sugiyama} {et~al.}(2023){Sugiyama}, {Yamauchi}, {Kobayashi},
  {Fujita}, {Arai}, {Hirano}, {Saito}, {Beutler}, \&
  {Seo}}]{2023MNRAS.523.3133S}
{Sugiyama}, N.~S., {Yamauchi}, D., {Kobayashi}, T., {et~al.} 2023, \mnras, 523,
  3133, \dodoi{10.1093/mnras/stad1505}

\bibitem[{{Sutter} {et~al.}(2014{\natexlab{a}}){Sutter}, {Lavaux}, {Hamaus},
  {Wandelt}, {Weinberg}, \& {Warren}}]{Sutter2014a}
{Sutter}, P.~M., {Lavaux}, G., {Hamaus}, N., {et~al.} 2014{\natexlab{a}},
  \mnras, 442, 462, \dodoi{10.1093/mnras/stu893}

\bibitem[{{Sutter} {et~al.}(2012){Sutter}, {Lavaux}, {Wandelt}, \&
  {Weinberg}}]{Sutter2012b}
{Sutter}, P.~M., {Lavaux}, G., {Wandelt}, B.~D., \& {Weinberg}, D.~H. 2012,
  \apj, 761, 187, \dodoi{10.1088/0004-637X/761/2/187}

\bibitem[{{Sutter} {et~al.}(2014{\natexlab{b}}){Sutter}, {Lavaux}, {Wandelt},
  {Weinberg}, {Warren}, \& {Pisani}}]{Sutter2014}
{Sutter}, P.~M., {Lavaux}, G., {Wandelt}, B.~D., {et~al.} 2014{\natexlab{b}},
  \mnras, 442, 3127, \dodoi{10.1093/mnras/stu1094}

\bibitem[{{Takada} {et~al.}(2014)}]{PFS:whitepaper2014}
{Takada}, M., {et~al.} 2014, \pasj, 66, R1, \dodoi{10.1093/pasj/pst019}

\bibitem[{{Talts} {et~al.}(2018){Talts}, {Betancourt}, {Simpson}, {Vehtari}, \&
  {Gelman}}]{Talts2018}
{Talts}, S., {Betancourt}, M., {Simpson}, D., {Vehtari}, A., \& {Gelman}, A.
  2018, arXiv e-prints, arXiv:1804.06788, \dodoi{10.48550/arXiv.1804.06788}

\bibitem[{{Tanikawa} {et~al.}(2013){Tanikawa}, {Yoshikawa}, {Nitadori}, \&
  {Okamoto}}]{2013NewA...19...74T}
{Tanikawa}, A., {Yoshikawa}, K., {Nitadori}, K., \& {Okamoto}, T. 2013, \na,
  19, 74, \dodoi{10.1016/j.newast.2012.08.009}

\bibitem[{{Tanikawa} {et~al.}(2012){Tanikawa}, {Yoshikawa}, {Okamoto}, \&
  {Nitadori}}]{2012NewA...17...82T}
{Tanikawa}, A., {Yoshikawa}, K., {Okamoto}, T., \& {Nitadori}, K. 2012, \na,
  17, 82, \dodoi{10.1016/j.newast.2011.07.001}

\bibitem[{Taruya {et~al.}(2010)Taruya, Nishimichi, \& Saito}]{Taruya:2010mx}
Taruya, A., Nishimichi, T., \& Saito, S. 2010, Phys. Rev. D, 82, 063522,
  \dodoi{10.1103/PhysRevD.82.063522}

\bibitem[{{Thiele} {et~al.}(2023){Thiele}, {Massara}, {Pisani}, {Hahn},
  {Spergel}, {Ho}, \& {Wandelt}}]{Thiele2023}
{Thiele}, L., {Massara}, E., {Pisani}, A., {et~al.} 2023, arXiv e-prints,
  arXiv:2307.07555, \dodoi{10.48550/arXiv.2307.07555}

\bibitem[{{To} {et~al.}(2024){To}, {DeRose}, {Wechsler}, {Rykoff}, {Wu},
  {Adhikari}, {Krause}, {Rozo}, \& {Weinberg}}]{Cardinal}
{To}, C.-H., {DeRose}, J., {Wechsler}, R.~H., {et~al.} 2024, \apj, 961, 59,
  \dodoi{10.3847/1538-4357/ad0e61}

\bibitem[{Tucci \& Schmidt(2023)}]{Tucci:2023bag}
Tucci, B., \& Schmidt, F. 2023.
\newblock \doarXiv{2310.03741}

\bibitem[{Valogiannis {et~al.}(2023)Valogiannis, Yuan, \&
  Dvorkin}]{Valogiannis:2023mxf}
Valogiannis, G., Yuan, S., \& Dvorkin, C. 2023.
\newblock \doarXiv{2310.16116}

\bibitem[{{Valogiannis} {et~al.}(2023){Valogiannis}, {Yuan}, \&
  {Dvorkin}}]{2023arXiv231016116V}
{Valogiannis}, G., {Yuan}, S., \& {Dvorkin}, C. 2023, arXiv e-prints,
  arXiv:2310.16116, \dodoi{10.48550/arXiv.2310.16116}

\bibitem[{{van den Bosch} {et~al.}(2005){van den Bosch}, {Weinmann}, {Yang},
  {Mo}, {Li}, \& {Jing}}]{vdBosch_et_al_2005}
{van den Bosch}, F.~C., {Weinmann}, S.~M., {Yang}, X., {et~al.} 2005, \mnras,
  361, 1203, \dodoi{10.1111/j.1365-2966.2005.09260.x}

\bibitem[{Vasudevan {et~al.}(2019)Vasudevan, Ivanov, Sibiryakov, \&
  Lesgourgues}]{Vasudevan:2019ewf}
Vasudevan, A., Ivanov, M.~M., Sibiryakov, S., \& Lesgourgues, J. 2019, JCAP,
  09, 037, \dodoi{10.1088/1475-7516/2019/09/037}

\bibitem[{Verde {et~al.}(2002)}]{Verde:2001sf}
Verde, L., {et~al.} 2002, Mon. Not. Roy. Astron. Soc., 335, 432,
  \dodoi{10.1046/j.1365-8711.2002.05620.x}

\bibitem[{{Verza} {et~al.}(2023){Verza}, {Carbone}, {Pisani}, \&
  {Renzi}}]{Verza2023}
{Verza}, G., {Carbone}, C., {Pisani}, A., \& {Renzi}, A. 2023, \jcap, 2023,
  044, \dodoi{10.1088/1475-7516/2023/12/044}

\bibitem[{{Verza} {et~al.}(2022){Verza}, {Carbone}, \& {Renzi}}]{Verza2022}
{Verza}, G., {Carbone}, C., \& {Renzi}, A. 2022, \apjl, 940, L16,
  \dodoi{10.3847/2041-8213/ac9d98}

\bibitem[{{Verza} {et~al.}(2019){Verza}, {Pisani}, {Carbone}, {Hamaus}, \&
  {Guzzo}}]{Verza2019}
{Verza}, G., {Pisani}, A., {Carbone}, C., {Hamaus}, N., \& {Guzzo}, L. 2019,
  \jcap, 2019, 040, \dodoi{10.1088/1475-7516/2019/12/040}

\bibitem[{Villaescusa-Navarro {et~al.}(2020)}]{Villaescusa-Navarro:2019bje}
Villaescusa-Navarro, F., {et~al.} 2020, Astrophys. J. Suppl., 250, 2,
  \dodoi{10.3847/1538-4365/ab9d82}

\bibitem[{Wadekar {et~al.}(2020)Wadekar, Ivanov, \&
  Scoccimarro}]{Wadekar:2020hax}
Wadekar, D., Ivanov, M.~M., \& Scoccimarro, R. 2020, Phys. Rev. D, 102, 123521,
  \dodoi{10.1103/PhysRevD.102.123521}

\bibitem[{{Wadekar} \& {Scoccimarro}(2020)}]{Wadekar_2020}
{Wadekar}, D., \& {Scoccimarro}, R. 2020, \prd, 102, 123517,
  \dodoi{10.1103/PhysRevD.102.123517}

\bibitem[{{Wang} {et~al.}(2022){Wang}, {Mao}, {Zentner}, {Guo}, {Lange}, {van
  den Bosch}, \& {Mezini}}]{Wang_et_al_2022}
{Wang}, K., {Mao}, Y.-Y., {Zentner}, A.~R., {et~al.} 2022, \mnras, 516, 4003,
  \dodoi{10.1093/mnras/stac2465}

\bibitem[{Wang {et~al.}(2023)Wang, Jeong, Taruya, Nishimichi, \&
  Osato}]{Wang:2022itv}
Wang, Z., Jeong, D., Taruya, A., Nishimichi, T., \& Osato, K. 2023, Phys. Rev.
  D, 107, 103534, \dodoi{10.1103/PhysRevD.107.103534}

\bibitem[{{Wechsler} {et~al.}(2022){Wechsler}, {DeRose}, {Busha}, {Becker},
  {Rykoff}, \& {Evrard}}]{addgals}
{Wechsler}, R.~H., {DeRose}, J., {Busha}, M.~T., {et~al.} 2022, \apj, 931, 145,
  \dodoi{10.3847/1538-4357/ac5b0a}

\bibitem[{{Wechsler} \& {Tinker}(2018)}]{2018Wechsler}
{Wechsler}, R.~H., \& {Tinker}, J.~L. 2018, \araa, 56, 435,
  \dodoi{10.1146/annurev-astro-081817-051756}

\bibitem[{{White} {et~al.}(2014){White}, {Tinker}, \& {McBride}}]{White2014}
{White}, M., {Tinker}, J.~L., \& {McBride}, C.~K. 2014, \mnras, 437, 2594,
  \dodoi{10.1093/mnras/stt2071}

\bibitem[{{Xu} {et~al.}(2021){Xu}, {Zehavi}, \& {Contreras}}]{Xu_et_al_2021}
{Xu}, X., {Zehavi}, I., \& {Contreras}, S. 2021, \mnras, 502, 3242,
  \dodoi{10.1093/mnras/stab100}

\bibitem[{Yoshikawa \& Fukushige(2005)}]{Yoshikawa_2005}
Yoshikawa, K., \& Fukushige, T. 2005, Publications of the Astronomical Society
  of Japan, 57, 849–860, \dodoi{10.1093/pasj/57.6.849}

\bibitem[{{Yuan} {et~al.}(2024){Yuan}, {Abel}, \& {Wechsler}}]{knnpaper}
{Yuan}, S., {Abel}, T., \& {Wechsler}, R.~H. 2024, \mnras, 527, 1993,
  \dodoi{10.1093/mnras/stad3359}

\bibitem[{{Yuan} {et~al.}(2018){Yuan}, {Eisenstein}, \&
  {Garrison}}]{Yuan_et_al_2018}
{Yuan}, S., {Eisenstein}, D.~J., \& {Garrison}, L.~H. 2018, \mnras, 478, 2019,
  \dodoi{10.1093/mnras/sty1089}

\bibitem[{{Yuan} {et~al.}(2022){Yuan}, {Garrison}, {Eisenstein}, \&
  {Wechsler}}]{2022bYuan}
{Yuan}, S., {Garrison}, L.~H., {Eisenstein}, D.~J., \& {Wechsler}, R.~H. 2022,
  \mnras, \dodoi{10.1093/mnras/stac1830}

\bibitem[{{Yuan} {et~al.}(2021){Yuan}, {Garrison}, {Hadzhiyska}, {Bose}, \&
  {Eisenstein}}]{Yuan:2021hod}
{Yuan}, S., {Garrison}, L.~H., {Hadzhiyska}, B., {Bose}, S., \& {Eisenstein},
  D.~J. 2021, \mnras, 510, 3301, \dodoi{10.1093/mnras/stab3355}

\bibitem[{Yuan {et~al.}(2023)Yuan, Hadzhiyska, \& Abel}]{Yuan:2022ibz}
Yuan, S., Hadzhiyska, B., \& Abel, T. 2023, Mon. Not. Roy. Astron. Soc., 520,
  6283, \dodoi{10.1093/mnras/stad550}

\bibitem[{{Yuan} {et~al.}(2023{\natexlab{a}}){Yuan}, {Zamora}, \&
  {Abel}}]{Yuan:2023ddknn}
{Yuan}, S., {Zamora}, A., \& {Abel}, T. 2023{\natexlab{a}}, \mnras, 522, 3935,
  \dodoi{10.1093/mnras/stad1275}

\bibitem[{{Yuan} {et~al.}(2023{\natexlab{b}}){Yuan}, {Zhang}, {Ross},
  {Donald-McCann}, {Hadzhiyska}, {Wechsler}, {Zheng}, {Alam}, {Gonzalez-Perez},
  {Aguilar}, {Ahlen}, {Bianchi}, {Brooks}, {de la Macorra}, {Fanning},
  {Forero-Romero}, {Honscheid}, {Ishak}, {Kehoe}, {Lasker}, {Landriau},
  {Manera}, {Martini}, {Meisner}, {Miquel}, {Moustakas}, {Nadathur}, {Newman},
  {Nie}, {Percival}, {Poppett}, {Rocher}, {Rossi}, {Sanchez}, {Samushia},
  {Schubnell}, {Seo}, {Tarle}, {Weaver}, {Yu}, {Zhou}, \&
  {Zou}}]{Yuan_et_al_2023}
{Yuan}, S., {Zhang}, H., {Ross}, A.~J., {et~al.} 2023{\natexlab{b}}, arXiv
  e-prints, arXiv:2306.06314, \dodoi{10.48550/arXiv.2306.06314}

\bibitem[{{Zennaro} {et~al.}(2022){Zennaro}, {Angulo}, {Contreras},
  {Pellejero-Ib{\'a}{\~n}ez}, \& {Maion}}]{Zennaro_2022}
{Zennaro}, M., {Angulo}, R.~E., {Contreras}, S., {Pellejero-Ib{\'a}{\~n}ez},
  M., \& {Maion}, F. 2022, \mnras, 514, 5443, \dodoi{10.1093/mnras/stac1673}

\bibitem[{{Zennaro} {et~al.}(2021){Zennaro}, {Angulo},
  {Pellejero-Ib{\'a}{\~n}ez}, {St{\"u}cker}, {Contreras}, \&
  {Aric{\`o}}}]{ZennaroAnguloPellejero2021}
{Zennaro}, M., {Angulo}, R.~E., {Pellejero-Ib{\'a}{\~n}ez}, M., {et~al.} 2021,
  arXiv e-prints, arXiv:2101.12187.
\newblock \doarXiv{2101.12187}

\bibitem[{{Zhai} {et~al.}(2023{\natexlab{a}}){Zhai}, {Percival}, \&
  {Guo}}]{Zhai_et_al_2023}
{Zhai}, Z., {Percival}, W.~J., \& {Guo}, H. 2023{\natexlab{a}}, \mnras, 523,
  5538, \dodoi{10.1093/mnras/stad1793}

\bibitem[{{Zhai} {et~al.}(2023{\natexlab{b}}){Zhai}, {Tinker}, {Banerjee},
  {DeRose}, {Guo}, {Mao}, {McLaughlin}, {Storey-Fisher}, \&
  {Wechsler}}]{Zhongxu_et_al_2023}
{Zhai}, Z., {Tinker}, J.~L., {Banerjee}, A., {et~al.} 2023{\natexlab{b}}, \apj,
  948, 99, \dodoi{10.3847/1538-4357/acc65b}

\bibitem[{{Zheng} {et~al.}(2007){Zheng}, {Coil}, \& {Zehavi}}]{Zheng2007}
{Zheng}, Z., {Coil}, A.~L., \& {Zehavi}, I. 2007, \apj, 667, 760,
  \dodoi{10.1086/521074}

\bibitem[{{Zheng} {et~al.}(2005){Zheng}, {Berlind}, {Weinberg}, {Benson},
  {Baugh}, {Cole}, {Dav{\'e}}, {Frenk}, {Katz}, \& {Lacey}}]{Zheng_et_al_2005}
{Zheng}, Z., {Berlind}, A.~A., {Weinberg}, D.~H., {et~al.} 2005, \apj, 633,
  791, \dodoi{10.1086/466510}

\end{thebibliography}
